%% file: MAIN.tex
\documentclass[epjST]{svjour}
\usepackage[toc,page,titletoc]{appendix}
\usepackage{subfigure}
\usepackage{graphicx}
\usepackage[intlimits]{amsmath}
\usepackage{amssymb}
\usepackage{amsfonts}
\usepackage[scanall]{psfrag}
\usepackage{multido}
\usepackage{psfrag,amsfonts,ifpdf}
\usepackage{amssymb}
\usepackage{color}

\def\mean#1{\big\langle#1\big\rangle}

\newcommand{\bbox}[1]{{\bf #1}}

\newcommand{\bea}{\begin{eqnarray}}
\newcommand{\eea}{\end{eqnarray}}
\newcommand{\appel}{\frac{\rm d}{{\rm d}t}}

\newcommand{\txr}[1]{\textrm{#1}}

\newcommand{\mN}{\mathcal{N}}
\newcommand{\mA}{\mathcal{A}}

\newcommand{\DBM}{D_\txr{BM}}
\newcommand{\DMM}{D_\txr{MM}}

\newcommand{\be}{\begin{equation}} 
\newcommand{\ee}{\end{equation}} 
\newcommand{\ba}{\begin{eqnarray}} 
\newcommand{\ea}{\end{eqnarray}}

\newcommand{\intl}{\int\limits}

\newcommand{\liml}{\lim\limits}

\newcommand{\lr}[1]{\langle #1 \rangle}

\newcommand{\bi}[1]{Fig.~\ref{fig:#1}}
\newcommand{\e}[1]{Eq.~(\ref{eq:#1})}

\begin{document}

\title{{Active Brownian Particles}}
\subtitle{From Individual to Collective Stochastic Dynamics}

\author{Pawel Romanczuk\inst{1}
  \fnmsep\thanks{\email{prom@pks.mpg.de}} \and M. B\"ar
  \inst{2} \and W. Ebeling \inst{3} \and B.  Lindner \inst{3}
  \and Lutz Schimansky-Geier\inst{3}}

\institute{
  Max Planck Institute for the Physics of Complex Systems Dresden, N\"othnitzer Str. 38, D-01187 Dresden, Germany  \and 
  Physikalisch-Technische Bundesanstalt, Abbestr. 2-12, D-10587 Berlin, Germany 
  \and
  Institute of Physics, Humboldt Universit{\"a}t zu Berlin, Newtonstr. 15, 12489 Berlin, Germany
  }
\abstract{ We review theoretical models of individual motility as well
  as collective dynamics and pattern formation of active particles.
  We focus on simple models of active dynamics with a particular emphasis 
  on nonlinear and stochastic dynamics of such
  self-propelled entities in the framework of statistical mechanics.
  Examples of such active units in complex physico-chemical and biological systems
  are chemically powered nano-rods, localized patterns in reaction-diffusion system, motile cells or macroscopic animals.
  Based on the description of individual motion of point-like active particles by stochastic differential equations, we discuss different velocity-dependent friction functions, the impact of various types of fluctuations and calculate characteristic observables such as stationary velocity distributions or diffusion coefficients.
Finally, we consider not only the free and confined individual active dynamics but also different types of interaction between active particles.  The resulting collective dynamical behavior of large assemblies and aggregates of active units is discussed and an overview over some recent results on spatiotemporal pattern formation in such systems is given.
}
  
\maketitle
\newpage
\tableofcontents





\newpage
\ 
\newpage
\thispagestyle{empty}
\ \vspace{15ex}
\begin{center} {\Large Dedicated to the late Frank Edward Moss  (1934-2011).}  \end{center}

\newpage
\input{SECTION_intro}
\clearpage
\newpage
\input{SECTION_beyond}

\clearpage
\newpage
\input{SECTION_self_driven}

\clearpage
 \newpage
 \input{SECTION_single}

\clearpage
 \newpage
\input{SECTION_confined}

\clearpage
 \newpage
 \input{SECTION_swarming}

\clearpage
 \newpage
 \input{SECTION_pattern}

\input{SECTION_concl}


\bibliography{BIB/benji_ALL,BIB/Review,BIB/Misc,BIB/ReviewAddOns}

\bibliographystyle{plain}

\end{document}

%% file: SECTION_intro.tex
\section{Introduction}
\label{sec:intro}

In recent years there has been a strong growth of research activities regarding the statistical description of systems far from equilibrium. 
A whole class of biological and physical systems which may be referred to as {\em active matter} have been studied theoretically and experimentally. The term ``active'' refers here to the ability of individual units to move actively by gaining kinetic energy from the environment.
Examples of such systems range from the dynamical behavior of individual units such as Brownian motors \cite{reimann_brownian_2002}, motile cells
\cite{friedrich_chemotaxis_2007,selmeczi_cell_2008,boedeker_quantitative_2010},
macroscopic animals 
\cite{kareiva_analyzing_1983,komin_random_2004} or artificial self-propelled particles 
\cite{paxton_catalytic_2004,howse_self-motile_2007} to large ensembles of interacting active particles and their large scale collective dynamics 
 \cite{vicsek_novel_1995,chate_simple_2006,baskaran_statistical_2009}. 
 A major driving force of the {\em active matter} research are the continuously improving experimental techniques such as for example automated digital tracking 
\cite{sokolov_concentration_2007,selmeczi_cell_2008,ballerini_interaction_2008} 
 or the realization of active granular and colloidal systems 
(see e.g. \cite{paxton_catalytic_2004,aranson_patterns_2006,kudrolli_swarming_2008,tierno_autonomously_2010,deseigne_collective_2010}). 

This review is devoted to the analysis of simple dynamical models of
active systems. 
In particular we will focus on stochastic models of individual active particles or agents as well as on large scale collective phenomena arising in systems of such interacting active particles. We define active motion as motion of particles or agents due to an internal driving, which may have different causes such as biological activity or non-equilibrium dynamics in artificial driven systems. It is fundamentally different from standard purely passive dynamical behavior of particles in gases, liquids or solid states at thermal equilibrium. In contrast,
active particles or agents are assumed to have an internal propulsion mechanism (``motor''), which may use energy from an external source and transform it under non-equilibrium conditions into directed acceleration motion. 

The great number of publications on active particle systems makes a complete review, describing all the different approaches across the different scientific disciplines, impossible. Thus, we will focus here on the review and analysis of generic models of individual active particles as well as their large scale collective dynamics, which may be considered as an extension of concepts well known in physics, such as ordinary Brownian motion, ferromagnetic or nematic media.  In particular, we will focus on the mathematical description and analysis of such systems from a (statistical) physicist's point of view. 

There are many fascinating research areas, which we will not discuss in this review, such as the dynamics of microswimmers at low Reynolds numbers (see e.g. \cite{dreyfus_microscopic_2005,pooley_hydrodynamic_2007,alexander_dumb-bell_2008,downton_simulation_2009,putz_cuda_2010}, or a recent review in \cite{lauga_hydrodynamics_2009}), or the biological implications of collective behavior of organisms and animals, for example, flocks of birds or schools of fish \cite{krause_living_2002,couzin_effective_2005,sumpter_information_2008,guttal_social_2010}.

An important feature of most \emph{active matter} systems, are the non-negligible random fluctuations in the motion of individual active units. This apparent randomness may have different origins,  for example, environmental factors or internal fluctuations due to the intrinsic stochasticity of the processes driving individual motion. In animals, they may be also associated with abstract decision processes which govern the direction and/or the speed of individual motion and which may appear as random to an external observer. A simple way to account for such fluctuations without being able to resolve the underlying mechanisms is to introduce stochastic forces into the equations of motion of individual units. Thus our general modelling approach will be based on the concept of stochastic differential equations (Langevin equation, SDE) and the corresponding Fokker-Planck equations for the evolution of the probability densities of the involved (stochastic) variables \cite{gardiner_handbook_1985,van_kampen_stochastic_1992,risken_fokker-planck_1996}. 
          
A large part of this review evolves around the concept of ``Active Brownian Particles'' introduced more than a decade ago. The term was first introduced by \cite{schimansky-geier_structure_1995}, referring to Brownian particles with the ability to generate a field, which in turn can influence their motion. In the following Ebeling, Schweitzer and others used this term in the context of self-propelled particles far from equilibrium  (see e.g. \cite{schweitzer_complex_1998,ebeling_active_1999,erdmann_collective_2003,schweitzer_brownian_2003,romanczuk_beyond_2008}). In general we will refer to ``Active Brownian Particles'' in the latter context as Brownian particles performing active motion, which may be accounted for by an internal energy depot and/or a (nonlinear) velocity-dependent friction function.

Throughout this review, we will focus on topics and research questions, which we have been actively working on over the past years. Hereby, an emphasis is put on presentation of the mathematical framework together with the discussion of its application to various problems. 

\subsection{Individual dynamics}
When the botanist Robert Brown observed the erratic motion of small pollen grain particles immersed in a liquid in 1827 \cite{brown_brief_1828},
he considered them first as living entities. However Brown, being also
the discoverer of the cellular nucleus, was a thorough scientist.  He
repeated the experiments with granular and glassy material, and
discovered its purely physical nature. Although his name is associated with the physical phenomenon of Brownian motion, he was certainly not the first one to observe it. The phenomenon of irregular motion of coal dust particles immersed in a fluid had already been reported by Jan Ingen-Housz in 1784 \cite{ingen-housz_bemerkungen_1784}.

Later on, in 1863, the origin
of the never-vanishing motion was experimentally traced back to the
 motion of the molecules of the surrounding liquid by { C. Wiener} \cite{wiener_erklaerung_1863}, who
was familiar with Maxwell's kinetic theory.  Afterwards { M. Gouy} \cite{gouy_note_1888}
found that the motion is amplified if viscosity of the liquid is
lowered. After the turn of the century, { A. Einstein} \cite{einstein_motion_1905,einstein_zur_1906}, { M. Smoluchowski} \cite{von_smoluchowski_zur_1906}, { P. Langevin} \cite{langevin_theory_1908}, and others showed theoretically that
the behavior of Brownian particles are due to the permanent molecular
agitation of the solution on the immersed particle.  Eventually, {J.
  Perrin} \cite{perrin_brownian_2005} was awarded the Nobel Prize in 1926 for the experimental observation of
Brownian motion, which confirmed the theoretical findings and permitted
the first realistic determination of Avogadro's number as indicator of
the molecular structure of matter. 

Without doubt, the theory of Brownian motion not only assumed a central role in the foundation of thermodynamics and statistical physics, but is still a major interdisciplinary research topic. 
Although Brownian motion is purely physical in origin, many models of
bacterial and small animal motion have employed the framework of Brownian
motion or its discrete counterparts, which are random walks. In parallel to Einsteins work, K.~Pearson
\cite{pearson_problem_1905,pearson_problem_1905-1,pearson_mathematical_1906} in 1905 invented the term ``random walks''
and used the concept for a statistical description of insects migration. 

Probably the first experiments on random motion of (living) microorganisms -- which consitute clearly a system far from equilibrium -- influenced by the theory of Brownian motion were performed by Przimbam in the second decade of the 20th century \cite{przibram_ueber_1913,przibram_ueber_1917}. Przibram has shown that the mean squared displacement of the protozoa in water increases linearly in time in analogy to Brownian motion but with a larger diffusion coefficient then predicted by the equilibrium kinetic theory of Brownian motion. Przimbrams work is the first experimental evidence of active Brownian motion. In fact, in his second paper Przimbram reported increasing diffusion coefficients of rotifiers with their increasing concentration, which appears to be the first report on hydrodynamically interacting active Brownian particles \cite{przibram_ueber_1917}. 

Przimbam's work was followed up by F{\"u}rth \cite{fuerth_brownsche_1920}, who based on his own experimental findings, introduced the notion of persistent random walk in the description of the motion of biological agents. F{\"u}rth arrived independently at the same result as Ornstein \cite{ornstein_brownian_1919} who considered inertial Brownian motion in thermal equilibrium.  

The mathematical description of the apparently random motion of biological agents and the corresponding diffusion processes are fundamental to understanding the ability of individuals to explore their environment and to describe the large scale dispersal of populations \cite{okubo_diffusion_2001}. Since the first pioneering works of Przimbam and F\"urth there have been a great number of publications on the theory of random walks and their application to biology and ecology.
A prominent example is the work of Howard C. Berg \cite{berg_random_1993}, who made random walk
theory an intrinsic part of modeling; new concepts along these lines are developed
even nowadays \cite{metzler_random_2000,ebeling_statistical_2005}. From the huge literature,  we
would like to highlight here few examples, such as the work by { H.~Gruler } and { M.~Schienbein} \cite{schienbein_langevin_1993}, by { H. Othmer} and coworkers \cite{othmer_models_1988,hillen_diffusion_2000} or the recent integrated theoretical and experimental approach by { H.~Flyvberg} and
collaborators \cite{selmeczi_cell_2005,selmeczi_cell_2008}.

Because of the strong connections between the theory of Brownian
motion and random walk theory to models of self-propelled or active
Brownian particles, we will start this review with a summary of general findings of the classical theory 
and some important mathematical concepts.

In the following sections, we will focus on the concept of self-propulsion in the framework of active Brownian particles. We will discuss various models of active motion not only differing in their deterministic equations of motion but also subject to different types of fluctuations: e.g. internal and external Gaussian fluctuations, (energetic) shot noise or dichotomous Markov noise. 
We will analyze the characteristics observables of individual active motion such as speed and velocity probability distributions or the mean squared displacement. In addition, we will discuss the behavior of active particles under external constraints, such as e.g. external potentials.

Despite the intense interdisciplinary research on active Brownian motion, there is still a lack of theoretical foundations. For example, one issue that has been neglected is the distinction between passive fluctuations (e.g. thermal fluctuations) and stochastic forces which have their origin in the active nature of the system and their different impact on the experimentally accessible observables such as stationary velocity and speed distributions. Only recently it was shown how internal (or active) fluctuations may lead to a complex behavior of the mean squared displacement of active particles with multiple crossovers or to characteristic deviations of stationary velocity and speed distributions in comparison to ordinary Brownian morion \cite{peruani_self-propelled_2007,romanczuk_brownian_2011}.

\subsection{Collective dynamics}
In the last two sections of this review, we will extend our scope to self-organization phenomena in systems of interacting active particles and the resulting modes of collective motion. 
In this context we introduce the term ``swarms'' of active particles.
We use it here to refer to a confined systems of
particles (animals in two dimensions and more general objects)
performing collective motions under far from equilibrium conditions.
The dynamics of swarms of animals is a traditional object of
biological and ecological investigations \cite{okubo_diffusion_2001} but still a
rather young field of physical studies (see e.g. Refs.
\cite{helbing_traffic_2001,vicsek_fluctuations_2001,mikhailov_cells_2002,schweitzer_brownian_2003}).  Here we will develop rather simple
models which are based on the idea that swarm motion possesses some
kind of universality \cite{okubo_diffusion_2001}. 
The existence of universal features makes it possible
to study swarm motion using simple models which are in the same
universality class.  Assuming that this hypothesis is true, we have the choice between
different models of swarm motion which are in the universality class
preferring the simplest ones. The idea which we will
follow here is: Swarms may be modelled as active particles using
the concepts of active Brownian motion \cite{anishchenko_nonlinear_2002,ebeling_statistical_2005}.

Over the past decades, collective dynamics of swarms of driven
particles has captured the growing interest of various theoretical
groups. Many interesting effects of the self-organization of swarms
have been revealed and in part already explained. We mention here the
comprehensive survey of Okubo and Levin \cite{okubo_diffusion_2001} on swarm
dynamics in biophysical and ecological context. Studies of Helbing
\cite{helbing_traffic_2001} relate to traffic phenomena and related self-driven
many-particle systems. Broad context of swarming dynamics in natural
science is also brought up in the comprehensive books and reviews by T. Vicsek
\cite{vicsek_fluctuations_2001,vicsek_collective_2010}, A.S. Mikhailov and V. Calenbuhr \cite{mikhailov_cells_2002}, F. Schweitzer \cite{schweitzer_brownian_2003}, I.D. Couzin \cite{couzin_collective_2009}, D.J.T. Sumpter \cite{sumpter_collective_2010} and Toner {\em et. al} \cite{toner_hydrodynamics_2005}.

A major shortcoming of the research in collective motion was the lack of empirical data on the structure and dynamics of real swarms. However, the situation is improving, due to technological advances in digital tracking and data processing, the empirical study of large scale collective motion in the field has become possible \cite{ballerini_interaction_2008,lukeman_inferring_2010,cavagna_scale-free_2010}. Furthermore, an increasing number of controlled laboratory experiments is being performed on collective motion of such different organisms as fish \cite{abaid_fish_2010}, insects \cite{buhl_disorder_2006,bazazi_nutritional_2010}, bacteria \cite{sokolov_concentration_2007,sokolov_enhanced_2009} or keratocytes (active tissue cells) \cite{szabo_phase_2006}. The resulting data provide the foundation to address the question on actual properties of social interactions in real world swarms and flocks. 

Beyond the dynamics of cohesive swarms, we will discuss systems of active particles with a symmetry breaking alignment interaction, which may lead to large scale collective motion.  

The most prominent (minimal) model of collective motion of this type was introduced by Vicsek and collaborators in 1995\cite{vicsek_novel_1995}.  Vicsek {\em et al.} have shown how an initially disordered state without collective motion at large noise intensities becomes unstable if the noise is decreased below a critical value and how large scale collective motion emerges via a spontaneous symmetry breaking.

We establish a link to publications on mesoscopic equation of motion for density and velocity fields of self-propelled particles  \cite{toner_long-range_1995,toner_flocks_1998,simha_statistical_2002,simha_hydrodynamic_2002,mishra_fluctuations_2010}, where the mesoscopic equations of motion for the density and velocity fields were constructed using symmetry and conservation laws. Recently, a corresponding kinetic description was derived by a Boltzmann approach \cite{bertin_boltzmann_2006}. Here, we derive the field equations in a systematic way from the microscopic Langevin equations of active Brownian motion without a restriction to constant speeds. In addition to the density and velocity fields, we consider explicitly the effective temperature field of the active Brownian particle gas. 
We consider also the important special case of self-propelled particles with constant speed and discuss in this context local alignment with different interaction symmetries.

In the end of Sec. \ref{sec:swarming}, we mention briefly other swarming mechanisms, such as chemotactic coupling or selective attraction/repulsion interactions (escape \& pursuit). 

At the end we will turn our attention pattern formation  of self-propelled particles with alignment and discuss clustering, phase separation, and emergence of large-scale coherent structures in such systems.

Finally, we will conclude this review with a summary of the contents together with a more detailed discussion of related expermental results as well as other modelling approaches and their relation to the theoretial framework presented in the paper.


%% file: SECTION_beyond.tex
\section{Brownian motion and beyond}
\label{sec:brown}
\subsection{Brownian motion revisited}

The behavior of ordinary Brownian particles is determined by the
(passive) stochastic collisions, the particles suffer from the
surrounding medium.  There is no active transfer of energy to the
particles. The energetic equilibrium between particles and surrounding
medium, which balances dissipation and fluctuations, is expressed
by the fluctuation--dissipation theorem.

In the most common  way (at first glance, also the simplest way),
Brownian motion is described by Newtonian dynamics including
friction and stochastic forces \cite{langevin_theory_1908}. The
motion of a Brownian particle subject to Stokes friction with
coefficient $\gamma$ in a space-dependent potential $U(\bbox r)$ can
be described by the Langevin equation
\begin{eqnarray}
  \label{langev-or}
  \frac{d {\bbox r}}{d t}={\bbox v}\,;\,\,
  m\frac{d {\bbox v}}{d t}= - \gamma{\bbox v} -\nabla U({\bbox r}) + {\bbox
    {\cal F}}(t)
\end{eqnarray}
with $\nabla U=0$ in the original publication by Langevin. He assumed
temporally short correlated random forces ${\bbox {\cal F}}(t)$,
independence between coordinate $\bbox r(t)$ and velocity $\bbox v(t)$,
and  the equipartition theorem  $\langle {\bbox{v}}^2 \rangle = 3k_{\rm B} T/m$ where
$k_{\rm B}$ is Boltzmann's constant.  Ornstein and Uhlenbeck
\cite{uhlenbeck_theory_1930} pointed out that the stochastic force
should be Gaussian distributed with independent components and
$\delta$-correlated time dependence
\begin{equation}
\label{stoch}
  \mean{{\bbox {\cal F}}(t)}=0 \,;\,\, \mean{{{\cal F}_i}(t){{\cal F}_j}(t')}=2D_p\delta_{i,j}\,\delta(t-t')\,, \,\,i,j=x,y,z\,. 
\end{equation}
The components  ${\cal F}_i(t)$ are referred to as 
Gaussian white noise with intensity $D_p$. Integrated over small time
intervals ${\rm dt}$, a stationary Wiener process is obtained. In
the Newtonian equation, these forces yield an increment of momentum in $\text{d}t$
with Gaussian distribution
\begin{eqnarray}
  \label{eq:wiener}
  {\rm d}W_{{\rm d}t,i} = \int_t^{t+{\rm dt}} {\rm d}s\,  {\cal F}_i(s)\,,\,\,i=x, y, z\,,
\end{eqnarray}
the average of which vanishes and second moment of which grows linear
in ${\rm d}t$ with slope $2D_p$.  Increments at different times are
independent \cite{gardiner_handbook_1985,risken_fokker-planck_1996}.

The noise strength $D_p$ for the momentum is connected with the noise
strength for the velocities $D$ by the simple relation $D_p = m^2
D$. In thermal equilibrium, according to Langevin's assumption,
the loss of energy due to friction compensates on average the gain of energy
resulting from the stochastic force.
In this case, the fluctuation-dissipation theorem  states:
\begin{equation}
  \label{fluct-diss}
  D_p\,=\,  D \,m^{2} \,= \,k_{\rm B}\,T \,\gamma\,,
\end{equation}
where $T$ is the absolute temperature.

We may rewrite the Langevin equation for the velocities as follows
\begin{eqnarray}
  \label{langev-v}
  \frac{d {\bbox r}}{d t}={\bbox v}\,;\,\,
  \frac{d {\bbox v}}{d t}= - \frac{\gamma}{m} \,\bbox{v} -\frac{\nabla
    U({\bbox r}))}{m} + \sqrt{2 D} \bbox{\xi}(t)\, , \label{langevv1}
\end{eqnarray}
where the stochastic source term obeys 
\begin{eqnarray}
  \label{eq:noise}
  \langle \bbox{{\xi}}(t)\rangle\,=\,0\,,\,\,\langle \xi_i(t)\xi_j(t^\prime)\rangle\,=\,\delta(t-t^\prime)\,\,\,i,j=x,y,z\,.
\end{eqnarray}
For $\nabla U(\bbox{r})=0$, the integration of Eq. (\ref{langev-v}) to
obtain the mean squared displacement $\langle{\bbox r}^2(t)\rangle$ starting
at $t=0$ at the origin gives in equilibrium using \eqref{fluct-diss}:
\begin{eqnarray}
  \label{eq:squareddispl}
  \langle {{\bbox r}(t)}^2 \rangle \,=\, 6 \,\frac{k_{\rm B} T}{\gamma}\left[t-\frac{1}{\gamma} (1-\exp(-\frac{\gamma}{m} t))\right]\,.
\end{eqnarray}
This expression yields a ballistic growth for times smaller than $t \le 1/\gamma $
and a linear growth for larger times 
\begin{align}\label{eq:diffcoeffBrown}
\langle  {{\bbox r}(t)}^2 \rangle = 2 d D_\text{eff} \,t\,,
\end{align}
with $d$ is the dimension of the Brownian motion. The diffusion coefficient obeys the Einstein-Sutherland
relation \cite{sutherland_dynamical_1905,einstein_motion_1905}
\begin{align}
D_\text{eff} = \frac{k_B T}{\gamma}\,.
\end{align}
The linear regime in $t$ is valid at length scales exceeding the brake path $l \propto \sqrt{k_{\rm B}T m }/\gamma $.

In the following we will always use dimensionless units and set the mass to $m=1$, resulting in $D_p=D$. Further on we use $D$ for noise intensity of purely additive Gaussian white noise in the velocity coordinate.

We are interested in the general statistical descriptions of
self-moving objects.  In the Markovian description, full information
is provided by the transition probability
$P(\bbox{r},\bbox{v},t|\bbox{r_0},\bbox{v_0},t_0)$ to find the
particle at location $\bbox{r}$ with velocity $\bbox{v}$ at time $t$
if started at $\bbox{r_0}$ with $\bbox{v_0}$ at initial time $t_0$.
As well known \cite{becker_theory_1967,risken_fokker-planck_1996}, the distribution function
density which corresponds to the Langevin equation (\ref{langev-v}), is
the solution of a Fokker-Planck equation of the form:
\begin{eqnarray}
  \label{eq:fpebrown}
  \frac{\partial P(\bbox{r},\bbox{v},t|\bbox{r_0},\bbox{v_0},t_0)}{\partial t} + {\bbox v}\,
  \frac{\partial P}{\partial \bbox{r}} + \nabla
    U(\bbox{r}) \,\frac{\partial P}{\partial \bbox{v}}\,=\,\frac{\partial}{\partial {\bbox v}}
  \left[\gamma \bbox{v}\, P + D\, \frac{\partial P}{\partial \bbox{v}}
  \right]
\end{eqnarray}
For harmonic potentials $U(\bbox{r})$, the solutions is a multi-modal
Gaussian distribution \cite{chandrasekhar_stochastic_1943} with
time-dependent moments. In the long-time limit, the stationary
solution of Eq. (\ref{eq:fpebrown}) becomes independent of the initial
distribution $P_{0}(\bbox{r},\bbox{v})$ and with Eq. (\ref{fluct-diss})
the ensemble of Brownian particles  obeys in equilibrium the
Maxwell-Boltzmann distribution:
\begin{eqnarray}
  \label{eq:maxw}
 P_0(\bbox{r},\bbox{v})& = & {\cal N} \, \exp{ \left\{
-\frac{1}{k_{B}T} \left[\frac{1}{2}\,{\bbox v}^{2}\,+\,U(\bbox{r})
\right] \right\}}
\end{eqnarray}

An important limiting case of Eq.~(\ref{langev-v}) is the so-called overdamped
Brownian motion. As pointed out by Purcell \cite{purcell_life_1977} and
Berg \cite{berg_random_1993}, this limit should be considered in the motion of bacteria and
other small micro-swimmers due to the low Reynolds number governing
their dynamics. Overdamped Brownian motion can be obtained from the Langevin
equation under the assumption of large friction where inertial
effects can be neglected, resulting in
\begin{eqnarray}
  \label{eq:overdamped}
  \frac{d {\bbox r}}{d t}={\bbox v}\,=\, -\frac{\nabla U(\bbox{r})}{\gamma}\,+\, \sqrt{2 D_r} \bbox{\xi}(t)   
\end{eqnarray}
with intensity $D_r=D /\gamma^2$. 

The corresponding Fokker-Planck equation for the overdamped dynamics reads
\begin{eqnarray}
  \label{eq:fpeover}
  \frac{\partial P(\bbox{r},t|\bbox{r_0},t_0)}{\partial t}\,=\,\frac{\partial }{\partial \bbox{r}} 
\left[\frac{\nabla U(\bbox{r})}{\gamma} P\right]\,+\,D_r\, \frac{\partial^2 P}{\partial \bbox{r}^2}\,
\end{eqnarray}
with the stationary solution
\begin{eqnarray}
  \label{eq:boltz}
  P_0(\bbox{r}) \,= \, {\cal N} \, \exp{ \left[ -\frac{U(\bbox{r})}{\gamma D_r} \right]}\,.
\end{eqnarray}
In thermal equilibrium, this becomes equal to the  Boltzmann
distribution by virtue of  $D_r=k_{B} T/\gamma$, which holds true  according to the
 relation Eq.~\eqref{fluct-diss}.

We note that the Langevin Eq. (\ref{langev-v}) has been generalized in many ways.
First of all, the description applies not only to mechanical degrees of freedom but also to
voltage fluctuations in electric circuits ('Johnson noise') and to fluctuations in the number of molecules undergoing  chemical reactions  ('Chemical Langevin equation'), to name only two prominent examples. Finite correlations in equilibrium fluctuations and dissipation ('memory damping') have been taken into account in the generalized Langevin equation \cite{Mor65,Kub66,Zwa73,haenggi_stochastic_1982}. The consequences of nonlinear dissipation in equilibrium systems were  studied as well \cite{Kli95}. More recently, also formulations of  Langevin and Fokker-Planck equations that are consistent with special relativity have drawn much attention  (for a comprehensive review on relativistic Brownian motion, see \cite{DunHan09}).  As for the many other generalizations and applications  in nonequilibrium systems, we just can refer the interested reader to recent collections of articles on these topics \cite{haenggi_reaction-rate_1990,100years_BrownianMotion,haenggi60_chemphys,lutz60_epjst}  and the references given therein.

\subsection{Polar representation of Brownian dynamics}

Let us now consider the dynamics in two spatial dimensions and without
external forces. A very useful representation is obtained in polar
coordinates of the velocities \cite{schimansky-geier_stationary_2005}. The Cartesian
components of the velocity $v_x,v_y$ may be written in terms of polar
coordinates as
\begin{eqnarray}
  \label{eq:polar_veloc}
  v_x\,=\,s(t)\,\cos\varphi(t),~~v_y\,=\,s(t) \,\sin\varphi(t)
  \end{eqnarray}
with $s(t)=|{\bf v}(t)| \ge 0$ being the speed of the particle and
$\varphi(t)$ the polar angle defining the direction of motion, i.e.
the angle between the velocity vector and the $x$-axis.

The two spatial components for the position follow the dynamics
\begin{align}
  \label{eq:pos}
  \frac{d}{dt} {\bf r}(t)= {\bf v}(t) = s(t) {\bf e}_v(t),
\end{align}
were ${\bf e}_v(t)=\{\cos \varphi(t),\sin\varphi(t)\}$ is the unit vector in the velocity direction at time $t$. The corresponding
stochastic equations for $s(t)$ and $\varphi(t)$ contain
multiplicative noise 
\begin{eqnarray}
  \label{eq:polar_diff}
  \frac{d}{dt}{s}\,=\,-\gamma s \,+\,\sqrt{2D}\, \xi_s(t)\, ,~~~~\frac{d}{dt}{\varphi}\,=\,\frac{1}{s} \sqrt{2D} \,\xi_{\varphi}(t)
\end{eqnarray}
and have to be interpreted in the sense of Stratonovich. The noise terms $ \xi_{v}(t), \xi_{\varphi}(t)$ read
\begin{eqnarray}
  \label{eq:noise_polar}
  \xi_s(t)\,=\, \big(\xi_x(t)\cos \varphi + \xi_y(t)\sin\varphi \big)\,,~~~\xi_{\varphi}(t)\,=\,\big(-\xi_x(t)\sin \varphi + \xi_y(t)\cos\varphi \big)\
\end{eqnarray}
and are statistically independent. However, in contrast to the angle
noise $\xi_{\varphi}(t)$ the mean of the speed noise
$\mean{\xi_s(t)}$ does not vanish. The corresponding Fokker-Planck
equation for the transition probability in the new variables
$\tilde P(s,\varphi,t|s_0.\varphi_0,t_0)$ reads
\begin{align}\label{eq:SG2d_FPEext}
  \frac{\partial \tilde P(s,\varphi,t|s_0.\varphi_0,t_0)}{\partial t} & = -\frac{\partial
  }{\partial s}\left\{\left(-\gamma s + \frac{D}{s}\right) \tilde P\right\} +
  D\frac{\partial^2 \tilde P}{\partial s^2} +\frac{D}{s^2}\frac{\partial^2
    \tilde P}{\partial \varphi^2}\,,
\end{align}
which  in the long time limit  becomes independent
of the angle and approaches the Rayleigh speed distribution
\begin{eqnarray}
  \label{eq:rayleigh}
  \tilde P_0(s, \varphi) \, = \, {\cal N}\,s\, \exp\Big(-\frac{\gamma s^2}{2 D}\Big).
\end{eqnarray}
We note here that the probability to have zero speed vanishes in
agreement with the Maxwellian velocity distribution. The stochastic
force agitating the particle permanently hinders it to come to full rest. 
This is one particular reason why the concepts of the simple Brownian motion framework have to be modified when applied to self-propelled objects.

\subsection{Internal coordinates frame for polar particles}
\label{sec:internalcoord}
Many active particles, such as for example biological agents, may have a distinct body axis defining their preferred direction of motion (head-tail axis). Whereas this asymmetry is obvious in higher organisms, it should be noted that the crawling motion of cells is also driven by a polar actin cytoskeleton \cite{mitchison_actin-based_1996,ridley_cell_2003,joanny_motion_2003,kruse_contractility_2006}.  Also, for artificial active particles, such as chemically-driven colloids, it might be natural to assume a preferred direction of motion based on their propulsion mechanism  \cite{paxton_catalytic_2004,howse_self-motile_2007,ruckner_chemically_2007}. 

\begin{figure}
\begin{center}
\includegraphics[width=0.5\textwidth]{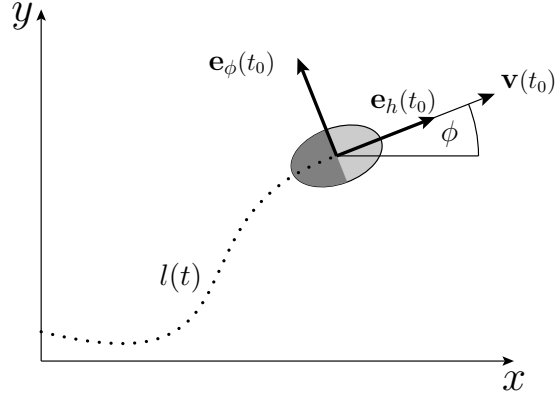}
\end{center}
\caption{Schematic visualization of the motion of a polar particle with unit vectors of the internal coordinates $\bbox{e}_h(t),\bbox{e}_\phi(t)$ at a given point in time $t=t_0$. The dashed lines indicates the trajectory of the particle $l(t)$. \label{fig:scheme_polar}}
\end{figure}

Without resolving the details of the origin of the asymmetry we simply assume that the polarity of particles introduces a distinct orientation, which we will refer to as {\em heading}. In general, we define the heading by a time dependent unit vector $\bbox{e}_h$, which in two dimensions is entirely determined by a single angular variable $\phi$:
\begin{align}
\bbox{e}_h(t) = (\cos\phi(t),\sin\phi(t)).
\end{align}
Thus the velocity of a point-like polar particle can be expressed by its velocity with respect to the heading and the corresponding heading vector ${\bbox v}=v {\bbox{e}_h}$. The velocity $v$ can be positive or negative, which can be identified with ``forward'' and ``backwards'' motion with respect to the heading.
In order to be able to span the two dimensional space we need a second unit vector, which can be associated with the angular direction perpendicular to the heading direction 
\begin{align}
\bbox{e}_\phi(t) = (-\sin\phi(t),\cos\phi(t)).
\end{align}
Here, we should emphasize that, despite the apparent similarities, the $(v,\phi)$-coordinates have to be distinguished from polar coordinates $(s,\varphi)$. In polar coordinates the speed is always positive $s(t)=|\bbox{v}|$ and the velocity unit vector $\bbox{e}_v$ is defined by the direction of the velocity vector at time $t$. In ($v,\phi$)-coordinates the velocity $v$ of a polar particle can be also negative, corresponding to backwards motion of the particle with respect to its heading. Furthermore, the heading vector does not depend on the momentary velocity but is defined by the intrinsic polarity of the particle. In Fig. \ref{fig:scheme_polar} a schematic visualization of a moving polar particle and the corresponding unit vectors is given.  

In general the following relations hold between the two coordinate systems:
\begin{align}
\bbox{e}_h \cdot \bbox{e}_v = 
\begin{cases} 
+1 & \text{for} \ v>0 \\
-1 & \text{for} \ v<0 
\end{cases} \qquad
\phi = \begin{cases}
\varphi & \text{for} \ v>0 \\
\varphi+\pi & \text{for} \ v<0 
\end{cases}
\end{align}
In cases, where the velocity $v$ is restricted to positive values, or where the particle is apolar, in the sense that its velocity dynamics are indistinguishable for forward and backwards motion, the internal coordinate frame cannot be distinguished from the polar coordinates. In this cases we will use throughout the review the corresponding polar coordinate unit vectors ($\bbox{e}_h\to\bbox{e}_v$, $\bbox{e}_\phi\to\bbox{e}_\varphi$).


\subsection{Markovian dichotomous process and white shot noise}
\label{sec:DMP}

In this review, we will often make use of the dichotomous Markovian
process (DMP), also called, the random telegraph noise. For this reason, we give here more details on  this process and its white-noise limit case, called white shot noise or Schottky-noise \cite{van_den_broeck_relation_1983,gardiner_handbook_1985}.

Because we are concerned with random motion, we illustrate the DMP by the
example of dispersing particles immersed in a flow through a
cylindrical tube with layers of different velocities. We follow here
the work by van den Broeck \cite{broeck_taylor_1990} who revisited Taylor
dispersion in 1990. The problem was first considered
experimentally and theoretically by G.I. Taylor in 1921
\cite{taylor_diffusion_1921} and in a series of papers in 1953/54
\cite{taylor_dispersion_1953,taylor_dispersion_1954,taylor_conditions_1954}. Similar theoretical problems have been formulated
also by S.~Goldstein and others
\cite{goldstein_diffusion_1951,kac_stochastic_1974,othmer_models_1988};
comprehensive reviews about DMP in dynamical systems can be found in
\cite{balakrishnan_continuous-time_1983,horsthemke_noise-induced_1984,bena_dichotomous_2006}.

Taylor found that  particles subject to layers of different flow velocities 
undergo additionally to the main stream flow a diffusion-like motion.  
To obtain this behavior, he considered scattering 
of the particles in the stream causing random jumps of the particle 
to other velocity layers. He even linked this behavior and his analysis to Pearson's problem of random insect
migration.

Let us consider a flow in one dimension  with two layers and with $x(t)$
denoting the position of the suspended particle in the frame co-moving
with the mean flow. It is assumed that the particle changes randomly
with a given constant rate $\lambda$ between the two layers in which it
moves with a constant relative velocity, either $+ v_0$ or $-v_0$. The velocity
is then a DMP $\xi_{DMP}(t)$, and the Langevin equation for the particle reads
\begin{eqnarray}
  \label{eq:dich_lang}
  \frac{{\rm d}}{{\rm d}t} x(t)\,=\, \xi_{DMP}(t).
\end{eqnarray}

For a temporally symmetric DMP, the life time of one of the velocities (before switching to the respective other one) is governed by an exponential distribution
\begin{eqnarray}
  \label{eq:exp}
  w(\tau)\,{\rm d} \tau \, = \, \lambda \exp (-\lambda \tau) {\rm d}\tau\,.
\end{eqnarray}

The transition probability densities $P_{\pm}(x,t|x_0,\xi_0,t_0,)$ for the particle 
to be at $r$ and the DMP to be at $\xi(t)=\pm v_{0}$ at time $t$ given the particle was at $r_0$ and the DMP was at $\xi_0$ at time $t_0$ is governed by equations like
\begin{eqnarray}
  \label{eq:telegraf1}
\frac{\partial P_+}{\partial t}\,&=&\,
  -\,v_0 \frac{\partial P_+}{\partial x} -\lambda P_+ + \lambda P_-\,,\\
\frac{\partial P_-}{\partial t}\,&=&\,
 + \,v_0 \frac{\partial P_-}{\partial x} +\lambda P_+ - \lambda P_-\,.
\end{eqnarray}
The conditions for $t_0$ enter by the initial conditions for $P_\pm$ at $t=t_0$. 
The probability averaged over the two velocity states and their initial states
$P(x,t)=P_+(x,t)+P_{-}(x,t)$ is given by the known second-order
telegraph equation \cite{shapiro_formulae_1978,horsthemke_noise-induced_1984,balakrishnan_continuous-time_1983,van_den_broeck_relation_1983,bena_dichotomous_2006}
\begin{eqnarray}
  \label{eq:telegraf}
  \frac{\partial^2 P}{\partial t^2} + 2\lambda \frac{\partial P}{\partial t} - v_0^2 \frac{\partial^2 P}{\partial x^2}\,=\,0\,. 
\end{eqnarray}
This equation can be solved in terms of modified Bessel
functions but has been supplemented by initial conditions for $P_{\pm
  \xi_0}$ or, respectively, $P$ and the flux $J=v_0(P_{+}-P_{-})$ in
$r_0$ and at $t_0$. Multiplication by $x^2$ and 
integrating over $x$  yields the same equation  which Langevin earlier obtained for Brownian motion \cite{langevin_theory_1908} with the respective re-assignments. By starting exactly at $x_0=0$ at time $t_0=0$, one finds that the mean square displacement evolves according to 
\begin{eqnarray}
  \label{eq:telegraf2}
  \langle x^2(t)\rangle \,=\,\frac{v_0^2}{\lambda}\left[t - \frac{1}{2\lambda}\, \left(1 -\exp(-2\lambda t) \right) \right]\, ,
\end{eqnarray}
which should be compared to  Eq.~\ref{eq:squareddispl}:  $2\lambda$ corresponds to the relaxation rate ($\gamma$ in Eq.~\ref{eq:squareddispl}) and $v_0^2/(2\lambda)$ is the spatial diffusion coefficient ($k_BT/\gamma$ in Eq.~\ref{eq:squareddispl}); an additional factor of three in Eq.~\ref{eq:squareddispl} is due to the difference in spatial dimensions.

A generalization to an asymmetric two state process with different
rates $\lambda_\pm$ and different values of the velocity $\xi_{DMP}(t)=v_\pm$  
can be easily formulated. The transition probability is
expressed by the equation
\begin{eqnarray}
  \label{eq:gen_tel}
  \frac{\partial^2 P}{\partial t^2} +
  (\lambda_+ + \lambda_-) \left(\frac{\partial P}{\partial t} - \mean{v} \frac{\partial P}{\partial t}\right) + v_+v_-\frac{\partial^2 P}{\partial x^2} + (v_++v_-)\frac{\partial^2 P}{\partial t \partial x}\,=\,0\,.
\end{eqnarray}
Here the mean velocity $\mean{v}= (\lambda_- v_+ \,+\,\lambda_+
v_-)/(\lambda_+\,+\,\lambda_-)$ was used. The diffusion
coefficient in the asymmetric case can be calculated as
\begin{eqnarray}
  \label{eq:dich_diff}
  D_{\rm eff}\,=\, \frac{\lambda_+\lambda_- (v_+-v_-)^2}{(\lambda_+ + \lambda_-)^3}.
\end{eqnarray}

We now outline the DMP's limit case of white shot noise
\cite{van_den_broeck_relation_1983}.  The asymmetric DMP, as discussed so far,
acts like a colored noise.  Its inverse correlation time is given by the sum of the switching rates
$\tau_c^{-1}=\lambda_-+\lambda_+$. The noise strength is determined by
$D_{DMP}=(v_+-v_-)^2\tau_c/4$. The limit to white noise can be taken if
shifting one of the possible velocity values to infinity combined with
a simultaneous vanishing of its live time. Specifically,  we let the upper level $v_+\to \infty$ and its live time $\tau_+=1/\lambda_+ \to 0$ while keeping the mean area   $v_+ \,\tau_+=h$ constant. 
In this way, the DMP $\xi_{DMP}(t)$ collapses to shot
noise $\xi_{SN}(t)$.  It consists of a sequence of $\delta$ peaks with
weights $h_i$:
\begin{eqnarray}
  \label{eq:shotnoise}
  \xi_{SN}(t)\,=\, \sum_i^{n(t)}\,  h_i\,  \delta(t-t_i)
\end{eqnarray}
The number of events $n(t)$ in the interval $(0,t)$ follows from a
Poisson distribution
\cite{masoliver_first-passage_1987,papoulis_probability_1991,caceres_generalized_1997,czernik_rectified_2000,eliazar_nonlinear_2005,kim_numerical_2007,zygadlo_relaxation_1993}.
\begin{eqnarray}
  \label{eq:poisson}
  P(n(t))\equiv Prob[n(t)=n] = \frac{(\lambda T)^n}{n!} \exp(-\lambda t).  
\end{eqnarray}
Here $\lambda$ is the rate of generating spikes, which equals the
inverse living time of the lower state $v_-$ of the DMP, i.e. $\lambda
=\lambda_-= 1/\tau_-$ . 

The weights $h_i$ are exponentially distributed, which is a property inherited  from the statistics of the stochastic duration of the DMP's upper state \cite{van_den_broeck_relation_1983} - the latter remains exponentially distributed even in the limit case of a vanishing mean live time. Its mean value approaches the fixed area in the limit $\lr{h_i}=h=\lim_{\tau_+\to0} v_+\tau_+$.

How similar is shot noise  to the more commonly used  Gaussian white noise? Like the latter, shot noise is a white noise, i.e. these fluctuations are $\delta$ correlated.  This can be also seen by means of  the correlation time  $\tau_c=(\lambda_++\lambda_-)^{-1}$, which vanishes if one of the live times ($\lambda_+^{-1}$ or $\lambda_-^{-1}$) goes to zero. Remarkably, $\tau_c$  is dominated by the smaller of the two mean live times (i.e. by the higher rate). 

While temporal correlations of shot noise and Gaussian white noise are similar, the distribution of shot noise is obviously not that of a mean-zero Gaussian distribution. First of all, shot noise has a  mean value $\mean{\xi_{SN}(t)}=v_-\, +\, h/\,\tau_-$ which is in general different from zero. Secondly, there is a strong asymmetry between the 
positive spikes  and the negative base line  $v_-$, which stands in marked contrast to the reflexion symmetry of any Gaussian distribution. This asymmetry  affects also  severely the statistics of dynamical systems driven by shot noise (see, for instance, studies by Richardson and Gerstner \cite{richardson_synaptic_2005,richardson_statistics_2006}). 

The difference from a Gaussian process can be quantified by the parameter
\cite{van_den_broeck_relation_1983} 
\begin{eqnarray}
  \label{eq:nongau}
  G\,=\, {|v_+\,-\,v_-| \,\tau_c}\,.
\end{eqnarray}
This value will vanish if simultaneously both lower and upper levels diverge according to  $v_\pm\to\pm\infty$ 
and the respective mean live times vanish $\lambda_\pm^{-1}\to 0$ while the two mean areas $v_+ \tau_+$ and $v_-\tau_-$ are kept constant and equal to each other. Obviously, this limit restores the symmetry of the process.
Furthermore, it can be shown that third- and  higher-order cumulants of the increments of this symmetric shot noise become negligible and that its effect on any dynamical system is completely equal to that of Gaussian white noise.
In this sense, the DMP approaches Gaussian white noise in the above limit, although it still attains only two discrete values in this limit.

%% file: SECTION_self_driven.tex
\section{Basic concepts of self-propulsion}
%
%
\label{sec:selfdriven}
\subsection{Dynamics of active Brownian particles}
\label{sec:selfdriven_ABP}
\subsubsection{Nonequilibrium, nonlinear friction and external
  forcing}
So far energy was supplied to the particle by molecular agitation
which has led to stochastic forces. In this Section, we want to
generalize the idea of Brownian particles by including an additional
energy input. In this way we will be able to derive
a simplified model of biological motion which we call Active
Brownian motion.

The major question we will address now is how this known picture
changes if we add an internal ``activity'' of particles. Our
main assumption is the additional inflow of energy leading to 
active motion can be described effectively by negative dissipation 
in the direction of motion. Hence, it will be modeled by negative friction
instead of a constant friction coefficient $\gamma$. We
introduce a nonlinear friction $\gamma(\bbox{r},\bbox{v})$ which
is a function of the position and velocity and, what is most
important, has regions in the phase space $(\bbox{r},\bbox{v})$,
where it assumes negative values. In addition, fluctuation-dissipation
relation becomes invalid for the non-equilibrium case of self-propulsion \cite{JulAjd97}.

Simple models of such active Brownian particles were studied already
in several earlier works (see e.g.  
\cite{schienbein_langevin_1993,steuernagel_elementary_1994,klimontovich_nonlinear_1994,schimansky-geier_structure_1995,riethmueller_langevin_1997,mikhailov_cells_2002,schweitzer_brownian_2003}). 
Here we will review and analyze in a more systematic way models of active Brownian particles with negative friction as well as the depot model of
particles which are able to store the inflow of energy in an internal
depot and to convert internal energy to perform different activities
\cite{schweitzer_complex_1998,ebeling_active_1999}. Other versions of active Brownian particle models
\cite{schimansky-geier_structure_1995,riethmueller_langevin_1997} consider more specific activities, such as
environmental changes and signal--response behavior. In these models,
the active Brownian particles (or active walkers, within a discrete
approximation) are able to generate a self-consistent field, which in
turn influences their further movement and physical or chemical
behavior. This non-linear feedback between the particles and the field
generated by themselves results in an interactive structure formation
process on the macroscopic level.  Hence, these models have been used
to simulate a broad variety of pattern formations in complex systems,
ranging from physical to biological and social systems \cite{schweitzer_clustering_1994,schimansky-geier_structure_1995,schweitzer_active_1997,helbing_active_1997,mikhailov_cells_2002,schweitzer_brownian_2003}.

Most of the time we will consider problems homogeneous in space with 
$\gamma({\bbox r},{\bbox v})=\gamma(\bbox{v})$. The situation of spatially
localized energy sources (food centers) generates more complicated dynamics
and was discussed in \cite{ebeling_active_1999}.  For such energetic pumps, the
basic ideas about nonlinear friction have been formulated by
Helmholtz and Rayleigh, whose aim was to model the complex energy input
in musical instruments. These models applied to walkers will be the
starting point, followed by a model of particles with an internal energy depot. It will be
assumed that the Brownian particles have the ability to
take up energy from the environment, to store it in an internal depot
and to convert internal energy into kinetic energy. Further on, we will discuss some specific models of active Brownian particles with a particular focus on the role of active fluctuations.  We will conclude this section with a brief discussion of a generalized model of active particles, where the direction of propulsion (motor direction) is considered as an additional degree of freedom.  

In other words we use Langevin equations which include
acceleration terms resulting from the energy inflow. We will show, that in
comparison with simple Brownian particles, the dynamics of active particles becomes
much more complex, which result in new dynamical features as e.g.:
\begin{itemize}
\item  new diffusive properties with large mean squared displacements,
\item  unusual velocity distributions with crater like shape,
\item  formation of limit cycles under confinement corresponding to motion on circles in space.
\end{itemize}
Some of these features may resemble active biological motion.  Hence,
the basic idea can be formulated as follows: how much physics is
needed to achieve a degree of complexity which gives us the impression
of motion phenomena found in biological systems?

In order to avoid misunderstandings we would like to stress again, that we do not
intend here to model any particular biological or social object but
instead to analyze general physical systems far from equilibrium, which
exhibit active motion and new types of dynamics.

\subsubsection{Active Brownian particles: velocity-dependent friction}
\label{sec:act}
The motion of Brownian particles with general velocity- and
space-dependent friction in a space-dependent potential $U(\bbox r)$
can be described again by the Langevin equation (see
Eq. (\ref{langev-or}):
\begin{eqnarray}
  \label{langev-or1}
  \frac{d {\bbox r}}{d t}={\bbox v}\,;\,\,
  \frac{d {\bbox v}}{d t}= {\bbox F}_{diss} -\nabla U({\bbox r}) + {\bbox{\cal F}}(t)
\end{eqnarray}
The new feature is the dissipative force which is now given with a
position and velocity dependent coefficient
\begin{eqnarray}
  \label{eq:nonlfriction}
  {\bbox F}_{diss} = - \gamma({\bbox r},{\bbox v}) {\bbox v}\,.
\end{eqnarray}
The force acts in direction of the motion and the friction
$\gamma(\bbox{r},\bbox{v})$ may depend on space, velocity and time. The term $\bbox{\cal
  F}(t)$ is a stochastic force with strength $D$ and a
$\delta$-correlated time dependence, see Eq. (\ref{stoch}). But now, due to nonequilibrium,
this noise strength is independent of the parameters in the dissipative
force and the Einstein relation is considered invalid. 

We consider, without loss of generality, $m=1$ and $D_p$ becomes $D$, which gives us
the following Langevin equation with unscaled Gaussian white noise
$\xi(t)$ as defined in Eq. (\ref{eq:noise}). 
\begin{eqnarray}
  \label{langev-v1}
  \frac{d {\bbox r}}{d t}\,=\,{\bbox v}\,;\,\,
\frac{d {\bbox v}}{d t}\,=\, - \gamma(\bbox{r},\bbox{v}) \,-\,\nabla U({\bbox r}) \,+ \,\sqrt{2 D}{\bbox \xi}(t) \label{langevv}
\end{eqnarray}
It gives the basis for the derivation of an energetic balance. From
the time derivative of the full mechanical energy one obtains
\begin{eqnarray}
  \label{eq:balance}
  \frac{d}{dt} E\,=\,\bbox{v}\,\frac{d \bbox{v}}{dt}\,+\,\nabla U(\bbox{r})  \frac{d \bbox{r}}{dt} 
\end{eqnarray}
which averaged over the noise yields
\begin{eqnarray}
  \label{eq:balance1}
  \frac{d}{dt} \langle E \rangle \,=\, -\gamma(\bbox{v},\bbox{r})\,\bbox{v}^2 + D
\end{eqnarray}
One sees that negative values of the friction coefficient lead to an
increase of the mechanical energy.

Since the fluctuating source is Gaussian white noise the transition
distribution density
$P(\bbox{r},\bbox{v},t|\bbox{r}_0,\bbox{v}_0,t_0)$ obeys a
Fokker-Planck equation
\begin{eqnarray}
  \label{eq:fokker1}
  \frac{\partial P(\bbox{r},\bbox{v},t|\bbox{r}_0,\bbox{v}_0,t_0)}{\partial t} + {\bbox v}\, \frac{\partial
    P}{\partial \bbox{r}} + \nabla U(\bbox{r})
  \,\frac{\partial P}{\partial\bbox{v}} =\frac{\partial}{\partial {\bbox v}}
  \Big[\gamma(\bbox{r},\bbox{v})\,\bbox{v}\, P+ D\, \frac{\partial P}{\partial \bbox{v}}
  \Big]
\end{eqnarray}
In the important case that the friction coefficient is only velocity
dependent and external forces are absent ($\nabla U(\bbox{r})=0$) the velocity
distribution becomes stationary and it holds
\begin{eqnarray}
  \label{eq:velocity_dis_general}
  P_0(\bbox{v})\, = \, {\cal N} \, \exp\Big\{  -\frac{1}{D}\,\int^{{\bbox{v}}} {\rm d} \bbox{v^\prime} \,\gamma(\bbox{v}^\prime)\,\bbox{v}^\prime\, \Big\} = {\cal N} \, \exp\Big\{  -\frac{\Phi({\bbox v})}{D} \Big\},
\end{eqnarray}
where $\Phi({\bbox v})$ is the effective velocity potential.
Let us consider now several models of the self--propelling mechanism.
Velocity-dependent friction plays an important role e.g. in certain
models of the theory of sound developed by Rayleigh and Helmholtz.
Following them we assume a parabolic behavior of the friction
coefficient
\begin{eqnarray}
  \label{eq:helm_ray}
  \gamma({\bbox r,v}) \,=\, -\, \alpha\, +\, \beta {\bbox v}^2
  \, =\,\alpha\,\left(\frac{v^2}{v_0^2}\, -\,1\right)\,=\,\beta\,(v^2\,-\,v_0^2)
\end{eqnarray}
This Rayleigh-Helmholtz- model is a standard model studied in earlier 
papers on Brownian dynamics \cite{klimontovich_nonlinear_1994,erdmann_brownian_2000}. We note that $v_0^2 = \alpha/\beta$ defines a special
value of the velocities where the friction is zero. At low velocities ($v^2<v_0^2$) the friction is negative. Hence, the particle gains kinetic energy from the pump.
Alternatively motion with greater velocities will be damped. 

\begin{figure}
  \centering
  \includegraphics[width=0.49\textwidth]{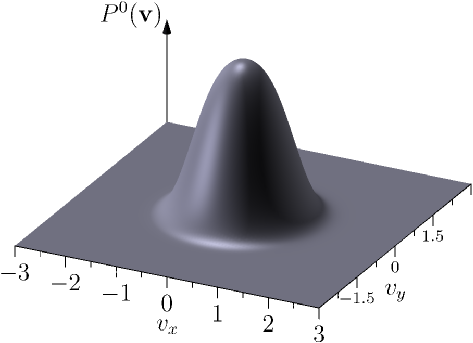}
  \includegraphics[width=0.49\textwidth]{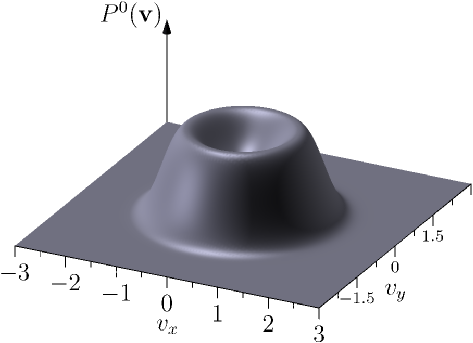}
  \caption{Stationary Cartesian velocity distribution in case of
    Rayleigh-Helmholtz friction. Left panel: passives regime with $\alpha=-0.1$
    ; Right panel: active regime with $\alpha=1.0$; Other parameter values:
    $\beta=1.0$, $D=0.5$.}
  \label{fig:craterRH}
\end{figure}

Without noise the direction of motion is defined by the initial
condition. With noise the particle moves in the long time limit in all
directions. This is seen from the shape of the stationary velocity distribution
\begin{eqnarray}
  \label{eq:velocity_dis_RH}
  P_0(\bbox{v})\, =
  \, {\cal N} \, \exp\Big\{ \frac{1}{D}\,\Big( \alpha \frac{\bbox{v}^2}{2}\,-\,\beta\,\frac{\bbox{v}^4}{4}\Big)\Big\}
\end{eqnarray}
Dependent on the sign of $\alpha$, the particle is passive
($\alpha<0$) or active ($\alpha>0$) extracting or pumping energy, out of or into the kinetic energy of the particle, respectively. The
corresponding single peaked and  crater-like distributions are
presented in Fig. \ref{fig:craterRH}.

The second standard model for active friction with a stationary velocity $v_0$ was derived from experiments with moving cells and analyzed by Schienbein and Gruler in 1993 
\cite{schienbein_langevin_1993,erdmann_brownian_2000}. 

It was originally formulated for the speed as a variable and may be seen as 
a linear simplification of the Rayleigh-Helmholtz friction. 
The dissipative friction force can be written in terms of speed $s=|{\bbox v}|$ and the corresponding unit vector ${\bbox e}_v={\bbox v}/s$ as  
\begin{align}
  \label{eq:sg_fri}
-\gamma({\bbox v}){\bbox v} = -\gamma_0 \left(s - v_0\right){\bbox e}_v.
\end{align}
It was shown by the mentioned authors that this model allows to
describe the active motion of several cell types as e.g.  granulocytes.
A disadvantage of this model is the discontinuity of the friction term at ${\bbox v} = 0$.
An advantage is the convergence of the  friction for large speeds $s$ to linear (Stokes) friction.
The stationary velocity distribution for the friction function in Eq. \eqref{eq:sg_fri} in Cartesian coordinates reads \cite{erdmann_brownian_2000}
\begin{align}\label{eq:SG_distro2d}
P_0 ({\bbox v}) = {\cal N} \exp\left\{-\frac{\gamma_0}{2 D}\left( |{\bbox v}| - v_0 \right)^2 \right\}  
\end{align}

An alternative formulation of the Schienbein-Gruler friction takes into account the polarity of an active particle (see Sec. \ref{sec:internalcoord}), which may be motivated by the distinct body axis of many organisms, defining their preferred direction of motion (head-tail axis).  Also, for artificial active particles it is often natural to assume a preferred direction of motion based on their propulsion mechanism.
The preferred direction of motion (heading) is given by the unit vector ${\bbox e}_h$ ($|{\bbox e}_h|=1$) and the Schienbein-Gruler friction can be written as (see Sec. \ref{sec:internalcoord}):
\begin{align}
\label{eq:sg_fri_polar}
-\gamma({\bbox v}){\bbox v} = -\gamma_0 \left(v-v_0\right){\bbox e}_h.
\end{align}

Please note that the two different Schienbein-Gruler variants are only equivalent if we define the heading vector as the velocity unit vector ${\bbox e}_h={\bbox e}_v={\bbox v}/|{\bbox v}|$.
In the general case, the second (polar) variant of the Schienbein-Gruler friction \eqref{eq:sg_fri_polar} is not symmetric with respect to $v=0$. It accounts also for the possibility of backwards motion with respect the heading direction, which Schienbein and Gruler neglected in their original work \cite{schienbein_langevin_1993}. As a consequence the first variant of the friction function is symmetric with respect to $v=0$ (apolar): there is no front or back for an active particle - the particle moves always to the ``front''. Related equations of motion with constant propulsion and linear friction have been used for example in the description of bacterial motion \cite{czirok_formation_1996,condat_randomly_2005}. 

In the following we will use, if not otherwise stated, the polar variant of the Schienbein-Gruler friction \cite{romanczuk_brownian_2011}.

With velocities in polar representation
(see Sec. \ref{sec:exter}) the stationary distribution can be
derived. It will be reached at times large compared with
$1/\gamma_0$ and is independent on the direction of motion
and reads for the polar Schienbein-Gruler friction:
\begin{eqnarray}
  \label{eq:SGhead_distro2d}
  P_0(\bbox{v})\, =
  \, {\cal N} \,\exp\Big\{ -\frac{1}{2 D}\,\gamma_0\,\big(|\bbox{v}|\,-\,v_0\big)^2\Big\} \Big[1\,+\, \exp\left\{ -\frac{2}{D}v_0|\bbox{v}|\right\}\Big]\,.
\end{eqnarray}
Again a crater-like distribution is established if $v_0>0$ \cite{romanczuk_brownian_2011}. 

\subsubsection{Depot model}
\label{sec:det_depot}
Now we will consider a friction function which is well behaved in the
full velocity range. This friction function is based on the idea of particles with an energy depot 
\cite{schweitzer_complex_1998,ebeling_active_1999,schweitzer_brownian_2003}, and has been recently used in the context of bacterial motion \cite{condat_diffusion_2002,garcia_testing_2011}.
We assume that the Brownian particle
itself should be capable of taking up external energy storing some of
this additional energy into an internal energy depot, $e(t)$. This
energy depot as an internal property of the considered objects may be
altered by three different processes:
\begin{enumerate}
 \item take-up of energy from the environment; where $q({\bbox r})$
  is a space-dependent pump rate of energy
\item internal dissipation, which is assumed to be proportional to the
  internal energy. Here, the rate of energy loss, $c$, is assumed to be
  constant.
\item conversion of internal energy into motion, where $h({\bbox v})$ is
the rate of conversion of internal to kinetic degrees of freedom.
This means that the depot energy may be used to accelerate motion
on the plane.
\end{enumerate}

The extension of the model is motivated by investigations of active
biological motion, which relies on the supply of energy, which is
dissipated by metabolic processes, but can be also converted into
kinetic energy.  The resulting balance equation for the internal
energy depot, $e$, of a pumped Brownian particle is then given by:
\begin{eqnarray}
  \frac{d}{dt} e(t) = q({\bbox r}) - c\;e(t) - h({\bbox v})\;e(t)
  \label{dep0}
\end{eqnarray}
A simple ansatz for $q({\bbox r})$ and $d({\bbox v})$ reads:
\begin{eqnarray}
  q({\bbox r})\equiv q_0 \; \qquad  h({\bbox v}) = d {\bbox v}^2 \label{dv}
\end{eqnarray}
where $d > 0$ is the conversion rate of internal into kinetic energy.
Under the condition of stationary depots we get
\begin{eqnarray}
  e_0 = \frac{q_0}{c+d {\bbox v}^2}
\end{eqnarray}
The energy conversion may result in an additional acceleration of the
Brownian particle in the direction of movement. This way we get for
the dissipative force including the usual passive friction and the
acceleration on the cost of the depot
\begin{eqnarray} 
  \bbox{F}_{diss} = - \gamma_0 {\bbox v} \,+\, d e(t) \bbox{v}
\end{eqnarray}

Correspondingly, we find a Langevin equation which contains an
additional driving force, $d e(t) {\bbox v}$:
\begin{eqnarray}
  \frac{d}{dt} \bbox{v} + \gamma_{0} \bbox{v} + \nabla U({\bbox r}) 
  = d e(t) {\bbox v} + {\cal F} (t) 
  \label{langev_depot}
\end{eqnarray}
Hence, the Langevin Eq. \eqref{langev_depot} is now coupled with the
equation for the energy depot, Eq. (\ref{dep0}). 

The energy loss of the depot is fully converted into kinetic energy of
motion of the Brownian particle. This is confirmed by the balance of
the kinetic energy:
\begin{equation}
 \dot{E}_{kin}=v_x \dot{v}_x + v_y \dot{v}_y=(de - \gamma_0) {\bbox v}^2 + \sqrt{2D} \ {\bbox v} \cdot {\boldsymbol \xi}
\end{equation}
where the first term on the r.h.s., the input of kinetic energy, is equal to the negative of the last term in Eq. (\ref{dep0}).

\begin{figure}[t]
  \centering
  \includegraphics[width=.5\linewidth]{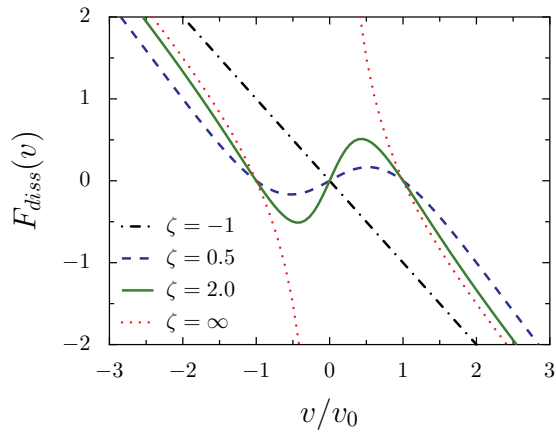}
  \caption{\label{diforc92} Friction force driving active particles
    corresponding to the depot model (SET-model): (i) Passive friction
    force $q = 0, \zeta = - 1$(dash-dotted straight line crossing the
    center). (ii) Depot model for positive values of the strength of
    driving: $\zeta = 0.5$ (dashed line); $\zeta = 2$ full line;
    $\zeta = \infty$ (dash-dotted line with a step at zero).}
  \label{gamma54}
\end{figure}

In most cases we will assume in the following that the energy depot is
stationary ${\dot e}(t) = 0$. This allows the adiabatic elimination of
the energy and leads to an effective dissipative force (see Fig.
\ref{gamma54}):
\begin{eqnarray}\label{depot_diss_force}
  {\bbox F}_{diss}({\bbox v}) = -
  \left[\gamma_0-\frac{d q}{c+d {\bbox v}^2}\right] {\bbox v} 
\end{eqnarray}
The corresponding friction function is
\begin{align}\label{eq:set_friction}
  \gamma({\bbox v})= \gamma_0-\frac{d q}{c+d {\bbox v}^2}\;.
\end{align}
The behavior of the force and the friction changes qualitatively in
dependence on the bifurcation parameter \cite{erdmann_brownian_2000} 
\begin{align}\label{eq:zeta}
\zeta = \frac{d q}{c \gamma_0} - 1.
\end{align} 
For positive $\zeta$--values we observe that the force disappears for three
values of the velocity, corresponding to an unstable velocity fixed point for vanishing velocity $v=0$ and two stable fixed points at a finite speed $v^2=v_0^2>0$.

Let us now consider several special cases in
more detail: In the case that the velocities are small (${\bbox v}^2\ll c/d$) we get
for the friction law
\begin{eqnarray}
  \gamma({\bbox v})= \left(\gamma_0- \frac{d q}{c}\right)- \frac{q\;
    d}{c^2} {\bbox v}^2 + {\cal O}\left({\bbox v}^4\right),
\end{eqnarray}
which corresponds with 
\begin{eqnarray}
  \alpha=\frac {d q}{\gamma_0}-\gamma_0; \qquad \beta=\frac {q d}{c^2} 
\end{eqnarray}
to the Rayleigh-Helmholtz model discussed above \eqref{eq:helm_ray}.

\begin{figure}
  \begin{center}
    \includegraphics[width=.49\linewidth]{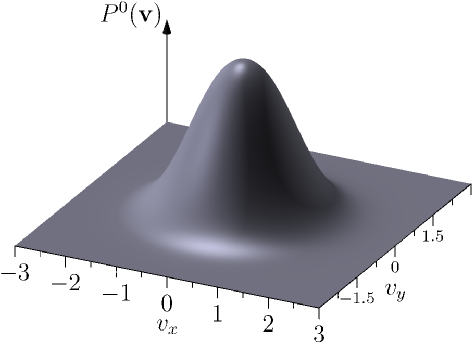}\includegraphics[width=.49\linewidth]{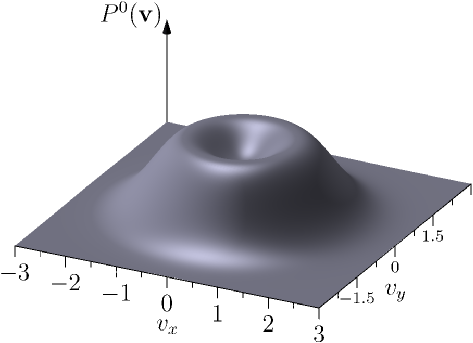}
    \caption{\label{pv02} Stationary Cartesian velocity distribution function of active
      Brownian particles for the depot model with under-critical
      values of the parameters (left panel; passive regime, $q_0=0.5$) and with
      over-critical parameter values (right panel; active regime, $q_0=2.0$). Other parameter values: $d=c=\gamma_0=1.0$, $D=0.5$}
  \end{center}
\end{figure}

The dissipative force for different values of $\zeta$ is shown in Fig. \ref{gamma54}. 
For positive $\zeta$, due to the pumping with free energy, slow
particles are accelerated and fast particles are damped. At certain 
conditions our active friction functions have  a zero corresponding
to stationary velocities $v_0$, where the friction function and the
friction force disappear. The deterministic trajectory of our system
moving on a plane is in both cases attracted by a cylinder in the
4d-space given by
\begin{eqnarray}
v_1^2 + v_2^2 = v_0^2
\end{eqnarray}
where $v_0$ is the value of the stationary velocity. For the
Rayleigh-Helmholtz model (RH) and the depot model (DM) the stationary velocities read:
\begin{equation}\label{eq:stat_v0}
  v_0^2=\frac{\alpha}{\beta} \  \text{(RH)}; \qquad v_0^2=\frac{q_0}{\gamma_0} \  -
  \frac{c}{d} \ \text{(DM)} \,. 
\end{equation}
We insert the effective force of the depot-model
\eqref{depot_diss_force} into the Fokker-Planck equation. The
stationary solution of the latter reads approximately
\begin{equation}
\label{distr_velo}
  P_0({\bbox v}) = {\cal N}
  \left(1+\frac{d}{c}\,{\bbox v}^2\right)^{\frac{q_0}{2D}}
  \exp\left[-\frac{\gamma_0}{2D} \,{\bbox v\,}^2\right]\,.
\end{equation}
The figure~\ref{pv02} shows a cross section of the probability distribution
for Rayleigh-Helmholtz and Schienbein-Gruler-Helbing friction function.
In case of strong noise and low pumping the particles approach high
velocities and broad distributions. Except in the close vicinity of
vanishing velocities the pre-factor in equation~(\ref{distr_velo}) can
be neglected.

\subsubsection{The limiting case of constant speed}
\label{sec:self_driven_const_s}
The velocity distributions, Eqs. \eqref{eq:velocity_dis_RH}, \eqref{eq:SG_distro2d},
\eqref{eq:SGhead_distro2d} and \eqref{distr_velo}, collapse to a $\delta$-peaked distribution at $v_0$ in the limit of vanishing noise with respect to the pumping term. The limit has to be taken such that $\beta/D$ or $\gamma_0/D \to
\infty$, respectively, and with $v_0=const. \ne 0$ one gets 
\begin{eqnarray}
  \label{eq:deltaspeed}
  P_0(|{\bbox v}|)\, \propto \,\delta (|{\bbox v}|-v_0)\ .
\end{eqnarray}
It yields the case of constant speed in which the
dynamics of the particles can be described by the angular direction
$\varphi(t)$ and spatial coordinates. The dynamics is given by Eq.
\eqref{eq:pos} and the corresponding Langevin equation for the angle, i.e.
the velocity dynamics Eq.~\eqref{eq:polar_diff} reduces to
\begin{eqnarray}
  \label{eq:speed}
  s(t)\,=\,v_0=\, const.\,,~~~~\frac{d}{dt} \varphi\, = \,\frac{1}{v_0} \,\sqrt{2 D}\,\xi(t)
\end{eqnarray}
A more detailed discussion of specific aspects of these models will be
given throughout the next two sections. Here we point out that the
important class of models with constant speed emerge from the
Active Brownian particles framework in the discussed limit. Thus, we emphasize, that
the concept of Active Brownian particles goes beyond this restriction and 
accounts naturally for fluctuations of the speed.

Assuming constant speed $s(t)=v_0=const.$ ($v_0>0$) the Fokker-Planck-Eq.
\eqref{eq:SG2d_FPEext} describes transitions in the angle dynamics by
$P(\varphi,t| \varphi_0,t_0)$ and reduces to
\begin{align}\label{eq:SG2d_FPEphi}
  \frac{\partial P(\varphi,t|\varphi_0,t_0)}{\partial
    t}\,=\,\frac{D}{v_0^2}\frac{\partial^2 P}{\partial \varphi^2}.
\end{align}
With the initial condition $P(\varphi,t_0|\varphi_0,t_0)\,=\,
\delta(\varphi-\varphi_0)$, this equation is solved by a Gaussian with
phase diffusion coefficients 
\begin{align}\label{eq:Dphi}
D_{\varphi}\,=\,D/v_0^2\,. 
\end{align}
Taking into
account the periodicity of the heading $P(-\pi,t)= P(\pi,t)$, the time
dependent solution of $\eqref{eq:SG2d_FPEphi}$ reads
\begin{align}
\label{eq:wrapped}
P(\varphi,t|\varphi_0,t_0)\,=\,\frac{1}{\pi}\left(\frac{1}{2}+\sum_{n=1}^{\infty}
  \cos n(\varphi-\varphi_0) \, \exp \Big(-n\frac{D}{v_0^2}(t-t_0)
  \Big) \right)\,.
\end{align}
For large times ($t-t_0\to\infty$), the distribution
$P(\varphi,t|\varphi_0,t_0)$ converges towards the homogeneous
distribution $1/(2\pi)$, which corresponds to a complete loss of
information of the initial direction.  The relaxation rate of the
angular function in Eq.~\eqref{eq:SG2d_FPEphi} is given by the relaxation
rate of the first Fourier mode ($n=1$), i.e. $\tau_r=v_0^2/D$.

It might be of interest to inspect the full dynamics of
Eq.~\eqref{eq:speed} in Cartesian velocity components. The Fokker-Planck
equation for the probability distribution of positions and velocities
assumes the form
\begin{eqnarray}
  \label{eq:cart_angle}
  \frac{\partial P(\bbox{r},\bbox{v},t|\bbox{r_0},\bbox{v_0},t_0)}{\partial t}\,=\, - {v_x}\,\frac{\partial P}{\partial{x}}- {v_y}\,\frac{\partial P}{\partial{y}} + 
  \,{D \over \bbox{v}^2}\left[\frac{\partial}{\partial{v_x}}\,v_y\,-\,\frac{\partial}{\partial {v_y}} v_x\,\right]^2 P,
\end{eqnarray}
where we have use the Stratonovich interpretation.  This equation is the starting point for the derivation of equations for the moments of the probability distribution. Later, we will use them to formulate hydrodynamic equations for the density, velocity,
and temperature of an ensemble of active particles. Here, it is instructive to look at the dynamics of the mean velocity $\bbox{u}(t)$. 
To obtain the dynamics for $\bbox{u}$, Eq.
\eqref{eq:cart_angle} is multiplied by ${\bbox{v}}$ and integrated over
possible velocity values. This expression is normalized by the local
density. Further, for simplicity we assume a homogeneous situations
and put all spatial derivatives to zero. This yields
\begin{eqnarray}
  \label{eq:mean_ve}
  {{\rm d} \over {\rm d}t}\,\bbox{u}(t) \,=\, - {D\over v_0^2} \bbox{u}\,. 
\end{eqnarray}
The mean velocity relaxes to zero and the mean relaxation time
$\tau_r= v_0^2/D$ is inverse to the intensity $D$ of the noise
which acts on the angle leading to a loss of directed motion.

\subsubsection{Angular dynamics of Active Brownian
  particles}
\label{sec:turning}
Some of recent measurements \cite{ordemann_motions_2003} indicate that during
short time intervals the hops leading to the overall random motion of
Daphnia may follow with some preferred angle to the left or right with
respect to their previous direction of motion. In the following we
will consider generic models which describes such kind of motion. We
will present some details within a model of active Brownian motion and
formulate also random walk model for a hopping dynamics.

The aim is to develop powerful techniques for the determination of the
diffusion coefficient. This diffusion coefficient if of central
interest \cite{okubo_diffusion_2001} in certain problems e.g. in
ecology and has received renewed attention recently
\cite{mikhailov_self-motion_1997,komin_random_2004,schimansky-geier_advantages_2005,Lin07,Lin08,peruani_self-propelled_2007}.
It was proposed to be the central interest since an optimally selected
diffusion coefficient might give an advantage in the search for food
\cite{garcia_optimal_2007,dees_patch_2008,dees_stochastic_2008}. Food consumption in a fixed time and in
bounded geometries depend significantly on the value of the spatial
diffusion and can be maximized by optimal selection of its value
\cite{schimansky-geier_advantages_2005}.

Here we look first at active Brownian particles performing the motion
with a preferred rotation direction, quantified by the angular
velocity $\Omega$. Motivated by the example of Daphnia mostly
performing an in-plane motion, we confine ourselves to a
two-dimensional situation.  In what follows, we first consider the
properties of the angular motion of such active particles. In the next
chapter, we will discuss how they affect their spatial diffusion.
Here, we concentrate on the quite realistic case when the speed
(absolute value of the velocity) can be taken as a constant but
fluctuations in the direction of motion cannot be neglected.

The inclusion of turning angles is best represented in a polar coordinates description. Simply requiring that the angle will change per unit time by a fixed amount is described as
\begin{equation}
  \label{eq:rotation}
  {\rm {d\over dt }}\, \varphi (t) \,=\, \Omega (t) g(s) \,+\, {\sqrt{2D}\over s}   \xi(t) \, ,
\end{equation}
where $\Omega(t)$ is the instantaneous angular velocity. The function
$g(s)$ shall describe a principal variation of the angular velocity in
dependence on the speed $s=|{\bbox v}|$. It is obvious that this function can
be also a function of $\varphi$ if the problem is not isotropic, say
in case of applied external forces.  

The second term on the r.h.s. of Eq.  (\ref{eq:rotation}) stands for the fluctuating
force with intensity $D$.  An important case is that $\Omega$ is
constant or switches between two different constant values, which
represent two possible turning behaviors of the particle (dichotomous
switching). As a transition rule for the switching between the two
turning velocities one might assume that the dynamics follows a
dichotomous Markov process, or as shown in the next section, the
problem can be formulated as a renewal process with arbitrary waiting
times densities in both states.  Last but not least, $\Omega$ can be
also a second Gaussian random process, which we will introduce later on as
active internal noise. It has a component perpendicular to the
direction of motion 
with $g(s)=1/s$ (see Eq. \ref{eq:Fstoch_a}) and stands for a
random variation of the angular velocity.

In case of active Brownian particles we formulate the problem as a
Newtonian dynamics with specific forces. The arising question what
forces correspond to the introduced angular velocity can be easily
answered via a transformation back to Cartesian velocities. In the two
dimensional case we find in addition to the active pump and the noise
term a force with
\begin{equation}
  \label{eq:lang1a} {\rm {d\over dt} } \bbox{v}(t) \,\propto \,  \mathbf{F}_{turning}\, = \,g(|{\bbox v}|)\,\left( \begin{array}{rr} {\Omega (t)
        v_y}\\{-\Omega (t) v_x}\end{array}\right)\,.
\end{equation} 
It resembles a Lorentz or Coriolis force with a vector ${\bbox
  \Omega}=(0,0,\Omega)$ pointing in the $z$-direction perpendicular to
the plane of motion. For the Lorentz-force, if the vector is the
magnetic field and $g(|{\bbox v}|)=1$, a charged particle performs a Larmor
precision motion on a circle with radius depending on the initial
conditions. Choosing $g(|{\bbox v}|)=1/|{\bbox v}|$ the turning force becomes independent
on the speed, which has been verified in experiments on artificial active particles \cite{paxton_catalytic_2004}. We point out that the consideration of white
noise (shot or Gaussian) will lead to a Stratonovich term, which will
be discussed later on.

We proceed with the assumption that the velocity approaches quickly a
stationary speed ($s\to v_0$, Eq. \ref{eq:deltaspeed}). From
(\ref{eq:rotation}) one obtains the Fokker-Planck-equation (for the case
that the angular velocity is not a white noise)
\begin{equation}
  \label{eq:sol} \frac{\partial}{\partial t} P(\varphi,t)= -\Omega(t)
  \frac{\partial P}{\partial \varphi} + D_\varphi \frac{\partial^2 P}{\partial
    \varphi^2}
\end{equation} 
with $D_\varphi$ defined in Eq. \eqref{eq:Dphi}.
We note that $D_\varphi$ is not necessarily small: For the
Rayleigh-Helmholtz model (\ref{eq:helm_ray}) with $\alpha^2/\beta
\gg D$ the value of $v_0$ is of the order $\sqrt{\alpha
  / \beta}$ and $D_\varphi \simeq D \beta / \alpha$ can be rather
large for moderate values of $\alpha$.

For a constant value of the angular velocity $\Omega$ this motion can be
described by the following unwrapped transition probability density
($t \ge 0$) :
\begin{equation} 
  P(\varphi,t|\varphi_0, 0)=\frac{1}{\sqrt{2 \pi D_\varphi t}} \cdot 
  exp\left(-\frac{(\varphi-\varphi_0-\Omega t)^2}{4 D_\varphi t}\right).
  \label{Gauss1}
\end{equation}

\subsubsection{Active Brownian dynamics resulting from coupling of molecular motors}
\label{sec:molmot_abp}
An example for active Brownian motion arising as a collective effect  is the intracellular transport by coupled molecular motors. Motor proteins like myosin or kinesin run along filaments (actin, microtubules) within the cell, transport vesicles, or provide active forces \cite{JulAjd97,How01}. Single motors have a preferred direction of motion along a given filament, a motion that is powered by ATPase. In many situations motors are coupled, for instance, when pulling collectively at a large vesicle \cite{GuePro10}. Collective effects in motor assemblies can be studied {\em in  vitro} in motility assays in which the roles of transporter and track are reversed: on a glass surface covered with motors fixed in their position, a filament is transported involving typically a large number of motors (up to a few hundreds). A genuine collective effect which has been found in such experiments, is the occurrence of bidirectional motion. A filament running for a while in one direction suffers apparently spontaneously a reversal of the direction of motion and runs "backwards" \cite{EndHig00}. This and other features of coupled molecular motors have been studied in the model by J\"ulicher and Prost, that we briefly discuss and for which we review the relation to an active Brownian particle's dynamics in the following.

\begin{figure}
 \centerline{\parbox[b]{0.5cm}{\rm\bf(a)\\[2em]}\parbox{6cm}{\includegraphics[width=6cm,angle=0]{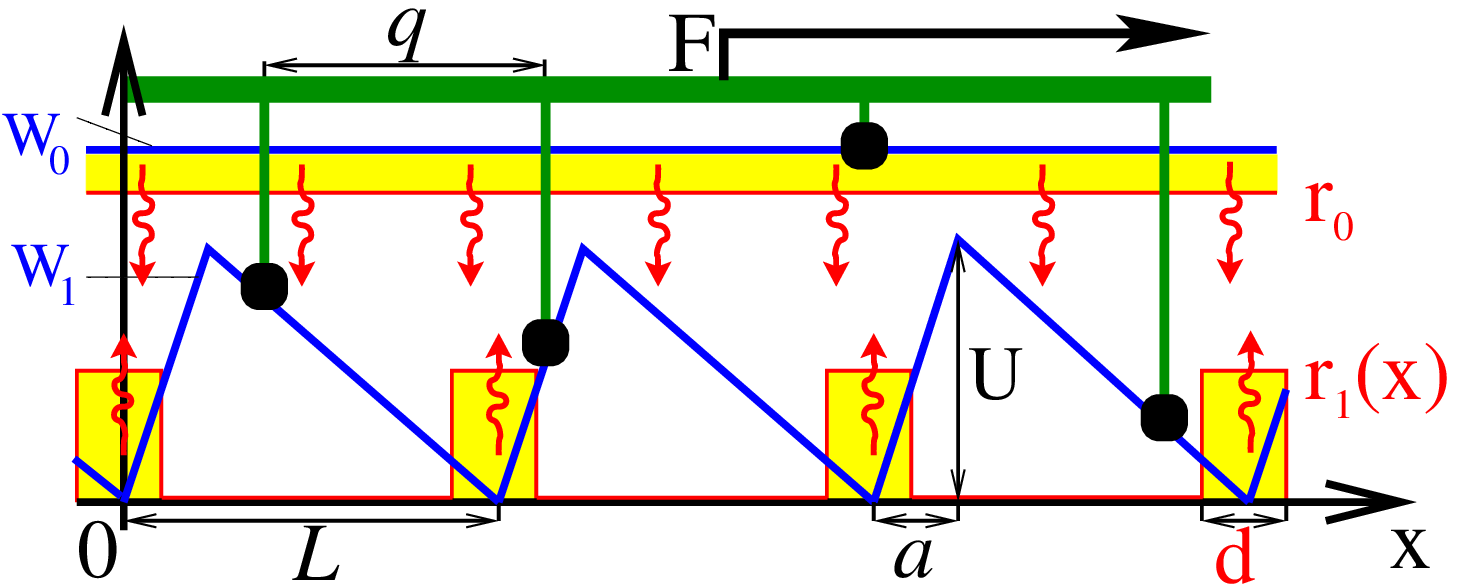}}\hspace{2em}\parbox{6cm}{\includegraphics[width=6cm,angle=0]{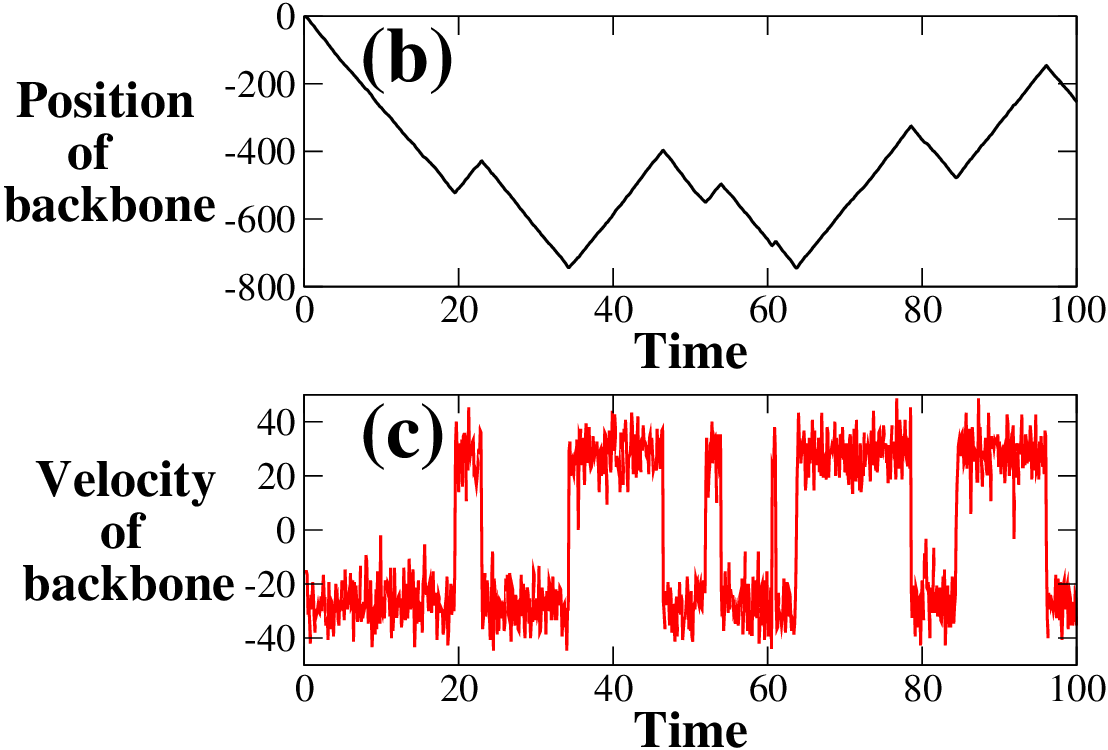}}}
\caption{\small  
Model of coupled molecular motors (CMM)  (a), a typical trajectory of the backbone (b), and the corresponding velocity (c) for $N=300$ and $a=L/2$.  For large $N$, the dynamics of the CMM model can be approximated by the dynamics of an active Brownian particle. Panel (a)  modified from \cite{LinNic08}. 
\label{fig:cmm_model}}
\end{figure}

The interaction between motor proteins and filament can be captured by different free energy landscapes between which the system switches by binding and release of ATP/ADP.  Essential features of the motor's dynamics can be already captured in a model system  switching between only two potentials where one of them is flat while the other one possesses a periodic structure with broken spatial symmetry (ratchet potential). Typically, the unbinding of the motor depends on its position along the backbone while  the binding of ATP can be assumed to be independent of the motor position. The motion of the motor is affected by two kinds of noise: the thermal motion and the switching between the two states which is likewise stochastic. In a simple approach to the dynamics of motor assemblies, the coupling of motors can be regarded as rigid leading to only one spatial degree of freedom: the position of the backbone (e.g. its center of mass). In this approximation,  the force acting on each motor contributes to the total force on the filamentous backbone and, in turn, the velocity of the single motor is given by that of the backbone. The equations for the overdamped dynamics of the backbone read
 \ba
\lambda\dot{x}&=&F_{ext}-\frac{1}{N}\sum_{j=1}^N\sigma_j(t)
W^\prime(x+ j\cdot q)+\sqrt{\frac{k_BT\lambda}{N}}\eta(t),\nonumber\\
&&\sigma_j=0\overset{r_0}{\underset{r_1(x+ j\cdot q)}{\rightleftharpoons}}\sigma_j=1.
\label{eq:cmm}
\ea
Here, $x$ is the central coordinate of the backbone, $\lambda$ is a an effective friction coefficient per motor, $F_{ext}$ an external force applied to the backbone, and $W(x)$ is the piecewise linear ratchet potential shown in \bi{cmm_model}a, the asymmetry of which determined by the parameter $a$ (see Fig. \ref{fig:cmm_model}a; for $a=L/2$ the potential is symmetric).  The state of a given motor $j$ at time $t$ is determined by the variable
$\sigma_j(t)$ which takes the values $0$ or $1$. The switching rate $r_0$ for the transition $0 \to 1$ does not depend on the spatial position. The rate for the transition   $1 \to 0$ attains the value $\hat{r}_1$ only if the motor is within the neighborhood of size $d$ centered around the potential minimum and is zero otherwise.  Parameters in the following are $L=1$,  $d=0.2L$, $\hat{W}=1$, $\lambda=0.01$, $r_0=40$, and $\hat{r}_1=500$. For simplicity,  thermal fluctuations are neglected.  
 
Simulations of the model for a large number of motors reveal a bidirectional motion of the assembly (cf. \bi{cmm_model}b) corresponding to a bistable velocity dynamics (\bi{cmm_model}c). 
It has been suggested \cite{BadJul02,LinNic08,TouSch10} that the velocity of a large  motor assembly ($N\gg 1$) can be well described by 
\begin{equation}
\dot{x}=v,\qquad\dot{v} = f(v)+ g(v)\xi(t),
\label{eq:langv}
\end{equation}
i.e. by an active Brownian particle (ABP) model, in which we also allow for a speed-dependent noise intensity.  
This means essentially that for large $N$ (i) the space dependence of the problem vanishes and (ii) the many dichotomous degrees of freedom together with the interaction of motors and potential results in one continuous degree of freedom (the velocity). The original problem does not possess any inertia, the new time scale introduced in \e{langv} comes from the switching times of the motor  and the relaxation of the probability density of motors in the spatial ratchet potential.

We can write down a rather lengthy Master equation with one spatial degree and $N$ dichotomous degrees of freedom for the system  \e{cmm} and try to approximate this equation by a Fokker-Planck  equation with one spatial and one velocity degree of freedom corresponding to \e{langv}. Whether this is possible and if so how this mapping can be carried out analytically is still an open problem. In the following, we present some numerical evidence that a mapping from CMM to ABP model is  a meaningful approximation. 

Assume, that \e{langv} is a reasonable approximation of the CMM model and suppose,  a long time series of the assembly's velocity is given, how could we determine the functions $f(v)$ and $g(v)$?  There are two methods to determine these functions \cite{YatErb09,TouSch10}. First, we could measure how the velocity changes on a short time-scale. Mean and variance of the instantaneous velocity changes should be proportional to $f(v)$ and $g(v)$, respectively. Put differently, one measures the Kramers-Moyal coefficients of $v(t)$ \cite{risken_fokker-planck_1996}. Secondly, we could also measure the stationary distribution $P(v)$  (long-time statistics), which is uniquely determined by the two functions we are seeking. Together with one additional piece of information involving the time scale of the system (e.g. the rate of velocity reversals or the spatial diffusion coefficient), one can determine both functions   $f(v)$ and $g(v)$.
\begin{figure}
 \centerline{\parbox{6cm}{\includegraphics[width=6cm,angle=0]{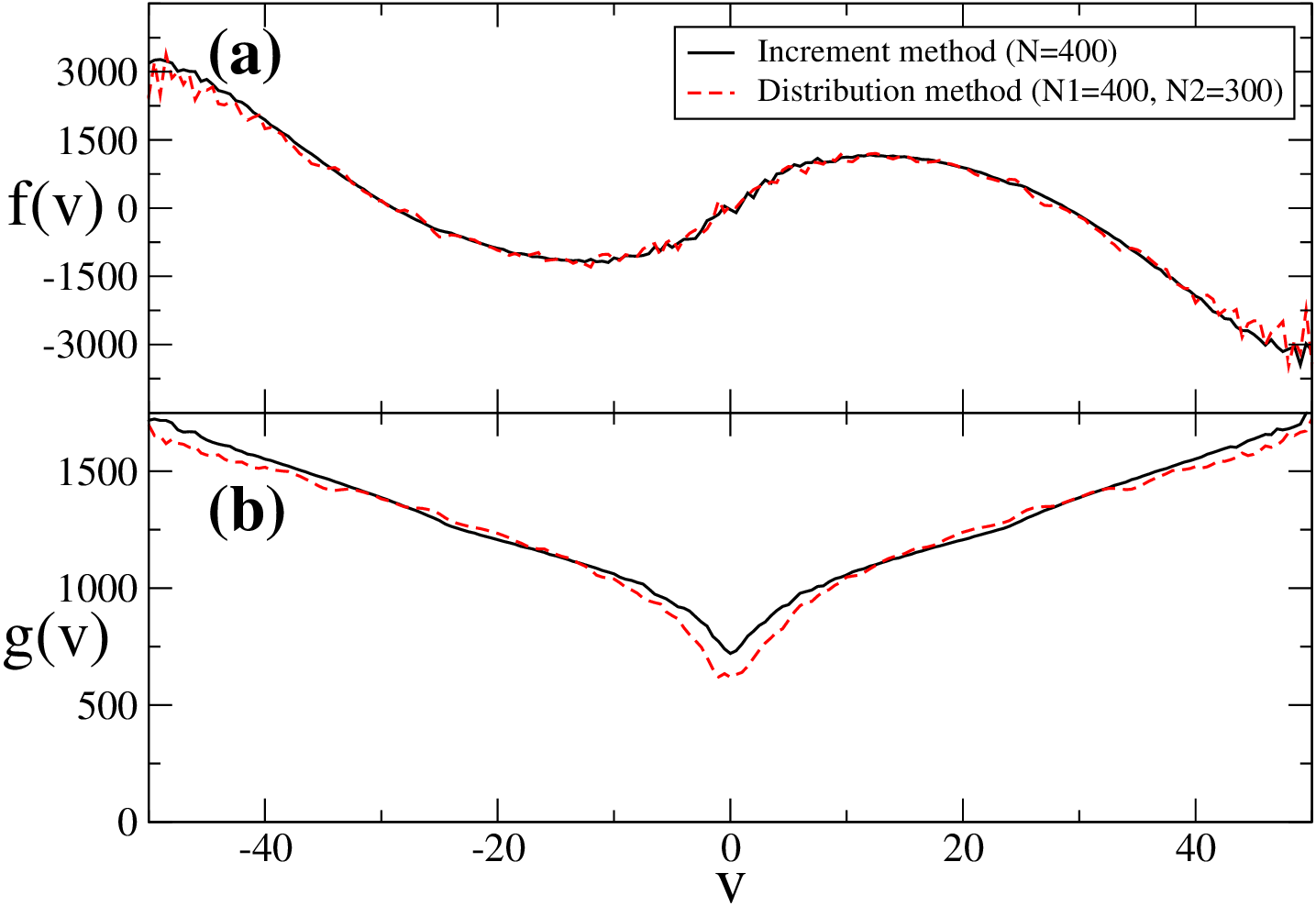}}\hspace{2em}\parbox{6cm}{\includegraphics[width=6cm,angle=0]{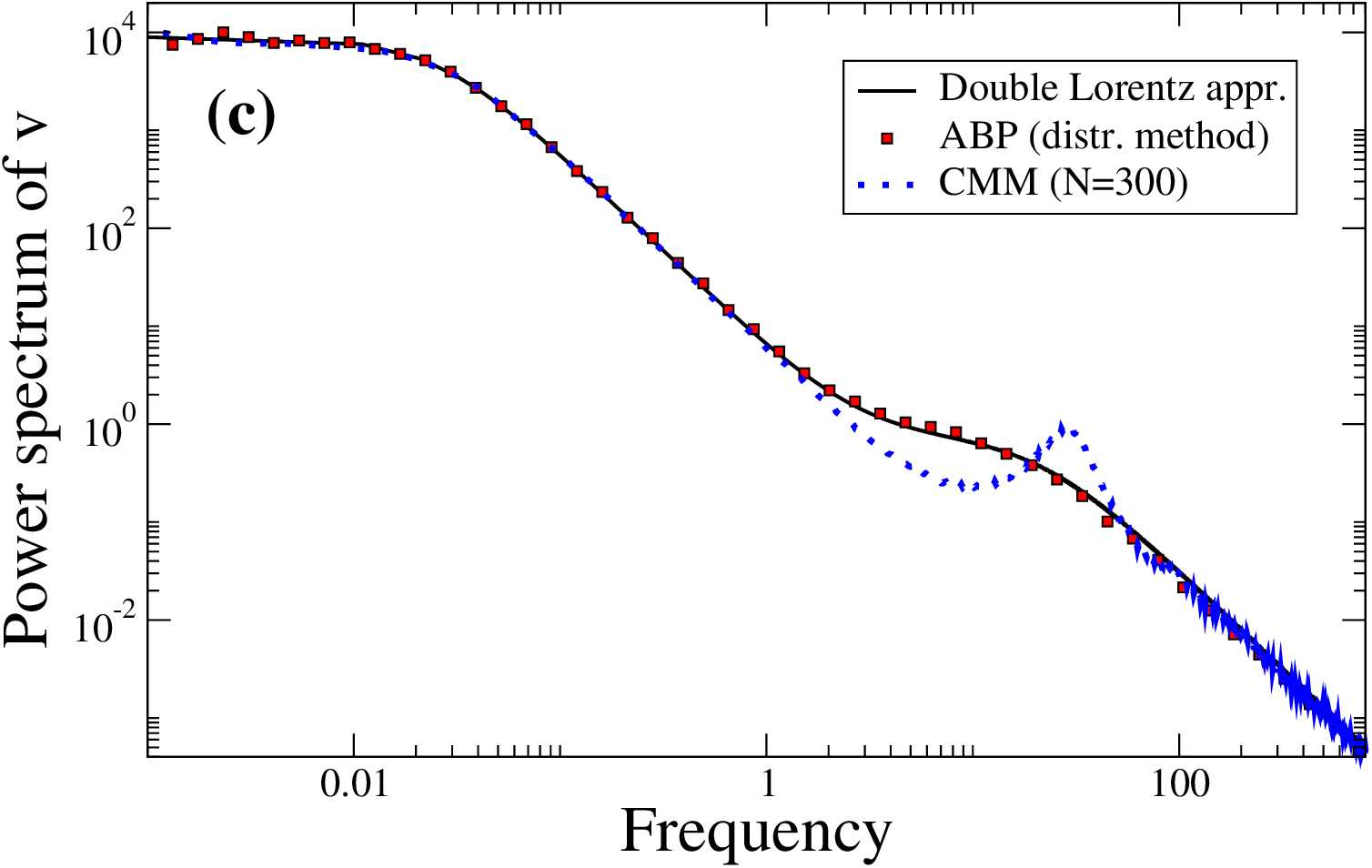}}}
\caption{\small  
Drift (a) and noise amplitude (b) of the CMM model estimated by the increment (solid) and distribution (dashed lines) methods for the symmetric system with $a=L/2$.  In (c) the velocity power spectrum of the CMM model (dashed line) is compared to that of  the approximating ABP model; for the latter, $f(v)$ and $g(v)$ were estimated via the distribution method. Also shown in (c) is the analytical approximation \e{spec_app} (solid line) to the ABP's power spectrum consisting of two Lorentzian spectra. Modified from \cite{TouSch10}. 
\label{fig:f_g_v}}
\end{figure}

The two methods have been applied to the CMM model in Ref.~\cite{TouSch10},  and the resulting drift and diffusion functions in the symmetric case are shown in \bi{f_g_v}.   The agreement between the functions determined by the independent increment and distribution methods is remarkable and thus shows that the ABP model is a reasonable approximation of the CMM model. The function $f(v)$ displays the shape of an inverted N, in particular, it possesses a negative slope for $v=0$, i.e. a vanishing velocity is dynamically unstable. 
The noise in the approximate ABP dynamics is indeed multiplicative as  the noise amplitude $g(v)$ increases for increasing speed. The latter feature is in marked contrast  to the system studied by Yates et al. \cite{YatErb09} in which the noise amplitude decreases with increasing speed. 


However, despite the consistent nature of the ABP approximation at short and long time scales, there exists also a discrepancy  between the CMM and ABP models  on an intermediate time scale. The power spectra of both systems, shown in \bi{f_g_v}c display a good agreement at low to intermediate frequencies and at very high
frequencies. In either limits the spectrum is well described by Lorentzian spectra, which correspond to those of a two state process of velocity reversals (range of low frequencies)  and an approximate Ornstein-Uhlenbeck process
for small-scale fluctuations around the two metastable velocities. For the ABP model, summing up these two contributions \cite{TouSch10}
\be
\label{eq:spec_app}
S(f)\approx \frac{2D_{\rm eff}}{1+(\pi f/r)^2}+\frac{2g^2(v_{min})/N}{f'(v_{min})^2+(2\pi f)^2}
\ee 
gives a reasonable approximation of the total spectrum (solid line in \bi{f_g_v}c). The CMM model, however, shows one additional feature, which is not captured by the ABP approximation: there is a pronounced peak around 
$f=25$. This corresponds to a small-scale damped oscillation in the velocity around its metastable states, which cannot be captured by a first-order velocity dynamics as in \e{langv}. In summary, the velocity process of the CMM model in all its details is more complicated  than the simple ABP dynamics even in the limit of a very large number of motors. Nevertheless, with respect to the statistics of the velocity and its reversals, to the spatial diffusion coefficient, and to the high-frequency fluctuations, the ABP model is an excellent approximation.

\subsubsection{Depot model for internal motors and broken friction symmetry}
\label{sec:depot_intosc}
So far, we have considered active motion induced by negative dissipation either through an internal energy depot or an effective velocity-dependent friction.  
In principle, such active Brownian particle models can always be considered as an effective, or ``coarse grained'', theoretical description of more complex microscopic dynamics leading to self-propelled motion as discussed in the previous section (Sec. \ref{sec:molmot_abp}) in the context of cooperative motion of many coupled molecular motors.

The physics of active transport on the sub-cellular level by such molecular motors and of related ratchet systems is a fascinating field of research, which has been under intensive investigation for decades now. A detailed review of the field is far beyond the scope of this work and we refer the interested reader to other comprehensive reviews on the subject \cite{juelicher_modeling_1997,reimann_brownian_2002,haenggi_artificial_2009}. Nevertheless, it leads us to a fundamental question, which we would like to address here in the context active particles: How can directed, self-propelled motion emerge in simple mechanical systems? In the case of directed transport in a ratchet, the directionality is based on breaking of the spatial symmetry of the underlying effective potential and in the presence of symmetric periodic forces or non-equilibrium fluctuations.

\begin{figure}
\centering
\includegraphics[width=.38\linewidth]{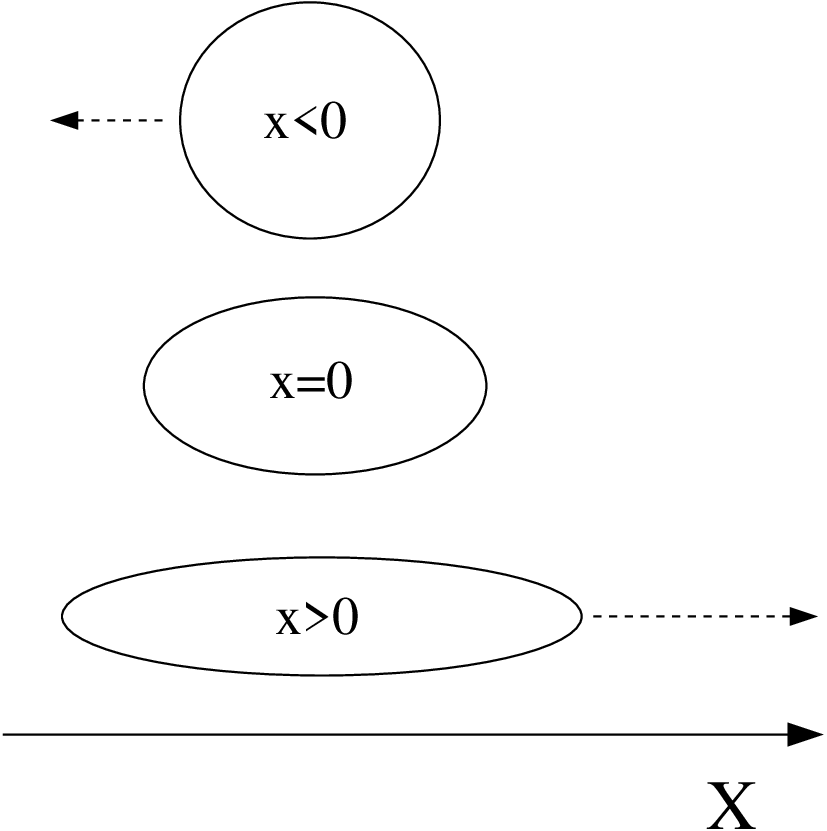}\hspace{5ex}
\includegraphics[width=.49\linewidth]{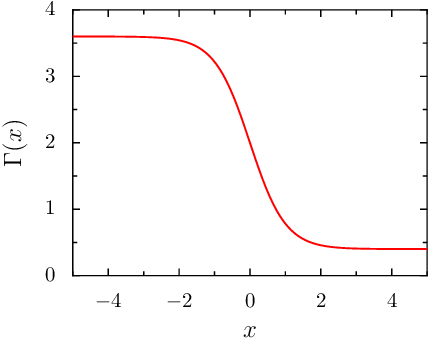}
\caption{{\bf Left panel:} A possible scenario for the dependence of the friction coefficient along the direction of motion $X$ on the internal coordinate $x$ coupled to the propulsion force indicated by the dashed lines. {\bf Right panel:} Nonlinear friction depending on the internal distance of the
  dimer in dependence on the amplitude of oscillations according to the
  KRR model with $\Gamma_0 = 2$ and $C = 0.8$ or $\gamma_h = 3.6,
\gamma_l = 0.4$.}
\label{fig:asy_friction}
\end{figure}

However, as seen by examples in living systems and technical systems,
directed motion is possible also without any imposed external spatial asymmetry.
The world around us is full of creatures like insects, birds and fishes, which are capable to move in a uniform medium by using special mechanisms developed in the process of evolution. Most of these creatures use some kind of periodic motion in order to propel themselves using specialized ``devices'', such as wings or fins.  The corresponding motion of individuals consists of interchanging intervals of acceleration and slowing down, and in a simplified picture we may think of an internal oscillating motor driving this periodic motion.  
Typically the various mechanisms are connected to the head-tail polarity of the animals and to periodic changes of the shape of the individual and the related changes in the friction with respect to the external medium (see e.g. Fig. \ref{fig:asy_friction}).  This leads to a symmetry breaking in the velocity space, leading to asymmetry with respect to forward and backward motion of a polar particle, which allows an effective directed motion even if the oscillatory dynamics of the motor are symmetric. Recently, several models for directed motion have been proposed which employ similar mechanisms of a broken symmetry with respect to an internal degree of freedom \cite{cilla_mirror_2001,norden_ratchet_2001,denisov_particle_2002,kumar_active_2008,barnhart_bipedal_2010}.

We consider now the situation where the energy depot is coupled to such an internal degree of freedom. Here, we assume that it drives the internal oscillatory dynamics in the inchworm model introduced by Kumar, Rao and Ramaswamy (KRR-model) \cite{kumar_active_2008}. This new ansatz eliminates some shortcomings of the original depot model, such as, the absence of periodic accelerations observed in animal motion, or the difficulty to account for the case of large friction (overdamped limit) \cite{romanczuk_active_2011}. Let us explain the latter in more detail: Within the original depot model described in Sec. \ref{sec:det_depot} the dynamics of the energy depot $e(t)$
depend on three constants: $q$ - input rate, $c$ - decay rate and $d$ - rate of transmission to energy of motion, which determine the functioning of the motor mechanism.
From the natural condition $v_0 > 0$, it follows that the depot model works only if
\begin{equation}
\label{workcond}
  \gamma_0 < \frac{q_0 d}{c}.
\end{equation}
This inequality means that the mechanism of the depot model breaks down at large friction values and, in particular, in the overdamped limit $\gamma_0\to\infty$.
However, in many biological examples, such as the motion of microorganism at low Reynolds numbers, this is the relevant limit. Here, we should mention the famous ``Scallop theorem'' formulated Purcell \cite{purcell_life_1977}, which states that a net-displacement of swimmers at low Reynolds numbers is only possible if the symmetry of the motion is broken with respect to time-reversal, which shows the fundamental role of broken symmetries for directed motion \cite{flach_directed_2000}.

Instead of considering particular swimming or propulsion mechanisms in the overdamped case, we discuss here a simple but generic variation of the depot model, which can be easily considered in the respective limit.  

The basic idea of the model is the following:
\begin{enumerate}
\item The energy depot drives oscillations of an internal degree of freedom $x$ (internal motor).
\item The internal oscillations are transformed into translational motion of the active object via a friction function dependent on the internal degree of freedom.
\end{enumerate}

The varying friction can be realized for example in the following way: We consider a two-dimensional ellipse-like object with the axes $a$ and $b$. We assume that the objects moves on a line parallel to the larger axis ($b>a$). Let $c$ be the equilibrium distance between the two focal points of the ellipse and the internal coordinate $x(t)$ describe the temporal deviations of the equilibrium distance. If $x(t)>0$ than the ellipse is more stretched and for $x(t)<0$ it assumes a more spherical shape. Thus a periodic change in $x(t)$ leads to a periodic change in the friction of the object. If further the force leading to a displacement of the center of mass of the object $X(t)$ is coupled to the internal variable $x$ (simplest case: elastic coupling), than directed motion of the object can  be observed despite symmetrical internal oscillations.  

As a model of such a mechanism we consider an object with a large mass $M\gg1$, located at position $X(t)$ which is elastically coupled to an internal degree of freedom $x(t)$ \cite{romanczuk_active_2011}: 
\begin{equation}
\label{Mdyn}
\dot X = V, \qquad  M {\dot V} = - M \Gamma (x) V + k x + Noise.
\end{equation}
Further we assume the internal degree of freedom (the motor) to have the following dynamics driven by the energy depot (see Eq. \ref{langev_depot})
\begin{equation}
\label{mdyn}
\dot x = v, \qquad  \dot{v}  = (d e v - \gamma_0) v - k x  - \omega_0^2 x + noise .
\end{equation}
The dynamics of the internal degree of freedom correspond to an active Brownian particle of unit mass in a confining harmonic potential $U(x)=(k+\omega_0^2)x^2/2$. 
The depot energy observes the standard balance equation
\begin{equation}
\label{depotdyn}
  \dot{e} =  q_0 - c e  + d v^2 e.
\end{equation}

We assume in our model, that the friction acting on the big mass $M$ in the medium depends on the internal motor variable $x(t)$ as in the KRR-model \cite{kumar_active_2008} with
\begin{equation}
\Gamma(x)=\Gamma_0 (1 - C \tanh (B x)).
\end{equation}
The friction is $\Gamma_0$ for $x=0$ and decreases with increasing $x$ up a minimal value $\Gamma_\text{min}=\Gamma_0(1-C)$ for $x\to\infty$. It increases with decreasing $x$ with $\Gamma_\text{max}=\Gamma_0(1+C)$ for $x\to -\infty$.
The internal degree of freedom $x$ corresponds in the original KRR-model to the relative distance of the two masses constituting the dimer. 

We see that the motor plays the role of an effective bridge between the energy depot and the dynamics of the mass $M$.
The above mechanism of self-propulsion, only seemingly violates the principle of mechanics that internal forces do not affect the motion of the center of mass. However, internal forces may affect the external dissipation, and may therefore lead to directed motion \cite{blekhman_vibrational_2000,gerasimov_self-similarity_2002,cilla_mirror_2001,norden_ratchet_2001}.

\begin{figure}
  \begin{center}
    \includegraphics[width=0.7\linewidth]{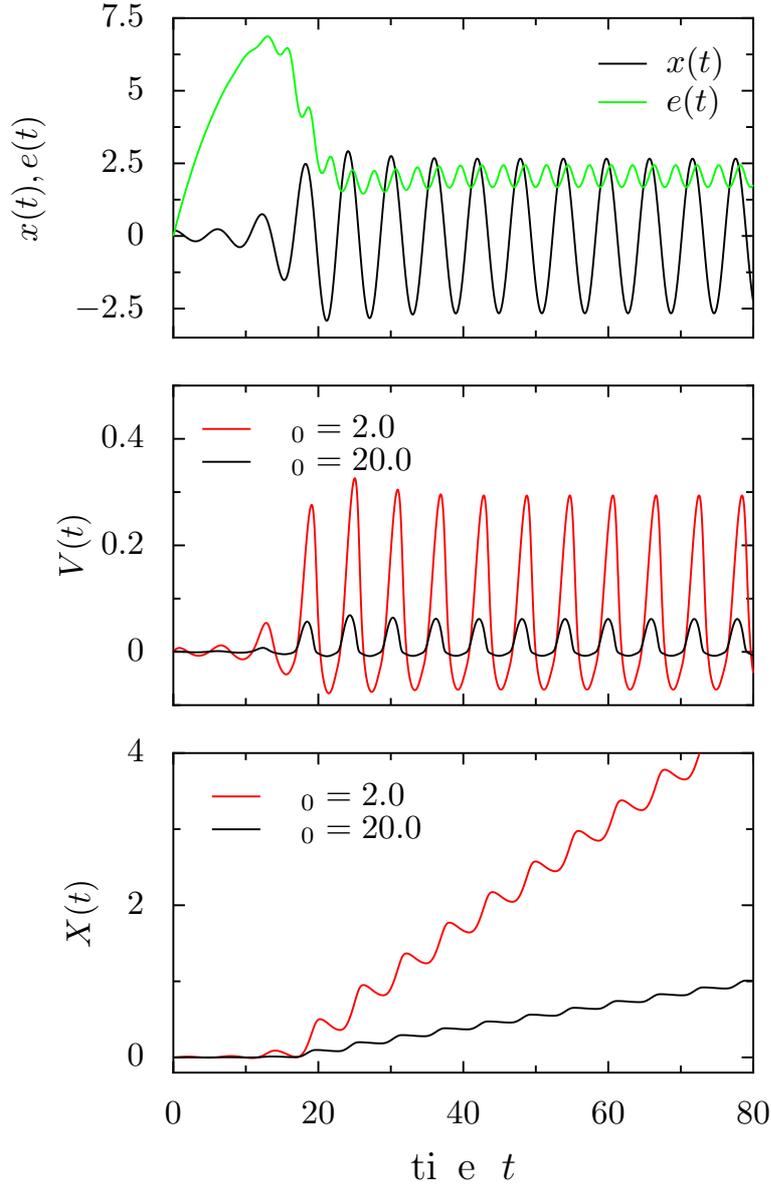}
    \caption{Propulsion via an oscillating internal motor: Internal
      coordinate $x(t)$ and energy depot $e(t)$ (top); external
      velocity $V(t)$ (center) and position $X(t)$ for two different
      friction coefficients $\Gamma_0 = 2$ and $\Gamma_0=20$. Other
      parameters: $\gamma_0 = 0.2$, $k = 1$, $\omega_0 = 0.1$, $M = 10$, $q =
      1$, $d=c=0.1$, $b=1$, $C=0.8$ }
  \end{center}
\label{dyn1}
\end{figure}
In Fig. \ref{dyn1} we shows an example of the temporal dynamics of different model variables. A characteristic feature of the dynamics is the periodic structure of acceleration, velocities and trajectories. Similar structures of the trajectories are quite typical mode of translational motion of animals. For example, the motion of different organisms such as Daphnia \cite{garcia_optimal_2007}, Chlamydomonas \cite{garcia_random_2011,drescher_direct_2010} consist of periodically repeating intervals. Each interval contains at the  beginning a subinterval of strong displacement, corresponding to a large accelerating force and a second subinterval where the displacement is nearly zero. Correspondingly, the velocity is a periodic function which alternates between high and low values. In our model this time structure is based on the action of a periodically working internal motor which generates the periodic accelerations.

Please note that the above model is only valid if the feedback of the time-dependent external friction $\Gamma(x)$ on the motor dynamics can be neglected. This approximation may lead to problems with the energy balance for example at small k-values and big amplitudes of the friction which are neglected here.


\subsection{Particles driven by active fluctuations}
 \label{sec:exter}
Another important difference between ordinary Brownian motion and active
moving objects consists in the possible directions of the
fluctuations and their statistical properties. This will be the central point of this section and we
will distinguish between passive noise terms from external sources acting as random (undirected) forces, 
e.g. the molecular agitation in Brownian motion and active from internal noise sources connected with the active dynamics of the particle such as its propulsion mechanism. 
The latter is connected to the actual state of the particle including its 
direction of motion. In the next Section (\ref{sec:depot}) we introduce a
fluctuating energy supply which will affect the particle's motion in a
similar manner.

We look again at two spatial dimensions and consider polar active particles with the heading unit vector ${\bbox e}_h$ as defined in Sec. \ref{sec:internalcoord}.
The time derivative of the velocity vector yields
\begin{align}\label{eq:polar_dotv}
\dot {\bbox v}=\dot v {\bbox e}_h + v\dot \phi {\bbox e}_\phi
\end{align}  
which gives two Langevin equations for the evolution of the velocity
$v$ and the orientation $\phi$\footnote{Note that the angular
  dynamics diverge for $v=0$. This is due to the fact that we are
  considering point-like polar particles, which for vanishing velocity may turn infinitely fast. This divergence can be eliminated by considering finite
  sized particles.}  :
\begin{align}
\label{eq:gv}
\dot v & = \,-\,\gamma(v) v+\, {\bbox{\cal F}}(t) {\bbox e}_v     \\
\dot \phi & = \,\frac{1}{v} \, {\bbox{\cal F}}(t) {\bbox e}_\phi.
\end{align}

Let us first concentrate on the random force $\bbox{\cal F}(t)$ in
Eqs.(\ref{eq:gv}). Further on, we will distinguish two different types
of fluctuations, which we will refer to as {\em passive} (or {\em external}) 
and {\em active} (or {\em internal}) fluctuations, respectively.  
Passive fluctuations are assumed to have their origin in an 
fluctuating environment in which the particle moves. 
In a homogeneous environment the passive
random force ${\bbox{\cal F}}_\txr{p}(t)$ has to be independent on the
direction of motion. The prominent example of particles subjected to
passive fluctuations is ordinary Brownian motion as presented in
Sec. \ref{sec:brown}, where the random force is associated with the
random collisions of the particle with the molecules of the
surrounding fluid.  Thus, we introduce the external fluctuations in the
same way as we did for Brownian motion in two dimensions, i.e. as a random
noise vector with the components of the vector given by two
uncorrelated, Gaussian white noise terms with the same noise intensity
$D$. The noise vector reads:
\begin{equation}
  \label{eq:Fstoch_p} 
{\bbox{\cal F}}_{p}(t)\,=\sqrt{2D}(\xi_{x}(t){\bbox e}_{x}+\xi_{y}(t){\bbox e}_{y}).
\end{equation}
Here, $\xi_i(t)$ ($i=x,y$) are $\delta$-correlated, normally distributed random
variables with zero mean (see Eq. \ref{eq:noise}, white Gaussian noise):
\begin{align}\label{eq:xi2d}
  \langle \xi_i(t) \rangle \,= \,0 \ , \quad \langle \xi_i(t)
  \xi_j(t')\rangle\, = \,\delta_{ij}\delta(t-t')\,, \,i\,=\,x,y\,..
\end{align}

A second possible model of fluctuations is assumed to have its origin in the internal dynamics of active particles.
This active fluctuations are a pure far-from equilibrium phenomenon and are relevant in the motion of an biological agents or artificial self-propelled particles. The origin of these fluctuations can be for example variations in the propulsion of chemically powered colloids \cite{paxton_catalytic_2004,howse_self-motile_2007,ruckner_chemically_2007}, complex intra-cellular processes in cell motility \cite{selmeczi_cell_2008,boedeker_quantitative_2010} or unresolved internal decision processes in animals \cite{niwa_self-organizing_1994,komin_random_2004,bazazi_nutritional_2010}.
For example, the motion of a macroscopic animal moving in a homogeneous environment, where the fluctuations
due the environment can be assumed as negligible, may nevertheless appear random to an external observer. 
The apparent randomness of the motion stems from internal decisions of the biological agent to
change its direction of motion $\phi$ and/or its velocity $v(t)$ (internal coordinates).
On the other hand, for artificial active particles, fluctuations in a microscopic power engine might occur due to its smallness. They may be associated, for example, with fluctuations in the concentration of fuel molecules driving self-propelled colloids (see e.g. \cite{paxton_catalytic_2004,ruckner_chemically_2007}). 
The important point is that the fluctuating forces are intrinsically connected with the internal propulsion mechanism of the active particle. Thus, the corresponding direction of the active fluctuating forces turn with the particles orientation.

We do not intend to resolve the internal processes, and assume for simplicity fluctuations in the direction of motion and in the velocity of the agent as independent stochastic processes, with possible
different statistical properties (see also \cite{peruani_self-propelled_2007}). 

Here, we make a simple ansatz for the active fluctuations, as
independent Gaussian white noise in the direction of motion ${\bf e}_h$ and in the angular direction ${\bf e}_\phi$:
\begin{eqnarray}
  \label{eq:Fstoch_a} 
  {\bbox {\cal F}}_a (t)\, =\,\sqrt{2 D_v} \xi_{v}(t) {\bf e}_{h} + \sqrt{2 D_\phi} \xi_{\phi}(t) {\bf e}_{\phi}
\end{eqnarray}
with $\xi_i(t)$ as defined in Eq. \ref{eq:xi2d} ($i=v,\phi$). In the general case we have to assume different noise intensities of the active angular and velocity noise: $D_\phi\neq D_v$.

\begin{figure}
\begin{center}
\includegraphics[width=0.6\linewidth]{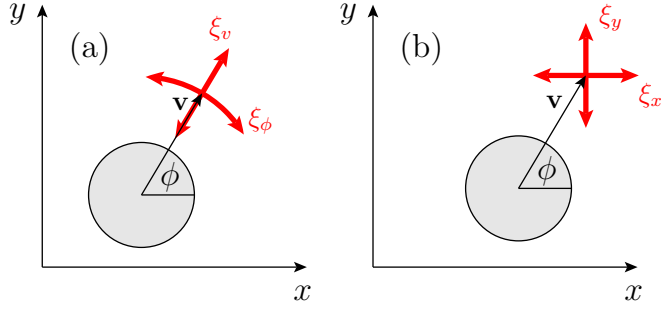} 
\end{center}
\caption{Visualization of the difference of passive (external) and active (internal) fluctuations (thick red/gray arrows). {\bf (a)} Passive fluctuations $\xi_x(t)$ and $\xi_y(t)$ uncorrelated to the direction of motion. 
{\bf (b)} Active fluctuations $\xi_v(t)$ and $\xi_\phi(t)$ (thick red/gray arrows). Parallel and perpendicular to the direction of motion (heading) \label{fig:scheme2dnoise} }
\end{figure}

We mention that smaller objects will experience always molecular
agitation or the stochastic character of the external forces. Therefore
the general situation is the combination of the introduced passive \eqref{eq:Fstoch_p} 
and active \eqref{eq:Fstoch_a} fluctuations. A visualization of the different fluctuation types is shown in Fig. \ref{fig:scheme2dnoise}.

\subsubsection{Quasi-Brownian particles with active fluctuations}
\label{sec:active_stokes}
Before we investigate (stochastically) pumped particles, we sketch shortly the
difference between the two different fluctuating forces by
considerating a simple model with linear friction discussed  in
Sec. \ref{sec:brown} in the context of ordinary Brownian motion.

Let us start with particles which are subject to Stokes friction and noise. 
The case of passive fluctuations (ordinary Brownian motion) was discussed in
Sec. \ref{sec:brown}. The stationary velocity distribution according
to Eq. \ref{eq:maxw} is a two dimensional Gaussian distribution
centered at the origin with width $D/\gamma$. 
Please note that the simple Stokes friction is apolar by definition (symmetric with respect to $v=0$). Thus, 
the heading vector is equivalent to the velocity unit vector ${\bbox e}_h={\bbox e}_v={\bbox v}/|{\bbox v}|$. 

\begin{figure}[t]
  \begin{center}
    \includegraphics[width=0.9\linewidth]{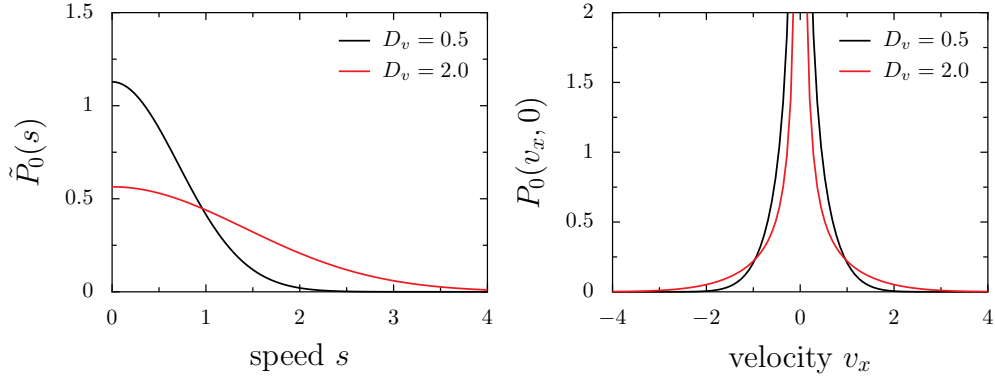}
    \caption{Stationary distributions of a linearly damped
      particle with internal noise for two different noise
      intensities $D_{v}$. Left panel: stationary speed distribution $\tilde p_0(s)$; Right panel: Cross section of the corresponding Cartesian velocity distribution $P_0(v_x,v_y)$
  \label{fig:Brown_inter}}
\end{center}
\end{figure}

The dynamics for purely active noise read: 
\begin{align}\label{eq:brown_w_AF}
  \frac{d}{dt } \bbox{v} \,=\, - \gamma \bbox{v}(t) +\sqrt{2D_v}\,\bbox{e}_{v}(t) \xi_v(t) \,+\, \sqrt{2D_\phi}\,\bbox{e}_{\phi}(t) \xi_{\phi}(t)\,,
\end{align}
where the stochastic sources are defined in Eq. \ref{eq:Fstoch_a}. In $(v,\phi)$-coordinates, we obtain two decoupled differential equations for the velocity and the angle:
\begin{align}
  \frac{d}{dt}{v}\,=\, -\gamma v +\sqrt{2D_v}\, \xi_v(t)\, ~~~~\frac{d}{dt}{\phi}= {1 \over v} \sqrt{2D_\phi}\, \xi_{\phi}(t)\,.
\end{align}
As $t \to \infty $ the velocity distribution becomes stationary
without any preferred direction. It is a Gaussian in the
velocity $v$
\begin{eqnarray}
\label{eq:FPE_inter}
 P_0(v)\,=\,{\cal N} \,exp\left(- {\gamma \,v^2\over 2 D_v}\right)\,,~~~~ P_0(\phi)\,=\, \frac{1}{2\pi}.
\end{eqnarray}
For the speed $s=|v|$ we derive 
\begin{align}
 \tilde  P_0(s)= \frac 12 \Big( P_0(v)\, +\, P_0(-v) \Big) \,=\,{\cal N} \, \exp\left(- {\gamma \,s^2\over 2 D_v}\right),
\end{align}
which in contrast to the Rayleigh distribution \eqref{eq:rayleigh} assigns maximal probability density to the state with zero speed. The reverse 
transformation to Cartesian velocities gives us the corresponding velocity distribution
\begin{eqnarray}
\label{eq:P_cart_act}
P_0(v_x v_y)\, = \, {\cal N}\, {1 \over \sqrt{v_x^2\,+\,v_y^2}}\,\exp\left( - {\gamma \,(v_x^2\,+\,v_x^2)\over 2 D_v}\right) \,,
\end{eqnarray}
which exhibits a singularity at the coordinate origin ($x=0,y=0$). The
distribution is clearly a nonequilibrium one.  The peak at the origin
is due to the many stochastic transitions created by $\xi_v(t)$ which
pass through the state with vanishing velocities. In
Fig. \ref{fig:Brown_inter} two examples of the stationary velocity
distribution are shown.

The most common situation is the presence of both, active and passive noise sources, with different intensities. 
Therefore we consider
\begin{equation}
  \frac{d}{dt}\, \bbox{v}\,=\, -\,\gamma \, \bbox{v}\,+\, {\bbox {\cal F}}_a(t) \,+ \,{\bbox {\cal F}}_p(t)\, 
\end{equation}
where the different noise terms are defined in Eqs. \ref{eq:Fstoch_p} and \ref{eq:Fstoch_a}.
Changing to internal coordinates,  we obtain the Fokker-Planck equation
for $P=P(v,\phi,t|v_0.\phi_0,t_0)$ with the typical $D/v^2$ term
in analogy to polar representation, see (\ref{eq:SG2d_FPEext}),
\begin{align}
  \label{eq:FPE_both}
  \frac{\partial P}{\partial t} \, =\, -\frac{\partial }{\partial
    v}\left\{\left(-\gamma v + \frac{D}{v}\right) P\right\} +
  (D+D_v)\,\frac{\partial^2 P}{\partial v^2}
  +\frac{D+D_\phi}{v^2}\frac{\partial^2 P}{\partial \phi^2}\,.
\end{align}
$D$, $D_v$ and $D_\phi$ are intensities of the different noise
sources: passive and active velocity and angular noise. 
One can solve for the stationary distribution which factorizes
again. Without a preferred direction we get again the uniform
distribution for the angular dependence which is due to increased
noise intensity approached at a faster time scale. The velocity distribution is symmetrical with respect to $v=0$ and reads
\begin{eqnarray}
  \label{eq:FPE_vel_both}
  P_0(v)\,=\,{\cal N}\, |v|^{\frac{D}{D+D_v}}\,\exp\Big\{ -\frac{\gamma}{2(D+D_v)}\,v^2\Big\},
\end{eqnarray}
which for the limiting cases $D \to 0$ and $D_v \to 0$ gives the
known distributions (\ref{eq:FPE_inter}) and (\ref{eq:rayleigh}). Due
to the symmetry the speed distribution corresponds directly to the velocity distribution, with $v\to s = |{\bbox v}|$ and an additional factor $1/2$ in the
normalization constant ${\cal N}$. Finally we obtain the Cartesian velocity distribution
to
\begin{eqnarray}
  \label{eq:FPE_vel_both_cart}
  P_0(v_x, v_y)\,=\,{\cal N}\, (v_x^2+v_y^2)^{\frac{D_v}{2(D+D_v)}}\,\exp\Big\{ -\frac{\gamma}{2(D+D_v)}\,\big(v_x^2+v_y^2\big)\Big\}
\end{eqnarray}
It reduces the Maxwellian velocity distribution for $D_v=0$ and the case
 (\ref{eq:P_cart_act}).

\subsubsection{Active shot noise fluctuations}
Active, or internal, noise as discussed is not restricted to Gaussian white
noise. We assume that an active particle is subjected to white shot noise
as introduced in \cite{van_den_broeck_relation_1983} as a limit of a dichotomous Markov process. The random ``force shots'' occur at exponentially distributed times $t_i$ with the mean interval
$\tau$ or rate $\lambda=1/\tau$ The weights $h_i$, which the delta
spikes in the shot noise are multiplied, are due to the distribution
with $\rho(h)$ with average $\mean{h}$.
\begin{eqnarray}
  \xi_{SN}(t) &=& \sum_i h_i\delta(t-t_i)\,, ~~~\mean{t_{i+1}-t_{i}}=\tau,~~~\mean{h_i}=h\,.
\end{eqnarray}
As the shot noise is the limiting case of a dichotomous Markov process the weights are exponentially distributed. Subsequently the white shot noise has the properties:
\begin{eqnarray}
  \mean{\xi_{SN}(t)}= \frac{h}{\tau},~~~~\mean{\xi_{SN}(t_1)\xi_1(t_2)}-\mean{\xi_{SN}}^2 = \frac{h^2}{\tau}\, \delta(t_1-t_2)\,.
\end{eqnarray}

\begin{figure}
\begin{center}
\includegraphics[width=0.6\linewidth]{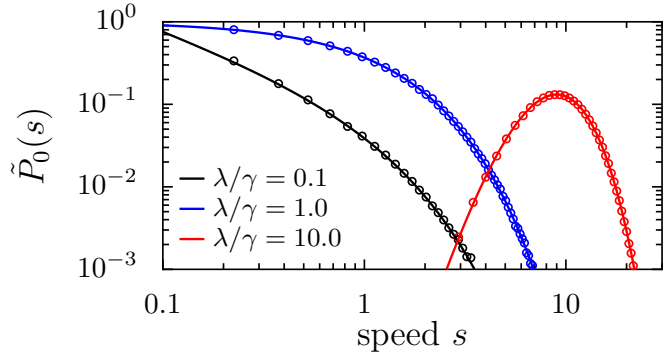}
\end{center}
\caption{Stationary speed distributions ${\tilde p}_0(s)$ for Stokes friction with active shot-noise fluctuations for different values of $\lambda/\gamma=1/(\tau\gamma)$. Solid lines show the result obtained in Eq. \ref{eq:ps_shot}, whereas symbols show numerical results \cite{romanczuk_brownian_2011}. \label{fig:Brown_shot} }
\end{figure}
The internal driving is assumed to act along the heading direction according to
\begin{eqnarray}
  \frac{d}{dt} \,\bbox{v} \, =\,  - \gamma \bbox{v} \,+ \,\bbox{e}_{h}(t) \xi_{SN}(t) \,+\, \sqrt{2D_{\phi}}\,\bbox{e}_{\phi}\, \xi_{\phi}(t)\,.
\end{eqnarray}
The dynamics are not overdamped, i.e. the shots do not affect
the coordinate immediately but act as forces pulses on the velocity. The
angular noise is assumed to be Gaussian and white as defined in
(\ref{eq:noise}).

Distributions of the velocities obtained from simulations are presented in
Fig. \ref{fig:Brown_shot}. For fast appearance of subsequent shots with short
mean interval $\tau$ the velocity distribution is approximately Gaussian around
the mean value of the speed which is $h/\tau$ (note that the velocity scale at the abscissa starts at $v=0$).
It results in a crater-like distribution in the Cartesian frame. Otherwise, in
case of rare shots the particles spend much time at the origin
leading to the peak in the Cartesian distribution. We do not show here
the situation where instantaneously the crater and the peak at zero
exist.

In polar representation with $v_x=v \cos(\phi)$ and
$v_x=v\sin(\phi)$ the equations of motion become
\begin{eqnarray}
  \dot{v}\,=\, -\gamma v\, +\, \xi_1(t)\, ~~~~\dot{\phi}= {1 \over v} \sqrt{2D_2}\, \xi_2(t)\,.
\end{eqnarray}
The equation for the velocity distribution is independent of the angle
dynamics
\begin{align}
  \label{eq:shot_vel}
  \frac{\partial}{\partial t} P(v.t|v_0,t_0) = \frac{\partial}{\partial v} \gamma v P -\lambda \,P(v,t|v_0,t_0) +\lambda\, \int_0^\infty {\rm d}h\,  \rho(h) P(v-h,t|v_0,t_0)   
\end{align}
The velocity is strictly non-negative and the difference to the speed
disappears. In Fig. \ref{fig:Brown_shot} we present results of
computer simulations for three different sets of parameters. Despite its
simplicity the model features typical properties of active motion like directed motion with non-vanishing mean velocity.

In the stationary limit the $\phi$-distribution becomes uniform again. The
$v$-distribution can be calculated taking the shot-noise limit of the
balance equation for the probability density function of a dichotomous Markov process.  
It reads
\begin{eqnarray}\label{eq:ps_shot}
  P^0(v)\,=\,{\cal N}\, v^{\Big(\frac{\lambda}{\gamma} - 1 \Big)}\, \exp \Big(-\frac{v}{h}\Big), ~~~v\,\geq 0 \,.
\end{eqnarray}
Please note that as $v\geq0$ the above velocity distribution corresponds directly to the stationary speed distribution $P_0(v)=\tilde P_0(s)$.

\begin{figure}[tbh]
\begin{center}
 \includegraphics[width=11.5cm,angle=0]{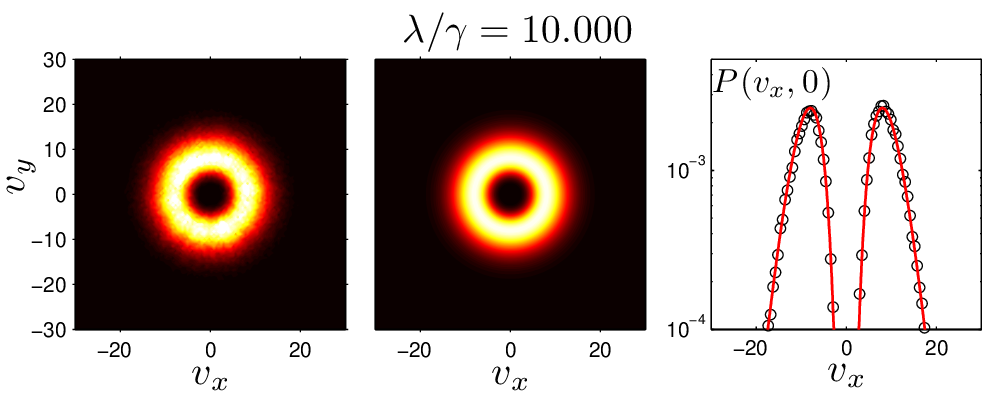}
 \includegraphics[width=11.5cm,angle=0]{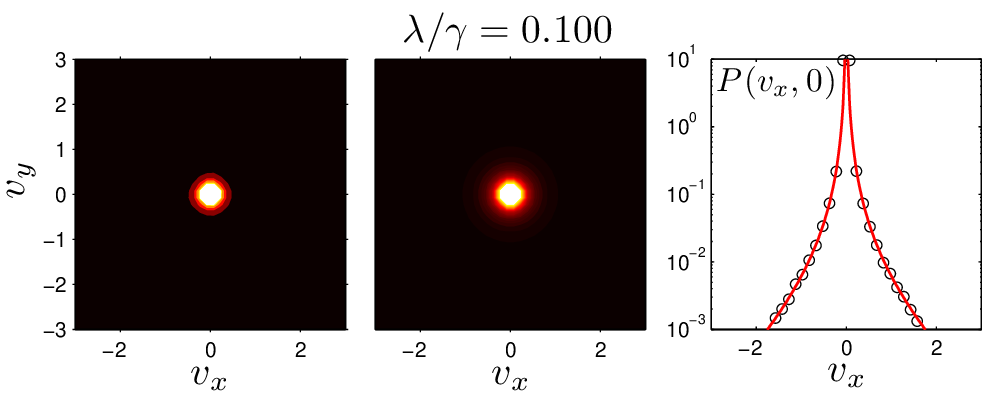}
\end{center}
\label{fig:shot_pvxvy}
\caption{$P_0(v_x,v_y)$ with internal shot noise, top: The
  case of fast occurrences of spikes $\lambda=10$ and damping
  $\gamma=1$, bottom: the case of rare generation of spikes
  $\lambda=0.1$ and damping $\gamma=1$, {Left:} Simulations; {Central:}
  Analytics; { Right:} cross-sections $P_0(v_x,0)$ analytics (solid)
  vs numerics (symbols). other parameter: $a=1$. }
\end{figure}

In Cartesian coordinates the probability distribution becomes sharply peaked at the origin for $1/(\gamma\tau)<2$: 
\begin{eqnarray}
P^0(v_x,v_y)\,=\,{\cal N} \, |{\bbox v}|^{\Big(\frac{1}{\gamma \tau} -2\Big)}\, \exp \Big(-\frac{|{\bbox v}|}{h}\Big), ~~~|{\bbox v}|=\sqrt{v_x^2+v_y^2} .
  \label{shotnoisspeed}
\end{eqnarray}
Examples of the distribution functions obtained from numerical simulations shown in
Fig. (\ref{fig:shot_pvxvy}) confirm our analytical findings. We point out that with constant weights the probability distribution functions obtained from simulations (not shown) differ strongly from the result obtained in Eq. \eqref{shotnoisspeed}.

\noindent

\subsubsection{Nonlinear friction and asymmetric driving}
In the polar representation the shape of the velocity distribution 
does not depend on the angle dynamics, as
$P(v,\phi,t)\,=\,P(v.t)\,P(\phi,t|v)$. On the other hand the angle
dynamics depends on the velocity since the derivative of $\phi(t)$
scales with $1/v$. Therefore a full factorization becomes
only possible in the case of constant velocities or as an approximations. Furthermore the distribution function is isotropic in the long time limit (independent on $\phi$) only in the absence of external forces.

Therefore the consideration of two- and also three dimensional problems
with internal noise acting on the scalar velocity $v(t)$ simplifies
the stationary analysis and many different models become tractable.
Our third example is motivated by vibrational dynamics and deals
with nonlinear friction and an asymmetric driving with vanishing mean. Surprisingly this dynamics may lead to a directed motion of particles resulting in a
non-vanishing mean velocity $v$.

As in vibrational dynamics we will assume nonlinear friction \cite{blekhman_vibrational_2000}
and, in contrast to the shot-noise case, stochastic forces
acting on the particle with a vanishing mean
\begin{eqnarray}
\label{eq:asym}
  \frac{d}{dt} \bbox{v} \,=\, - \,\gamma |\bbox{v}|^{n-1}\,\bbox{v}(t) \,+\,\bbox{e}_v(t) \xi_{DMP}(t) \,+\, \sqrt{2D_{\phi}}\,\bbox{e}_{\phi}(t) \xi_2(t)\,.
\end{eqnarray}
Here, the stochastic term acting on the velocity $\xi_{DMP}(t)$ is assumed to be stochastic force given by a dichotomous Markov process (see Sec. \ref{sec:DMP}), whereas the angular noise is again white and Gaussian.  

Prominent types of friction are Stokes friction ($n=1$) or the ``quadratic
drag force'' ($n=2$) occurring for objects moving at
large velocities through fluids, e.g.  in aerodynamic engineering. Another prominent type of
friction is the so called ``dry friction'' with $n=0$. 

\begin{figure}
  \begin{center}
    \includegraphics[width=0.7\textwidth]{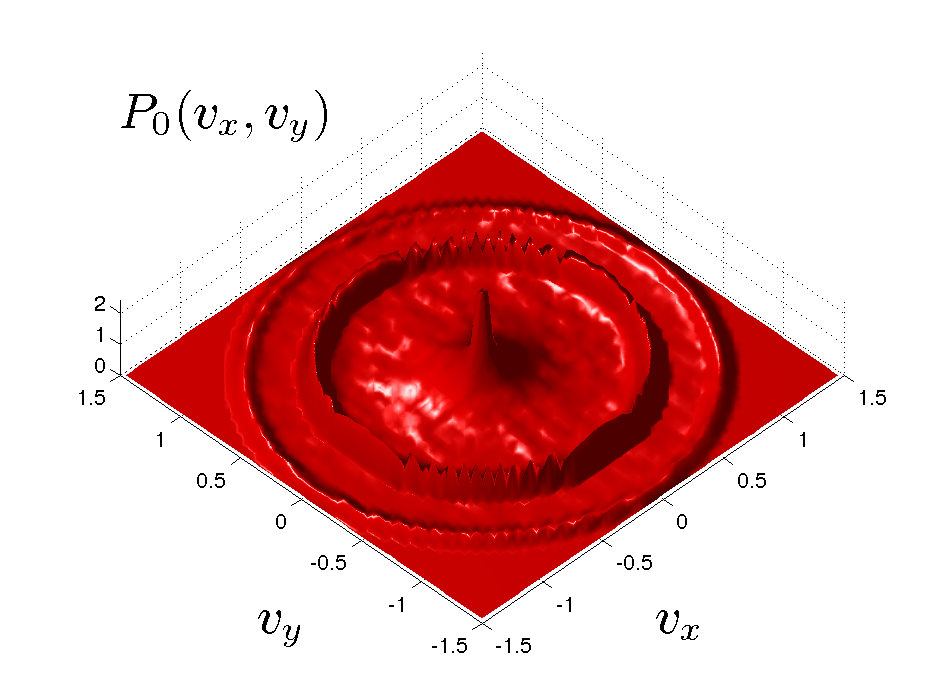}
    \caption{Stationary velocity distributions (Cartesian frame) of
      particle driven by an internal DMP and Gaussian white noise in
      the angle-dynamics obtained from simulations. The ``double-crater'' arise due to the asymmetry of the forward and backward motion.}
    \label{fig:Brown_velo}
  \end{center}
\end{figure}

It was shown by Cebiroglu {\em et. al.} \cite{cebiroglu_rectification_2010}, that for a nonlinear friction the particle starts to move into the direction of ${\bbox e}_h$ even if the time average of the stochastic force vanishes  ($\langle \xi_{DMP}(t) \rangle=0$) as long as $\xi_{DMP}(t)$ is asymmetric. This was considered by Blekhman \cite{blekhman_vibrational_2000} in an
investigation of trajectories, whereas in \cite{cebiroglu_rectification_2010} stochastic methods
have been used to find the vanishing average. The dichotomous Markov process
$\xi_{DMP}(t)$ can assume two values $A_\pm$ and $\lambda_\pm$ are
the rates giving the probability per unit time to leave the corresponding
states. The time average of the applied stochastic force vanishes if
\begin{eqnarray}
  \label{eq:asy_dich}
  \mean{\xi_{DMP}(t)}\,= {\, A_+\,\lambda_- \,+\,A_-\,\lambda_+ \over \lambda_+ \,+\,\lambda_- } \,=\,0\,.
\end{eqnarray}
The asymmetry of $\xi_{DMP}$ is described by he ratio
\begin{eqnarray}
  \label{eq:asymmetry}
  p=|A_-/A_+|, \quad \text{with} \quad 0\leq p \leq 1,
\end{eqnarray}
where we assume, without loss of generality, $|A_-| \le |A_+|$.

Here we present only the stationary velocity probability
density function 
\begin{align}
  P_{st}\left(v\right)&={N  |i_+(v)i_-(v)|}\nonumber\\
  &\times\exp\biggl[-\kappa_+\int^v i_+(v')dv'-\kappa_-\int^v i_-(v')dv'\biggr],\nonumber
\label{pdis22}
\end{align}
where the functions $i_{\pm}(v)$ are defined as
\begin{equation}
 i_{\pm}(v)=\frac{1}{-\gamma|v|^n \textmd{sign}(v)+A_\pm} \quad .\nonumber  
\label{supp1}
\end{equation}
which due to the applied angular noise is independent on $\phi$.

\begin{figure}
  \begin{center}
    \includegraphics[width=0.49\textwidth,angle=270]{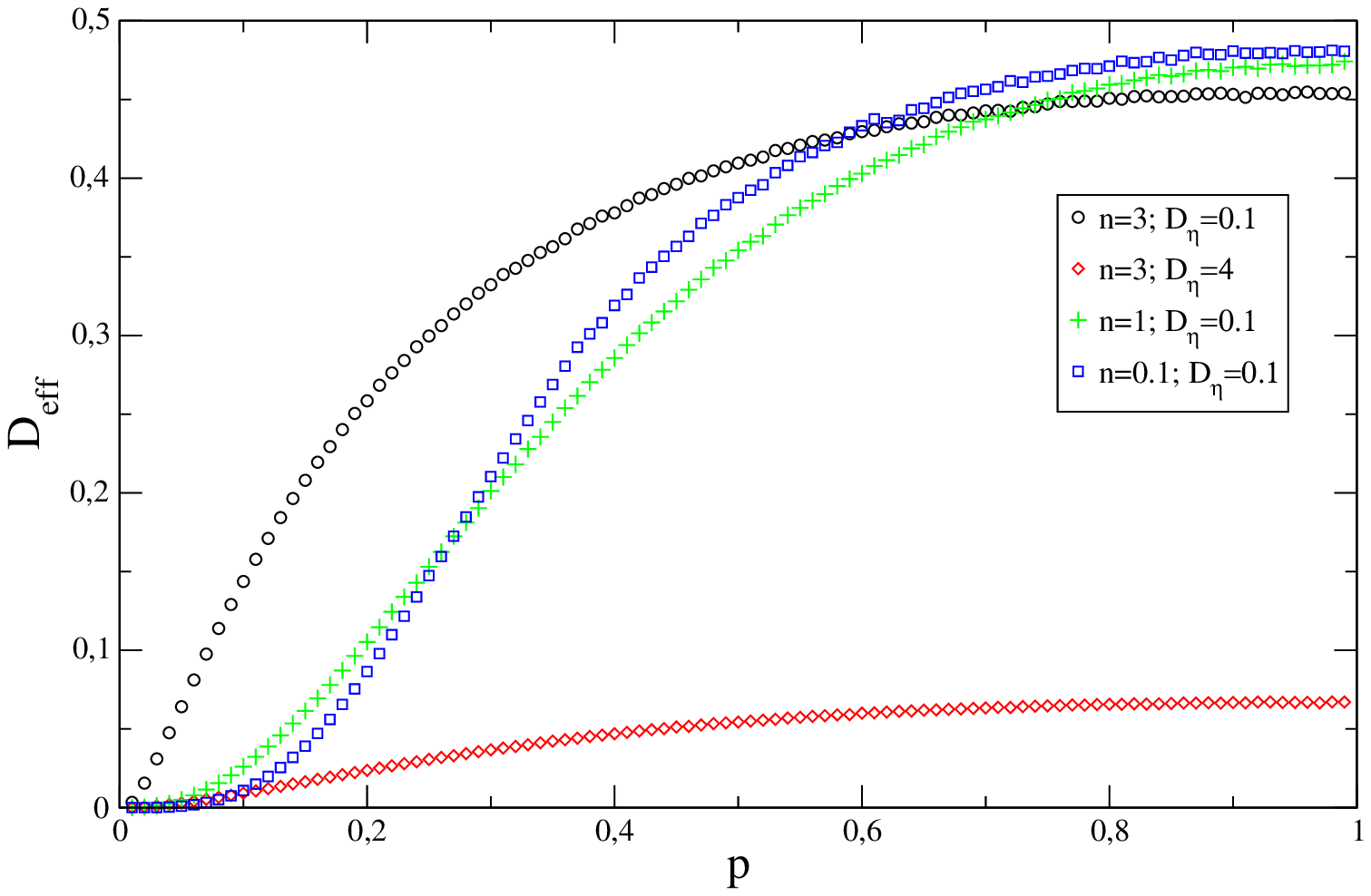}
    \includegraphics[width=0.49\textwidth,angle=270]{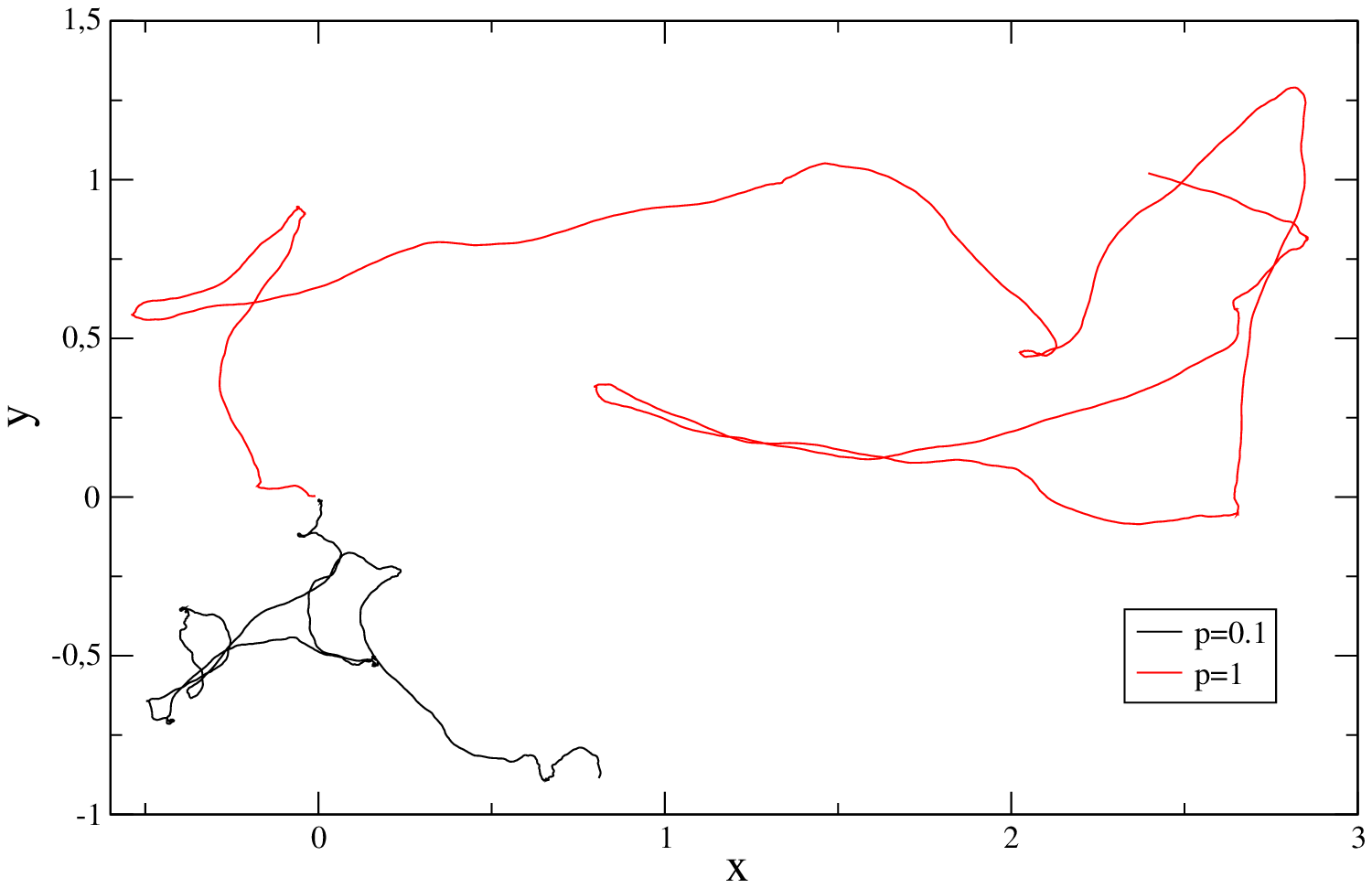}
    \caption{Particles with nonlinear friction subject to DMP fluctuations with vanishing mean. Left: Two paths of particles in two dimensions for different asymmetry of strokes. Right: Diffusion coefficient
      for different exponents in the friction coefficient (\ref{eq:asym}) as function of the asymmetry \cite{cebiroglu_rectification_2010}}
    \label{fig:DMP_nonlin}
  \end{center}
\end{figure}

The velocity distribution mapped
back to Cartesian velocities is shown in Fig. \ref{fig:Brown_velo}
with a central peak at the origin and a ``double-crater'' like distribution. The ring-like maxima of the probability density function correspond to the preferred speed values in any
direction. The peak at the origin stems from the finite probability of vanishing speed, as for particles with Stokes friction with active Gaussian velocity fluctuation (see Sec. \ref{sec:active_stokes}). 

We have simulated \cite{cebiroglu_rectification_2010} the two dimensional situation
\begin{equation}
   \label{eq:veloc}
   \frac{{\rm d}}{{\rm d} t } \bbox{r}(t)=\bbox{v}(t)\,~~\mbox{while}~~  \bbox{v}(t)=v(t)\cdot \bbox{e}_h(t)\quad.
 \end{equation}
 and two paths of realizations are shown in Fig. \ref{fig:DMP_nonlin}.

The particles have a nonzero average speed. The diffusive
motion is becoming stronger if the asymmetry vanishes $p\to 1$, which is
reflected in the diffusion coefficient shown in
Fig.\ref{fig:Brown_shot}. This growth is caused mostly by the variance
of the velocities which scales the diffusion coefficient and which
increases if the strokes becomes more symmetric \cite{cebiroglu_rectification_2010}.

\subsubsection{Fluctuating energy depots}
\label{sec:depot}

Another example where noise depends on the orientation can be
discussed within the depot model.  We so far assumed a permanent food
supply. Here we study the situation that the food supply occurs
randomly and underlies stochastic influences. Doing so, we extend the
model with energy depot (see Sec. \ref{sec:det_depot}) to the case
that the energy is provided at discrete times with packets of chemical
energy which is subsequently converted into acceleration of motion
\cite{strefler_dynamics_2009}.  In contrast to mechanical external noise
which has not preferred direction, such energetic noise is directed
likewise an internal noise and acts into the direction of motion. We
assume, that alterations of the direction have a different origin than
the energetic noise supply.

We propose here an oversimplified model \cite{strefler_dynamics_2009} and
assume that the food supply is a Poissonian shot noise process as it
was introduced in Sec. (\ref{sec:DMP}).  More precisely we assume
that the particles are supplied at discrete times with packets of
chemical energy which is subsequently converted into acceleration of
motion.  This should model the situation that animals, birds, insects
or bacteria pick up nutrients at different, randomly distributed times.
For simplicity we assume that the time intervals between the
food-supplies follow the pattern of a Poisson point process
\cite{masoliver_first-passage_1987,papoulis_probability_1991,caceres_generalized_1997,czernik_rectified_2000,eliazar_nonlinear_2005,kim_numerical_2007,zygadlo_relaxation_1993}.
Possible application to ratchet problems with energy supply
modeled as shot noise imposed on Brownian motion have also been
presented \cite{ebeling_statistical_2008,fiasconaro_tuning_2009,fiasconaro_active_2008}.

We relate a shot-noise driven energy depot with the Brownian dynamics
of individual particles. We adapt the picture in which the energy
depot accelerates the particle along the direction of motion with the
strength $d$
\cite{ebeling_statistical_2008,fiasconaro_active_2008,schweitzer_complex_1998,ebeling_active_1999,strefler_dynamics_2009} as
introduced in Section \ref{sec:det_depot}. The new assumption is that
the carried energy in the depot $e(t)$ now obeys the stochastic
balance equation
\begin{eqnarray}
  \label{eq:dyn}
  \frac{{\rm d}}{{\rm d}t}\, {e}\, = \,  q(t) - (c + d {\bbox v}^2) e.
\end{eqnarray}
Therein the function $q(t)$ corresponds to a shot noise consisting of
energy packets $h_i$ which are exponentially distributed with mean $h$
and arriving at discrete times $t_i$ with rate $\lambda=1/\tau$. The
time average of the shot noise process is denoted as $\mean{q(t)} =
q_0 = h/\tau$. 

In the case of vanishing mechanical noise, we obtain two stationary
solutions for the velocities averaged over the stochastic food supply.
The first is resting with $\mean{\bbox{v}_{1}} \,= \,0,
\mean{e_1}\,=\, q_0/c$ and the second stands for running with
$\mean{\bbox{v}_2^2} \, =\, q_0/\gamma_0 - c/d,~ \mean{e_2} =
\gamma_0/d\,$.  Therefore dependently on the energy supply, the system
exhibits two different regimes.  When the energy input is high enough,
the particle moves at a non-zero velocity. If the energy input is too
small, the energy fluctuates but the velocity remains zero.

\begin{figure}
  \centering
  \includegraphics[width=.49\linewidth]{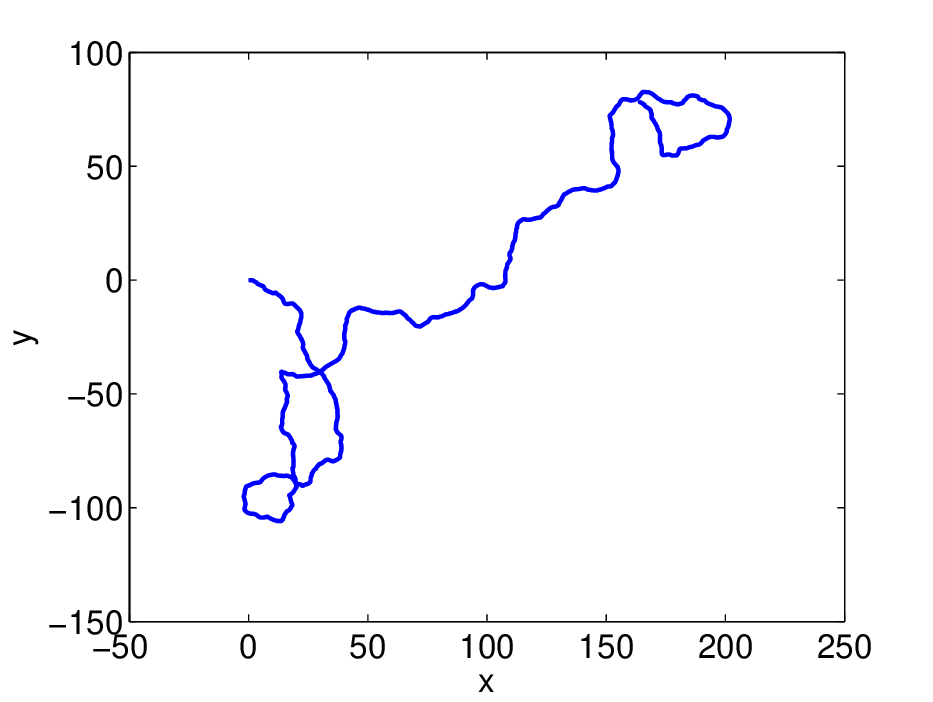}\hfill\includegraphics[width=.49\linewidth]{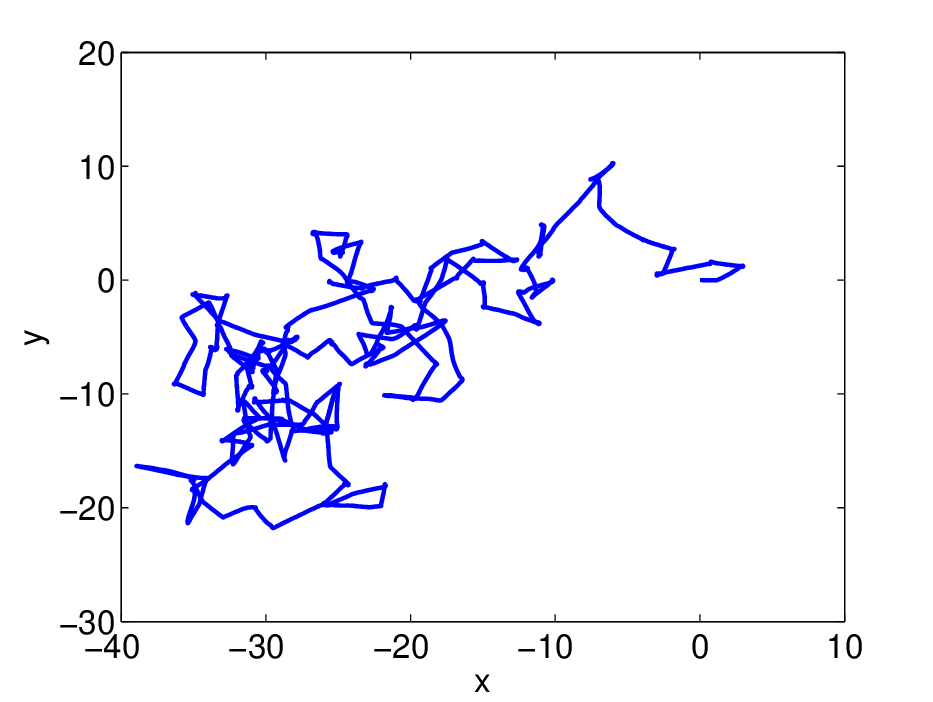}
  \caption{Typical trajectories in the smooth regime (left,
    $\tau=0.05$) and shot regime (right, $\tau=5$). The simulation
    times are equal for both figures, however the length scales are
    different. Other parameter values: $q_0 = 1, D = 0.01,
    \gamma_0=10, c=0.1, d=10$ \cite{strefler_dynamics_2009}.}
\label{fig:trajektorie}
\end{figure}

Further on, we can distinguish two limiting cases \cite{strefler_dynamics_2009}: (i) The smooth
regime: Here the shots are so dense that the mechanical system sees
practically a continuous flow of energy.
This is the case, when the mean time interval between two shots
$\langle t_{shot} \rangle = \tau$ is much shorter than the timescale of the decay of
the energy depot $t_e = 1 / (c+d \bbox{v}^2)$. Now we insert the
stationary solution for the velocity, Eq. (\ref{eq:stat_v0}, DM).  Then we
speak of a smooth regime when the rate of the shots $1/\tau \gg d q_0
/ \gamma_0$. In this situation, the stochastic food supply can be well
approximated by Gaussian noise. (ii) The shot regime: The shots arrive
so seldom that every shot is a special event which accelerates the
particle.  This is the case when the mean
time interval between two shots is much longer than the timescale of
the decay of the energy depot, i.e. the rate $1/\tau \ll d q_0 /
\gamma_0$.

\begin{figure}
  \centering
  \includegraphics[width=.49\linewidth]{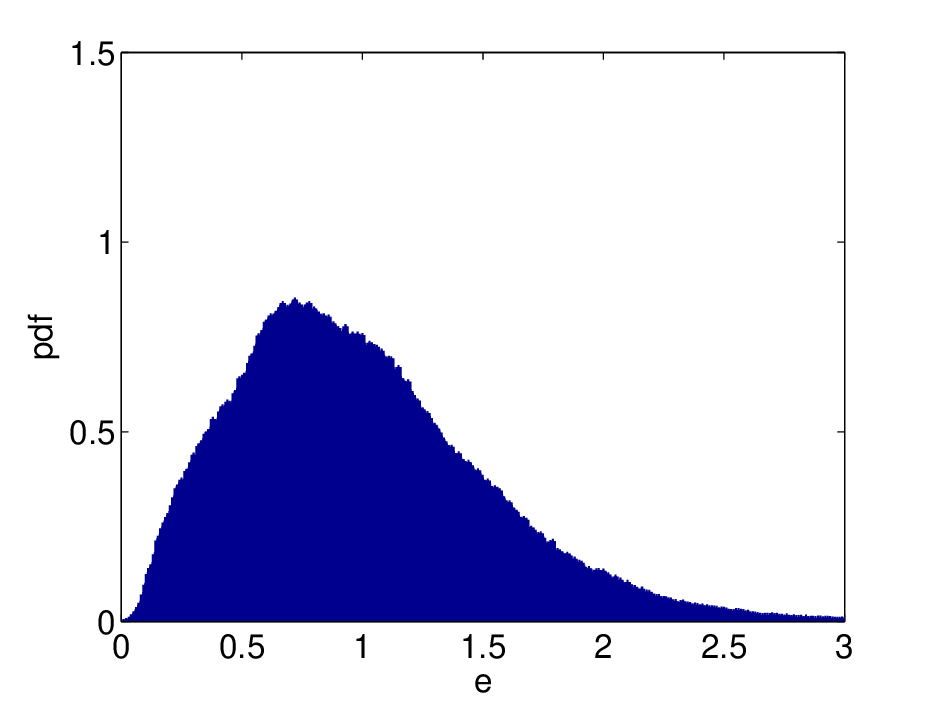}\hfill\includegraphics[width=.49\linewidth]{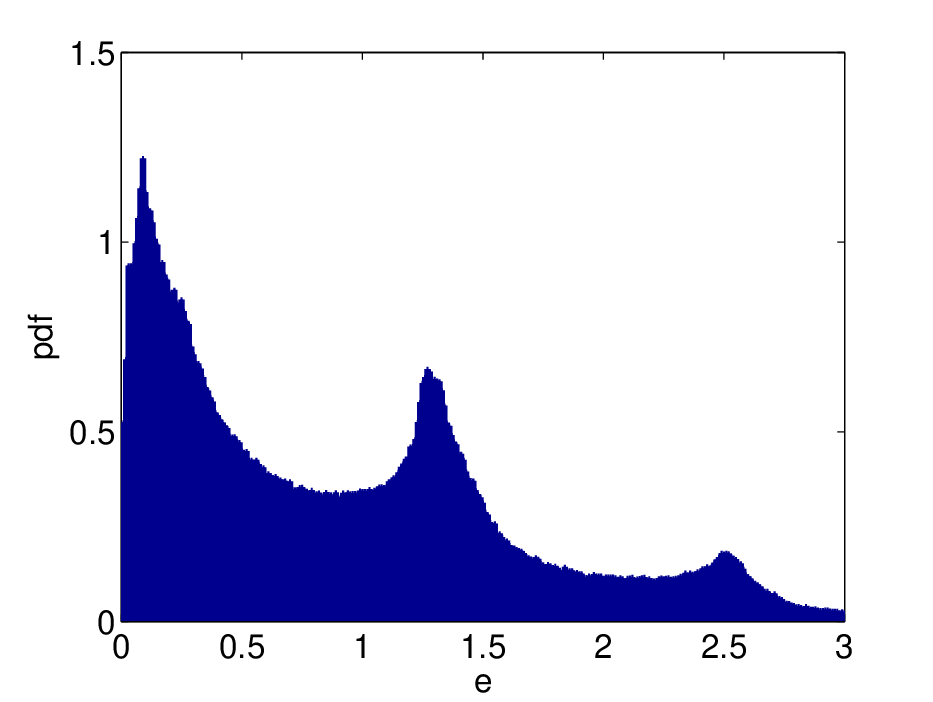}
  \caption{Energy distributions for the smooth regime: (left
    $\tau=0.05$) and shot regime (right, $\tau =5q$).  $q_0 = 10$,
    other parameter values as in Fig. \ref{fig:trajektorie} \cite{strefler_dynamics_2009}.}
  \label{fig:edist}
\end{figure}

Typical trajectories and energy distributions of the smooth and the
shot regime are shown in Fig. \ref{fig:trajektorie} and Fig.
\ref{fig:edist}. Note that the mean energy input is equal for the two
regimes. In the smooth regime, the energy shows a maximum at $e=
\gamma_0 / d$. In the shot regime we observe a maximum at very low
energies and additional smaller maxima at multiples of $h$. The
maximum at low energies occurs because the particle converts the
available energy from the depot to kinetic energy much faster than the
average time between two shots. The maxima at multiples of $h$ occur
when multiple shots of energy arrived before the particle converted
the energy of the depot to kinetic energy and accelerated its motion.

\label{section:exp}
Now we will concentrate on the impact of the shot noise and set the
external Gaussian white noise acting on the velocity $D$ to zero.
Like in Sec. (\ref{sec:DMP}) we will treat the white Poisson shot
noise as a limit of the dichotomous Markovian process (DMP)
\cite{czernik_rectified_2000,zygadlo_relaxation_1993,van_den_broeck_relation_1983,fulinski_non-markovian_1994}.  
This will allow us to derive
an expression for the stationary probability distributions for
velocities and energy.

Because of our shot noise source does not change the direction of the
velocity it acts like an internal noise as introduced previously.
Therefore we may restrict the consideration to the motion along the
direction of motion.  We change to the polar representation and get
\begin{align}
   \dot{s} \,= \,(d e - \gamma_0) s \,,~~~~\dot{e} \,=\, q(t) - e
   \left[c + d s^2\, \right]. \label{eq:v_shot}
\end{align}
Please note that the polar representation is equivalent to the internal coordinates as the velocity can assume only positive values $s(t) = v(t)\geq0$.
In the following we will discuss two cases for which we can specify
the stationary distributions.

(i) Adiabatic approximation: For $q_0 \gg \gamma_0^2 / d$ we use the
adiabatic approximation $\dot{e} \approx 0$ and insert $e(t) = q(t) /
c+d s^2(t)$ into Eq.  (\ref{eq:v_shot}),
\begin{eqnarray}
  \label{eq:fg}
  \dot{s} \,= \,- \,\gamma_0 s \,+ \,\frac{d v}{c + d s^2}\,q(t) \,= \,f(s) \,+ \,g(s)\, q(t).  
\end{eqnarray}
Following the approach that $q(t)$ is a limit of a DMP we obtain
\begin{eqnarray}
  \label{eq:shot_prob_depot}
  \tilde P_0 (s)\, = \,{\cal N} \ s^{\left(\frac{q_0}{\gamma_0 h} - \frac{c}{d h} -
      1\right)} \exp\Big\{-\frac{s^2}{2 h}\Big\}\,.
\end{eqnarray}

(ii) Static regime: For $q_0 < c \gamma_0 / d$ the system is in the
static regime, so that $s = 0$. Therefore the dynamics reduces to
\begin{eqnarray}
  \label{eq:shot_static_depot}
  \dot{e} \,= \,q(t)\, -\, c\,e.
\end{eqnarray}
 we obtain
\begin{eqnarray}
  \label{eq:prob_stat_shot}
  P_0(e)\, = \, {\cal N} e^\frac{1}{\tau\,c-1}\,\exp\Big\{\,-\,\frac{e}{h} \Big\}. 
\end{eqnarray}
The second term of this expression yields a monotonous decreasing for
all parameter values. The first term is monotonously increasing if $c
\tau <1$ and decreasing otherwise. Therefore the probability
distribution of the internal energy shows a maximum in the static
regime when $1/\tau >c$.

In another limit, we consider small mean time interval $\tau$ between
the spikes and low mean amplitudes $h$. We assume that the energy
input contains a constant and a fluctuating part,
\begin{equation}
\label{q(t)}
q(t) = q_0 + \eta(t).
\end{equation}
where $q_0$ is the mean value of the energy input. We approximate the
fluctuating part as Gaussian white noise $\eta(t) = \sqrt{2 D_q} \
\xi_q(t)$. According to \cite{van_den_broeck_relation_1983}, white shot noise with
exponentially distributed weights $h_i$ converges to Gaussian white
noise in the limit $\tau \rightarrow 0, h \rightarrow 0$ with constant
noise strength $D_q = h^2 /\tau = q_0^2 \tau$. The parameter $G$ of
\eqref{eq:nongau} than vanishes.  

If the energy follows changes in the velocity very fast, we can assume
that the energy takes quickly values such that $\dot{e} \approx 0$ in
\eqref{eq:dyn}. We solve \eqref{eq:dyn} for $e(t)$ and insert the
expression into the equation for the speed. Afterwards, we obtain
in Cartesian coordinates an additive external noise and a
multiplicative internal noise caused by the fluctuating energy supply

\begin{figure}
  \centering
  \includegraphics[width=.6\linewidth]{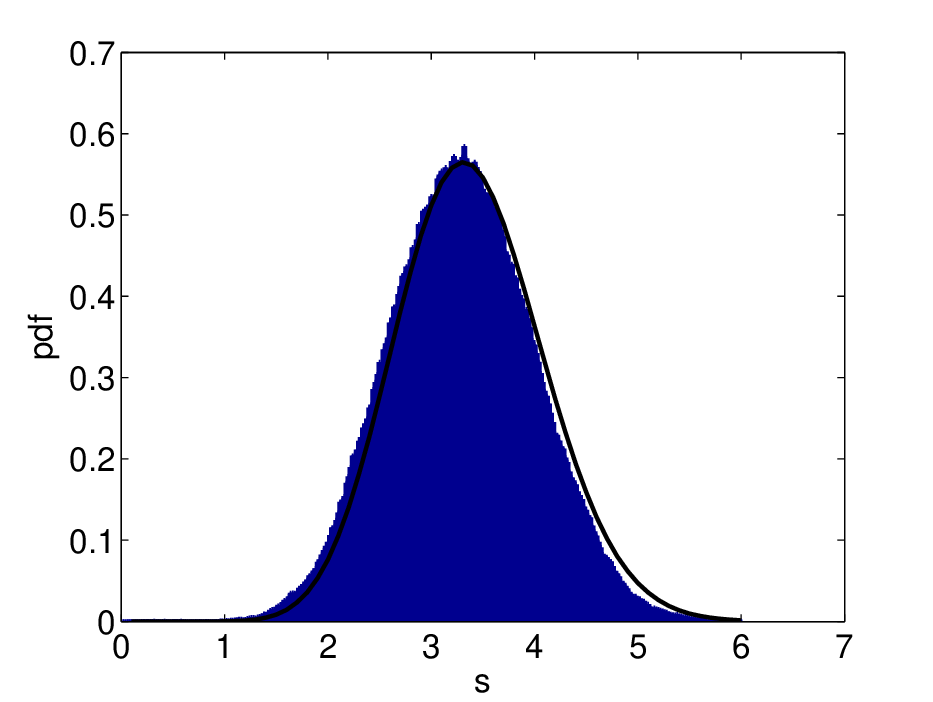}
  \caption{Speed distribution and theoretical estimate (solid line,
    Eq. \ref{eq:solFP}). Parameter values: $\gamma_0 = 0.1, c=0.01,
    d=1, D=0.1, q_0=1, \tau=0.01, D_q=0.01$ \cite{strefler_dynamics_2009}}
\label{fig:vdist_th}
\end{figure}
In polar coordinates we obtain the following equations of motion:
\begin{align}
&  \dot{s} \,=\, \left(\frac{d q_0}{c + d s^2} - \gamma_0 \right) s + \sqrt{2 D}\, \xi_s\, +\,\frac{d s}{c + d s^2} \, \sqrt{2 D_q} \xi_q\,,\label{eq:polar1} \\ 
& \dot{\varphi} \,=\, \frac{\sqrt{2 D}}{s}\,  \xi_{\varphi}(t) \label{eq:polar2}\,. 
\end{align}
Therein $\xi_s(t)$, $\xi_{\varphi}(t)$ are due to
Eqs. (\ref{eq:noise_polar}).

The total noise in both equations consists of two terms, (1) the
standard white mechanical noise, which has no preferred direction, and
(2) the driver noise which is directed pointing to the direction of
the velocity $\bbox{e}_v={\bbox v}/s$.  As a result, the shot-noise affects only the speed.

From the two equations we obtain the corresponding Fokker-Planck
equation and look for the stationary solution. It reads
\begin{eqnarray}\label{eq:solFP}
  P_0(s) \,=\, {\cal N} s\left[D + D_q \left( \frac{d s}{c + ds^2} \right)^2 \right]^{- \frac{1}{2}} \exp[-\Phi(s)]
\label{velo}
\end{eqnarray}
where the potential
\begin{eqnarray}
  \label{eq:FP_depot}
  \Phi(s)\,=\, - \,\int^s\,{\rm d}s'\, \frac{\left( \frac{d q_0}{c
        + ds'^2} - \gamma_0 \right) s' - D_q \left( \frac{d s'}{c +
        ds'^2} \right)^2}{D + D_q \ \left( \frac{d s'}{c + ds'^2}
    \right)^2} 
\end{eqnarray}
In the limits of this approximation the analytical result fits well
the numerical simulations as seen in Fig. \ref{fig:vdist_th}.
\subsection{Active Brownian particles: active versus passive
   fluctuations}
 \label{sec:SG}
Now we consider active Brownian particles with an intrinsic polarity defined by the heading vector ${\bbox e}_h$ (see Sec. \ref{sec:internalcoord}). In case of the Schienbein-Gruler friction (\ref{eq:sg_fri_polar}), the equations of motion
with both passive and active noise (Sec \ref{sec:active_stokes}, Eqs. \ref{eq:Fstoch_p}, \ref{eq:Fstoch_a}) in $v,\phi$-coordinates read
\begin{align}\label{eq:sg2d_w_PF_AF}
  \frac{d}{dt}\,v & = \gamma_0\,( v_0 -v ) +
  \sqrt{2 D}(\xi_x(t)\cos\phi\,+\,\xi_y(t)\sin\phi) + \sqrt{2 D_v}\xi_v(t)\,, \\
  \frac{d}{dt}\,\phi & = \,\frac{1}{v}\left(\sqrt{2
      D}\left(-\xi_x(t)\sin\phi + \xi_y(t)\cos\phi\right)\right)+\sqrt{2 D_\phi}\xi_\phi(t)\,.
\end{align}

The above Langevin equations equations have multiplicative noise terms
and the corresponding Fokker-Planck equation for
$P(v,\phi,t|v_0,\phi_0,t_0)$ again gets the Stratonovich shift
and reads:
\begin{align}
  \frac{\partial P(v,\phi,t|v_0,\phi_0,t_0)}{\partial t}  = & 
  -\frac{\partial }{\partial v}\left\{\gamma_0 \left(v_0-v
    \right)P + \frac{D}{v}P - (D+D_v) \frac{\partial
      P}{\partial v}\right\}  +\frac{D+D_\phi}{v^2}\frac{\partial^2
    P}{\partial \phi^2}
\end{align}

Please note that the angular diffusion given by the last term of the
above equation depends directly on $v(t)$.  In the case of vanishing external forces (${\bbox
  F}_\txr{ext}=0$) there is no distinguished angular direction. The
stationary distribution with respect to $\phi$ is homogeneous and
we may write $P_0(v,\phi)=P_0(v|\phi)/(2 \pi)$ whereas $P_0(v|\phi)$ has to
fulfill the following equation:
\begin{align}
0 = -\frac{\partial}{\partial v}\left\{\left(-\gamma(v)v+\frac{D}{v}\right)P_0-(D+D_v)\frac{\partial}{\partial v} P_0\right\}
\end{align}
By inserting the Schienbein-Gruler friction \eqref{eq:sg_fri_polar} and
solving the above equation we obtain the stationary velocity
distribution for any given heading angle as
\begin{align}\label{eq:SG2d_gen_pvel}
P_0(v|\phi) = \mN |v|^{\frac{D}{D+D_v}} \exp \left\{-\frac{\gamma_0(v-v_0)^2}{2 (D+D_v)}\right\}.
\end{align}
This is the general solution for the probability density function of the velocity with respect to the heading in the presence of both noise types. In the following we discuss the stationary distributions for the two limiting cases of only passive and only active fluctuations. 

\label{sec:SGpassive}
\begin{figure}[p]
\includegraphics[width=0.9\linewidth]{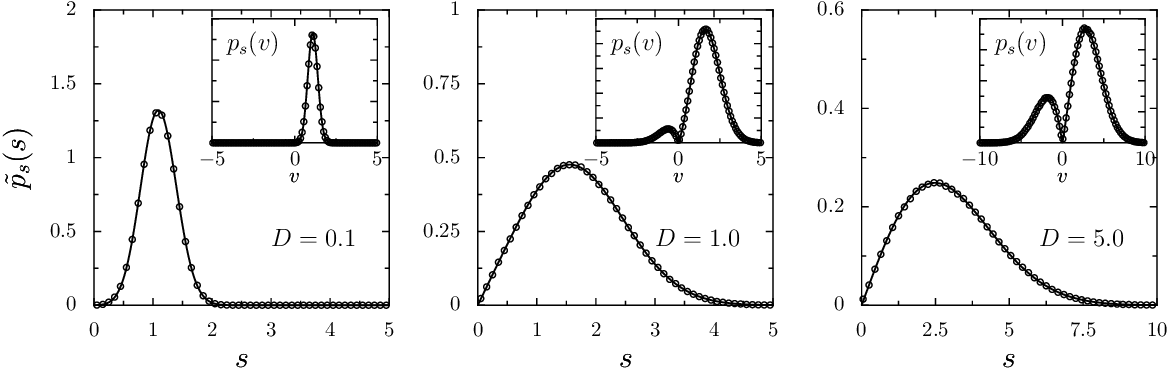}
\caption{Stationary speed distribution $\tilde P_0(s)$ of the Schienbein-Gruler model with external noise for different noise strengths: Analytical solution obtained from \eqref{eq:SG2d_ext_ps} (solid line) and numerical results (symbols). The insets show the corresponding plots for the velocity distributions $P_0(v)$ \eqref{eq:SG2d_ext_pvel}.   Other parameters $\gamma_0=1.0$, $v_0=1.0$ \cite{romanczuk_brownian_2011}. }
\label{fig:SG2d_pvel}
\end{figure}
\begin{figure}[p]
\begin{center}
 \includegraphics[width=0.8\linewidth]{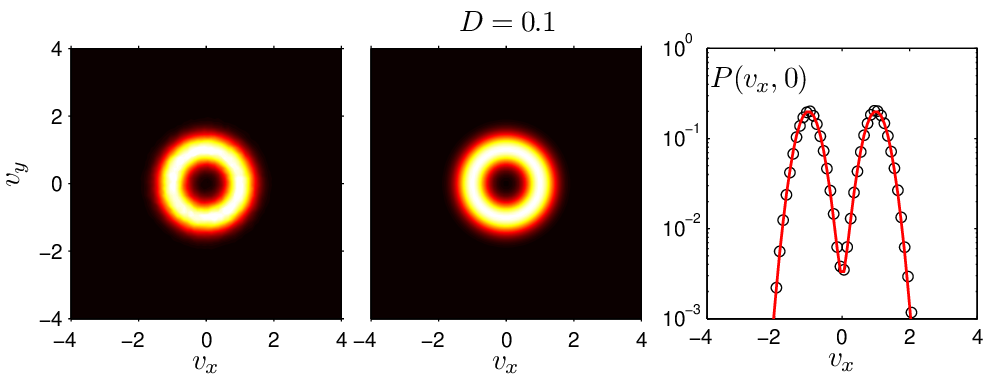}\\
 \includegraphics[width=0.8\linewidth]{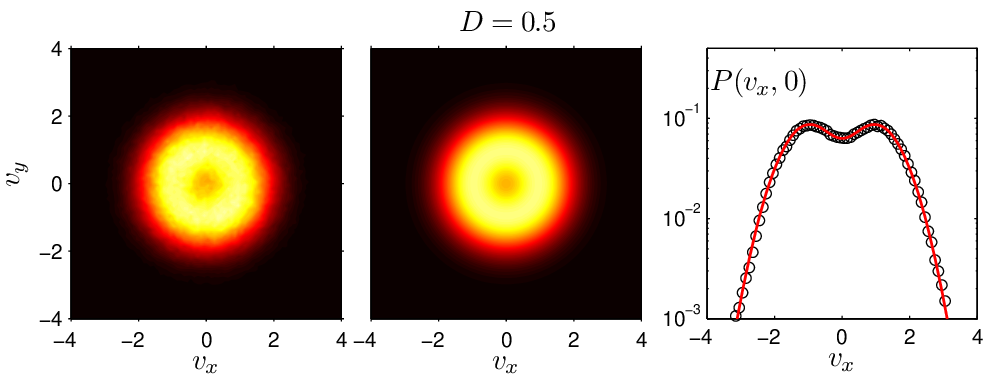}\\
 \includegraphics[width=0.8\linewidth]{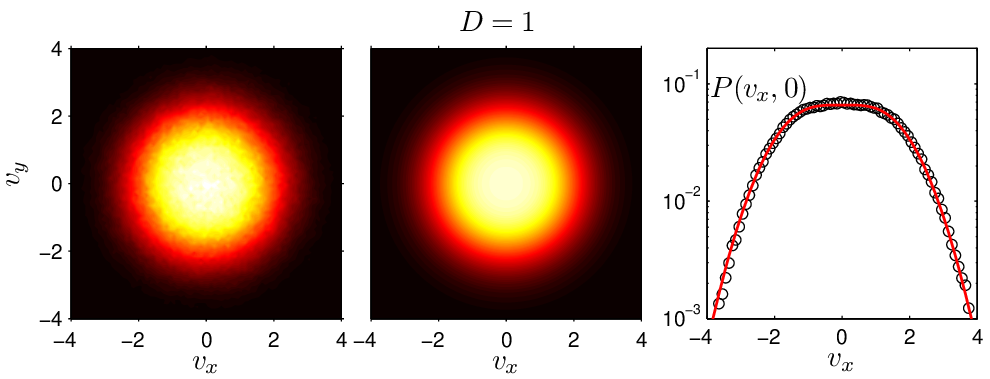}
\end{center}
\caption{Stationary velocity distributions $P_0(v_x,v_y)$ of the SG2d-model with external noise for different noise strengths: $D=0.1$, top; $D=0.5$, center; $D=1.0$. {\bf Left column:} Results obtained from Langevin simulations; {\bf Central column:} Analytical solution given in Eq.~\eqref{eq:SGhead_distro2d}; {\bf Right column:} One dimensional cross-sections $P_0(v_x,0)$ comparing analytical solutions  (solid lines) with numerics (symbols); Other parameters $\gamma_0=1.0, v_0=1.0$ \cite{romanczuk_brownian_2011}.}
\label{fig:SG2d_pvxvy}
\end{figure}
The stationary velocity density for only passive fluctuations can be directly obtained from \eqref{eq:SG2d_gen_pvel} by setting $D_v=0$ to
\begin{align}\label{eq:SG2d_ext_pvel}
P_0(v|\phi) = \mN_p |v| \exp \left\{-\frac{\gamma_0(v-v_0)^2}{2 D}\right\}.
\end{align}
The inverse normalization constant in this case can be calculated using $\int_{-\infty}^{\infty} P(v|\phi) dv=1$ to:
\begin{align}
\mN_p^{-1} = \frac{2 D}{\gamma_0}\text{exp}\left(-\frac{\gamma_0 v_0^2}{2 D}\right)+\frac{\sqrt{2\pi D}}{\gamma_0}v_0\text{Erf}\left(\sqrt{\frac{\gamma_0 v_0^2}{2 D} }\right).
\end{align}
This result is confirmed by the velocity
distribution obtained from numerical simulations of the
Schienbein-Gruler model in two dimensions (SG2d-model) with only external
fluctuations as shown in Fig. \ref{fig:SG2d_pvel}. At vanishing noise
intensities $D/\gamma_0\to0$ the distribution converges towards a
$\delta$-peak at $v_0$. With increasing noise intensity it is
approximately given by a narrow Gaussian around $v_0$. With a further
increase in $D$ clear deviations from the Gaussian distribution become
evident by the appearance of a second maximum of the probability
density at negative velocities, caused by backwards motion of the
particle with respect to its heading. The action of the external
fluctuation leads to a vanishing probability of $v=0$. Finally at
$D/\gamma_0$ the distribution approaches a symmetric distribution which
corresponds to the Rayleigh-distribution mirrored along $v=0$.

In polar representation $v\to|v|=s$ the corresponding speed distribution along any given
angle $\phi$ is a symmetric superposition of particles moving forward and backward with respect to the corresponding heading direction:
\begin{align}\label{eq:SG2d_ext_ps}
  {\tilde P}_0 (s,\varphi )\,& =\, \frac 12 \Big( P_0(v,\phi)\,+\, P_0(-v,\phi) \Big) \nonumber \\
  & = \,\mA \, \mN_p\,s\,\exp \Big\{-\frac{\gamma_0\,(s-v_0)^2}{2D_v}\Big\}
  \left[ 1+ \exp\Big\{-\frac{2\gamma_0 s
      v_0}{D_v}\Big\}\right],
\end{align}
Please note  that ${\tilde P}_0(s,\varphi)$ differs from $P_0(v,\phi)$ as it is only defined for positive speed values.
The constant factor $\mA=(2\pi)^{-1}$ is determined by the normalization with respect to the angular variable. 

In analogy to $P_0(v)$ for $D/\gamma_0\to\infty$ ($v_0\to0$) the
distribution converges to the Rayleigh-distribution of
ordinary Brownian motion (see Eq. \ref{eq:rayleigh}), whereas for
$D/\gamma_0\to0$ the limiting distribution is a
$\delta$-distribution at $v_0$. Finally we obtain the stationary
distribution in Cartesian coordinates $P_0(v_x,v_y)$ from the
stationary distribution in polar coordinates $\tilde
P_0(s,\phi)\,= \,\tilde P_0(s)/(2\pi)$ by
corresponding coordinate transformation and we obtain Eq. (\ref{eq:SGhead_distro2d}).  Examples of $P_0(v_x,v_y)$ are shown Fig.
\ref{fig:SG2d_pvxvy}.
\begin{figure}[p]
\includegraphics[width=0.9\linewidth]{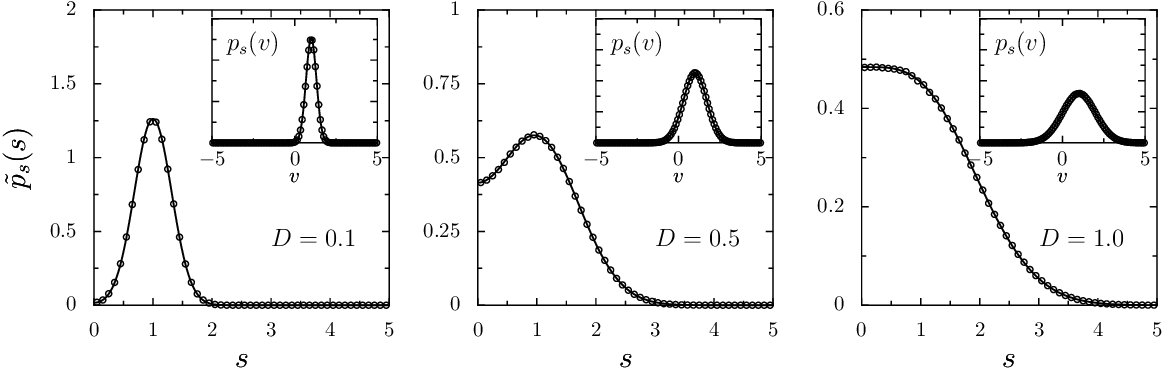}
\caption{Stationary speed distribution $\tilde P_0(s)$ of the Schienbein-Gruler model with internal noise for different noise strengths: Analytical solution obtained from \eqref{eq:SG2d_int_ps} (solid line) and numerical results (symbols). The inset shows the corresponding velocity distribution $P_0(v)$ given by a Gaussian \eqref{eq:SGveldis}. Other parameters $\gamma_0=1.0$, $v_0=1.0$ \cite{romanczuk_brownian_2011}. }
\label{fig:SG2d_int_ps}
\end{figure}
\begin{figure}[p]
\begin{center}
 \includegraphics[width=0.78\linewidth]{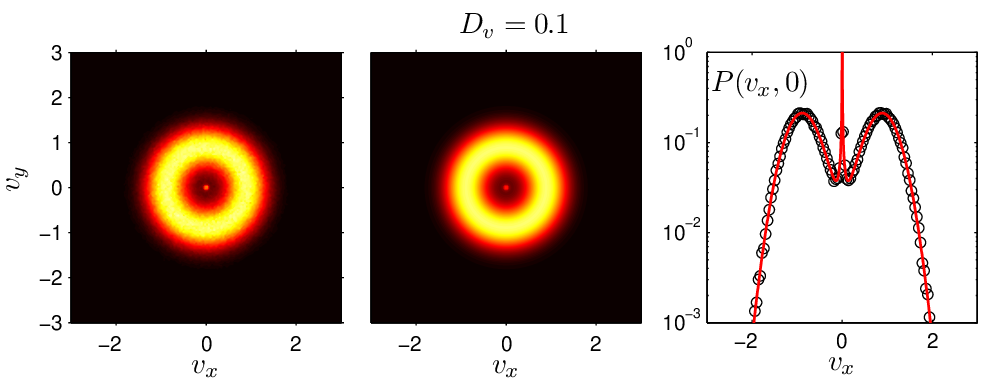}\\
 \includegraphics[width=0.78\linewidth]{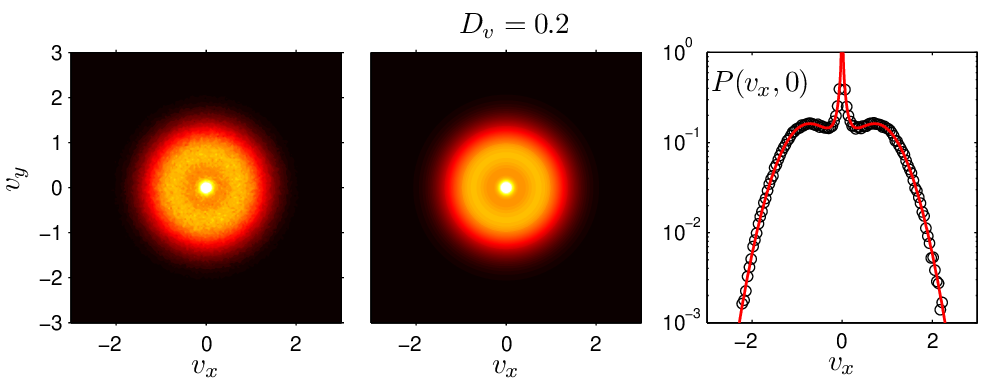}\\
 \includegraphics[width=0.78\linewidth]{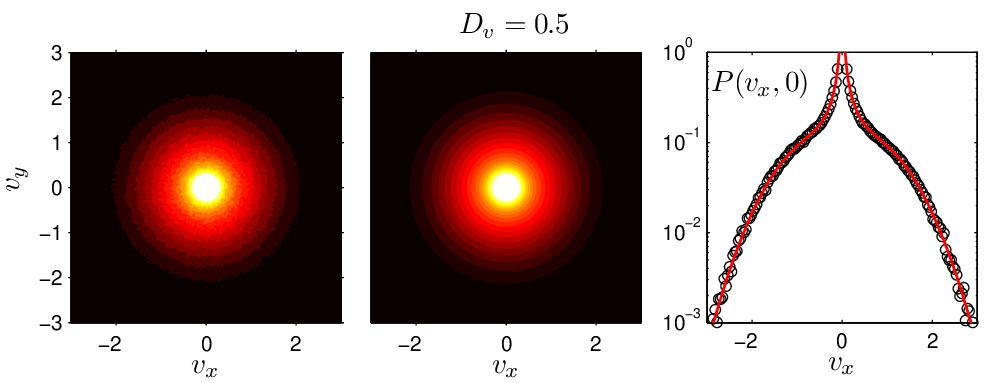}
\end{center}
\caption{Stationary velocity distributions
  $P_0(v_x,v_y)$ of the Schienebein-Gruler model with internal noise for different
  noise strengths:Top $D=0.1$, bottom: D=0.2. {\bf Left column:}
  Results obtained from Langevin simulations; {\bf Central column:}
  Analytical solution given in Eq.~\ref{eq:sg2d_int_pvxvy}; {\bf Right
    column:} One dimensional cross-sections $P_0(v_x,0)$ comparing
  analytical solutions (solid lines) with numerics (symbols); Other
  parameters $\gamma_0=1.0, v_0=1.0$ \cite{romanczuk_brownian_2011}.}
\label{fig:SG2d_int_pvxvy}
\end{figure}

For only active fluctuations ($D=0$) the stationary velocity probability density \eqref{eq:SG2d_gen_pvel} becomes a Gaussian centered at $v_0$ with width $D_v$:
\begin{align}\label{eq:SGveldis}
  P_0(v|\phi) & = \mN_a \exp\Big\{-\frac{\gamma_0
    (v-v_0)^2}{2D_v}\Big\},
\end{align}
with $\mN_a=\sqrt{\gamma_0/2\pi D_v}$. The normalized probability density function in polar coordinates reads
\begin{align}\label{eq:SG2d_int_ps}
   {\tilde P}_0 (s,\varphi )\, & = \frac{1}{(2\pi)^{3/2}}\,\sqrt{\frac{\gamma_0}{D_v}}\,\exp \Big\{-\frac{\gamma_0\,(s-v_0)^2}{2D_v}\Big\}
  \left[ 1+ \exp\Big\{-\frac{2\gamma_0 s
      v_0}{D_v}\Big\}\right].
\end{align}

We emphasize the non-vanishing probability density at $s$ and the
corresponding absence of an increase of ${\tilde P}_0$ for small
$s$ as shown in Fig. \ref{fig:SG2d_int_ps}. This atypical
behavior indicates that there is no limit where ${\tilde
  P}_0(s)$ converges towards the Rayleigh-distribution.

The probability density in the Cartesian velocity coordinates
$v_x,v_y$ can be directly obtained through the corresponding coordinate transformation 
to 
\begin{align}\label{eq:sg2d_int_pvxvy}
  P_0(\bbox{v})=\frac{1}{(2\pi)^{3/2}}\sqrt{\frac{\gamma_0}{D_v}}
  \frac{1}{|\bbox{v}|}\left[ 1+ \exp \Big\{-\frac{2\gamma_0 |\bbox{v}|
      v_0}{D_v}\Big\}\right] \exp
  \Big\{-\frac{\gamma_0(|\bbox{v}|-v_0)^2}{2D_v}\Big\}
\end{align}
with $|\bbox{v}|=\sqrt{v_x^2+v_y^2}$. Please note that the velocity probability density function in Cartesian
coordinates diverges for $v_x=0$, $v_y=0$ and as a consequence exhibits a sharp peak close to the origin as shown in
Fig~\ref{fig:SG2d_int_pvxvy}.


\subsection{Particles with Internal Motor Control}
\label{sec:motorcontrol}
\subsubsection{Propulsion of  particles in arbitrary direction}
Here we discuss briefly several concepts describing the
two-dimensional dynamics of particles with motor, where the motor is
included in the dynamical description. We note at first that polar
particles have an internal direction. It was defined in Section
\ref{sec:act} by the tail-head structure expressing the spatial
structure of the particle. In a similar way, one could speak about
polarity in the presence of a dipole moment, magnetic moments or other
features playing a role in the interaction of particles, or in the
dynamical response to external forces. This may introduce a polar
head-tail or left-hand asymmetry and define a distinguished direction characterizing the particle.

We have connected this polarity axis with the direction of motion
${\bbox e}_v(t)$. It corresponds for positive velocities to the heading
vector ${\bbox e}_h(t)$, whereas for negative velocities with respect to the heading both vectors are anti-parallel. It is a situation similar to a
moving ship where the positive and negative directions of velocity are
parallel to the bottom of the ship from the bow to the rear, but the
velocity points differently to the heading for a ship moving ``backwards''. Both vectors have associated angular unit vectors, we recall that ${\bbox
  e}_\phi(t)$ describe rotations of the heading and ${\bbox
  e}_\varphi(t)$ rotations of the velocity, respectively.

So far in most cases, we also put the non-random propulsion parallel
to the velocity (Exceptions are torques which act perpendicularly). This parallelism now will be dropped, in contrast to
the previous sections, we will now assume that there exist no strict
and fixed relation between the velocity (or polarity) and the
propulsive direction. We will explicitly allow {for a} propulsion
mechanism that acts in a direction which is different from the
direction of motion (see e.g. \cite{szabo_phase_2006,enculescu_active_2011}). 

We define a new direction, which we will connect with the action of
the propulsive motor. Let ${\bbox a}_p(t)$ the acceleration associated
to the propulsive force, which an active particle experiences from the
action of an internal motor. The associated unit vector is ${\bbox
  e}_p(t)$ with direction cosines $\{\cos \chi(t), \sin \chi(t)\}$.
{Here}, $\chi(t)$ is the angle between the axis of abscissas and the
considered force.  We also introduce the magnitude of acceleration
$a_p(t)$ by
\begin{eqnarray}
  \label{eq:acceler}
  {\bbox a}_p(t)\,=\,a_p(t)\, {\bbox e}_p(t)\,.
\end{eqnarray}
{An active particle with unit mass ($m=1$) obeys the dynamical equations:}
\begin{eqnarray}
  \label{eq:dyn_prop1}
  \frac{d {\bbox r}}{d t}\,=\,{\bbox v}\,~~~~~\frac{d {\bbox v}}{d t}\,=\, {\bbox a}_p (t) \, - \gamma_0 \, {\bbox{v}} \,    
  +\,{\bbox F}_{\mbox{ext}} \,+ \,\sqrt{2 D}\,{\bbox \xi}(t) \,.
\end{eqnarray}
Here the propulsion acts in direction of the unit vector ${\bbox
  e}_p(t)$. Furthermore, the particle is driven by external forces $
{\bbox F}_{\mbox{ext}}$, by linear Stokes friction and by noisy
agitations, which might be passive or active. The treatment of the
latter was described in the previous sections.

Because propulsion and velocity vectors are not parallel anymore, the
vector ${\bbox e}_h(t)$ and its associated angle $\phi$ are not
uniquely determined as they were in the previous sections. We define
in this section that the particle starts at time $t=0$ always with
positive velocity and that in this case ${\bbox e}_h(0)={\bbox
  e}_v(0)$ independent of the acting force. By the temporal evolution
no further ambiguity occurs.

We expand the propulsive acceleration again along the ${\bbox
  e}_{h}(t)$ and the perpendicular unit vector ${\bbox e}_{\phi}(t)$.
Consequently, we formulate
\begin{eqnarray}
  \label{eq:propuls}
  {\bbox a}_p (t)\,=\, a_h(t) \, {\bbox e}_{h}(t)\,+\, a_\phi(t)\,{\bbox e}_{\phi}(t)\,. 
\end{eqnarray}
Then, without noise and if the external force acts along the $x$-axis
${\bbox F}_{\mbox{ext}}\, =\,F_0\,{\bbox e}_{x}$, it follows for the
dynamics of the particle
\begin{eqnarray}
  \label{eq:dyn_prop2}
    \frac{d {v}}{d t}\,=\,{a_h}(t)\,-\gamma_0 \, v\, +\, F_0 \cos \phi(t)\,  ~~~~~v\, \frac{d {\phi}}{d t}\,=\, { a}_{\phi} (t) \, - \, F_0 \sin \phi(t) \,.
\end{eqnarray}
where $v(t)$ is again the projection of the velocity-vector ${\bbox
  v}(t)$ on the heading axis ${\bbox e}_h(t)$, which is defined by the
actual angle $\phi(t)$. If we insert explicitly the direction of the
propulsion, the corresponding dynamics becomes
\begin{eqnarray}
  \label{eq:dyn_prop3}
  \frac{d {v}}{d t}\,=\,{a}_p(t) \cos (\chi - \phi )\,-\gamma_0 \, v\, +\, F_0 \cos \phi(t)\,  ~~~~~v\,\frac{d {\phi}}{d t}\,=\, { a}_p (t) \sin (\chi-\phi)\, - \, F_0 \sin \phi(t) \,.
\end{eqnarray}
For the simplest case of a constant propulsion ${\bbox
  a}_p(t)\,=\,{\bbox a}_0 $ pointing along a stationary angle $\chi_0$
and without external force $F_0=0$, the angle $\phi(t)$ turns into the
direction of the acceleration. Asymptotically, we get
\begin{eqnarray}
  \label{eq:asym_a}
  \phi(t\to \infty)\,=\,\phi_0 \to \, \chi_0\, ,~~~~~~ v(t \to \infty) \,\to\,
  a_0/\gamma_0 \,.  
\end{eqnarray}
Therefore, we find that the particle behaves asymptotically like it
would follow a Schienbein-Gruler propulsion, i.e.  the propulsion
coincides with the direction of motion and in the simple case in which
we obtained a stationary speed.

For a non-vanishing external force $F_0\ne 0$, the particle is pulled
in direction of the $x$-axis. If now the propulsion has constant
magnitude $a_0$ and direction $\chi_0$ the stationary solution can be
found as well. The asymptotic angle of direction $\phi_0$ obeys
\begin{eqnarray}
  \label{eq:direc_a_F} 
  \tan \phi_0\,=\,\frac{a_p \, \sin \chi_0}{F_0\,+\,a_p \,\cos \chi_0} 
\end{eqnarray}
which follows also easily from a geometric construction in response {to}
the action of two constant forces. In result, the stationary
direction of motion is always sensing between the propulsive force and
the $x$-axis. For the magnitude of the stationary speed, one obtains
\begin{eqnarray}
  \label{eq:vel_a_F}
  v_0\,=\,\frac{1}{\gamma_0}\, \sqrt{F_0^2\,+\,a_p^2 + 2\,a_p\,F_0 \cos \chi_0} 
\end{eqnarray}
which behaves as the value of a scalar product.

{Within our more general framework, we briefly summarize the
propulsion functions  which we have used so far or will 
use later:} 
\begin{itemize}
\item Schienbein and Gruler \cite{schienbein_langevin_1993} have found empirically 
the following simple "ansatz"  for the cell dynamics \eqref{eq:sg_fri_polar}
  \begin{eqnarray}
  \label{eq:sg_prop}
  a_h\,=\gamma_0 v_0\ \,\,~~~~a_{\phi}\,=\,0\,.
\end{eqnarray}
The polarity axes coincides with directions of motion. In
perpendicular direction normally angular noise is applied.
\item In case of the Rayleigh-Helmholtz pump with $\tilde \alpha=\alpha-\gamma_0\geq 0$ in
  \eqref{eq:dyn_prop1} we defined \eqref{eq:helm_ray} 
  \begin{eqnarray}
  \label{eq:ray_hel}
  a_h\,=\,(\tilde \alpha v\,-\,\beta v^3)\,,~~~~a_{\phi}\,=\,0\,.
\end{eqnarray}
Again the pump induces a acceleration parallel to the current motion
and there is no systematic torque applied and in $\phi$ direction we
apply noise with vanishing mean.

{Note that} the propulsion in this case has no preferred action along
the polarity axis. { It is impossible to distinguish} between forward
and backward motion along the polarity axes of the particle.  The
propulsion acts symmetrically in both positive and negative
directions.

\item In the frame of the the energy depot model, the stored energy
  $e(t)$ was transformed into kinetic energy with rate $d$ as (see Eq.
  \ref{langev_depot})
  \begin{eqnarray}
  \label{eq:depot_pop}
    a_h\,=\,d\,e(t)\,v(t)\, \,,~~~~a_{\phi}\,=\,0\,.
  \end{eqnarray}
  The internal motor increases kinetic energy in direction of motion.
\item The case with constant speed ${\bbox v}(t)\,=\,v_0\,{\bbox
    e_h}(t)$ and random or systematic turning angle: The propulsion in
  directions of motion has to compensate at least the friction force.
  The angular dynamics can be systematic, for example, circling with
  constant torque $\Omega_0 \ne 0 $. It can be as well a function of
  the orientation if additional forces affect the motion
  anisotropically.  Of course, random torques with Gaussian nature or
  shot noise and other time-dependent noise acts on the orientation.
  This case can be characterized with large $\gamma_0 \to \infty $ by
\begin{eqnarray}
  \label{eq:_ang_prop}
  a_h\,=\,\gamma_0 v_0 \,,~~~~a_{\phi}\,=\,  \Omega(v_0, \phi(t),\xi(t), t)\,, 
\end{eqnarray}
where $\Omega (v_0, \phi(t),\xi(t), t)$ shall describe the mentioned
influences. It will be presented in more details in next sections.
\end{itemize}

\subsubsection{Navigation in space following a given protocol}

{The human motion control of objects, say the steering of a ship or
  airplane or the driving of cars} usually involves some moving task
or protocol of navigation. There is {a} leader, the driver or captain,
who has to follow a given optimal route in spite of stochastic forces
and factors as wind, nonuniformities of space, etc..  In animal
mobility, the tasks are more modest but anyhow they do already exist
in elementary form.  An animal has to find food and shelter in a
landscape with given geometry, with some distribution of food and
shelter.

In order to describe such phenomena in a two-dimensional {spatial set
  up}, we have to assume at first a coordinate system {in which} the
protocol can be {defined and applied.}  We propose that the protocol
contains the information about the current state of the propulsion
mechanism or of the motor of the active particle. We will assume that
this state can be characterized by a velocity-like vector ${\bbox
  w(t)}$.  At every instant of time, it has a direction given by the
unit vector ${\bbox e}_w$ with components ${\bbox e}_w = \{ \cos
\nu(t), \sin \nu(t)\}$ in Cartesian presentation and is defined by the
magnitude or by the length of the vector $w(t)$.

Both values $w(t)$ and $\nu(t)$ {together make up} the protocol {and}
will define the performance of the motor over the considered time. In
addition to the position and velocity, they are new variables
describing the momentary state of the running motor. Changes of the
motor in magnitude and direction induce a propulsive force which is
exerted on the particle.  The corresponding acceleration has two
components
\begin{eqnarray}
  \label{eq:acc_mot}
  {\bbox a}_p (t)\,=\, a_w(t) \, {\bbox e}_{w}(t)\,+\, a_\nu(t) \,{\bbox e}_{\nu}(t)\,. 
\end{eqnarray}
resulting either from a change of magnitude or of direction 
\begin{eqnarray}
  \label{eq:change_mot}
  a_w(t)\, =\, {{\rm d}\over {\rm d}t}\, w(t)\,,~~~~~~a_\nu(t)\, =\,w(t)\,  { {\rm d} \over {\rm d} t }\,\nu(t)\,=\, w(t)\,\Omega(t)\,.
\end{eqnarray}
and we have set $\dot{\nu}(t)= \Omega$. Both components of ${\bbox
  a}_p$ vanish for a constant {activity} of the motor. The direction
of the common action from the two items differs from the direction of
motion ${\bbox e}_v(t)$ and from this of the protocol ${\bbox e}_w(t)$,
in general.

The function ${\bbox w}(t)$ would define a ideal course of the
particle {parameterized} in time. {In the absence of other forces,} the performance of
the particle would follow the protocol:
\begin{eqnarray}
  \label{eq:course}
  {\bbox v}_{\mbox{id}}(t) \,=\,{\bbox w}(t)\,.  
\end{eqnarray}
With a respective initial condition, one would get the ideal course
\begin{eqnarray}
  \label{eq:path}
  {\bbox r}_{\mbox{id}}(t)\,=\,{\bbox r}_0\,+\,\int^t {\rm d}t^\prime
  {\bbox w}(t^\prime)  
\end{eqnarray}
which is the wanted path in the considered coordinate system.

\subsubsection{Navigation in space: The general case}
In more complicated cases, the time-dependent protocol is unknown and
the temporal evolution of the motor {variables} $w(t)$ and $\nu (t)$
has to be defined as the solution of a given dynamical problem.  In
general, {it is reasonable to} assume that the two functions {obey}
first order differential equations
\begin{equation}
  \label{eq:motor_dyn}
  \frac{d { w}}{d t}\,=\,f_w({\bbox r },{\bbox v},{\bbox w}, t )\, + \,\sqrt{2D_w}\xi_w(t)\,,~~~~~~~~~~~~
  \frac{d {\nu}}{d t}\,=\, f_{\nu}({\bbox r },{\bbox v},{\bbox w}, t )\,+\,\sqrt{2D_{\nu}}\xi_{\nu}(t)\,.
\end{equation}
The $\xi_w$ and $\xi_{\nu}$ stand for motor noise with corresponding
intensities. In particular, the solution of these equations can be
also defined as {a} control problem, {in which} one has to
find the correct temporal forcing in order to follow a given route
under given loading.

The existence of {the} ``internal vector'' { ${\bbox w}(t)$
  determining} the direction of the propulsion by Eq.
\eqref{eq:course} {turns} the point-like particle {into} a Brownian
particles with time-dependent propulsion.  This new dynamical feature
in the description of the Active Brownian particle leads to an
extended state space which includes the new state variables ${\bbox
  w}(t)$, respectively $w(t)$ and $\nu(t)$. The situation is similar
to the case of active polar particles discussed recently by Szab{\'o}
{\emph et al.}  \cite{szabo_phase_2006} or Enculescu and Stark
\cite{enculescu_active_2011} . {In our new frame work}, the active
Brownian particle is defined by the position ${\bbox r}(t)$, by the
velocity ${\bbox v}(t)$, and by the vector ${\bbox w}(t)$ or by the
Langevin equations for position and velocity supplemented by Eqs.
\eqref{eq:motor_dyn} for the motor.

{The Langevin dynamics includes} the acceleration exerted by the
motor, linear friction, an external force ${\bbox F}$, and noise
(again we set $m=1$, leading to $\gamma=\gamma_0$):
\begin{eqnarray}
  \label{eq:motor_new}
  \frac{d {\bbox r}}{d t}\,=\,{\bbox v}\,;\,\,~~~~~~~~~\frac{d {\bbox v}}{d t}\,=\,\dot{w}(t)\, {\bbox e}_{w}(t)\,+\, w(t)\, \Omega(t)\,{\bbox e}_{\nu}(t)\,  - \gamma \, {\bbox{v}} \,-\,{\bbox F} \,+ \,\sqrt{2 D}{\bbox \xi}(t) \,.
\end{eqnarray}
{Here}, the motor creates an acceleration with magnitude
$a(t)$ acting in the direction of action $\chi(t)$.

{It is instructive to first consider a simplified system.} We assume
that the acceleration in the direction of the motor is some constant
and take for simplicity $\gamma w_0$. {This choice} is equivalent to a
Schienbein-Gruler propulsion.  Furthermore, we replace in the angular
acceleration the velocity in front of the angular frequency by the
asymptotic constant value $w_0$.  Doing so, the propulsion term reads
\begin{equation}
  {\bbox a}_p(t)\, = \gamma w_0 {\bbox e}_w \,+ w_0\, \Omega (t) {\bbox e}_{\nu} \,.
  \label{at2}
\end{equation}
These assumptions lead to 
\begin{eqnarray}
  \label{eq:motor1}
  \frac{d {\bbox v}}{d t}\,=\, v_0 \gamma  {\bbox e}_w \, - \gamma \, {\bbox{v}} \, 
  + v_0 \Omega (t) {\bbox e}_{\nu}   
  +\,{\bbox F} \,+ \,\sqrt{2 D}\, {\bbox \xi}(t) \,,
\end{eqnarray}
which is similar to the problem in Eq.\eqref{eq:dyn_prop2}. {The
  only difference} is the third term, {which describes} a rotation
of the particle as consequence of a change of direction of the motor
$\dot{\nu}(t)\,=\,\Omega \,\ne \,0$.

Next, we consider a constant external force, which acts in
$x$-direction ${\bbox F} = F_0 {\bbox e_x}$. We project the Langevin
equation onto the vectors ${\bbox e}_h(t)$ being parallel to the
direction of motion and the perpendicular angular direction ${\bbox
  e}_{\phi}$. In result, we obtain
\begin{eqnarray}
  \label{eq:control_polar}
  \frac{d {v}}{d t} &=& \gamma w_0 \, \cos (\nu - \phi) - \gamma \,v \,+\,F_0 \cos \phi \,+\,w_0\, \Omega(t) 
  \sin (\phi - \nu) \,+\,\sqrt{2D_v} \xi_v(t)\nonumber\\\\
  v\,  \frac{d { \phi}}{d t} &=&  \gamma v_0 \sin (\nu - \phi) + F_0 \sin \phi \,+\,  w_0 \Omega(t)\, \cos (\nu - \phi)\, 
  +\,\sqrt{2D_{\phi}} \xi_{\phi}(t) \,,\nonumber
\end{eqnarray}
One easily sees, that without external forces and with a constant
direction of the motor $\nu_0$ the direction $\phi$ of motion of the
particle will asymptotically follow $\nu_0$. Since items with $\Omega$
which is zero disappears the resulting equations are identical to the
situation explained following Eq.\eqref{eq:dyn_prop2}.

In other cases, the angular driving $\Omega(t)\,=\,\dot{\nu}(t)$ is
not defined as protocol and has to be defined.  The description
comprises the dynamical equations for the motor. For example, a wanted
time and space dependent route $\nu_0 ({\bbox r},t)$ may be given e.g.
by the relaxation dynamics
\begin{equation}
  \Omega (t) \,=\, \dot{\nu}(t) = - {1 \over \tau}\,  \left(\nu(t) - \nu_0 ({\bbox r}, t)\right)\,, 
\end{equation}
with some relaxation time $\tau$. For the particular $\nu_0 ({\bbox
  r}, t)$ one may find from this equation $\nu(t)$ as a function of
time and insert the solution into Eq.  \eqref{eq:control_polar}
{in order to solve} for $\phi(t)$ and for $v(t)$ and {hence to
  obtain} the complete trajectory.

As a further example, we {consider} a circling body. For this purpose we
suppose constant motor load $\gamma w_0$ and constant torque
$\Omega_0$, i.e.
\begin{equation}
  \label{eq:const_acc}
  \dot{w}\,=\,\gamma \, w_0 \,,~~~~~~~~~~~~~~\dot{\nu}\,=\, \Omega_0 \,.
\end{equation}
We simplify to the case without external force $F_0=0$. Then the
particle follows the rotating motor with the same angular velocity
$\dot{\phi}=\Omega_0$.  The constant difference of the angles
$\Delta\, =\,\phi\,-\,\nu $ and the stationary speed $v_0$ is
determined by the two equations 
\begin{eqnarray}
  \label{eq:explain}
  \gamma \,v_0\,=\,\gamma\, w_0 \, \cos \Delta \,+\, w_0\, \Omega \sin \Delta\,,~~\Omega \,v_0\,=\,\gamma\, w_0 \, \sin \Delta \,+\, w_0\, \Omega \cos \Delta\,.
\end{eqnarray}
for given motor load $w_0$ and $\Omega$. They are solved by
\begin{equation}
  \label{eq:phase_diff}
  \cos \Delta \,= \frac{v_0}{w_0}\,.
\end{equation}
The stable solution {should} obey $\Delta=\nu - \phi >0$.  For the
speed of the circling particle one obtains
\begin{equation}
  \label{eq:circling}
  v_0\,=\, w_0 \,  \frac{\Omega_0}{\sqrt{\gamma^2\,+\,\Omega_0^2}}\,.
\end{equation}
which becomes maximal if {the} dissipation is {weak}, i.e. $\gamma$ is small. {In this case},  the velocity
becomes approximately $w_0$ and the difference of the angle between
motor and particle vanishes.

{Generally,} the motor works with maximal efficiency (maximal
speed) {if the} direction of the motor agrees with the current
velocity vector.  Therefore, an important requirement is a
minimization of the deviation between $\nu(t)$ and $\phi(t)$ also in
the case {when} $\nu(t)$ changes permanently.  {One
  possibility to implement this requirement would be an adaptation
  dynamics, in which $\nu(t)$ approaches $\phi(t)$ with a positive
  rate $q>0$}
\begin{equation}
  \dot \nu(t) = - q [\nu(t) - \phi(t)]\,. 
\end{equation}
More generally, the dynamics for $\nu$ may be described in terms of a potential 
\begin{equation}
  \label{eq:motor_pol}
  \frac{d {\nu}}{d t}\,\propto \,-\frac{\partial U(\nu-\phi)}{\partial \nu}\,.
\end{equation}
With the particular choice $ U(\nu-\phi)= - q \cos(\nu-\phi)$ the
motor will also follow the current velocity, in case $q \to \infty$
both directions converge and the motor dynamics reduces to forces
which can be interpreted as negative friction. In general, the
outlined search for obtaining a maximal speed defines a problem of
optimization. { This problem is then  similar to that of the
navigation problem of a captain who selects} the direction of his ship by
turning the rudder and the speed by changing the motor power. {Finally,} we mention that one can also assume that the direction of motion in the dynamics is  {delayed by a fixed period}, i.e. $\phi(t-\tau)$, with $\tau > 0$. 

{Particularly important in our review} will be the case that the motor is
driven by random forces. We will discuss this point in more detail in
the next Section, where we {consider the} diffusion of active particles.
Already a first inspection of the Eqs. \eqref{eq:control_polar} shows
that the action of isotropic random alterations of the motor will not
result in a preferred direction. Nevertheless, the random action of
the motor will induce a diffusional spread over the free space. The
respective diffusion coefficients and distribution functions in
confined geometries which quantify this behavior will be the topic of
the next two Chapters (\ref{sec:sing} and \ref{sec:rot_ext}).

%% file: SECTION_single.tex
\section{Diffusion of active particles}
\label{sec:sing}

In this Section, we discuss several aspects of
the undirected spatial spreading of active particles in absence of external forces,
i.e. diffusion of free active particles. Diffusion is not only a fundamental feature of random motion and the mean squared displacement and the corresponding effective diffusion coefficient are characteristic observables easily accessible in experiments.
 
The properties of diffusive motion have important biological implications if applied to animal searching behavior as they have a major impact on the ability of individuals animals to exploit spatially distributed patches of nutrients (see e.g. \cite{komin_random_2004,schimansky-geier_advantages_2005,bartumeus_animal_2005,bartumeus_influence_2008}). 
As a starting point, before discussing different aspects of diffusion of active particles, we will give in this context one example for the importance of the value of the effective diffusion coefficient \cite{komin_random_2004,schimansky-geier_advantages_2005}. 
Other examples from ecology have been discussed by Okubo and Levin \cite{okubo_diffusion_2001}. 
We intend to show that simple organisms might gather different
amount of food in dependence on the value of the spatial diffusion
coefficient. It will be of special importance that the food is localized, 
say in bounded regions which extend over a finite distance only. 
If then the organism has only a finite time for foraging his strategy 
to locate a maximal amount of food is obviously an object of optimization 
as recently outlined by Moss and
coworkers
\cite{garcia_optimal_2007,dees_stochastic_2008,dees_patch_2008}.

Assuming that the density of the independent organisms
$\rho(\bbox{r},t)$, thus the single individual probability distribution,
obeys a diffusion equation with a diffusion
coefficient $D_r$
\begin{equation}
\label{eq:diff_daphnia}
  \frac{\partial \rho}{\partial t} 
  = D_r \Delta\rho.
\end{equation}
In two dimension the resulting probability density in space is
\begin{equation}
  \label{eq:density}
  \rho(\bbox{r},t)=\frac{1}{4\pi\,D_r\,t}
  \exp{\left[-\frac{{\bbox{r}}^2}{4D_rt}\right]} \, .
\end{equation}
We assume that the particle consumes with
constant rate $k$ a food $C$, which is given by its density
$c(\bbox{r},t)$, during its random motion. The food is stationary in space. Hence the consumption
of food is described by
\begin{equation}
  \label{eq:food}
  \frac{\partial}{\partial t} c(\bbox{r},t)\,=\, -\,k\,c(\bbox{r},t)\,\rho(\bbox{r},t)
\end{equation}

The latter equation can be solved exactly if the solution
(\ref{eq:density}) is inserted. Simple quadrature gives
\begin{equation}
  \label{eq:food2}
  c(\bbox{r},t)\,=\,c_0(\bbox{r})\,\exp\left[ - \int_{t_0}^t \,k\, \rho(\bbox{r},t^\prime)\, {\rm d}t^\prime\right].
\end{equation}
With the definition of the exponential integral \cite{AbrSte70}
\begin{equation}
  \label{eq:int}
  {\rm Ei}(1,a)\,=\, \int_a^\infty {\exp-t \over t }{\rm d}t
\end{equation}
we obtain in compact form
\begin{equation}
  \label{eq:compact}
  c(\bbox{r},t)\,=\,c_0(\bbox{r})\,\exp\left[ - {k \over 4 \pi D_r} \,{\rm Ei} \left(1,{\bbox{r\,}^2 \over 4D_r t} \right)\right].
\end{equation}
where we used $t_0=0$.

We consider now bounded distributed food patch. As a prototypical
distribution we take a circular patch with radius $R$ where $C$ is present.
Outside of the patch no food is located. For simplicity we assume the
circle positioned at the origin, i.e.
\begin{equation}
  \label{eq:patch}
  c_0(\bbox{r})\,=\,c_0\,=\,const., ~~~ {\rm if} \quad |\bbox{r}| < R.
\end{equation}
and vanishing food concentration, elsewhere. Then after a fixed time $T$ we determine the
overall food which is still left. It results from integration \cite{schimansky-geier_advantages_2005}
\begin{equation}
  \label{eq:food3}
  C(T)\,=\, 2\pi\,c_0 \int_0^R\exp\left[ - {k \over 4 \pi D_r} \,{\rm Ei} \left(1,{r^2 \over 4D_r T} \right)\right] \,r\,{\rm d}r.
\end{equation}

Inspection of this expression shows that there exist a minimum with
respect to the diffusion coefficient $D_r$. Thus the variation of the
effective diffusion coefficient gives rise to a change in the overall food
consumption. One might speculate that in nature the effective diffusion coefficient has been optimized for maximal foraging success \cite{garcia_optimal_2007,dees_stochastic_2008,dees_patch_2008}

\begin{figure}
  \centering
  \includegraphics[width=0.7\linewidth]{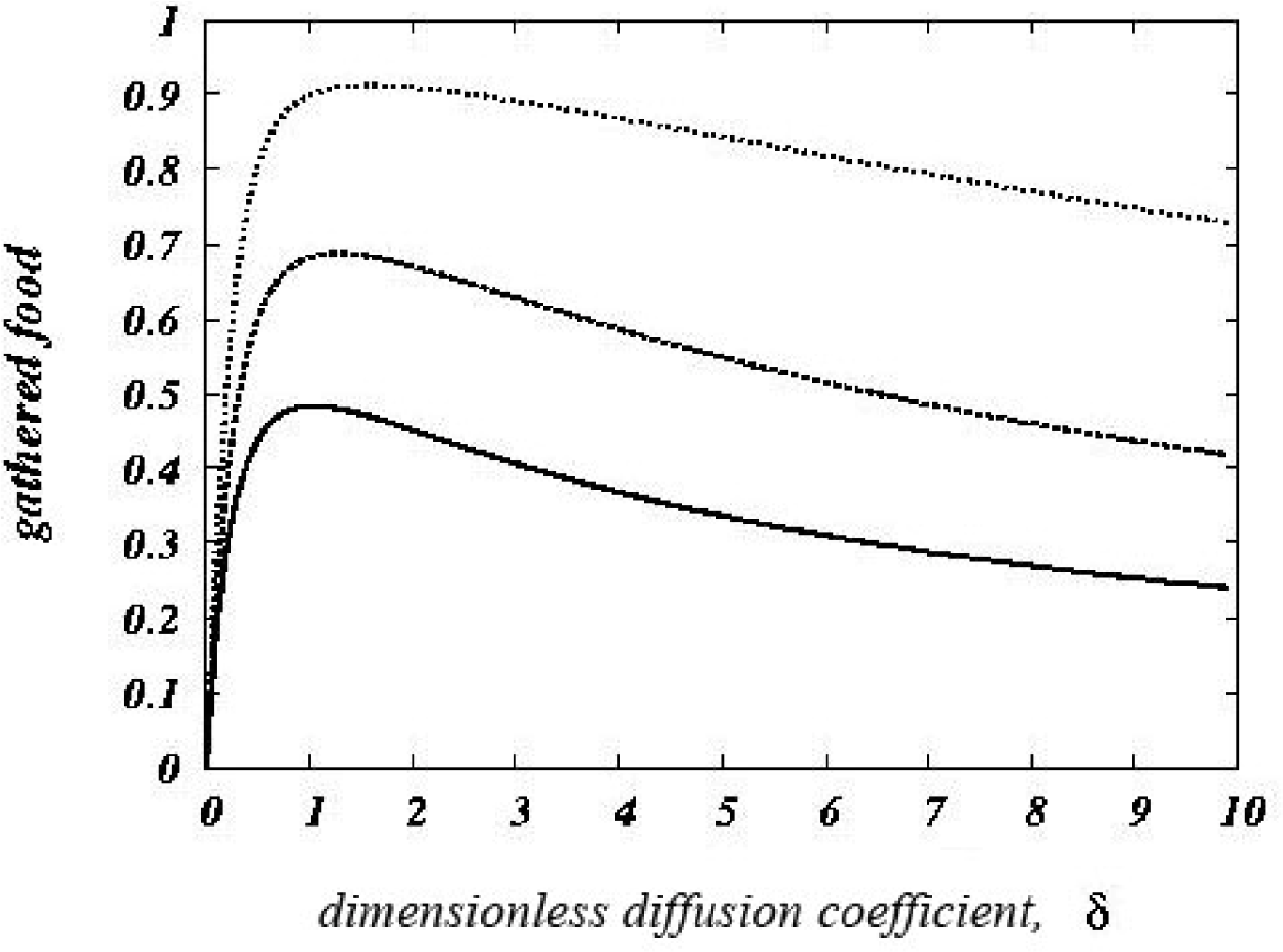} 
  \caption{Fraction of food gathered, which can be obtained using Eq. \eqref{eq:food3},
    as function of the dimensionless diffusion coefficient for three
    values of the clearance rate: $k =1$ (solid line), $k =1$
    (dashed line), $k =5$ dotted line. The gathered food is
    expressed as a fractions of the original total
    amount \cite{garcia_optimal_2007}.}
  \label{fig:consumed}
\end{figure}

The fraction of consumed food $(C(0)-C(T))/C(0)$ as a function of the diffusion coefficient is shown in
Figure \ref{fig:consumed}. We have chosen the spatial diffusion
coefficient $D_r$ with respect to the normal diffusion coefficient
$D_R\,=\,R^2/4T$ to cover the given sphere with radius $R$ in time $T$
and assigned $\delta=D_r/D_R$.  A simple explanation can be given as
follows. If the diffusion of particles is very large, e.g. they move almost exclusively straight forward, they will quickly leave the food
patch. On the other hand, staying localized due to low diffusion coefficient, e.g. jumping permanently forwards and backwards, is also disadvantageous as the particle spends a lot of time  in the region where it already consumed the food. Hence
there exists an optimal diffusion coefficient, for which the particles spend the most time within the
food patch and are able to consume the maximal amount of food. It does not matter whether they start at the center
of the food patch or not.

\begin{figure}[htb]
  \centering
  \includegraphics[width=0.6\linewidth]{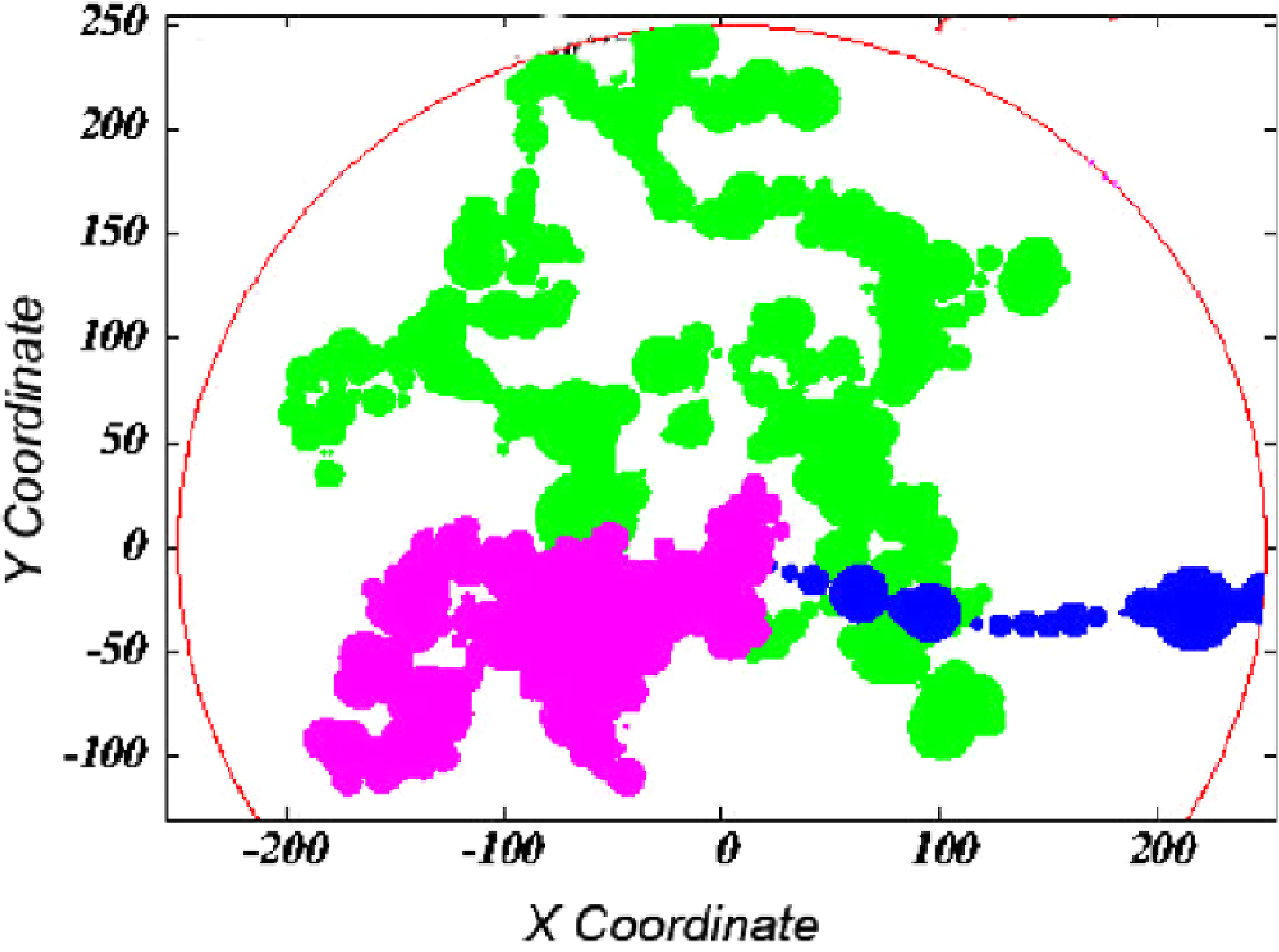} 
  \caption{Traces of food gathered by active particles for different
    widths of an exponentially distributed turning angle. The
    particles hop according to the turning angle distribution a finite
    step and rest a finite time in every new position gathering food;
    Blue: $\sigma=0.1$, green: $\sigma=1.2$ and magenta: $\sigma=10$,
    for ballistic, correlated and uncorrelated motion, respectively
    \cite{garcia_optimal_2007}.}
  \label{fig:gather_2}
\end{figure}

This situation is illustrated in Fig. \ref{fig:gather_2} where
particles hop with exponentially distributed turning angles
\begin{eqnarray}
  \label{eq:expon_turn}
  P(\Delta \varphi)\,=\, \sigma \,\exp\left(- \sigma |\Delta \varphi| \right)
\end{eqnarray}
and a constant hop length \cite{garcia_optimal_2007}. Between two hops they rest a
random time (with mean $\mean{\Delta t}=0.25$) and consume food.  In
Fig \ref{fig:gather_2},  the consumed food for three different width's $\sigma$ of
the turning angle distribution is shown. The large circle depicts the
finite radius of the food patch. It is seen that an optimal width leads to a
maximal consumption of food from the patch.

\subsection{Taylor-Kubo relations for the diffusion coefficient}
An interesting yet nontrivial question about the dynamics of active
particles concerns their diffusive behavior in the simple situation in
which there is neither interaction with other active particles nor
nonlinear external fields. In these situations it is of interest
whether the Active Brownian particles exhibits a mean drift with non-vanishing mean velocity
$\lr{\bbox v}$, possibly due to an asymmetry in its friction function
or due to some interaction between the friction and the driving
noise. Furthermore, it is not clear {\em a priori} whether the active particle 
will execute a normal Brownian motion or whether for certain friction
functions anomalous diffusion may be observable. In many cases, it
turns out that active Brownian particles perform just normal Brownian motion with a linear
asymptotic growth of the mean square displacement. In these cases,
however, it may still be surprising how the diffusion coefficient
depends on system parameters as, for instance, the noise intensity.

Generally, if the velocity dynamics obeys a stationary statistics, we may express the mean square displacement
(m.s.d.) $\Delta {\bbox r}^2(t) = ({\bbox r}(t)-{\bbox r}(0))^2(t)$ by the velocity auto-correlation function via the Green-Kubo relation. For convenience, we set the initial position to ${\bbox r }(0)={\bbox 0}$ and can thus express the position at $t$ by the integral over the velocity vector $\int_0^t dt' {\bbox  v(t')}$. First we assume vanishing stationary mean velocity $\langle {\bbox{v}} \rangle=0$. By inserting this into the m.s.d., exchanging integration and averaging, using stationarity,  and simplifying the integrals, one arrives at
\begin{align}\label{eq:msd_gen_taylor_kubo}
\langle \Delta {\bbox
  r}^2(t) \rangle = \langle {\bbox r^2}(t) \rangle = \int_0^t\int_0^t dt' dt'' \langle {\bbox v}(t') {\bbox v}(t'')\rangle.
\end{align}

If the stochastic process governing the evolution of ${\bbox v}$ is
stationary, the expectation value of $ \langle {\bbox v}(t') {\bbox
  v}(t'')\rangle$ depends only on the absolute value of the time
difference $\tau=|t'-t''|$ and we may rewrite the above relation as an
integral over the velocity-velocity correlation function $C_{{\bf v
  v}}=\langle {\bbox v}(\tau) {\bbox v}(0)\rangle$:
\begin{align}
  \label{eq:kubo1}
  \langle { \Delta \bbox{r}}^2(t)\rangle \,=
  \,=\, 2\;t\,\int_0^t d\tau \left(1-\frac{\tau}{t}\right)\,C_{{\bf v v}}(\tau)
\end{align}
Assuming a sufficiently fast decaying $C_{{\bf v v}}(\tau)$ so that the
integral over $\tau$ converges for $t\to\infty$ we may or large times
$t$ replace the upper limit of the integration by infinity which
yields:
\begin{align}\label{eq:diff_vv}
\langle {\Delta \bbox{r}^2}(t) \rangle = 2  \left[\int_0^{\infty} d\tau  C_{{\bf v v}}(\tau)\right] t.
\end{align}
Thus we define the diffusion coefficient according to the Taylor-Kubo formula in arbitrary spatial dimensions $d$ as:
\begin{align}\label{eq:deff_Dd}
D_{\rm eff} = \frac{1}{d} \int_0^\infty d\tau \langle {\bbox v}(\tau){\bbox v}(0)\rangle.
\end{align}
Note that the pre-factor with $d$ being the spatial dimension was chosen in analogy to ordinary
Brownian diffusion  (Eq. \eqref{eq:diffcoeffBrown}), and may differ from other definitions of the
(effective) diffusion coefficient. All of the outlined formulas can be generalized to the case with a finite mean speed.
The diffusion coefficient  then describes the diffusive spread around the mean motion:
\be
\label{eq:benji_deff_kubo}
D_{\rm eff}=\frac{1}{d} \liml_{t\to\infty} \frac{\langle {\bbox{r}^2(t)} \rangle - \langle {\bbox r(t)}\rangle^2}{2  t}
	=\frac{1}{d}\intl_0^\infty d\tau [\lr{{\bbox v}(0){\bbox v}(\tau)}-\lr{\bbox{v}}^2].
\ee
There exists a simple yet rather useful reformulation of this formula. Using the variance $\lr{\Delta {\bbox v}^2}$ 
and one particular definition of a velocity correlation time, namely the integral over the normalized auto correlation function, we can write
\be
\label{eq:benji_deff_corrtime}
D_{\rm eff}=\frac{\lr{\Delta {\bbox v}^2}\tau_{corr}}{d}
\ee 
So, the diffusion coefficient depends not only on the variance of the velocity fluctuations but also on their typical duration which is characterized by the correlation time. We stress that this interpretation holds only for non-oscillatory correlation functions.

For the cases studied in the following, the velocity correlation function is generally not known because of the nonlinearities introduced by the friction function. In one spatial dimension, however, one can solve for the integral of the correlation function  over the entire time interval $(0,\infty)$ and in this way obtain an analytical expression 
for the diffusion coefficient.  This can be achieved not only for the dynamics of active Brownian particles with white noise but also for a nonlinear velocity dynamics driven by a specific correlated (colored) fluctuations, namely, the dichotomous noise. In higher spatial dimensions, approximations for the calculation of the diffusion coefficient may be employed which exploit the specific symmetries of the speed-dependent friction (phase dynamics in 2d and 3d,
scaling behavior for friction functions obeying a power law).

In the following, we review the diffusive behavior of active particles with different friction functions, under the influence of  asymmetries, and driven by white or colored noise.  A particularly simple kind of nonlinear friction functions, namely those given by a power law, is treated separately because it makes certain limit cases for the diffusion of active particles more comprehensible. We then proceed with new features occurring in the diffusion in higher spatial dimensions.

\subsection{Diffusion in one dimension}

\subsubsection{Diffusion for a symmetric friction function and white noise}
Let us first consider $\gamma(-v)=\gamma(v)$ and a driving by a white Gaussian noise. In this spatially symmetric case  we obtain  a vanishing 
stationary mean velocity
\be
\lr{v}=0
\ee 
and the diffusion coefficient is the remaining statistics of interest. We mention that different approximations for the diffusion coefficient have been suggested for the one-dimensional case \cite{mikhailov_self-motion_1997,ConSib02,Ebe04,Lin06b};
the derivation of the exact solution will be  shown in the following.

Useful for the calculation of the diffusion coefficient is 
the introduction of a velocity potential given by
\be
\Phi(v)=\intl_0^v dv' \gamma(v') v'.
\ee
For a typical friction  function of an active particle, this potential is  bistable and thus has two metastable states with finite speed  whereas the zero velocity is dynamically unstable. Specifically, for the Rayleigh-Helmholtz friction function introduced in \e{helm_ray}, the potential reads
\be
\Phi(v)=\beta \frac{v^4}{4}-\alpha \frac{v^2}{2};
\label{eq:benji_rh_pot}
\ee
in this section, we set  $\alpha=\beta=1$.  For the friction function of the depot model in \e{set_friction}, one obtains the potential
\be
\Phi(v)=\frac{\gamma_0}{2}\left[v^2 -\frac{q}{\gamma_0}\ln(1+ v^2 d/c)\right].
\label{eq:benji_set_pot}
\ee
In the following, we use the numerical  values $\gamma_0=20, d/c=40, q=5$. 

The Fokker-Planck equation for the velocity process reads 
\be
\label{eq:benji_fpe}
\partial_t P(v,t)=\partial_v[\Phi'(v)+D \partial_v] P(v,t).
\ee
We recall \cite{risken_fokker-planck_1996} that (i) this equation can be easily solved in the stationary state, yielding a Boltzmann-like distribution $P_0\sim \exp[-\Phi(v)/D]$; (ii) the transition probability $P(v,t|v_0,0)$ is governed by the FPE with the initial condition $P(v,0)=\delta(v-v_0)$; (iii)  there is no general solution for $P(v,t|v_0,0)$ for a general nonlinear force $-\Phi'(v)$. If we would know  $P(v,t|v_0,0)$ we could calculate the velocity correlation function by
\be
\lr{v(t)v(0)}=\int dv \int dv_0 v v_0 P(v,t|v_0,0) P_0(v_0)
\ee
where the last two factors express the joint probability density $P(v,t; v_0,0)$ by means of Bayes' theorem. 
Inserting the correlation function into the Green-Kubo formula, we may write
\be
D_{\rm eff}=\int dv v   \intl_0^\infty d t  \int dv_0 v_0 P(v,t|v_0,0) P_0(v_0)= \int dv v G(v)
\ee
An ordinary differential equation for $G(v)$ can be obtained from the Fokker-Planck \e{benji_fpe} by integrating over time, multiplying with the known stationary probability density in $v_0$,  and integrating over the latter variable.
The resulting equation reads:
\be
-vP_0(v)=(\Phi'(v) G(v))'+D G''(v)
\ee
which can be solved in terms of quadratures. After a few simplifications (using also the symmetry of the potential),
the result for the diffusion coefficient reads \cite{Lin07}
\be
\label{eq:benji_deff_symm}
D_{\rm eff}=\frac{\intl_0^{\infty} dv_2\; e^{\Phi(v_2)/D} \left[ \intl_{v_2}^{\infty} dv_1\;
e^{-\Phi(v_1)/D} v_1\right]^2}{D^2 \intl_0^{\infty}  dv_3\; e^{-\Phi(v_3)/D}}
\ee
We note that the same techniques has been used by Jung and Risken \cite{JunRis85} and in particular by Risken in the second edition of his well-known textbook \cite{risken_fokker-planck_1996} to obtain an analytical expression for the correlation time of a stochastic process governed by a nonlinear Langevin equation (the connection to $\tau_{corr}$ is evident in \ref{eq:benji_deff_corrtime}).

Before coming to the case of active Brownian motion, let us first consider a simpler nonlinear speed-dependence of the friction which is given by a power law. In particular, we assume that 
\be
\gamma(v)=\gamma_0 v^{2\alpha}, \alpha\ge 0.
\ee
(note that $\alpha$ has nothing to do with the parameter of the RH friction function, which we  set to zero).
This problem has been studied in detail in Ref.~\cite{Lin08} where it has been shown that the diffusion coefficient in this case scales like a power law with the system's parameters:
\be
\label{eq:benji_powlaw_scaling}
D_{\rm eff}\sim  \gamma^{-\frac{2}{1+\alpha}} D^{\frac{1-\alpha}{1+\alpha}}
\ee
This result can be obtained from the exact expression by changing the integration variables appropriately (the factor of proportionality is then given by an integral that still depends on $\alpha$ but not on $D$ or $\gamma_0$ anymore). It can be, however,  also obtained by renormalization of the original Langevin equation
(see \cite{Lin07}) a techniques by means of which one can show that \e{benji_powlaw_scaling} holds also true in  arbitrary  dimensions although of course with different pre-factor than in the one-dimensional case.

On closer inspection, \e{benji_powlaw_scaling} is a little surprising. It tells us, for instance, that for a pure cubic friction  ($\alpha=1$), the diffusion coefficient  does not depend on the noise intensity $D$ at all. Furthermore, for stronger than cubic friction ($\alpha >1$), the diffusion coefficient decreases monotonically with increasing $D$ and diverges in the limit of vanishing noise intensity. In the case of normal Brownian motion ($\alpha=0$), the diffusion coefficient is proportional to the noise intensity as expected. All of these conclusions are illustrated and confirmed by results of numerical simulations in \bi{benji_powlawfric}.

\begin{figure}[h!]
 \centerline{\parbox{7.5cm}{\includegraphics[width=7.5cm,angle=0]{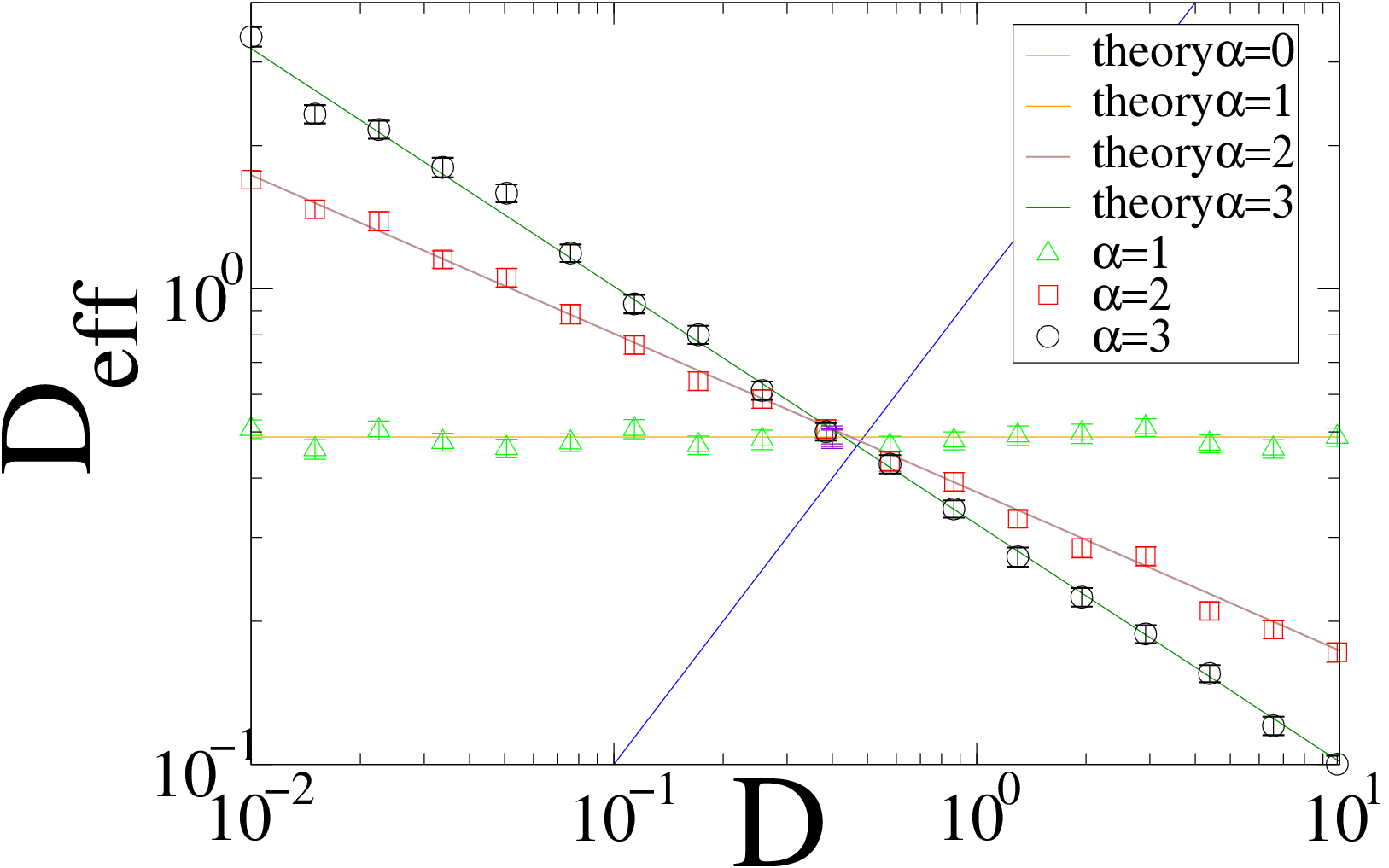}}}
\caption{\small  
Diffusion coefficient vs noise intensity $D$ for the one-dimensional case
($\gamma=1$) and various values of the exponent. Theory (numerical evaluation of \e{benji_deff_symm})  is 
compared to simulation results. Modified from \cite{Lin08}. 
\label{fig:benji_powlawfric}}
\end{figure}

In order to explain this behavior we have to recall that according to \e{benji_deff_corrtime} the diffusion coefficient is 
given by the product of the variance and the correlation time of the velocity fluctuations. Increasing the noise intensity will in all considered cases increase the variance. The effect on the correlation time, however, is more complicated. For strong noise and a strong nonlinear friction ($\alpha\ge 1$), the velocity attains typically large values at which  the  dissipation term $v^{2\alpha+1}$ is stronger than in the linear case, hence the correlation time is shorter and becomes even shorter for further increasing noise. In the opposite limit of vanishing noise, the velocity is typically at values close to zero where the dissipation term is now much weaker than in the case of normal Brownian motion. For a cubic friction the tendency of increasing velocity variance and decreasing correlation time balance exactly and their product remains constant upon variations of $D$. For a stronger friction, the sensitivity of the correlation time with respect to changes in the noise is even stronger and dominates the dependence of the diffusion coefficient on $D$. 

Let us now turn to the case of an active Brownian particle. First we consider the dependence of the diffusion coefficient for the depot model introduced in Sec.~\ref{sec:det_depot}. In this case the quadrature formula has been evaluated numerically and the result is shown in \bi{benji_deff_set}.
\begin{figure}[h!]
 \centerline{\parbox{7.5cm}{\includegraphics[width=7.5cm,angle=0]{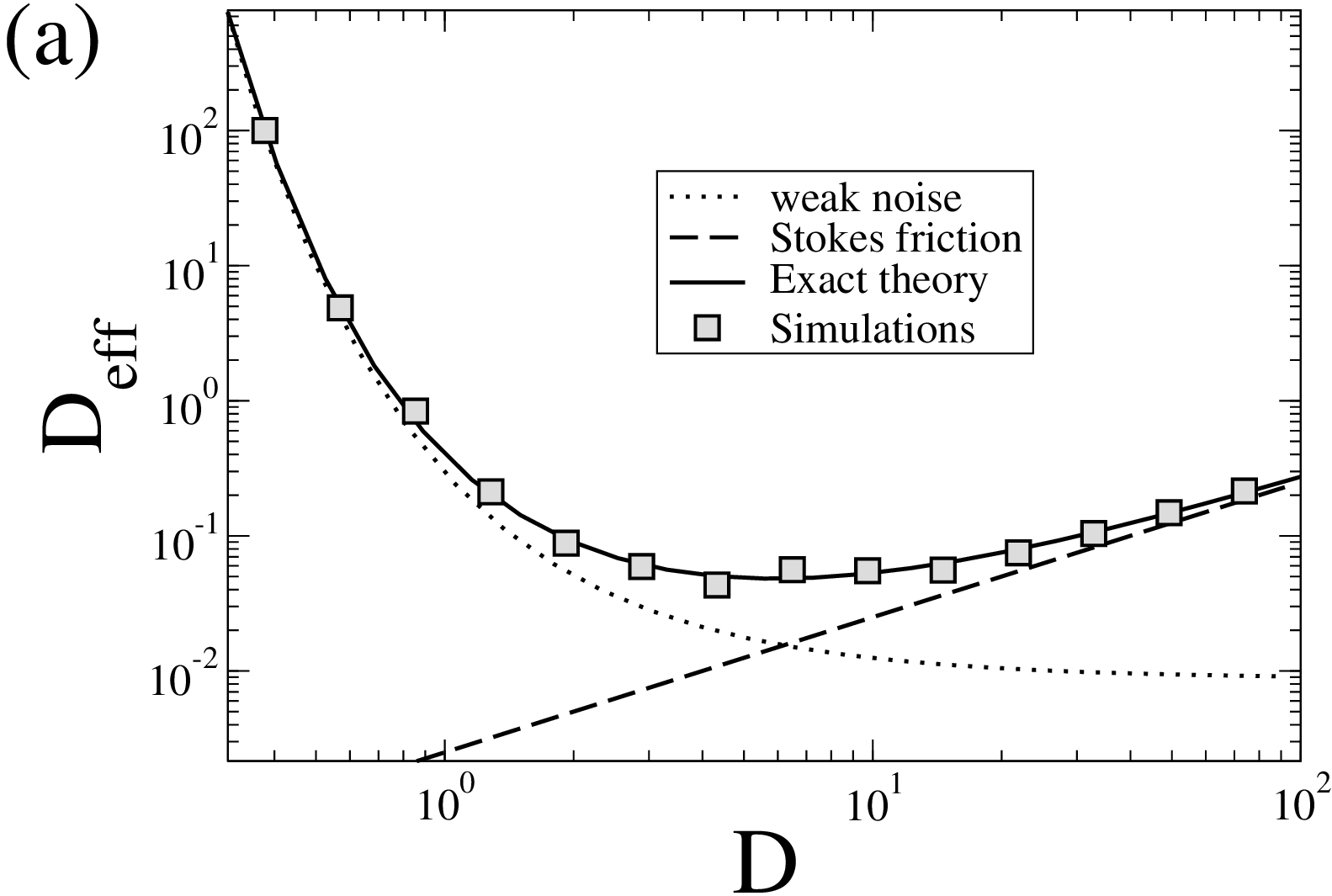}}}\centerline{\parbox{7.5cm}{\includegraphics[width=7.5cm,angle=0]{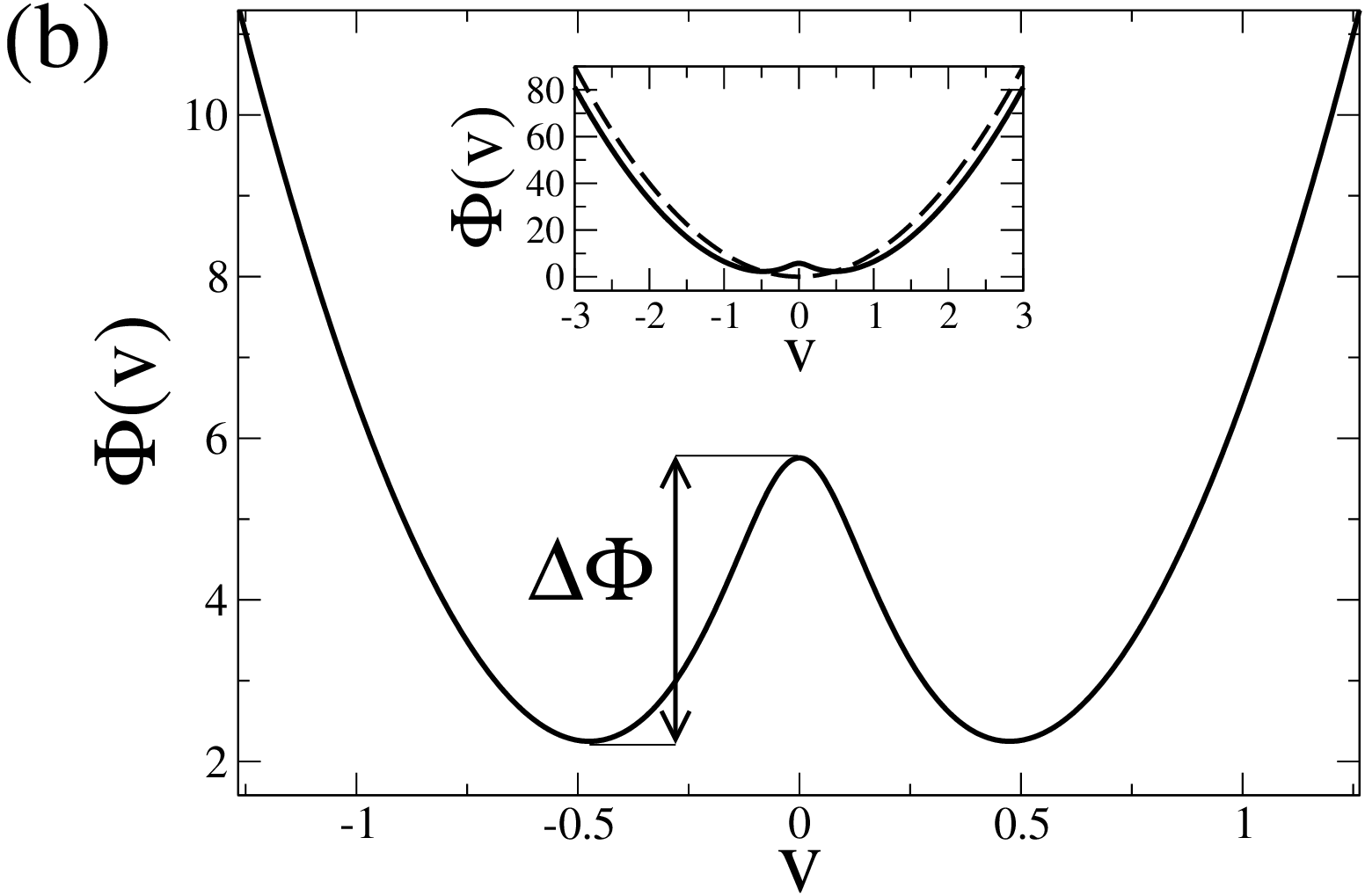}}}
\caption{\small  
Diffusion coefficient vs noise intensity $D$ (a) and velocity potential (b) for the friction function of the depot model.  In (a), theory (numerical evaluation of \e{benji_deff_symm} with the potential \e{benji_set_pot} is 
compared to simulation results. The inset in (b) illustrates that the velocity potential behaves asymptotically as a parabola $\Phi(v)\sim v^2$  (dashed line).  Modified from \cite{Lin07}. 
\label{fig:benji_deff_set}}
\end{figure}

We observe a nonmonotonic relation between the diffusion coefficient and the noise level: the diffusion is minimized at an intermediate noise level. We can understand the growth with $D$ seen at large noise intensity because  at large values of the velocity (typically attained at strong noise) the depot model approaches 
the Stokes limit of a speed-independent friction coefficient. At the level of the velocity potential this implies a parabolic potential at large speed (cf. \bi{benji_deff_set}b). In the opposite limit of weak noise the bistability of the velocity potential becomes essential. Here the velocity undergoes noise-induced transitions between the two metastable states corresponding to the minima of the potential. For such a process, the diffusion coefficient is inversely proportional  to the Kramers hopping rate \cite{mikhailov_self-motion_1997,Ebe04}
\ba
\label{eq:benji_deff_sd}
D_{\rm eff}&\approx& \frac{v_0^2 \pi}{Q\sqrt{\Phi''(v_0)|\Phi''(0)|}} \exp[\frac{\Delta
\Phi}{D}]= \frac{v_0^2}{2 r_{K}}   
\ea
The latter relation agrees with \e{dich_diff} for $\lambda_\pm=r_k, |v_\pm|=v_0$; it has been derived for a velocity following a telegraph process with rate $r_k$ in \cite{San84}.
Assuming weak noise $D$, \e{benji_deff_sd} can be also derived via a saddle-point approximation from the exact result \e{benji_deff_symm}.

The divergence of the diffusion coefficient can be understood by considering the spread of an ensemble of active Brownian particles. At equilibrium half of the particles are in the left well with a corresponding finite speed going to the left while the other half is situated in the right well and thus go with finite speed to the right. Without any 
transitions between the wells (a limit that is approached for vanishing noise), the growth of the m.s.d. would occur ballistically. Incidentally, this has to be taken into account when choosing a simulation time for an estimation of the diffusion coefficient. One has to wait at least a multiple of the correlation time of the process, in order to see a diffusive growth of the m.s.d. Very similar to the case of normal Brownian motion, there is a ballistic phase in the m.s.d. time course. The time of this $t^2$ growth can be estimated from the relation between diffusion coefficient, variance, and correlation time \e{benji_deff_corrtime}: $\tau_{corr}=D_{\rm eff}/\lr{\Delta v^2}$. because in our numerical example $ \lr{\Delta v^2}\simeq 1$ but $D_{\rm eff}$ is exponentially large, the correlation time is rather large. In order to get reasonable estimates for the diffusion coefficient, an ensemble of trajectories has to be simulated at least for a multiple of this large correlation time.
\begin{figure}[h!]
 \centerline{\parbox{7.5cm}{\includegraphics[width=7.5cm,angle=0]{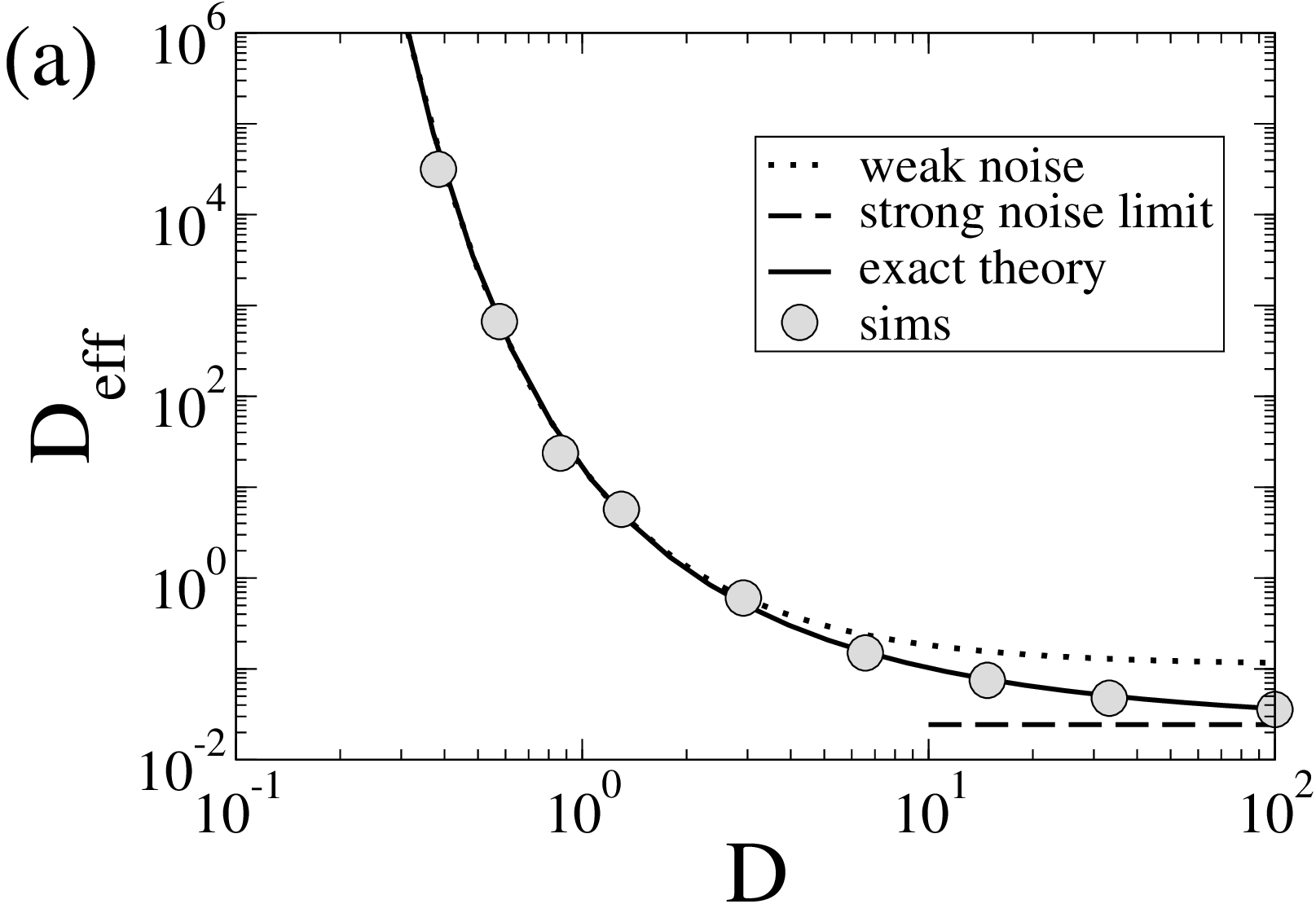}}}\centerline{\parbox{7.5cm}{\includegraphics[width=7.5cm,angle=0]{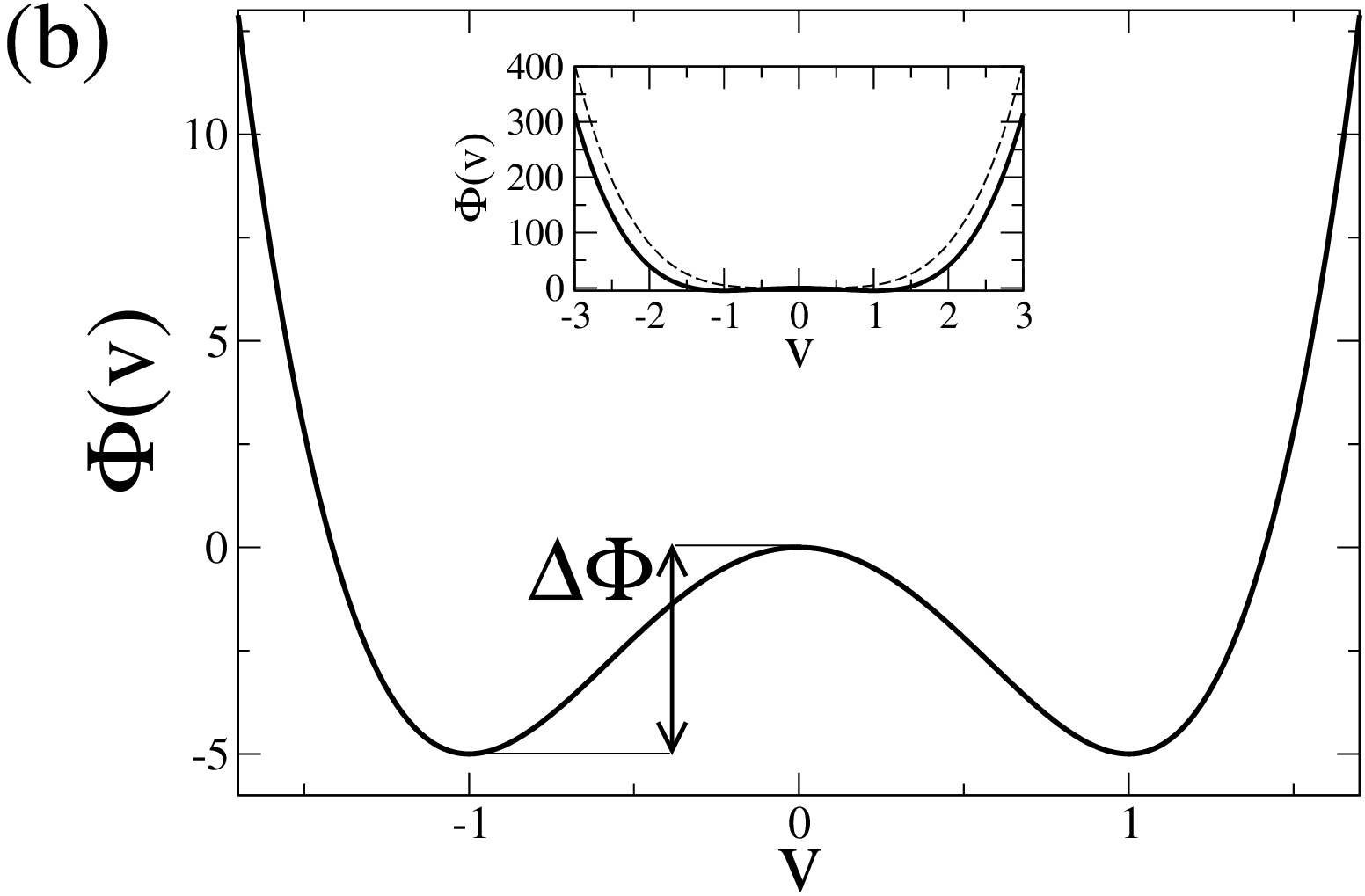}}}
\caption{\small  
Diffusion coefficient vs noise intensity $D$ (a) and velocity potential (b) for the RH model. 
In (a), theory (numerical evaluation of \e{benji_deff_symm} with the potential \e{benji_rh_pot}) is 
compared to simulation results. The inset in (b) shows the asymptotic behavior of the potential which is dominated for the RH model by the quartic term $\Phi(v)\sim v^4$ (dashed line). Modified from \cite{LinNic08b}. 
\label{fig:benji_deff_rh}}
\end{figure}

Turning back to the minimum, which we observed for the diffusion coefficient vs noise intensity: is such a minimum 
observed for all possible friction functions of active Brownian particles? The results on the power-law friction function tell us otherwise. If a friction function has a nonlinear asymptotic limit such that the friction increases like a power equal or stronger than cubic friction, the diffusion coefficient will either saturate (for cubic friction $\gamma(v)v \sim v^3$) or even decrease (for friction stronger increasing than $\gamma \sim v^2$) in the asymptotic limit of large noise $D$. So in these cases we cannot expect to observe a minimum. Indeed for the RH model which is at large speed dominated by the cubic term, the diffusion coefficient decreases monotonically with the noise intensity $D$. We note that the divergence of the diffusion coefficient in the limit $D \to 0$ is untouched by the asymptotics of the velocity potential because this divergence hinges solely upon the presence of a potential barrier.

\subsubsection{Diffusion in the spatially symmetric case with colored noise}
The assumptions we make on the phenomenological model of an active particle are often  based on simplicity.
One such assumption is the lack of correlations in the driving noise process $\xi(t)$.  After we have reached some
understanding of the active Brownian motion dynamics in one dimension with white noise, we may relax this condition and consider the case where the fluctuations have a finite correlation time, i.e. we consider
\be
\appel x=v,\;\; \appel v=-\gamma(v) v+\eta(t)
\ee
where $\eta(t)$ is a colored noise that we choose here to have an exponential correlation function
\be
\lr{\eta(t)\eta(0)}=\sigma^2 e^{-2\lambda t}.
\ee
There are two simple stochastic processes with such a correlation function but different probability density.
The first one is  the symmetric dichotomous process (DMP, also called telegraph noise) which jumps between two values $-\sigma$ and $\sigma$ with transition rates $\lambda$. The other one is a Gaussian noise, the  Ornstein-Uhlenbeck process (OUP) $\eta$, which  can be obtained by integrating the stochastic differential equation
 \be
\appel \eta=-2\lambda \eta(t) +2\sigma \sqrt{\lambda} \xi(t)
\ee
($\xi(t)$ is as usual white Gaussian noise with $\lr{\xi(t)\xi(t')}=\delta(t-t')$). The unusual scaling in the OUP 
ensures that variance and correlation time are the same  for OUP and DMP in terms of the parameters $(\sigma,\lambda)$.

If the driving noise introduces a finite correlation time, one question of interest is how this time scale affects the diffusion coefficient. The DMP offers a simple case in which this question can be explored analytically. By a very similar approach as discussed above for the white-noise case, one can derive an exact quadrature formula for the diffusion coefficient \cite{Lin10}. 

For the velocity process driven by a dichotomous noise, we may write down evolution equations for the probability 
densities  $P_+(v)$ and  $P_-(v)$ that the velocity attains the value $v$ and that the noise is at $\eta=\sigma$ and 
$\eta=-\sigma$, respectively:
\ba
\label{eq:benji_fpe_dicho}
\partial_t P_+ &=& \partial_v[\gamma(v)v-\sigma]P_+ -\lambda P_++\lambda P_-\\
\partial_t P_- &=& \partial_v[\gamma(v)v+\sigma]P_- +\lambda P_+-\lambda P_-
\ea
These two equations solved for different initial conditions with respect to the noise would yield four transition probabilities $p_{\pm,\pm}(v,t)$ ($p_{+,-}(v,t|v_0)$, for instance, is the probability to find the velocity at $v$ and the noise at $\eta=\sigma$ if initially at $t=0$ the velocity was at $v_0$ and the noise was $\eta(0)=-\sigma$)  by means of which the velocity correlation function can expressed:
\be
\label{eq:benji_corr_v}
 \lr{v(t)v(t+\tau)} =\intl_{-v_m}^{v_m} dv v \left[\intl_{-v_m}^{v_m} dv_0\; v_0\left\{(P_{++}+ P_{-+}) P_+^0+(P_{+-}+ P_{--}) P_-^0 \right\} \right]
\ee
We can use again the fact that for the diffusion coefficient we have to know only the integral over the correlation function.  By integrating Eqs.~(\ref{eq:benji_fpe_dicho}) we obtain a system of ordinary differential equations for two auxiliary functions (comparable to the function $G(v)$ in the previous subsection), the integral of which then yields  the diffusion coefficient. Further details on the lengthy but straightforward calculation can be found in \cite{Lin10}; the final result reads
\be
 \label{eq:benji_diffu_dicho}
 D_{\rm eff}=2 \lambda \sigma^2 \frac{\int\limits_0^{v_m} dx \frac{e^{\hat{\Phi}(x)}}{\sigma^2-f^2(x)} \left(\int\limits_x^{v_m} dy \frac{e^{-\hat{\Phi}(y)}}{\sigma^2-f^2(y)} y\right)^2}{\int\limits_0^{v_m} dz e^{-\hat{\Phi}(z)}[\sigma^2-f^2(z)]^{-1}},
\; \;\; \hat{\Phi}(v)=-2\lambda\intl_0^v dx \frac{f(x)}{\sigma^2-f^2(x)}.
\ee
The value of $v_m$ is set by the maximal speed by which the particle can go under dichotomous driving and is found from the equation $\gamma(v_m)v_m=\sigma$.  

In the white-noise  limit of the dichotomous fluctuations $\lambda\to \infty, \sigma \to \infty$ with $D=\sigma^2/(2\lambda)=const$, the expression approaches the result \e{benji_deff_symm}.  Further limit cases as well as the numerical evaluation of the integrals is discussed in \cite{Lin10}. Here we review the dependence of the diffusion coefficient on the new time scale in the problem given by the switching rate of the DMP.

\begin{figure}[h!]
 \centerline{\parbox{7.5cm}{\includegraphics[width=7.5cm,angle=0]{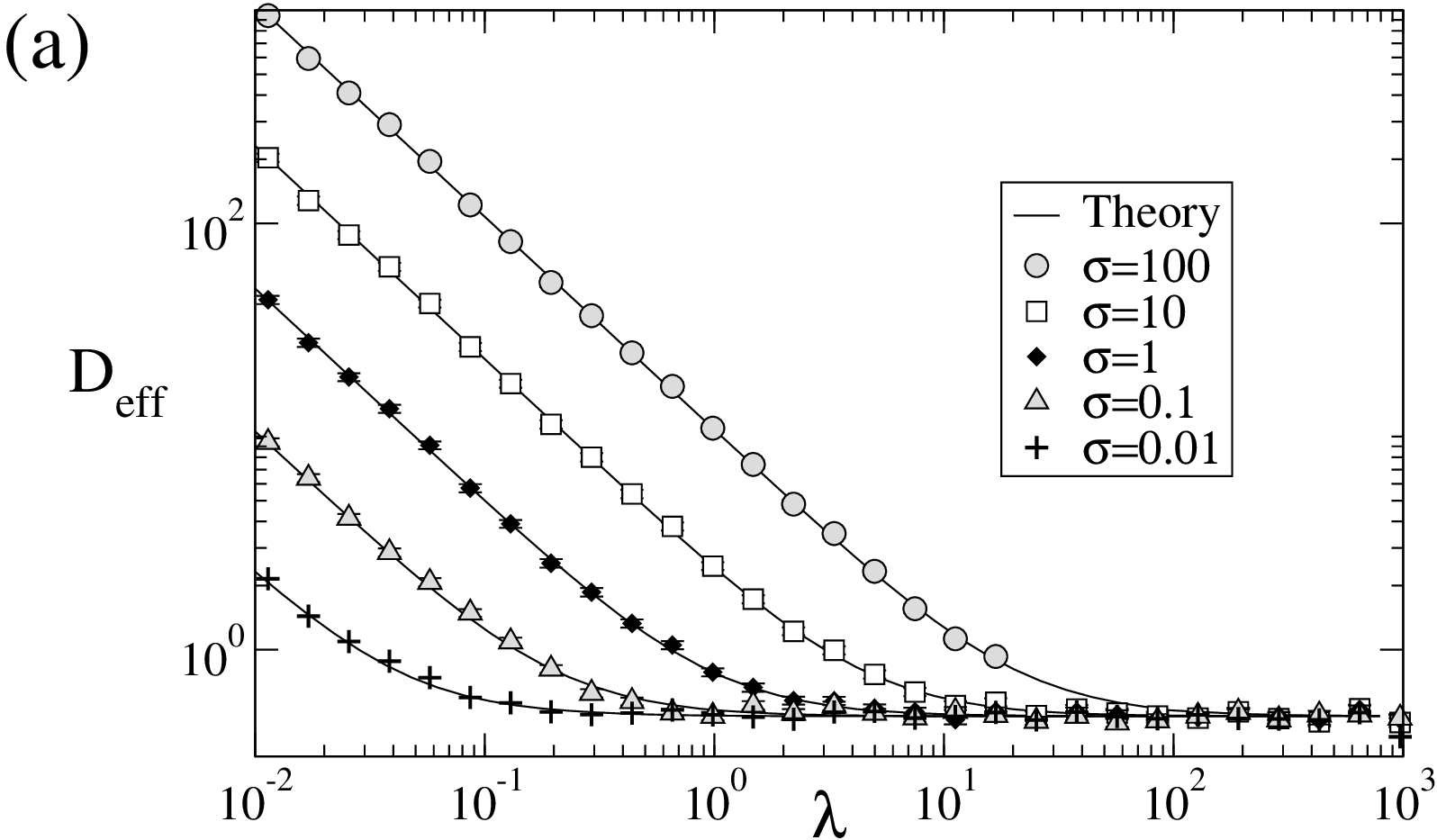}}} \centerline{\parbox{7.5cm}{\includegraphics[width=7.5cm,angle=0]{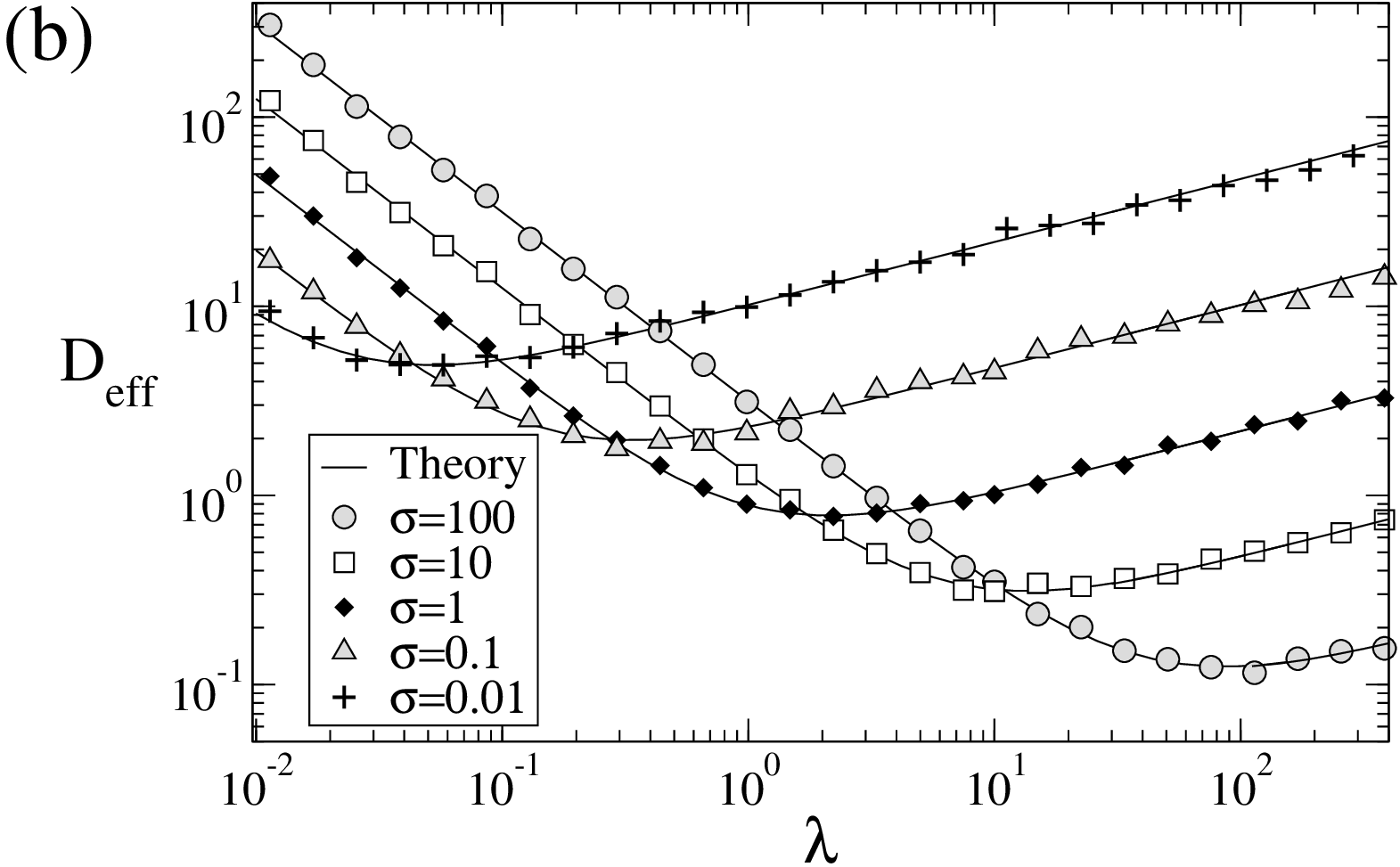}}}
\caption{\small  
Diffusion coefficient vs switching rate of the telegraph noise for a cubic (a) and a quintic friction function (b)
for different amplitudes of the driving as indicated. Theory (solid lines, \e{benji_diffu_dicho})  is
compared to simulation results. Modified from \cite{Lin10}. 
\label{fig:benji_deff_dicho_powlaw}}
\end{figure} 
Let us first consider the simple case of a power law friction. The diffusion coefficient for a cubic friction decreases 
monotonically with increasing switching rate. We can understand the limits of small and large rate based on the results from the previous  subsection. At small switching rate the velocity process relaxes towards the asymptotic 
value $\pm v_m$ - the dichotomous character of the input process carries over to the velocity with an amplitude 
which is determined by the nonlinear equation for $v_m$. We mentioned already that for a velocity following a telegraph noise, the diffusion coefficient is proportional to the inverse of the switching rate and this is exactly the behavior observed in \bi{benji_deff_dicho_powlaw} in the limit $\lambda\to 0$. In the opposite limit of a large rate, the DMP approaches white noise  with a vanishing intensity because $D=\sigma^2/(2\lambda)$ ($\sigma$ is constant). For a cubic friction, however, the diffusion coefficient does not depend on the noise intensity and thus $D_{\rm eff}$ approaches the constant value that we already obtained in the white noise case for $\gamma(v) v=\gamma_0 v^3$
(cf. in \bi{benji_powlawfric} the data for $\alpha=1$).

For a friction function with stronger nonlinearity, for instance, for a quintic dependence on $v$, the divergence at small rate remains the same (also here we observe a dichotomous velocity process dictated by the slow driving),
at strong rates, however, the diffusion coefficient also diverges in marked contrast to the cubic case (\bi{benji_deff_dicho_powlaw}b). The reason for the increase at large rate is also quite plausible. With increasing rate the white-noise limit is approached with a noise intensity that decreases with rate. We saw before that for a friction function with $\alpha>1$ the diffusion coefficient diverges for vanishing noise and it is exactly this divergence that 
we now observe for increasing rate. As a consequence of asymptotic behavior, the diffusion coefficient attains a minimal value at intermediate rate. The rate for the minimal diffusion coefficient depends on the amplitude and is approximately given by \cite{Lin10}
\be
\lambda_{min}\approx 2(A^4 \gamma)^{1/5}.
\ee
\begin{figure}[h!]
 \centerline{\parbox{7.5cm}{\includegraphics[width=7.5cm,angle=0]{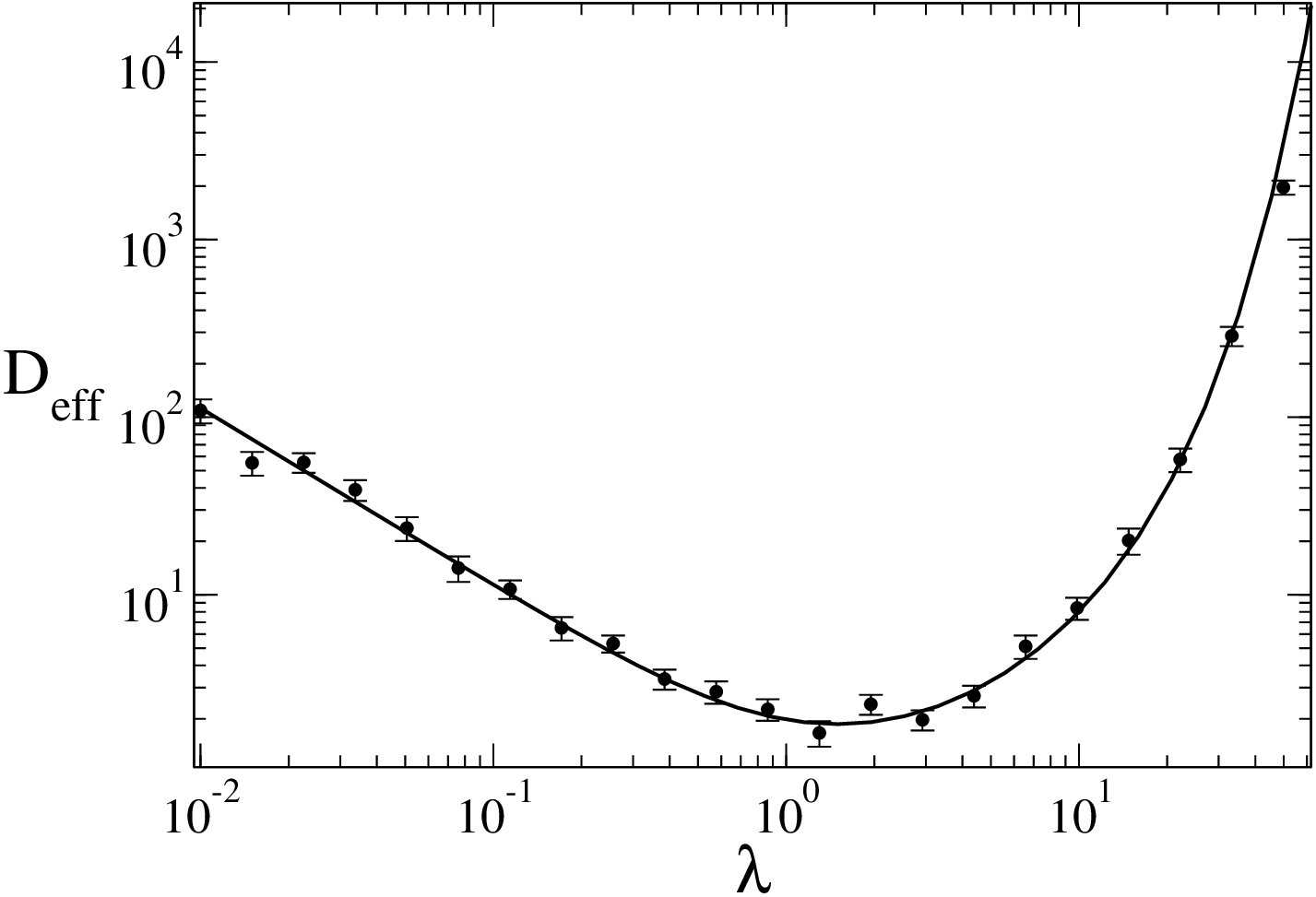}}}
\caption{\small  
Diffusion coefficient vs DMP switching rate a Rayleigh-Helmholtz friction function; amplitude of the driving is $\sigma=1.875$. Theory  (solid line, \e{benji_diffu_dicho})  is compared to simulation results. Modified from \cite{Lin10}. 
\label{fig:benji_deff_dicho_rh}}
\end{figure}  

For an active Brownian particle  we also find a minimum in the diffusion coefficient vs rate (\bi{benji_deff_dicho_rh}) because in this case we recover in the limit of large rate the weak-noise limit of the white-noise case associated with a bidirectional motion with rare reversals. Hence, here in both limits of small and large switching rate of the telegraph noise we observe a bidirectional motion although for two different physical reasons. 
We note that for the Rayleigh-Helmholtz friction function used in (\bi{benji_deff_dicho_rh}), our theory is constraint 
to the condition of a sufficiently large amplitude ensuring that transitions between the metastable states are possible at all.
\begin{figure}[h!]
 \centerline{\parbox{7.5cm}{\includegraphics[width=7.5cm,angle=0]{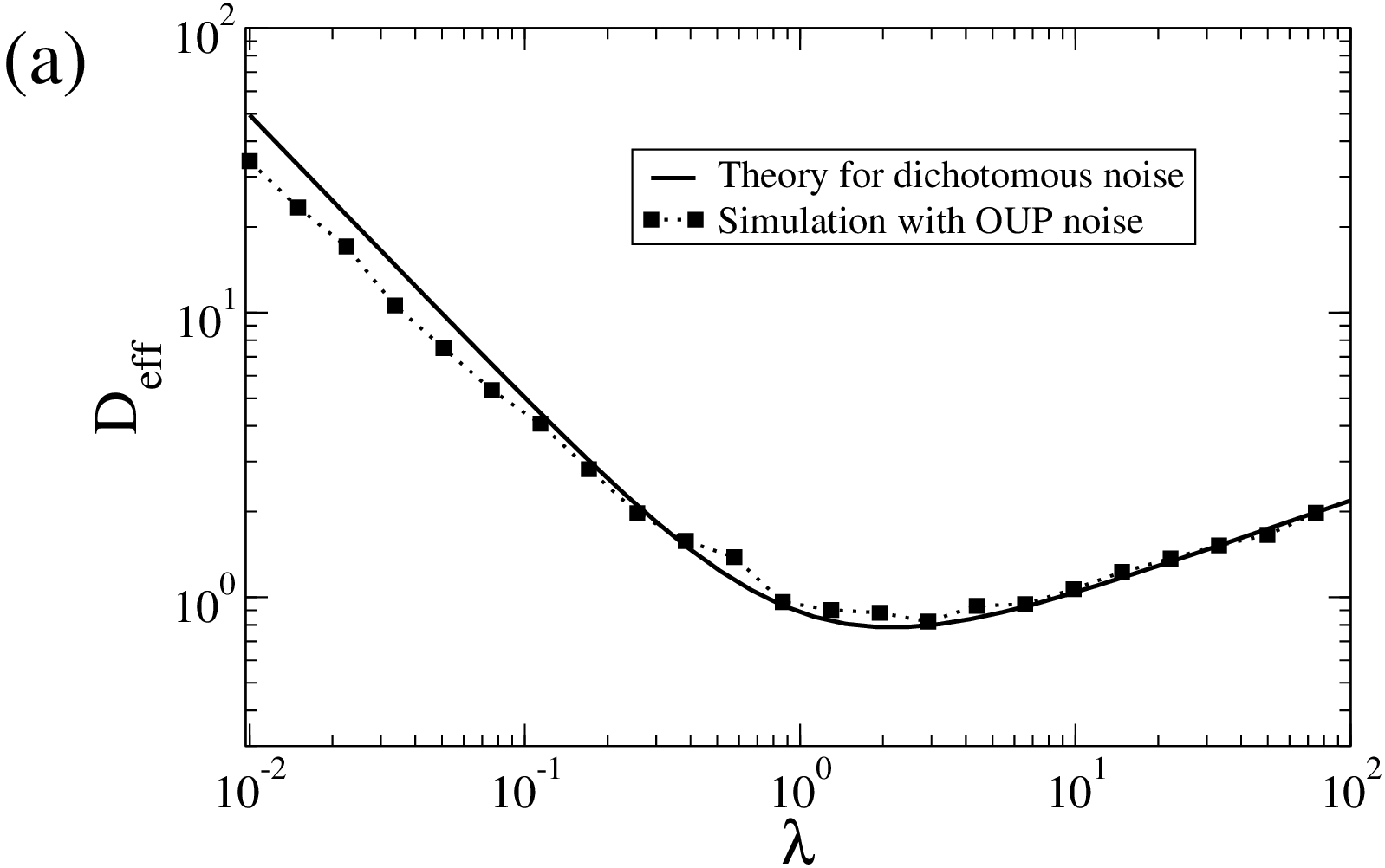}}} \centerline{\parbox{7.5cm}{\includegraphics[width=7.5cm,angle=0]{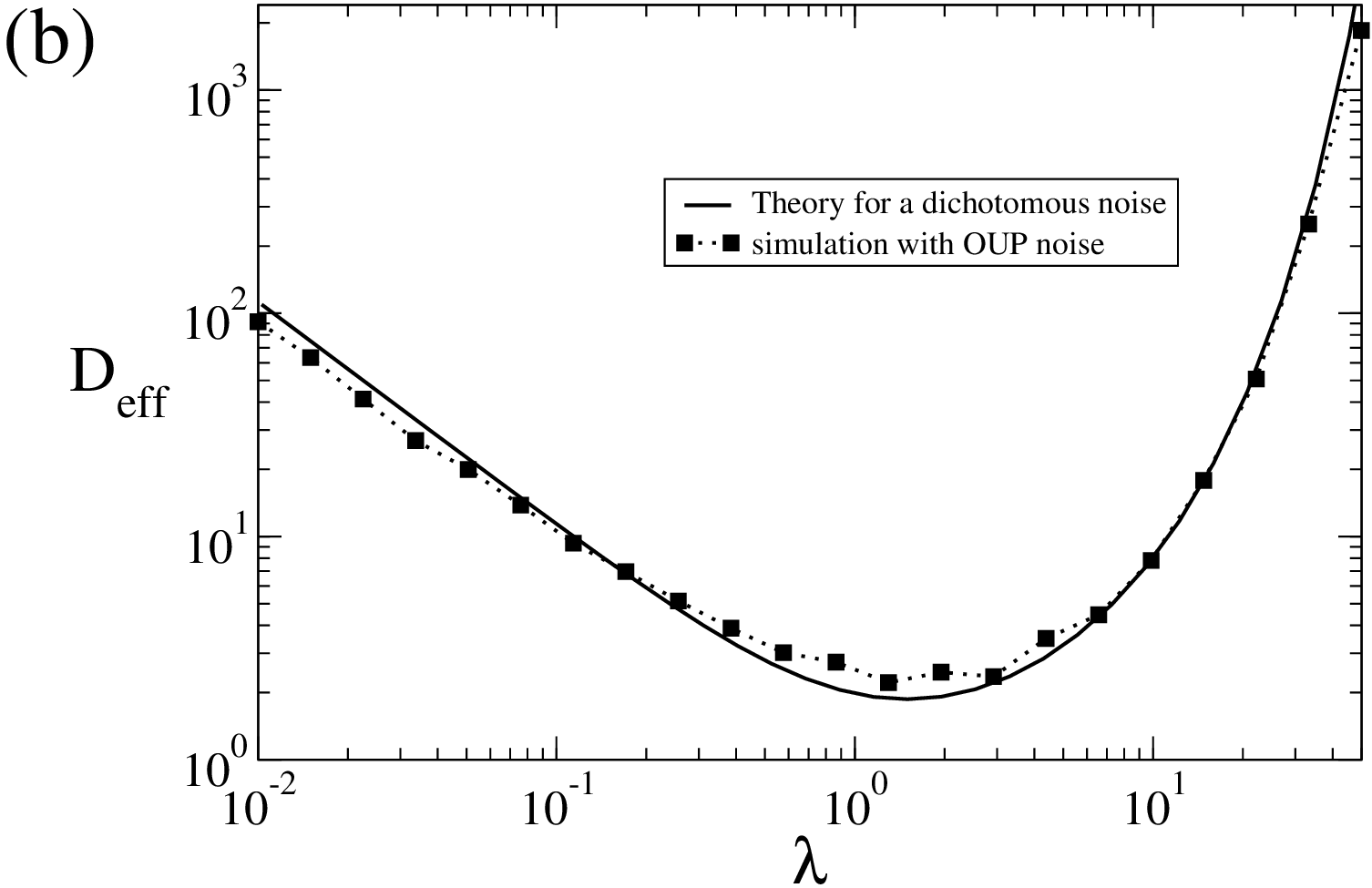}}}
\caption{\small  
Diffusion coefficient vs switching rate of the Ornstein-Uhlenbeck noise for a quintic (a) and a Rayleigh-Helmholtz friction function (b). The theory for the dichotomous process with the same variance and rate (solid lines, \e{benji_diffu_dicho})  is shown for comparison. Modified from \cite{Lin10}. 
\label{fig:benji_deff_oup}}
\end{figure}  

One may wonder how much the above results hinge upon the discrete support of the telegraph noise. 
Surprisingly, for all the cases discussed above it does not make much of a difference if we replace the colored dichotomous noise by a colored Gaussian noise, i.e. by the Ornstein-Uhlenbeck process. In this case,
simulations reveal a minimum in the diffusion coefficient for a quintic friction function (\bi{benji_deff_oup}a) and also  for a for a Rayleigh-Helmoltz friction of an active particle (\bi{benji_deff_oup}b). Even a naive comparison of the OUP simulations to the theory for dichotomous driving shows a rather good agreement  (cf. solid lines in \bi{benji_deff_oup}). At least for a sufficient amplitude of the driving, the variance and the correlation function of the driving noise (statistical features shared by DMP and OUP)  seem to be essential in determining the diffusion coefficient and the exact distribution of their values (a feature that distinguishes DMP and OUP) is immaterial. 
\subsubsection{Particles with asymmetric friction function under white-noise driving: critical asymmetry}
So far we have considered symmetric friction functions and symmetric noise sources for which the mean velocity
vanishes. A perfectly symmetric friction function is not expected in the dynamics of active particles. Assemblies of molecular motors interact with filaments of a certain polarity; if the assemblies are described by an active Brownian particle dynamics then this dynamics will not be symmetric  ($\gamma(-v)\neq \gamma(v)$). 

A simple way to introduce an asymmetry is an external bias. Thus we may consider the Langevin dynamics 
\be
\label{eq:benji_langevin_asymm}
\appel x=v,\;\; \appel v=-\gamma(v) v+F+\sqrt{2 D} \xi(t)
\ee
We can lump both nonlinear friction ("pumping") force and the external bias into the derivative of one effective 
velocity potential. This potential is not symmetric anymore but tilted and for a sufficiently strong force even its bistability will be lost (see \bi{benji_pots_asymm}) . In terms of this potential it is very clear that the asymmetry will have a profound effect on the diffusive properties of the particle. The strong diffusion we observed in the symmetric case and at weak noise  relies on the bimodality of the velocity which will, however, vanish if the velocity potential is not bistable anymore. Thus we can  expect  that the diffusion coefficient becomes very small in the limit of strong force whereas it diverges for finite but small forces. It is, however, not clear what the exact value of the force is where the divergence of the diffusion coefficient vanishes.  One first guess would be the force at which the potential loses its bistability. A closer inspection, however, disproves this conjecture.

\begin{figure}[h!]
 \centerline{\parbox{12.5cm}{\includegraphics[width=12.5cm,angle=0]{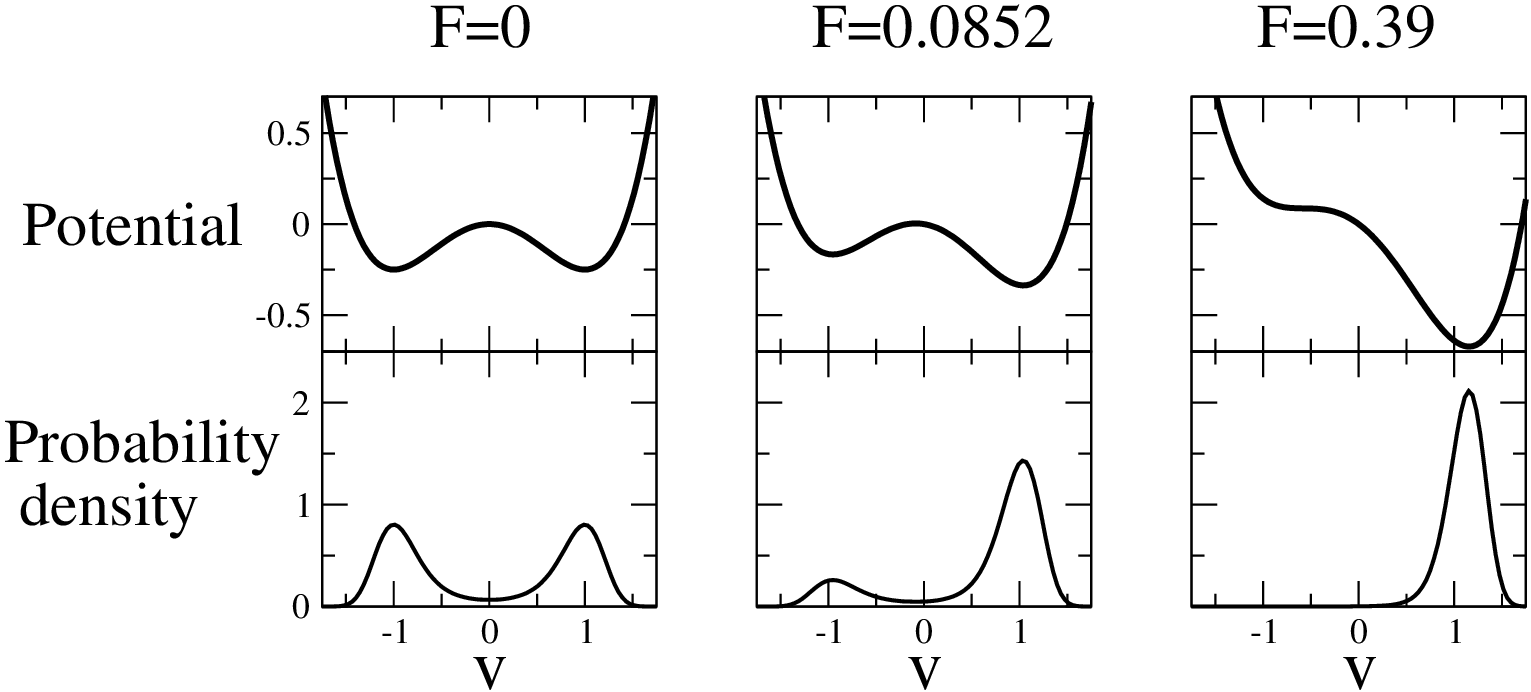}}}
\caption{\small  
Quartic velocity potentials $\Phi(v)=v^4/4-v^2/2-Fv$ (upper row) and the corresponding velocity probability densities $P(v)\sim \exp[-\Phi(v)/D]$ (lower row) for three different values of the external force $F$ as indicated and a noise intensity of $D=0.1$.   
\label{fig:benji_pots_asymm}}
\end{figure}  

First, we can repeat the previous calculation for the calculation of the diffusion coefficient based on the Fokker-Planck equation (see the related calculation for the correlation time in supplement S.9 in \cite{risken_fokker-planck_1996}). The more general result for the diffusion coefficient reads \cite{LinNic08}
\be
\label{eq:benji_deff_asymm}
D_{\rm eff}=\frac{\textstyle\int_{-\infty}^{\infty} dx \; e^{\Phi(x)/D}\!\!\left[\int_{-\infty}^x
dy\; \displaystyle [y-\lr{v}] e^{-\Phi(y)/D}\right]^2}{D \int_{-\infty}^{\infty} dv\; e^{-\Phi(v)/D}}
\ee
Here $\Phi(x)$ is the new effective potential which includes the bias. For weak noise and a bistable potential we can perform a saddle-point approximation of this result and obtain a much simpler formula. The latter can be alternatively derived by a two-state theory that yields a better physical insight.

\begin{figure}[h!]
 \centerline{\parbox{7.5cm}{\includegraphics[width=7.5cm,angle=0]{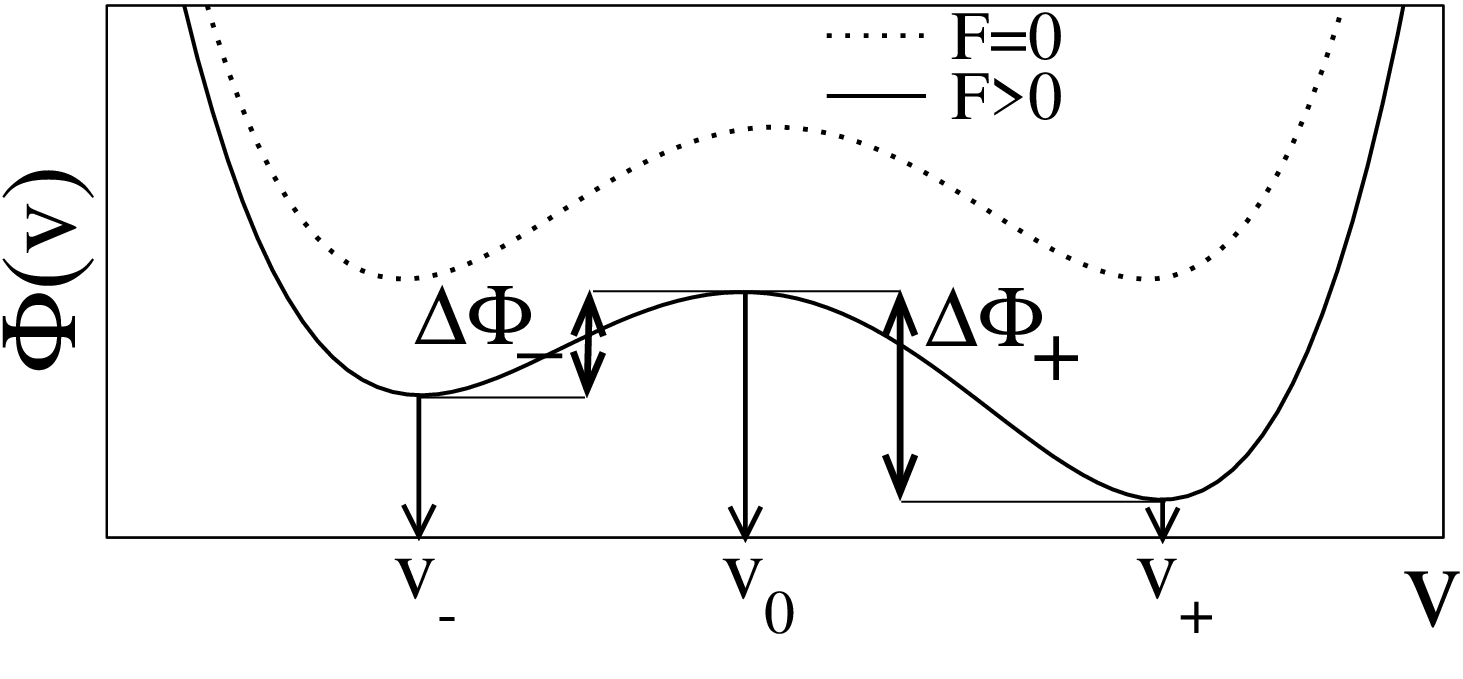}}}
\caption{\small  
Symmetric (dotted line) and asymmetric (solid line) potentials with potential barriers indicated.    
\label{fig:benji_pot_asymm}}
\end{figure} 
In the asymmetric case, the motion of the velocity in the bistable potential is governed by two Kramers rates 
$r_\pm$ for the escape over the larger (smaller) barrier and the velocity obeys the statistics of a dichotomous noise.  Correlation time and variance the product of which forms the diffusion coefficient 
are given by \cite{Gar85}
\be
\tau_{corr}=[r_-+r_+]^{-1},\;\;\;\;  \lr{\Delta v^2}=\frac{(v_+-v_-)^2 r_+ r_-}{(r_++r_-)^2}, 
\ee
in accordance with \e{dich_diff}. We are specifically interested in the limit of vanishing noise intensity for which both rates go to zero. In the general asymmetric case, the rates differ by an exponential pre-factor given by the difference between the two potential barriers, $r_+/r_-\sim \exp[-(\Delta \Phi_+-\Delta \Phi_-)/D]$ and, hence, the correlation time diverges but the 
variance vanishes. The latter finding implies that in the limit of vanishing noise the particles dwells with probability one within the deeper well. Because the diffusion coefficient is the product of variance and correlation time, it is not clear at all when the divergence of the correlation time and when the vanishing  of the variance dominates the diffusion coefficient. Inserting the Kramers formula into the product one obtains the specific formula
\be
\label{eq:benji_kramers}
D_{\rm eff} \! \approx \! \frac{2\pi (v_+-v_-) \omega_+\omega_-/|\omega_0|}{[\omega_-e^{-(2\Delta \Phi_--\Delta \Phi_+)/(3D)}\!+\!\omega_+e^{-(2\Delta \Phi_+-\Delta \Phi_-)/(3D)}]^3},
\ee
which is also obtained by a saddle-point approximation of the exact result \e{benji_deff_asymm}. Here, $\omega_\pm$ are the curvatures in the extrema of the potential. This approximation 
is valid in the limit of weak noise and yields the most interesting result in the thermodynamic limit of vanishing noise
($D\to 0$) that we explore now.  Let us assume without loss of generality that $\Delta \Phi_+$ is the {\em larger} barrier. Then the second exponential in the denominator of \e{benji_kramers} will vanish  rapidly for $D \to 0$. The first exponential, however,  can either go to infinity (resulting in a vanishing diffusion coefficient) or to zero (yielding a divergence in the diffusion coefficient) depending on the sign of the exponent $2\Delta \Phi_--\Delta \Phi_+$. In the case of a mild asymmetry, the larger barrier $ \Delta \Phi_+$  will be still smaller than the double of  the smaller barrier $\Delta \Phi_-$ --- in this case the diffusion coefficient diverges as in the symmetric case. A stronger asymmetry, which does not destroy the potential's bistability but makes the larger barrier greater than the double of the smaller barrier, leads to a positive exponent and thus makes the diffusion coefficient vanish in the limit of zero noise. The critical force by which these two distinct diffusive behaviors are separated is clearly determined by the condition
\be
\label{eq:benji_cond}
\Delta \Phi_+(F_{\rm crit})=2\Delta \Phi_-(F_{\rm crit})
\ee
\begin{figure}[h!]
 \centerline{\parbox{7.5cm}{\includegraphics[width=7.5cm,angle=0]{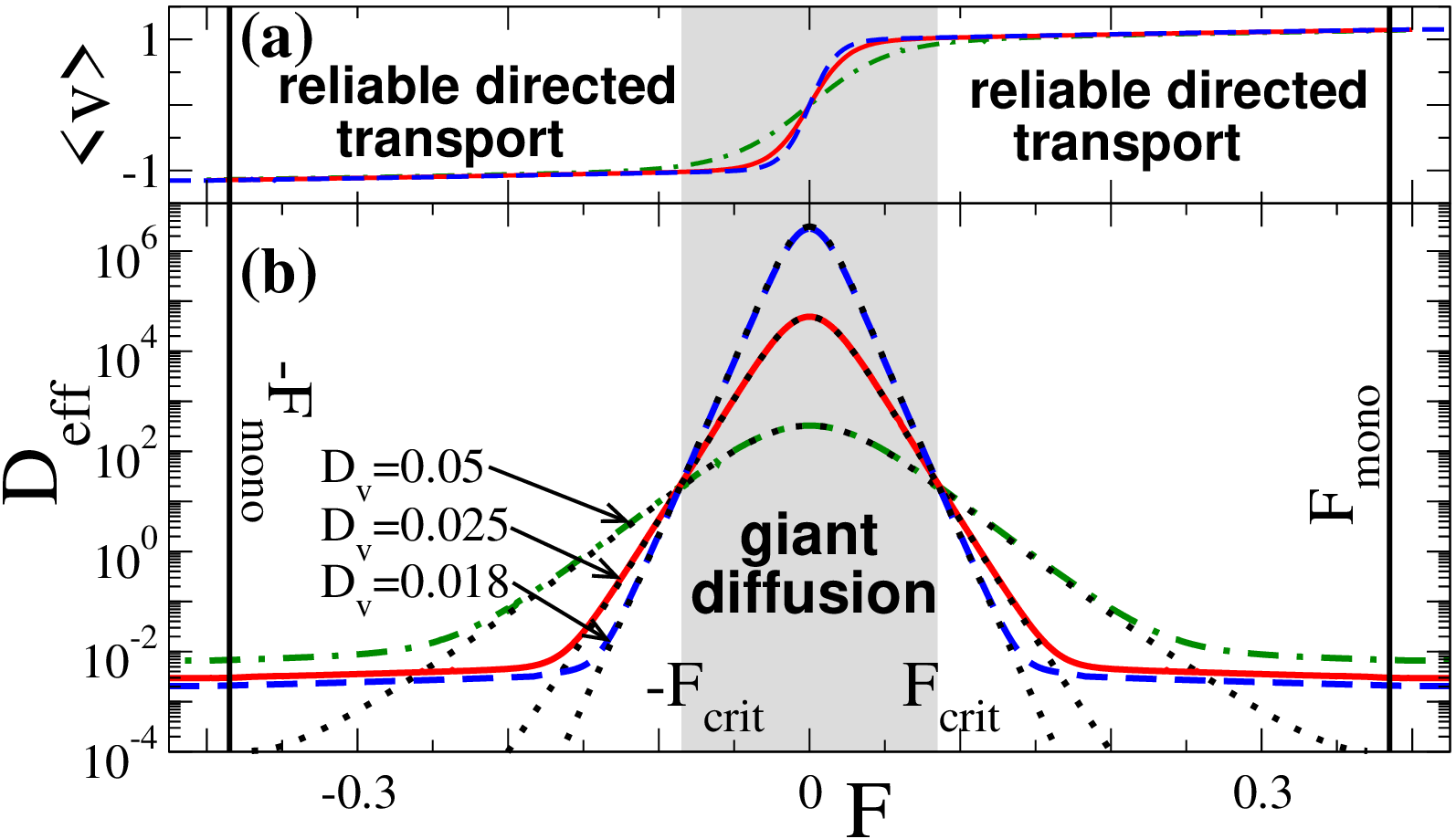}}}
\caption{\small  
Mean velocity (top) and effective diffusion  coefficient (bottom) vs external bias for three different noise intensities 
as indicated. Color lines give the theory from \e{benji_deff_asymm} while black dotted lines is the weak-noise approximation from \e{benji_kramers}.  Modified from \cite{LinNic08}.
\label{fig:benji_deff_asymm}}
\end{figure} 

The critical force (and its symmetric counter part $-F_{\rm crit}$) form the boundary of a region of giant diffusion (shaded region in \bi{benji_deff_asymm}b).  Generally, no transition  is  seen in the mean velocity (shown in \bi{benji_deff_asymm}a). Hence if for strong asymmetry the mean velocity is finite, we are entitled to talk about a parameter region of regular transport (outside the shaded region in     \bi{benji_deff_asymm}).
In  \bi{benji_deff_asymm} the comparison between the full numerical evaluation of the exact solution \e{benji_deff_asymm}
and the weak-noise expansion \e{benji_kramers} reveals good agreement in the region of foremost interest where both potential barriers still exist but can also significantly differ.
 
Note that we have not used  in our arguments above that  $F$ is an additive bias. Thus all conclusions regarding 
the critical asymmetry apply also to the case where the asymmetry controls the barrier heights in a more complicated way; also we may generalize the line of reasoning to the case of an active Brownian particles with multiplicative noise. In all cases we will have a behavior of the diffusion coefficient similar to this shown in \bi{benji_deff_asymm}: There is a finite region of weak asymmetry, for which the diffusion coefficient diverges in the limit $D \to 0$ whereas outside this region the same limit yields a small or vanishing diffusion coefficient. 

In Sec. \ref{sec:molmot_abp} we gave some evidence that the active Brownian dynamics can reproduce the bidirectional motion of  coupled molecular motors. Does this imply that for coupled molecular motors a similar critical asymmetry as found above exists? This question was studied in \cite{LinNic08} and the result of extensive simulations is shown in \bi{benji_deff_motors} for a symmetric (a,b)  or an asymmetric (c,d) motor-filament interaction.  

\begin{figure}[h!]
 \centerline{\parbox{6.5cm}{\hspace{0.5cm}\includegraphics[width=6cm,angle=0]{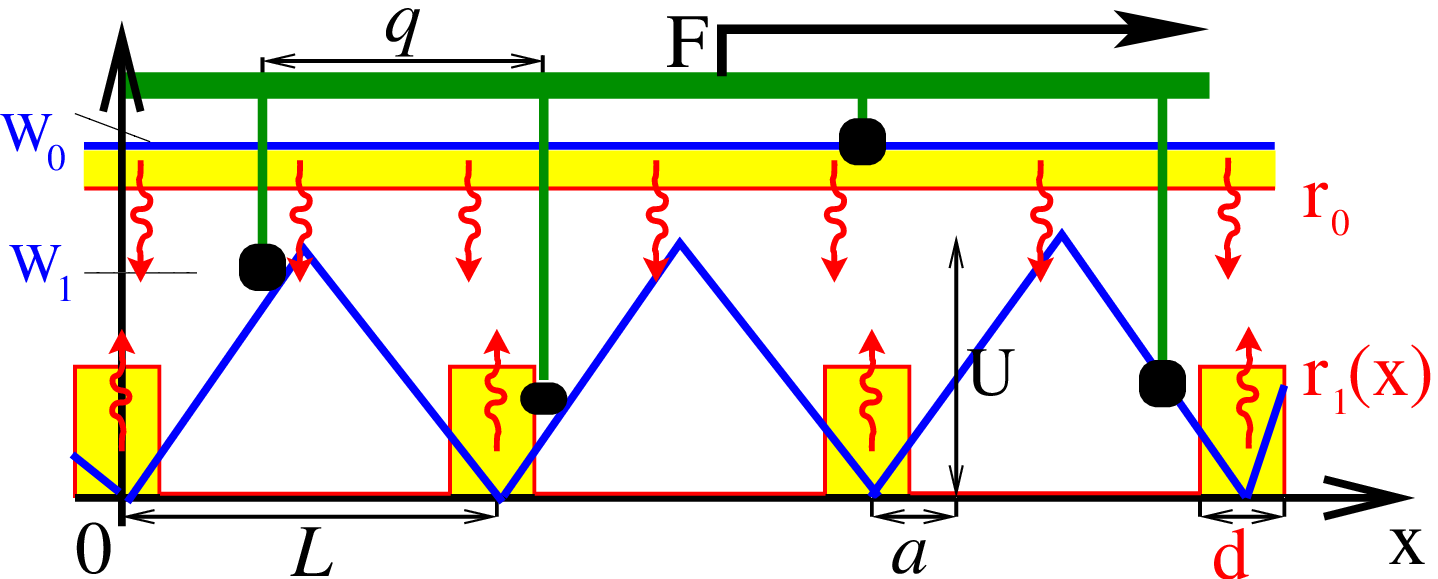}} \hspace{1cm}\parbox{6.5cm}{\hspace{0.5cm}\includegraphics[width=6cm,angle=0]{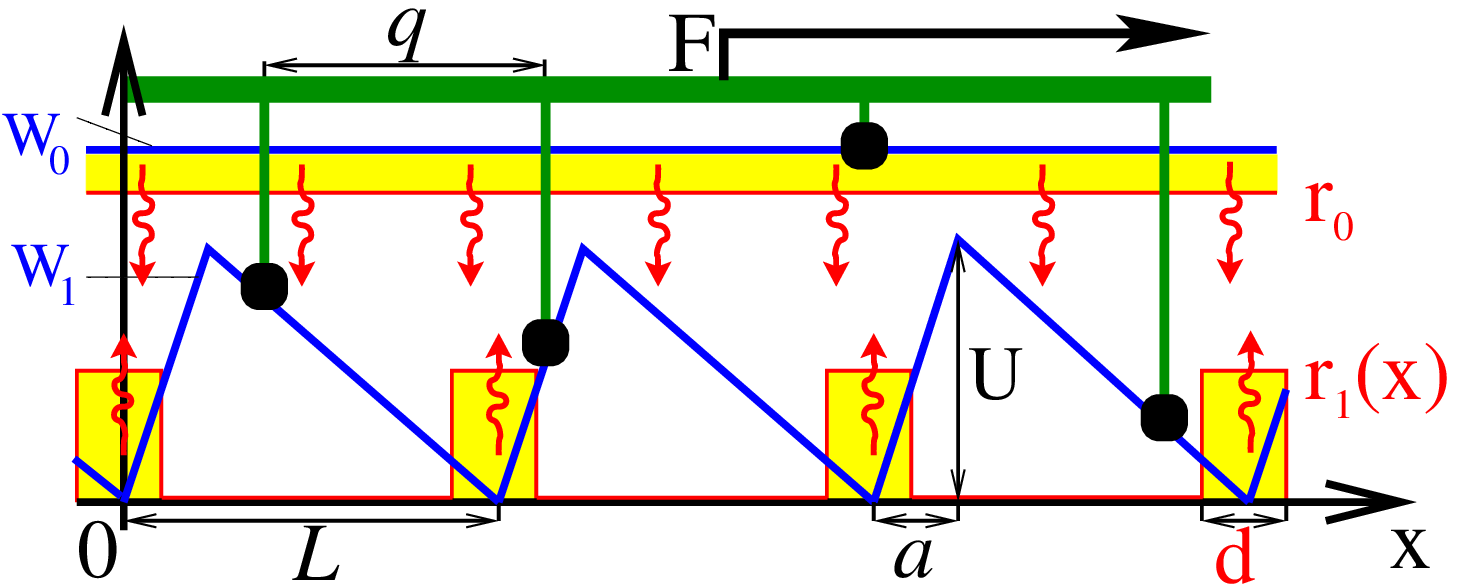}}}
  \centerline{\parbox{6.5cm}{\includegraphics[width=6.5cm,angle=0]{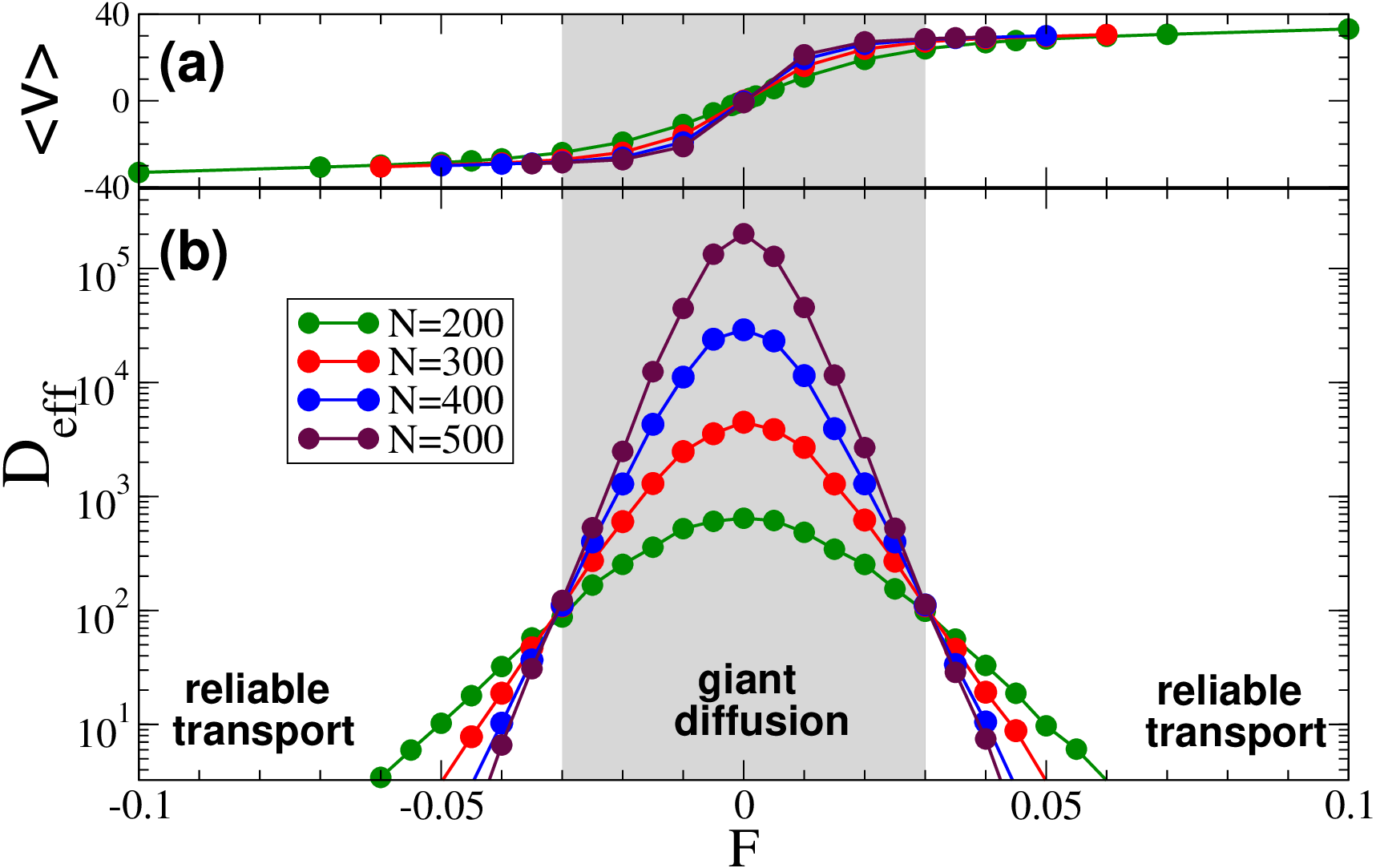}}\hspace{1cm}\parbox{6.5cm}{\includegraphics[width=6.5cm,angle=0]{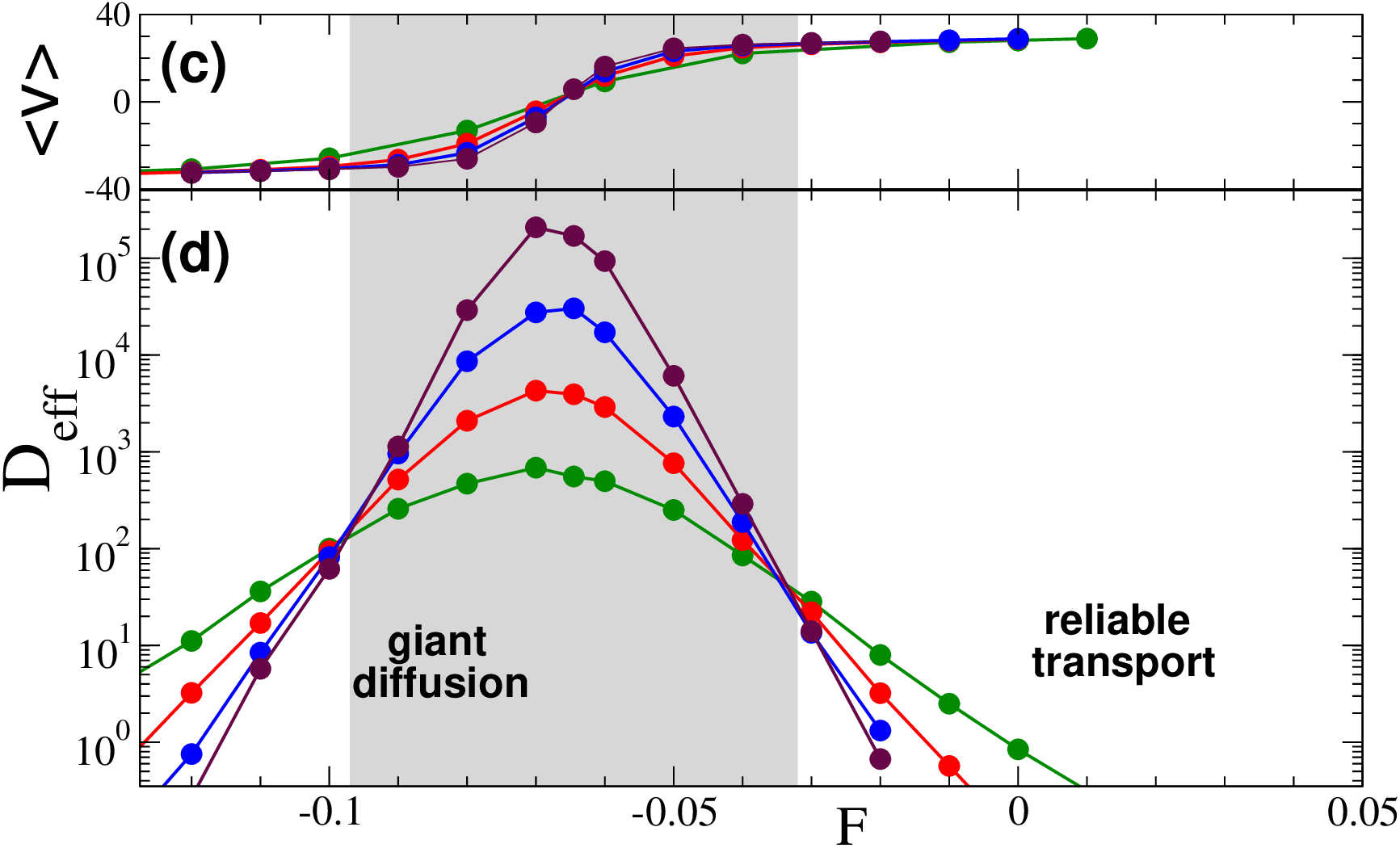}}}
\caption{\small  
Mean velocity (a, c)  and diffusion coefficient (b, d)  of assemblies of molecular motors as a function of the 
external force. (a) and (c) are for a symmetric motor-filament interaction (see the spatial potential  sketched above panels) with $a=0.5$; (b) and (d) are for the asymmetric case with $a=0.4$. Results for various numbers of motors 
as indicated. Modified from \cite{LinNic08}.
\label{fig:benji_deff_motors}}
\end{figure} 

For the motor system noise intensity is set by the inverse number of motors. Indeed, just as in the case of active Brownian particles, diffusion curves for different motor numbers (different noise levels) intersect at critical values 
of the external force. These values are the boundaries for the forcing region of giant diffusion.  If the unbiased system is  spatially symmetric as in \bi{benji_deff_motors}a,b, the dependence of mean velocity and diffusion coefficient  
on $F$ resemble strongly the results for the Brownian motion with Rayleigh-Helmholtz friction. Please note that the latter is also spatially symmetric and that any asymmetry in the system is due to the bias force $F$. If the system, however, shows a spatial asymmetry ($a=0.4$ for the piecewise-linear ratchet potential as shown in Fig. \ref{fig:benji_deff_motors}), both statistics are shifted. At zero force, there is already a finite mean velocity and the motion is rather regular because the diffusion coefficient is small. 
Biasing the motor assembly against its preferred (ratchet-induced) direction, however, results for a finite rage of forces again in a giant diffusion. Roughly speaking, the two kinds of asymmetries add up and maximal diffusion is achieved if the system is rather symmetric, i.e. both directions of motion are roughly equal.   Further support for the similarity of the underlying mechanism for the  critical asymmetry observed for active Brownian particles and for coupled molecular motors  comes from a measure based on the velocity's probability density which can be derived from \e{benji_cond}; for details, see \cite{LinNic08}. 

\subsection{Diffusion in two dimensions}

\subsubsection{Two dimensional random walk with correlated turning angle}
\label{sec:okubo}
Starting from this paragraph we consider diffusion of active particles
in two dimensions. The central aim remains the calculation of the
diffusion coefficient including correlations in motion, respectively,
if he trajectories of the particles exhibit some persistence.

We will start with a discrete random walker in two dimensions which
jumps with a fixed step length $l_0$. It could be modeled by a shot
noise where the sequence of $\delta$-peaks is supplied with constant
weight $l_0$.  As in Eq.(\ref{eq:overdamped}), we restrict in
the consideration to an overdamped situation. The equation of motion
reads.
\begin{eqnarray}
  \label{eq:shot}
  { {\rm d}\over {\rm dt}} \bbox{r}\,=\,  \bbox{\xi}_{S}(t)\,=\, \sum_i^{n(t)} \,l_0 \bbox{e}_{v_i} \, \delta (t-t_i).  
\end{eqnarray}
This $\delta$-kicks change the position vector by $l_0$ pointing in
the direction of the unit vector $\bbox{e}_{v_i}$. The $t_i$ are
generated due to the waiting time density between to strokes
$w(\tau)$. The mean of this shot noise is obviously $1 /< \tau>$ where
$<\tau>$ is the mean time between two spikes. For simplicity, the
occurring jumps shall be also clocked with a given time dispersionless
interval $\tau_0$ \cite{kareiva_analyzing_1983,okubo_diffusion_2001}, i.e.
$w(\tau)=\delta(t-\tau_0)$.

The mean speed is obviously $s=v_0=l_0/\langle \tau \rangle>0$. The unit vector of the
direction $\bbox{e_v}$ is determined by its angle respective to the
$x$-axis which we again label by $\varphi(t)$. We assume that the two
angles between two subsequent steps $i$ and $i+1$ are not
independent. The angle of the velocity vector are shifted by a turning
angle $\eta_i$ for which we assume a given turning angle distribution
$P(\eta)$. The existence of a structured distribution $P(\eta)$
different from a uniform one creates some persistence in the motion.
We will assume that the correlations are weak and that they extend
over one step, only. It implies that the sequence of turning angels
$\eta_1,\eta_2,\ldots, \eta_{i},\ldots $ shall be independent.

This situation was considered by Kareiva and Shigesada and is
discussed in detail in the monograph by Okubo and Levin
\cite{kareiva_analyzing_1983,okubo_diffusion_2001}.  Than the $i$th
displacement vector can be given by:
\begin{eqnarray}
  \Delta \bbox{r}_i \,=\, l_0\, \big(\bbox{e}_x \cos \varphi_i \,+\, \bbox{e}_y \sin \varphi_i\big)\,,  ~~~~~~~\varphi_i\,=\,\varphi_{i-1}+\eta_i
\end{eqnarray}
where $\varphi_i$ is the angle between the x- axis and the direction
of motion. After $n=[t/\tau_0]$ steps the walker has a squared
distance from the point of start of:
\begin{equation}
\label{eq:shig}
\bbox{r}_{n}^2 \,=\,\sum_{i=1}^n \,\Delta\bbox{r}_i \cdot \sum_{j=1}^n \Delta\bbox{r}_j 
\end{equation}
Averaging over an ensemble of walkers and expanding the sum yields:
\begin{eqnarray}
  \langle \bbox{r}^{2}_n \rangle
  &=&  \sum_{i=1}^n \,\langle\Delta \bbox{r}^2_i\rangle\,+\,2 \sum_{i=1}^{n-1}\sum_{j>i}^n \,\langle \Delta \bbox{r}_i\, \cdot\,\Delta \bbox{r}_j\rangle\nonumber\\
  &=& n\,l_0^2 \,+ \,2\,l_0^2 \,\sum_{i=1}^{n-1}\sum_{j>i}^n \,\langle cos(\varphi_i\,-\,\varphi_{j})\rangle\,.\label{taylor-kubo}
\end{eqnarray}
The second term on the r.h.s requires a distribution of the turning
angles $P(\eta)$ which we assume to be symmetric in $\eta$. Then the
mean squared displacement is expressed by the angular correlation
defined as the average $\Gamma$ of the cosine of the turning angle
\begin{eqnarray}
  \label{eq:turning}
  \Gamma\,=\,\mean{\cos \eta} \,=\, \int_{-\pi}^\pi {\rm d}\eta\, \cos \eta\, P(\eta)
\end{eqnarray}
This expression has the physical meaning of how much on average the
length of the unit vector $\bbox{e}_v$ is reduced if projected on the
direction of the previous jump. Since correlations have been considered
only between successive steps we get for two jumps 
\begin {eqnarray}
  \langle\cos(\varphi_i-\varphi_{i+2})\rangle\,=\,\int_{-\pi}^{\pi}
  {\rm d}\eta \,P(\eta)\,\int_{-\pi}^{\pi} {\rm
    d}\eta'\,P(\eta')\,\cos(\eta+\eta')\,=\,\Gamma^2\,,
\end{eqnarray}
It can be shown as well with $k>2$ that
\begin{equation}
  \langle\cos(\varphi_i-\varphi_{i+k})\rangle=\Gamma^k\,.
\end{equation}
So we can solve the sum in (\ref{taylor-kubo}) and get a formula for
the mean squared displacement after $n$ steps of a 2D correlated random walk like in
\cite{okubo_diffusion_2001} and \cite{wu_modelling_2000}, which was derived previously by Kareiva
and Shigesada \cite{kareiva_analyzing_1983}:
\begin{equation}
  \langle \bbox{r}^{2}_n \rangle \,=\, 
  l_0^2 \left( n \frac{1+\Gamma}{1-\Gamma} - 
    2 \Gamma \frac{1-\Gamma^n}{ \left( 1- \Gamma \right)^2 }
  \right) \, \propto \,l_0^2 \, \left(n \frac{1+\Gamma}{1-\Gamma}-\frac{2\Gamma}{\left(1-\Gamma\right)^2}\right)\,. \label{limit_kareiva}
\end{equation}
if $n \to \infty$ since $|\Gamma| < 1$. In this limit the second term
on the right hand side of (\ref{limit_kareiva}) becomes time
independent for large $n$.

Therefore the diffusion for short times anomalous at at long times is
normal with coefficient
\begin{equation}
  D_{\rm eff}  =\frac{1+\Gamma}{1-\Gamma}\,\frac{l_0^2}{4\,\tau_0} \,. \label{D_gamma}
\end{equation}
The second term in the sum can be interpreted as the square of a
characteristic length scale. If this length scale is much smaller than
the length scale of the process itself, normal diffusion approximation
holds.

As can be seen from Eq.~(\ref{eq:turning}) $\Gamma$ vanishes if $\eta$
is equally distributed. This corresponds to the motion of a freely
diffusing particle without any directional correlations between the jumps.  We see also
that a persistence of the trajectory with $\Gamma$ larger than zero
enhances diffusion. An negative sign of the correlation parameter
decreases the diffusion coefficient.

We will also have a look at the diffusion coefficient without
correlations $\Gamma=0$. One can replace the jump length by the 
duration of the jump and the velocity, i.e. $l_0=v_0 \tau_0$. Afterwards we set $\tau_0$ equal to the relaxation time for the angular fluctuations from Eq. \eqref{eq:cart_angle}, i.e. $\tau_0=\tau_{\varphi}= D
/v_0^2$. Put together we obtain that the effective diffusion
coefficients behaves as $\sim v_0^4/D$ as we will find several times
from the continuous theory of Active Brownian particles, later on
\cite{mikhailov_self-motion_1997}. It means, surprisingly, that
increasing noise decreases the value of the diffusion coefficient.

An application to particular situation of a hopping Daphnia motion was
given in \cite{komin_random_2004,garcia_optimal_2007}. Therein
specific angle distribution, like exponential and bimodal ones have
been assumed in order to control the diffusion coefficient.

One has to add that the same way of calculation of the diffusion
coefficient was used by R. F\"urth \cite{fuerth_brownsche_1920} who
considered persistent random walk in one dimension. He assumed the
probabilities $p$ and $1-p$ that the particles jumps after $t_0$ the
length $l_0$ preferentially in the same direction or in the opposite
one, respectively. Such way he formulated correlations between
successive hops. In consequence, one can also find the probabilities 
for all longer sequences of possible jumps and also the
correlations between over next hops are included.  If summing up the
all jumps in \eqref{taylor-kubo} one determines the mean squared
displacement in one dimension.

One limit is the case with $p=1/2$, i.e. no preference in the angle is
assumed and the diffusional motion is free and he diffusion
coefficient equals $l_0^2/(2\tau_0)$.  In contrast in the limit $p \to
1$ and $\tau_0 \to 0$ with $(1-p)/\tau_0=\gamma/2=const$ the diffusion
coefficient of a Brownian particles with inertia and $m=1$ likewise in
\eqref{eq:squareddispl} is obtained. The correlation time
$\tau_{corr}=1/\gamma$ quantifies the persistence length $l_{pers}=v_0
\tau$ with assumed fixed velocity $v_0=l_0/\tau_0$.

\subsubsection{Diffusion for independent velocity and heading dynamics}
\label{sec:diffhigh}
\label{sec:diff_independent}
In higher dimensions, the diffusion of  active Brownian particles  becomes more complicated and is only in a few cases analytically tractable. A popular and reasonable approximation in this case is the assumption that velocity and direction of motion are independent variables; this will be presented here, following the work by Peruani and Morelli \cite{peruani_self-propelled_2007}. 

In general the motion of an active Brownian particle in arbitrary dimensions $d$ can be described by its  a unit vector describing the direction of motion (heading) ${\bbox e}_h(t)$ at time  $t$ and the velocity with respect to the heading direction (see Sec. \ref{sec:internalcoord}).

In two spatial dimensions, with ${\bbox e}_h(t)=(\sin(\phi(t),\cos\phi(t))$,  assuming completely independent dynamics in $v$ and $\phi$ (heading angle) the evolution of the velocity vector ${\bbox v}=v(t){\bbox e}_h(t)$ is determined by the following stochastic differential equations:
\begin{align}\label{eq:general_langevin}
\dot  v   & =  -\gamma(v)v+{\eta}_{v}(t) \\
\dot \phi & =  {\eta}_{\phi}(t) \label{eq:ind_phidyn}. 
\end{align}
The first term in the velocity equation is again an arbitrary velocity dependent friction function which describes the deterministic evolution of the velocity. The second term is the stochastic part with $\eta_v$ being the component of random force acting on the velocity. The evolution of the heading angle is assumed to consist only of a stochastic torque acting on $\phi$.
In \eqref{eq:ind_phidyn}, it is assumed that the angular velocity (or turning rate) does not dependent on the velocity $v$, and we should note that it is in general not the case, in particular, if we consider turning behavior with inertia.  

In general the fluctuations ${\eta}_{v,\phi}(t)$  can obey arbitrary distributions and temporal statistics. Here, we consider independent and uncorrelated Gaussian (white) noise terms, based on our ansatz for active fluctuations (Sec. \ref{sec:exter})
\begin{align}
{\bbox {\eta}}_{v}=\sqrt{2 D_v}\xi_v, \quad {\eta}_{\phi}(t) = \sqrt{2 D_\phi} \xi_\phi,
\end{align}
which is, based on the center limit theorem, a reasonable approximation of many random processes occurring in nature. 

For independent $v$ and ${\phi}$-dynamics the joint probability distribution my be decomposed as $P(v,{\phi},t)=P_v(v,t)P_\phi({\phi},t)$. Thus the expectation value for the displacement $\langle {\bbox r}(t)\rangle$ reads
\begin{align}
\langle {\bbox r}(t) \rangle = \int_0^t dt' \langle v(t'){\bbox e}_h(t') \rangle = \int_0^t dt' \langle v(t') \rangle \langle {\bbox e}_h(t') \rangle
\end{align}
In the absence of a preferred direction of motion (uniform distribution of $\phi$), the time average and ensemble average with respect to ${\bbox e}_h$ vanishes and so does the mean displacement $\langle {\bbox r}(t)\rangle=0$.

The mean square displacement $\langle {\bbox r}^2 \rangle(t)$ in the absence of directional bias can be obtained according to Eq. \eqref{eq:msd_gen_taylor_kubo} from the velocity-velocity correlation function. Here $C_{\bf v v}(\tau)$ decomposes into independent $vv$ and ${\bbox e}_h{\bbox e}_h$-correlation functions:
\begin{align}\label{eq:msd_vvhh}
\langle {\bbox r}^2(t) \rangle = \int_0^t dt' \int_0^{t'} d\tau \langle {\bbox v}(0){\bbox v}(\tau) \rangle = \int_0^t dt' \int_0^{t'} d\tau \langle {v}(0){v}(\tau) \rangle \langle {\bbox e}_h(0){\bbox e}_h(\tau) \rangle
\end{align}
Assuming an exponential decay of the velocity correlations we may rewrite the velocity correlations in the stationary case as
\begin{align}\label{eq:vv-corr}
\langle {v}(0){v}(\tau) \rangle & =  \langle v^2 \rangle e^{-\kappa_v\tau} +  \langle v \rangle^2(1 - e^{-\kappa_v \tau})
\end{align}
with $\langle v \rangle$ and  $\langle v^2 \rangle$ being the first and second moment of the stationary velocity distribution $p_s(v)$ and $\kappa_v$ being the relaxation rate (inverse relaxation time). 

The angular dynamics in Eq. \eqref{eq:ind_phidyn} results in a free diffusive motion of ${\bbox e}_h$ on a 
unit circle, with an exponential decay of autocorrelations:
\begin{align}\label{eq:hh-corr}
\langle {\bbox e}_h(0){\bbox e}_h(\tau) \rangle & =  e^{-\kappa_h \tau},
\end{align}
with $\kappa_h$ being the correlation decay rate of the direction of motion. In the case of free diffusion of $\phi$ it is equal to the angular diffusion coefficient: $\kappa_h= D_\phi$.

Inserting Eqs. \eqref{eq:vv-corr} and \eqref{eq:hh-corr} in Eq. \eqref{eq:msd_vvhh} yields finally \cite{peruani_self-propelled_2007}:
\begin{align}\label{eq:peruanimorelli}
\langle {\bbox r}^2(t) \rangle  = & \ 2\frac{\langle v \rangle^2}{\kappa_h}\left[t+\frac{1}{\kappa_h}\left(e^{-\kappa_h t}-1\right)\right] \nonumber \\
 & + 2\frac{\langle v^2 \rangle-\langle v \rangle^2}{\kappa_v+\kappa_h}\left[t+\frac{1}{\kappa_v+\kappa_h}\left(e^{-(\kappa_v+\kappa_h) t}-1\right)\right]
\end{align}
The first term describes the contribution of the mean square displacement due to self-propelled motion with (constant) mean velocity $\langle v \rangle$ and a stochastic direction of motion, whereas the second term takes into account the impact of velocity fluctuations on the mean square displacement.

For particles moving with constant speed ($\langle v \rangle = v_0 =const.$) there exist two different scaling regimes of $\langle {\bbox r}^2 (t) \rangle$ with $t$ similar to ordinary Brownian motion: The so-called ballistic regime at short times with $\langle {\bbox r}^2 (t) \rangle \propto t^2$ and the diffusive regime with  $\langle {\bbox r}^2 (t) \rangle \propto t$ for $t\to\infty$. Between the two regimes there is a crossover at time $t\sim \kappa_h^{-1}$.

The introduction of an additional time-scale due to velocity fluctuations may lead to a mean square displacement with four distinct regimes and three crossovers in between.  For large time scale separation: $(\kappa_h+\kappa_v)^{-1} \ll \kappa_h^{-1}$ the angular correlation decays much slower than the velocity correlations ($\kappa_h/\kappa_v \ll 1$).  Thus starting from its initial position at $t=0$ the particle performs first a quasi one-dimensional motion along its initial direction of motion. The stochastic decorrelation of the velocity in this effective one-dimensional motion leads to a first crossover from ballistic to diffusive motion. After the velocity dynamics reach a stationary state, what matters is only the average motion with the mean velocity $\langle v \rangle$. This regime corresponds to a self-propelled particle with constant speed performing (still) effectively a one-dimensional motion with $\langle {\bbox r}^2 \rangle\propto t^2$. As a consequence a second crossover can be observed from the transient diffusive regime to a second ballistic regime. Finally, in the limit of large times $t\to\infty$, the slow decorrelation of the direction of motion leads to the third and last crossover towards the final diffusive regime. 

The diffusion coefficient of the limiting diffusive regime reads (Eq. \ref{eq:deff_Dd}):
\begin{align}\label{eq:diff_per_mor}
D_\text{eff} = \frac{1}{d}\left( \frac{\langle v \rangle^2}{\kappa_h} + \frac{\langle v^2 \rangle-\langle v \rangle^2}{\kappa_v+\kappa_h} \right).
\end{align}
This results, as well as Eq. \eqref{eq:peruanimorelli}, holds for arbitrary dimensions $d=1,2,3$ as long as the heading correlations decay exponentially and are governed by a single time scale given by $\kappa_h^{-1}$.

An additional bias (e.g. external force), may even introduce yet another time-scale and in this case even five crossovers are possible \cite{peruani_self-propelled_2007}. This complicated behavior of the mean squared displacement can be an explanation for experimental observations of superdiffusive motion of active particles \cite{wu_particle_2000,dieterich_anomalous_2008}: several crossovers and averaging over nonidentical active particles may lead to observations of mean squared displacements, which appear to be in between ballistic and diffusive motion.    

\subsubsection{Diffusion of active particles with passive fluctuations}
\label{sec:diff_gruler}
In contrast to the previous section, we consider now active particles with correlated fluctuations in the velocity and direction of motion, due to only passive (or external) fluctuations acting simultaneously on both degrees of freedom as introduced in Sec. \ref{sec:SGpassive} in Eq. \eqref{eq:sg2d_w_PF_AF} with $D_v,D_\phi=0$. At low external noise intensities $D$ the velocity distribution for an arbitrary friction function is given by a narrow peak around
the stationary velocity $v_0$ and as a first approximation the velocity can be assumed as
constant: $v=v_0=const$. The external (passive) fluctuation act only on the direction of motion. In this limit the velocity correlation function reads
\begin{align}\label{eq:corr_vconst}
  \langle v(0)v(\tau) \cos(\varphi(\tau)-\varphi(0))\rangle\,=\, v_0^2
  \, \mean{\cos(\varphi(\tau)-\varphi(0))}\,=\, v_0^2
  \exp\left(-\frac{D}{v_0^2}\tau\right)\,.
\end{align}
Inserting Eq. \eqref{eq:corr_vconst} in Eq. \eqref{eq:peruanimorelli} gives us
the mean square displacement:
\begin{align} 
\label{eq:msd_gruler}
  \mean{\Delta {\bbox r}^2(t)}=2\frac{v_0^4}{D}\left[
    t+\frac{v_0^2}{D}\left(\exp
      \big\{-\frac{D}{v_0^2}t\big\}-1\right)\right]
\end{align}
This is a well known result obtained previously by Meink\"ohn and
Mikhailov \cite{mikhailov_self-motion_1997}.

The corresponding long time effective spatial diffusion coefficient is
inversely proportional to $D$ and reads
\begin{align}
  \label{eq:DMM}
  \DMM=\lim_{t\to\infty} \frac{\mean{\Delta {\bbox r}^2(t)}}{4t} = \frac{v_0^4}{2
    D}.
\end{align}
Please note the different power-law dependence of $\DMM$ on $v_0=\langle v \rangle$ in comparison to \eqref{eq:diff_per_mor} for vanishing velocity fluctuations ($\langle v^2 \rangle - \langle v \rangle^2 =0$). This is due to the fact that in Eq. \eqref{eq:sg2d_w_PF_AF} active motion with inertia is considered.

In general, the diffusion coefficient of self-propelled particles ($v\approx v_0=const.$), decreases with increasing noise strength $D$ in contrast to ordinary Brownian motion. In the limit $D=0$ the fluctuations in the direction of motion vanish and the self-propelled particles moves with constant velocity along its initial direction of motion. The mean squared displacement increases as $\sim t^2$ and we observe only ballistic motion - no diffusion. Thus, for $D\to0$ the effective diffusion coefficient diverges. On the other hand, with increasing $D$ the fluctuations in the direction of motion increase --- the particle changes its direction of motion with increasing frequency. As the velocity does not increase with $D$ this leads effectively to a localization of the particle and decreasing diffusion. Thus, for active particles moving with a constant speed the diffusion coefficient vanishes $\DMM\to0$ as $D\to\infty$. 

For the linear Schienbein-Gruler model (see Sec. \ref{sec:act}) the above 
low-noise limit corresponds to $D/\gamma_0
\ll v_0^2$. Thus in the limit $D \to 0$ the effective diffusion
coefficient will converge to Eq.  \ref{eq:DMM}.  However, in
the limit of large noise intensities $D / \gamma_0 \gg v_0^2$ we may
neglect the active motion term in the Schienbein-Gruler friction and put $
v_0\,=\,0$. The dynamics reduces to ordinary Brownian motion with the
effective diffusion coefficient $\DBM=D/\gamma_0$ which increases
linearly with $D$.

Based on the two asymptotic limits it becomes evident that there must
exist a minimum of the effective diffusion coefficient. A crude
approximation for $D_\text{eff}$ can be obtained by a sum of the two
asymptotic diffusion coefficients:
\begin{align}\label{eq:SG2d_ext_DI}
D_\txr{I}=\DMM+\DBM=\frac{v_0^4}{2D}+\frac{D}{\gamma^2}.
\end{align}
This approximation has the right asymptotic behavior and reproduces
qualitatively the behavior of $D_\text{eff}$ at intermediate noise
strengths $D$. But a comparison with the numerical results reveals
that this approximation underestimates the diffusion coefficient close
to the minimum.

We attempt to improve the approximation of $D_\text{eff}$ by considering
the velocity drift term in the Fokker-Planck Equation
\eqref{eq:SG2d_FPEext} with $v_0>0$. The most probable velocities
$\tilde v$, corresponding to the maxima of the velocity probability distribution are given as
roots of the drift term according to the Stratonovich interpretation:
\begin{align}
  -\gamma_0(\tilde v-v_0)+\frac{D}{\tilde v} = 0.
\end{align}
By multiplying with $\tilde v$ we obtain a quadratic equation for
$\tilde v$ with the roots:
\begin{align}
  \tilde v_{+/-} =
  \frac{v_0}{2}\pm\sqrt{\frac{v_0^2}{4}+\frac{D}{\gamma_0}}.
\end{align}
The positive root corresponds to the maximum at positive velocities
close to $v_0$ whereas the negative root corresponds to the maximum at
negative velocities (backwards motion). At low $D$ the backwards motion
may be neglected and the most probable velocity is given by the
positive root $\tilde v_+$. Inserting $v=\tilde v_+$ in $\DMM$
\eqref{eq:DMM} gives us qualitatively the right behavior of the
diffusion coefficient with a minimum at intermediate $D$ but does not
reproduce the correct asymptotic for $D\to \infty$. In order to
eliminate this deviation we add a correction term $D/(2 \gamma_0^2)$ and
obtain a second approximation as:
\begin{align}\label{eq:SG2d_ext_DII}
D_\txr{II}=\frac{\left(\frac{v_0}{2}+\sqrt{\frac{v_0}{4}+\frac{D}{\gamma_0}}\right)^4}{2D}+\frac{D}{2\gamma_0^2}.
\end{align}
A comparison of $D_\txr{II}$ with numerical results shows that it
offers a better approximation then the $D_\txr{I}$ but overestimates
the diffusion coefficient close to the minimum.

\begin{figure}
  \begin{center}
    \includegraphics[width=\linewidth]{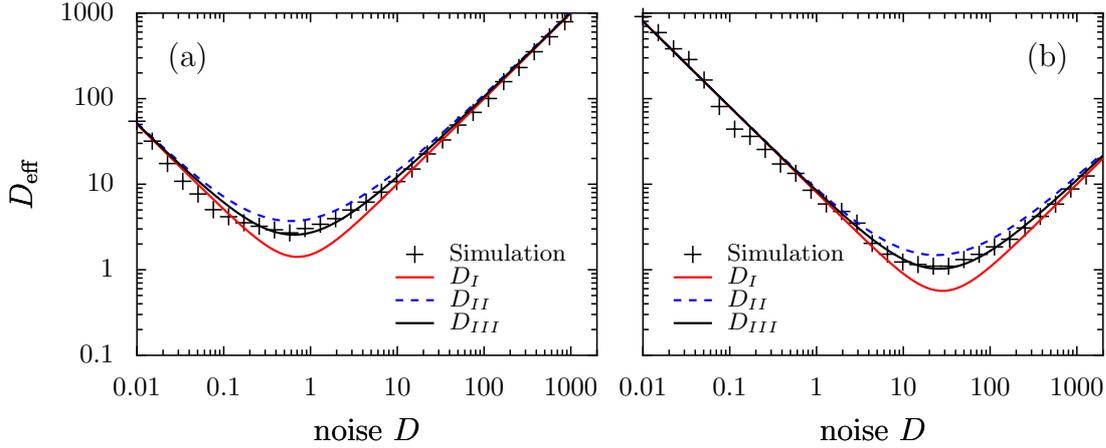}
  \end{center}
  \caption{Effective diffusion coefficient of Schienbein-Gruler model with external
    noise versus noise intensity $D$ for two different parameter sets:
    (a) $\gamma_0=1.0$ and $v_0=1.0$, (b) $\gamma_0=10.0$ and $v_0=2.0$.
    Comparison of numerical results with the analytical approximations
    $D_I$ \eqref{eq:SG2d_ext_DI}, $D_{II}$ \eqref{eq:SG2d_ext_DII} and
    $D_{III}$ \eqref{eq:SG2d_ext_DIII}.  }
\label{fig:SG2d_ext_Deff_vs_D}
\end{figure}

The two approximations appear to provide a lower and an upper bound of the
effective diffusion coefficient close to the minimum. We obtain a third approximation by taking
the average of $D_\txr{I}$ and $D_\txr{II}$.
This heuristic ansatz does not yield any additional qualitative insights but results in an analytical expression for $D_\text{eff}$ 
with a good agreement to numerical simulation:
\begin{align}\label{eq:SG2d_ext_DIII}
D_\text{eff} \approx D_\txr{III} & =\frac{1}{2}\left(D_\txr{I}+D_\txr{II}\right) \nonumber \\
& =\frac{3D}{4\gamma_0^2}+\frac{v_0^4}{4D}+\frac{1}{64 D}\left(v_0+\sqrt{v_0^4+\frac{4 D}{\gamma_0}}\right)^4.
\end{align}
In Fig.~\ref{fig:SG2d_ext_Deff_vs_D} we show a comparison of the three
analytical approximations for $D_\text{eff}$. Please note that at low noise
intensities all approximations seem to yield systematically larger
values of $D_\text{eff}$ then the numerical simulations. This can be associated 
with the coupling of the effective angular diffusion to the
velocity dynamics, which have not been taken into account correctly.

\subsubsection{Turning angles as Gaussian Ornstein-Uhlenbeck process}
Another model to include correlations of the trajectories starts with
equation of motion for the angle dynamics (See Eq. \ref{eq:rotation}).
So far in case of constant speed $v_0$ and a continuous angle changes
we widely have used that the turning angle is a Wiener process and,
respectively, its angular velocity Gaussian white noise.  The
corresponding solution of the FPE is simply formulated from Eq.
\eqref{Gauss1} putting $\Omega=0$ or from Eq. \eqref{eq:wrapped} if
periodically wrapped.

The situation without preferred turning angle (see Eq. \ref{eq:speed})
can be generalized to a correlated Ornstein-Uhlenbeck process
\cite{uhlenbeck_theory_1930,chandrasekhar_stochastic_1943,horsthemke_noise-induced_1984,haenggi_colored_1995}
as a model of possible correlations. Some exemplary models of
correlated dynamics are found in
\cite{sancho_analytical_1982,lekkas_stochastic_1988,schwalger_interspike_2008}.
Here, applied to angular dynamics, the corresponding angular velocities
have vanishing mean but possess Gaussian deviations correlated over a
characteristic time $\tau_C $. The corresponding dynamics for the
angular velocity is defined by the system of stochastic differential
equation
\begin{eqnarray}
  \label{eq:ornstein}
  \dot{\varphi}\,= \,{1 \over v_0} \, \Omega(t)\,, 
  ~~~\dot{\Omega}\,\,=\, - \,\kappa \Omega \,+\,\sqrt{2 D_{\Omega}}\,\xi(t)\,.
  \nonumber
\end{eqnarray}
The coefficients $\kappa$ plays the role of the inverse of the
correlation time $\tau_C$ of the angular velocity.
\begin{figure}[htb]
  \centering
  \includegraphics[width=0.7\linewidth]{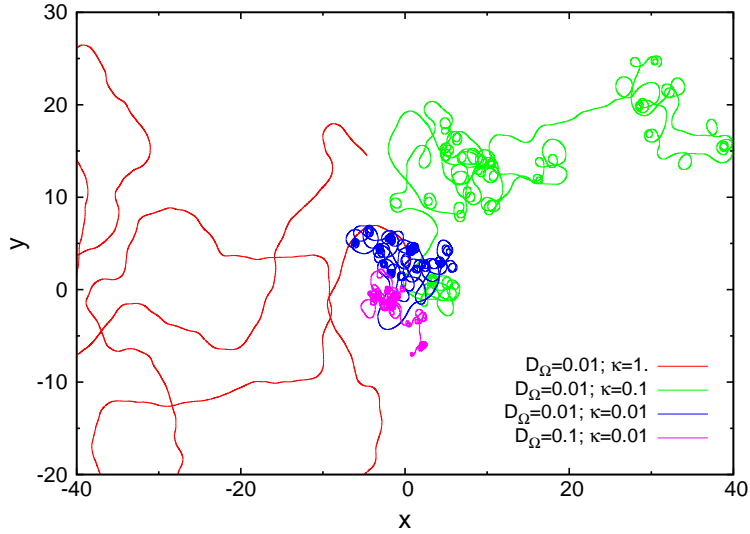}
  \caption{Trajectories of particles with constant speed and
    correlated turning angles for different values of correlation.
    Other parameters: $v_0=1$.\cite{weber_diffusion_2010}}
  \label{fig:traj_oup}
\end{figure}

Considering unwrapped angles with $\varphi\in[-\infty,\infty]$, the angular dynamics resemble the
equations of motion of a Brownian particle with Stokes friction
$\propto \kappa$ and noise intensity $D_{\Omega}$ (compare Eq.
\eqref{langev-v}).  The solution of this Langevin equation is known
since the seminal paper by Uhlenbeck and Ornstein
\cite{uhlenbeck_theory_1930} which was later generalized by
Chandrasekhar to motion in a higher dimensional space
\cite{chandrasekhar_stochastic_1943}. From the time-dependent solution
of the FPE, which is a Gaussian in the angle $\varphi$ and in the
angle velocity $\Omega$, we obtain, for a given initial angular
velocity $\Omega_0$, the transition probability of the angle as
\cite{uhlenbeck_theory_1930}
\begin{eqnarray}
  P(\varphi_2,\tau|\varphi_1,\Omega_0,0)\,&=& \,\sqrt{\frac{\kappa^3 v_0^2}{2\pi D(\tau)}}\\
&&\,\exp \Big(-\frac{ k^3 v_0^2}{2D(\tau)}\big(\varphi_2-\varphi_1-\frac{\Omega_0}{\kappa}(1-\exp(-\kappa\tau)) \big)^2\Big)\,. \nonumber
\end{eqnarray}
Therein the time dependent mean square increment of the angle reads
\begin{eqnarray}
  D(\tau)\,=\, D_{\Omega}\Big(2\kappa \tau -3 +4 \exp( -\kappa \tau) - \exp( -2\kappa \tau)  \Big) \nonumber
\end{eqnarray}
which exactly agrees in the unwrapped approximation with the spatial
diffusion coefficient calculated by Langevin for the mean squared
displacement \eqref{eq:squareddispl}.

And thus the increment in the angle is a Gaussian distributed random
variable. It simplifies the calculation of the mean projection along
the former path according to \eqref{eq:corr_vconst} which yields
\begin{eqnarray}
\label{eq:two_exp} 
  \langle \cos \Delta \varphi(\tau)\rangle \,=\, \exp \Big(-\frac{D_{\Omega}}{v_0^2\kappa^3}(\kappa \tau -1 +\exp( -\kappa\tau) )\Big)
\end{eqnarray}
\noindent
and results in the mean squared displacement of the particle
\begin{eqnarray}
  \label{eq:mean_oup}
  \langle \Delta \bbox{r}^2(t) \rangle \,=\, 2 \,v_0^2\, \int_0^t (t-\tau)
  \langle \cos \Delta \varphi(\tau)\rangle\,{\rm d}\tau. 
\end{eqnarray}
The expression with double exponentials \eqref{eq:two_exp} can be
evaluated either numerically in the general case or analytically in limiting cases \cite{weber_diffusion_2010}. 
\begin{figure}[htb]
  \centering
  \includegraphics[width=0.7\linewidth]{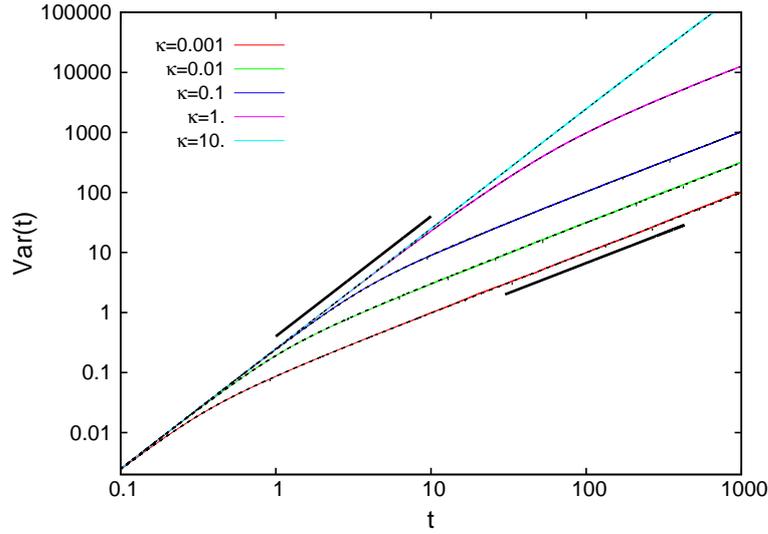}
  \caption{Mean squared displacement of a particle with constant speed
    and correlated angular velocity for different inverse correlation
    times $\kappa$.\cite{weber_diffusion_2010}}
  \label{fig:diff_oup}
\end{figure}
We observe that it starts with a ballistic growth $\propto t^2$ and
crosses over to a diffusional regime $\propto t$. One
can easily verify that in the two temporal limits
\begin{eqnarray}
  \lim_{\tau \to 0} {\rm {d \over dt}} \langle \, \langle \Delta \bbox{r}^2(t) \rangle \,=\, 2 \, v_0^2\, \tau\,. 
  \end{eqnarray}
and 
\begin{eqnarray}
  \lim_{\tau \to \infty} {{\rm d} \over { \rm d}\tau}  \langle \Delta \bbox{r}^2(t) \rangle \,=\, 2 v_0^2 \int_0^\infty \exp \left[ - \frac{D_{\Omega}}{v_0^2 \kappa^3}\Big(\kappa \tau -1\Big)\right] {\rm d}\tau \,=\, {\rm const}  
\end{eqnarray}
The crossover times between both regimes depend on the coupling. Let
$\tau_1 = v_0 \sqrt{\kappa/D_\Omega}$. For large correlation times
$\tau\,=\,1/\kappa \,\ge \, \tau_1$ the crossover time equals
$t_{cross}=\tau_1$. In contrast is $\tau\,=\,1/\kappa \, \le\,\tau_1$
then the mean squared displacement becomes diffusional beyond
$t_{cross}=\tau_1^2\,\kappa$.

An analytic tractable limit is $\kappa \to \infty$ with $D_{\Omega}={\rm
  const}$.  It corresponds to a frozen angle and the variance is
ballistic
 $ \langle\, \Delta \bbox{r}^2(t) \,\rangle \,=\, v_0^2 \tau^2 $
As well one might return to the case that the angular velocity is
white noise in the same limit $ \kappa \to \infty$.  But this time the
noise intensity shall scale with $\kappa$ as $D_{\Omega}=\kappa^2 D$.
This case was studied previously in
Section \ref{sec:diff_gruler}, where the mean squared displacement
is given by Eq.\eqref{eq:DMM} \cite{mikhailov_self-motion_1997}.

As shown before in Sec. \ref{sec:diff_independent} under the assumption of independent speed and angular fluctuations both contribute additively to the mean squared displacement. Assuming an energy pump of the Schienbein-Gruler type
\begin{eqnarray}
  \dot{v}\,= \,-\gamma (v-v_0)\,+\,\sqrt{2 D_v}\,\xi(t)\,
  \nonumber
\end{eqnarray}
the speed fluctuations lead to an replacement of the constant squared
velocities in \eqref{eq:mean_oup} by the time dependent
autocorrelation function
\begin{eqnarray}
  \label{eq:replac}
  v_0^2 \to v_0^2 + \frac{D_v}{\gamma} \exp\Big( -\gamma \tau \Big)
\end{eqnarray}
\begin{figure}[htb]
  \centering
  {\includegraphics[width=0.7\linewidth]{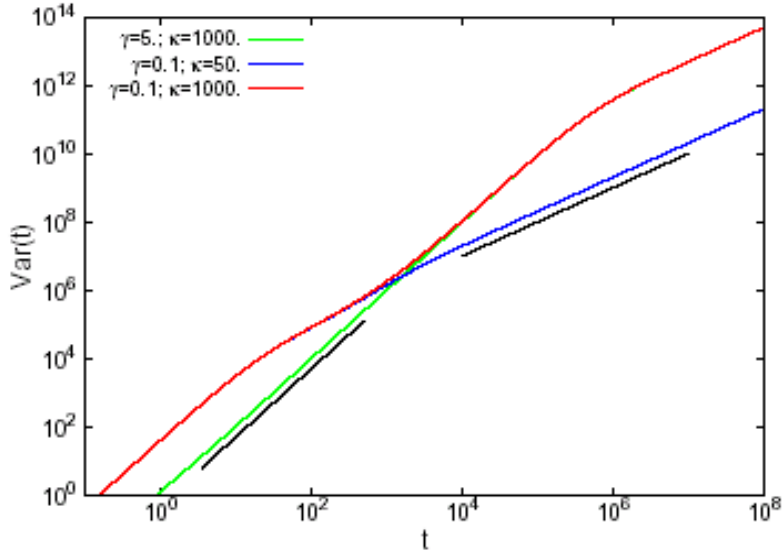}}
  \caption{Mean squared displacement of a particle with fluctuating
    speed and correlated angular velocities \cite{weber_diffusion_2010}. }
  \label{fig:msd_oup_withspeed}
\end{figure}
Parameters have been selected such, that the growth of the mean
squared displacement versus time undergoes three crossovers
\cite{peruani_self-propelled_2007} (see also Sec.
\ref{sec:diff_independent}). The corresponding times are given by the
relations \cite{weber_diffusion_2010}
\begin{eqnarray}
  t_1=\frac{1}{\gamma}\,,~~~ t_2= \frac{D_\Omega}{\gamma^2 v_0^2}, ~~~  t_3= \frac{v_0^2 \kappa^2}{D_\Omega}\,.
\end{eqnarray}
The first time $t_1$ gives the crossing of the speed fluctuations,
whereas the last one $t_3$ has its origin in the change of the angular
fluctuations. At the time $t_2$ both the value of ballistic
displacement caused by the angular noise starts to overcome the
diffusional displacement created by speed fluctuations. Therefore the
displacement returns to a ballistic growth.

\subsubsection{Stochastic, dichotomous angular dynamics}
\label{sec:stoch_dich_angles}
\subsubsection*{Markovian switch between two operating orientations}
In the following, we consider again particles moving with
constant speed $s=v_0$ and under the influence of a time-dependent torque 
\cite{van_teeffelen_dynamics_2008,friedrich_chemotaxis_2007,friedrich_stochastic_2008}.
However, this time we assume that the rotational motion switches dichotomously between leftward and
rightward turning
\cite{komin_random_2004,schimansky-geier_advantages_2005,haeggqwist_hopping_2008}.
In detail, we consider as switching protocol for $\Omega(t)$ the
symmetric dichotomous Markov process (see Sec.  \ref{sec:DMP}), which
is another colored noise taken as the driving torque.  $\Omega(t)$ may
possess the two values $\pm \omega$ and switches between them with a
constant rate $\lambda$.  The particles shall move with constant speed
$v_0$ and, additionally, angular white Gaussian noise with intensity
$D_\varphi=D/v_0^2$ is present and perturbs the angle.

\begin{figure}[htb]
  \centering
  \includegraphics[width=0.45\linewidth]{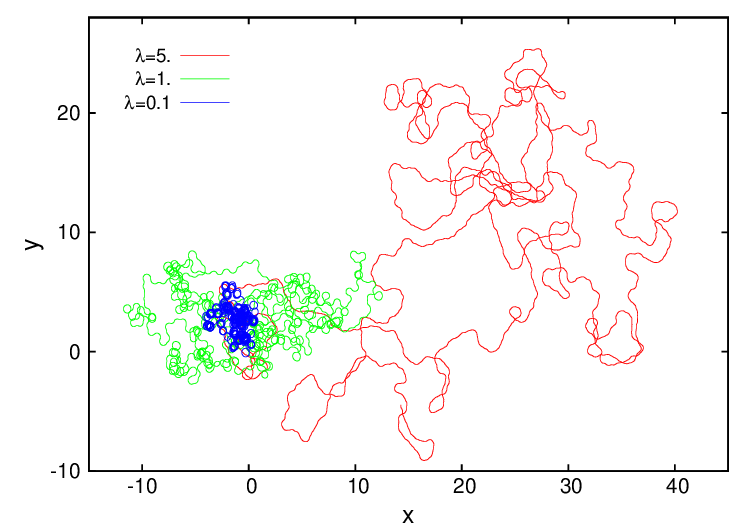}~\includegraphics[width=0.45\linewidth]{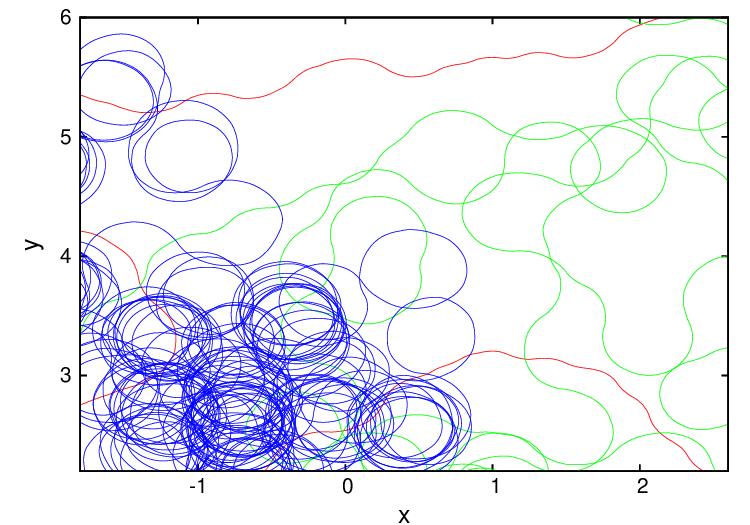}
  \caption{Trajectories of particles with constant speed and random
    torque. The angular velocities switches as a symmetric DMP between
    $\Omega=\pm 1.5$.  Different rates of transitions create different
    magnitude of spreading as presented in the left figure.  The right
    figure zooms into the left one to show the different trajectories.
    Other parameters: $v_0=0.5$ \cite{weber_diffusion_2010}}
  \label{fig:random_torque}
\end{figure}

The Langevin equation for this situation is easily formulated and
reads
\begin{eqnarray}
  \label{eq:torque}
  \dot{\varphi}\,= \,\Omega(t)\,+\,\sqrt{2D_\varphi}\, \xi_2(t)\,,
\end{eqnarray}
In Fig. \ref{fig:random_torque} we show simulations of typical
trajectories where the random torque $\Omega(t)$ is given by a
symmetric dichotomous Markov process \cite{weber_diffusion_2010,weber_active_2011}, as introduced
in Eq.  \eqref{eq:torque}. We assume that the angular velocity
vanishes in average, i.e. it holds $\mean{\dot{\varphi}}=0$ .

We define the probability density functions for both rotational modes in analogy to Sec. \ref{sec:DMP} and the description reduces to angular dynamics $\varphi(t)$. 
The particle has the joint
probability density function $P_{+}(\varphi,t|\varphi_0)$ to have the angle
$\varphi$ and angular velocity $\Omega(t)=+\omega$ at time $t$
conditioned by the initial value $\varphi_0$ at $t=0$. Equivalently, one
defines $P_{-}(\varphi,t|\varphi_0)$ for the state with $\Omega=-\omega$.
Both densities obey the equations \cite{anishchenko_nonlinear_2002}
\begin{align}
  \label{eq:dich}
  \frac{\partial}{\partial t} P_{\pm} (\varphi,t) & = \mp \omega
  \frac{\partial P_{\pm}(\varphi,t)}{\partial \varphi} + D_\varphi
  \frac{\partial^2 P_{\pm}(\varphi,t)}{\partial \varphi^2}-\lambda
  P_{\pm}(\varphi,t)+ \lambda P_{\mp}(\varphi,t)\, .
\end{align}
and we have omitted the initial values. The common probability density
to have an angle $\varphi$ at time $t$ then follows from
\begin{equation}
  \label{eq:common}
  P(\varphi,t) = P_{+}(\varphi,t)+ P_{-}(\varphi,t)
\end{equation}
which will be used for the determination of the effective spatial
diffusion coefficient.

We proceed to discuss the physical situation in two limiting cases.
First we neglect the Gaussian fluctuations. We put $D=0$ and
subsequently $D_{\varphi}=0$.  Then the evolution of $P(\varphi,t)$ is given
by the telegraph equation \eqref{eq:telegraf}
\begin{equation}
  \label{eq:tele}
  {1 \over 2 \lambda} \frac{\partial^2}{\partial t^2} P(\varphi,t)  +  \frac{\partial}{\partial t} P (\varphi,t) = {\omega^2 \over 2 \lambda} 
  \frac{\partial^2 P(\varphi,t)}{\partial \varphi^2}\,.
\end{equation}
For large times $t \gg 1/\lambda$ and $\varphi-\varphi_0 \ll \omega t$ the
term with the second derivative can be neglected. One gets again a
diffusion equation for the angel dynamics. In result, again a Gaussian
density is established as in Eq. \eqref{Gauss1}. The effective angle
diffusion coefficient follows as
\begin{equation}
  \label{eq:diff_easy}
  D_\varphi\,=\, {\omega^2 \over 2\lambda} \, .
\end{equation}

In the second case we look at the governing equation for the common
probability density function including the Gaussian noise.  It looks
complicated
\begin{align}
  \label{eq:prob-common}
  \frac{\partial}{\partial t} P (\varphi,t) = & - {1 \over 2 \lambda}
  \frac{\partial^2}{\partial t^2} P(\varphi,t)
  + {D_\varphi \over \lambda} \frac{\partial^3 P(\varphi,t)}{\partial t \partial \varphi^2}  \nonumber \\
  & + \Big({\omega^2 \over 2 \lambda} + D_\varphi \Big)\frac{\partial^2
    P(\varphi,t)}{\partial \varphi^2} - {D^2_\varphi \over 2 \lambda }
  \frac{\partial^4 P(\varphi,t)}{\partial \varphi^4}\,,
\end{align}
but this equation simplifies significantly in the limit of fast
switching $\lambda \to \infty$. Additionally we assume in this limit a
fixed $\omega^2/\lambda=const$.  Then, the evolution operator again
reduces to a simple diffusion equation
\begin{equation}
   \frac{\partial}{\partial t} P (\varphi,t) =  \Big({\omega^2 \over 2 \lambda}  + D_\varphi \Big)\frac{\partial^2 P(\varphi,t)}{\partial \varphi^2}\,.
\label{eq:EffDiff}
\end{equation}
Changes of the angel becomes diffusional with the effective diffusion
coefficient possesses 
\begin{equation}
  \label{eq:large1}
D_\varphi^{\rm eff}\,=\,{\omega^2 \over 2 \lambda}  + D_\varphi\,.   
\end{equation}
In addition to the intensity of the Gaussian angel noise a second
contribution arises from the switching.  Again the time dependent
solution has a Gaussian form \eqref{Gauss1} and the wrapped
$2\pi$-periodic solution corresponds to Eq.  \eqref{eq:wrapped} with
\eqref{eq:large1}.  We mention that both limits correspond to the
Gaussian white limits of the stochastic force $\Omega(t)$ in the
Langevin equation for the angle dynamics, with fixed velocity $v=v_0$
and $g(v)=1/v_0$. In this approximation $\Omega(t)$ was interpreted as
a second Gaussian white noise in \eqref{eq:rotation}, which is
uncorrelated with the stochastic term $\xi_(t)$.

Eq. \eqref{eq:EffDiff} and its Gaussian solution are valid in the limit
of large angle and long times. It turns to be out that the
corresponding approximation is insufficient to obtain the correct
value of the effective spatial diffusion coefficient
$D_\text{eff}$ describing the motion of the particle in a
plane. The latter is strongly determined by the small angle behavior
(long stretches without considerable turning). The largest
contribution to the spatial diffusion stems from trajectories
corresponding to long stretches along the same direction, i.e. is
determined by the small-angle behavior.  

Therefore, before turning to the calculation of the spatial diffusion
coefficient, we discuss a renewal approach which allows a
generalization to a non-Markovian case as well as leads us to the
effective angular diffusion \eqref{eq:large1} more directly.

\subsubsection*{Renewal model for two operating orientations}
The previous section dealt with a Markovian dichotomous switchings.
Hence, the times spent within two states with fixed direction of
rotation are distributed exponentially. Here, we generalize to
arbitrary distributions of the switching times
\cite{haeggqwist_hopping_2008}. A different techniques will be
applied, which is based on a continuous time generalization of a
persistent random walk, in particular our analysis applies methods
developed in Ref.
\cite{masoliver_continuous-time_1989,haeggqwist_hopping_2008}.

We consider the switching between the two states as an alternating
renewal process. The probability density function of the sojourn time
$\tau$ in each state is given by a function $w(\tau)$. In each of
these states the motion rotates with the constant angular velocity
$\pm \omega$. The overall process is expressed as a sequence of steps
in $\varphi_i$, each of which corresponds to the sojourn in one of the
rotational states; the overall

As in Eq. \eqref{eq:shig} for the position, the angular
displacement $\varphi(t)$ of the system is given as the sum of the
displacements $\varphi_i$ in $n$ completed steps probably followed by the
last, $(n+1)$st incomplete step:
\begin{align}
  \label{eq:seq_ang}
  \varphi(t) = \sum_{i=1}^n \varphi_i + \varphi_{n+1}\,.
\end{align}
The corresponding joint probability density function (pdf)
$h(\varphi,\tau)$ of the displacement $\varphi$ and the duration $\tau$
of a completed step in a corresponding state can be written as
\begin{align} 
\label{pdf:joint}
  h_\pm(\varphi,\tau) = R_\pm (\varphi,\tau) \,w(\tau)
\end{align} 
where $R_\pm(\varphi,t)$ is the conditional pdf of rotation for
given step duration $\tau$.  It is given by Eq. \eqref{Gauss1} with
$\Omega = \pm \omega$:
\begin{align}
  \label{eq:prob} 
  R_\pm(\varphi,\tau) = \frac{1}{\sqrt{2 \pi D_\varphi \tau}}\exp
  \left[-\frac{(\varphi\mp\omega \tau)^2}{4 D_\varphi \tau}\right]
\end{align}
As auxiliary functions in the approach of Ref.\cite{masoliver_continuous-time_1989} we need the
probability not to perform a step up to time $t$
\begin{align}
  W(\tau)= 1- \int_0^\tau w(\tau')\mbox{d}\tau',
\end{align}
which defines the two distributions
\begin{align} 
  H_\pm(\varphi,\tau) = R_\pm(\varphi,\tau) W(\tau)
\end{align}
corresponding to \eqref{pdf:joint}.

The probabilities $Q_\pm$ to rotate over the overall angle $\varphi$ just
having completed a step corresponding to $+$ or $-$ state at time $t$ read
\begin{align} 
  Q_\pm(\varphi,t) &=& \frac{h_\pm(\varphi,t)}{2}+\int_{-\infty}^\infty \mbox{d} \varphi
  \int_0^t\mbox{d} t' h_\pm(\varphi-\varphi',t-t')Q_\mp(\varphi',t') 
\end{align} 
It follows from the consideration that the temporal stay in one of the states ($+/-$), 
ending at time $t$ is either the first stay, with the 
probability $1/2$ of being in one of the states,
or the walker has already made several steps, and entered at time
$t'$, to its last $+$ ($-$)-state. The latter ends at time $t$ and
since the process is a renewal one, transition depend on the last jump
only.

The probabilities $P_\pm$ still to be at $\varphi$ at time $t$ are
\begin{align}
  \label{eq:jack} P_\pm(\varphi,t) &=
  \frac{H_\mp(\varphi,t)}{2}+\int_{-\infty}^\infty
  \mbox{d}\varphi'\int_0^t\mbox{d}t' H_\mp(\varphi-\varphi',t-t')Q_\pm(\varphi',t')
\end{align} 
By adding the probabilities for $\varphi$ at time $t$ in both states one
finally gets the common probability $P(\varphi,t)$ for the angle
distribution.

The integrals in Eq. \eqref{eq:jack} have the form of a convolution in
spatial and temporal variables.  Such equations transform into
algebraic ones in the Fourier-Laplace domain where solutions can be
easier found. In \cite{haeggqwist_hopping_2008} the solution was given as
\begin{eqnarray}
   \label{eq:linde1} 
   \tilde p(k,\tilde s) = \frac{\tilde H_+(k,\tilde s)\bigl[1+\tilde h_-(k, \tilde s)\bigr]+ \tilde H_-(k,\tilde s)\bigl[1+\tilde h_+(k,\tilde s)\bigr]} {2\bigl[1-\tilde h_-(k,\tilde s)\tilde h_+(k,\tilde s)\bigr]}\,,
\end{eqnarray} 
where $\tilde{p}(k,\tilde s)$, $\tilde{p}_\pm(k,\tilde s)$ assign the
Fourier-Laplace transforms of the corresponding probability density
function. 

The Fourier-Laplace transforms of the densities $h_\pm$ and $H_\pm$
follow from the transformation of a Gaussian and from the shift
theorem for the Laplace transform. They can be expressed as
\begin{eqnarray} 
  \tilde{h}_\pm(k,\tilde s) = \tilde{w}(\tilde s + D_\varphi k^2 \pm i \omega k),~~\tilde{H}_\pm(k,\tilde s)= \tilde{W}(\tilde s + D_\varphi k^2 \pm i \omega k)
\end{eqnarray} 
with $\tilde{W}(u) = [1- \tilde{w}(u)]/u$. 

As example we consider again that the two states of the random torque
$ \Omega(t)$ are due to a symmetric and Markovian protocol. In
consequence, we take exponential waiting time distribution with rate
$\lambda$ (see Eq. \ref{eq:dich}) $w(t)=\lambda
\exp{(-\lambda t)}$. After corresponding Fourier Laplace transforms we

obtain 
\begin{equation}
  \label{eq:rwt39} 
  \tilde p(k,\tilde s) =\frac{\tilde s+2\lambda+D_\varphi k^2}{\tilde s^2 +
    [\omega^2 +2\lambda D_\varphi]k^2+D_\varphi^2k^4+2\lambda \tilde s+2 \tilde s D_\varphi k^2}\,.
\end{equation}
It is interesting to underline that this solution from the renewal
model solves Eq.  \eqref{eq:prob-common} in the Fourier-Laplace domain.
It will be used in the next paragraph to obtain the effective spatial
diffusion coefficient which describes the asymptotic behavior of the
particle.

Let us shortly return to the diffusional approximation of the angular
dynamics. We let $\tilde s \rightarrow 0$. The first term of the denominator
(quadratic in $\tilde s$) can be neglected.  Additionally we look at the
limit of large overall turning angles
taking leading terms in the limit $k \rightarrow 0$, only. In this
limit it one re-obtains angular diffusion with the effective diffusion
coefficient defined in Eq.  \eqref{eq:large1}. We remark that one gets
the same result for any waiting time density $w(t)$ if its first
moment $\tau = \int_0^\infty \mbox{d}t' t' w(t')=1/\lambda$
\cite{haeggqwist_hopping_2008}.

\subsubsection*{Spatial diffusion coefficient under two random rotation} 
Eventually, we aim to calculate the spatial diffusion coefficient
$D_\text{eff}$ on a plane for the situation with two random rotations and
Gaussian angular noise.  We apply again the Taylor-Kubo relation
\cite{ebeling_statistical_2004,schimansky-geier_advantages_2005} taking $d=2$ in Eq. \eqref{eq:deff_Dd}.  For a
motion with constant speed $v_0$ the correlation function $C_{\bf v
  v}$ is governed by the behavior of the angular coordinate, only, so
that
\begin{equation} 
  C_{\bf v v}(t)= v_0^2 \left\langle \cos[\varphi(t)]\right\rangle \,.
\end{equation} 
Here we assumed $\varphi(0)=0$. The diffusion coefficient reads
\begin{equation} D_{\rm eff}= \frac{v_0^2}{2}\int_0^\infty
  \left\langle \cos[\varphi(t)] \right\rangle \mbox{d} t\,.
\label{eq:cvv}
\end{equation} 
Let
\begin{equation}
  \label{eq:chara}
  \tilde{P}(k,t) = \left\langle e^{ik\varphi} \right\rangle =\int_{-\infty}^\infty \mbox{d}\varphi e^{ik\varphi} P(\varphi,t) 
\end{equation}
be the characteristic function of the angular distribution at time $t$
(its temporal Laplace transform gives us $\tilde{p}(k,\tilde s)$). The mean
value in the expression Eq. \eqref{eq:cvv} is then $\left\langle
  \cos[\varphi(t)] \right\rangle = \mbox{Re} \, \tilde{P}(1,t)$.
Therefore the Laplace transform of the latter expression in its
temporal variable yields
\begin{equation} 
  \int_0^\infty \mbox{d}t e^{-\tilde s t} \left\langle \cos[\varphi(t)] \right\rangle = \mbox{Re}\, \tilde{p}(1,\tilde s)\,. 
  \label{eq:lala}
\end{equation} 
Since in our case the function $\tilde{p}(1,\tilde s)$ is real, the Re-symbol
can be omitted. From comparison of Eq. \eqref{eq:lala} and
Eq. \eqref{eq:cvv}, we obtain the diffusion coefficient for the considered
particle dynamics. It reads following Eq. \eqref{eq:rwt39}
\begin{equation} 
\label{d2d}
D_{\rm eff}\,= \,\frac{1}{2} v_0^2 \tilde{p}(1,0)\, = \, \frac{1}{2} v_0^2
\frac{2\lambda + D_{\varphi}}{\omega^2+2\lambda D_{\varphi}+ D_{\varphi}^2}.
\end{equation}
Using Eq. \eqref{eq:Dphi} we can rewrite the last expression as \cite{haeggqwist_hopping_2008}
\begin{equation}
  \label{eq:final2}
  D_{\rm eff}= {v_0^4 \over 2 D}\,\cdot {1\over 1+{\omega^2 v_0^4 \over D(D+ 2\lambda v_0^2)}}\,.
\end{equation}
\begin{figure}
  \centering
  \includegraphics[width=.6\linewidth]{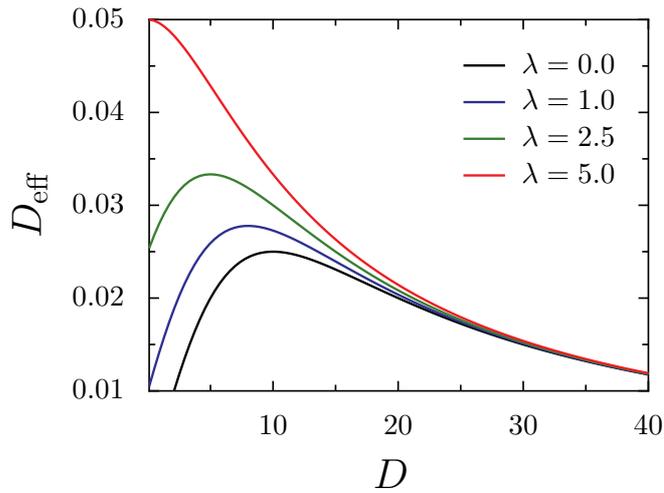}
  \caption{Diffusion coefficient for active particle with negative and
    positive angular velocity $\omega=10$ and $v_0=1$ vs noise
    intensity $D$. The upper curve $\lambda=5$ converges in good
    agreement with the dependence of $D_{\rm eff}$ for $\lambda \to
    \infty$.  The lower curve corresponds to $\lambda=0$ but averaged
    over both values of $\pm \omega$.}
  \label{fig:diff_non}
\end{figure}
The first factor is the already often cited diffusion coefficient of
active Brownian particles \cite{mikhailov_self-motion_1997} which appears, for example, also in Eq. \eqref{eq:msd_gruler}). It defines the upper limit in the situation with two random rotations. We see that a directed
turning motion decreases the diffusion process since the second
multiplicand is always smaller than $1$ if $\omega \ne 0$.  It is the
effect of the bounded motion along the circles and the precision along
these curved trajectories which obviously decreases the effective
diffusion. The dependence of the diffusion coefficient as function of
the noise intensity $D$ is presented in Fig \eqref{fig:diff_non} for
several rates of transitions $\lambda$ between the two angular
velocities.

Starting from Eq. \eqref{eq:final2} this behavior is also found
for $\omega=0$ or in the case of high noise $D \to \infty$. In this situation the second factor becomes unity. As $\lambda \to 0$ the particle will rotate always clockwise or counterclockwise without switching between
the two turning directions. Then one obtains
\begin{equation}
  \label{eq:final3}
  D_{\rm eff}= {v_0^4 \over 2 D}\,\cdot {1 \over 1+{\omega^2 v_0^4 \over D^2}}\,.
\end{equation}
This result was previously derived in \cite{schimansky-geier_advantages_2005}. The effect of the switchings between the two values of $\omega$ decreases the value standing in the denominator of the second factor. Hence the diffusion coefficient decreases if switches between both turning directions are allowed.

Another interesting case is the Gaussian limit of the dichotomous
phase velocities. As in the section 4 we suppose the common limit
$\lambda \to \infty$ and $\omega^2 \to \infty$ holding thereby fixed
their ratio. Then it follows
\begin{equation}
  \label{eq:final4}
  D_{\rm eff}= {v_0^4 \over 2 (D+ { \omega^2 v_0^2 \over 2 \lambda })}.
\end{equation}
It yield the result of Meink\"ohn and Mikhailov but with the increased
noise intensity $D \to D+ { \omega^2 v_0^2 /2 \lambda }$ coming
from the fast switches of the velocity-phase $\varphi$.

%% file: SECTION_confined.tex
\section{Stochastic dynamics of active particles in confining potentials}
\label{sec:rot_ext}
In this section we address the question of active motion in 
external confinements and analyze the typical trajectories and distribution functions. Throughout this section we use passive fluctuation, independent on the direction of motion.  

Rotational motion of biological agents has been observed as a result of an external confinement \cite{ordemann_motions_2003,ordemann_pattern_2003,gautrais_analyzing_2008}. Furthermore,
collective rotation is a very common mode of the motion of
swarms \cite{okubo_diffusion_2001}.  Motivated by these observations, in order to imitate the
rotational mode, we will therefore consider first a single particle in an external
harmonic potential
\begin{equation}
  \label{eq:eq:lin_poz}
  U_{\rm H}(\bbox{r}) = \frac{1}{2} \omega^2 \bbox{r}^2\,.  
\end{equation}
It is well known that this system possesses a stable solution
corresponding to a circular motion with radius $r_0 = v_0/\omega$
\cite{schimansky-geier_stationary_2005}. In the absence of noise the angular momentum $\bbox{L}
= \bbox{r} \times \bbox{v}$ is fixed, the direction of motion depends on
the initial conditions. 

In this section we include an external potential in the description. In
case of an constant force it will break the isotropy of the problem.
In a large or an infinite space the presence of dissipation will
create a situation with a constant velocity and we can cross to a co-moving frame. 

A more interesting situation is obtained for potentials which create a confinement.
New dynamical solutions are then limit cycles in the phase space. In the following, we will restrict mainly to the discussion of the the Rayleigh-Helmholtz friction.
Then the problem of active Brownian motion in external attracting potentials
transforms into the study of (higher dimensional) van-der-Pol
oscillators with noise.

\subsection{Basic solutions for the active motion in harmonic
  potentials}
\label{sec:harm}
\begin{figure}[htbp]
\begin{center}
  \includegraphics[width=.7\linewidth,height=.5\linewidth]{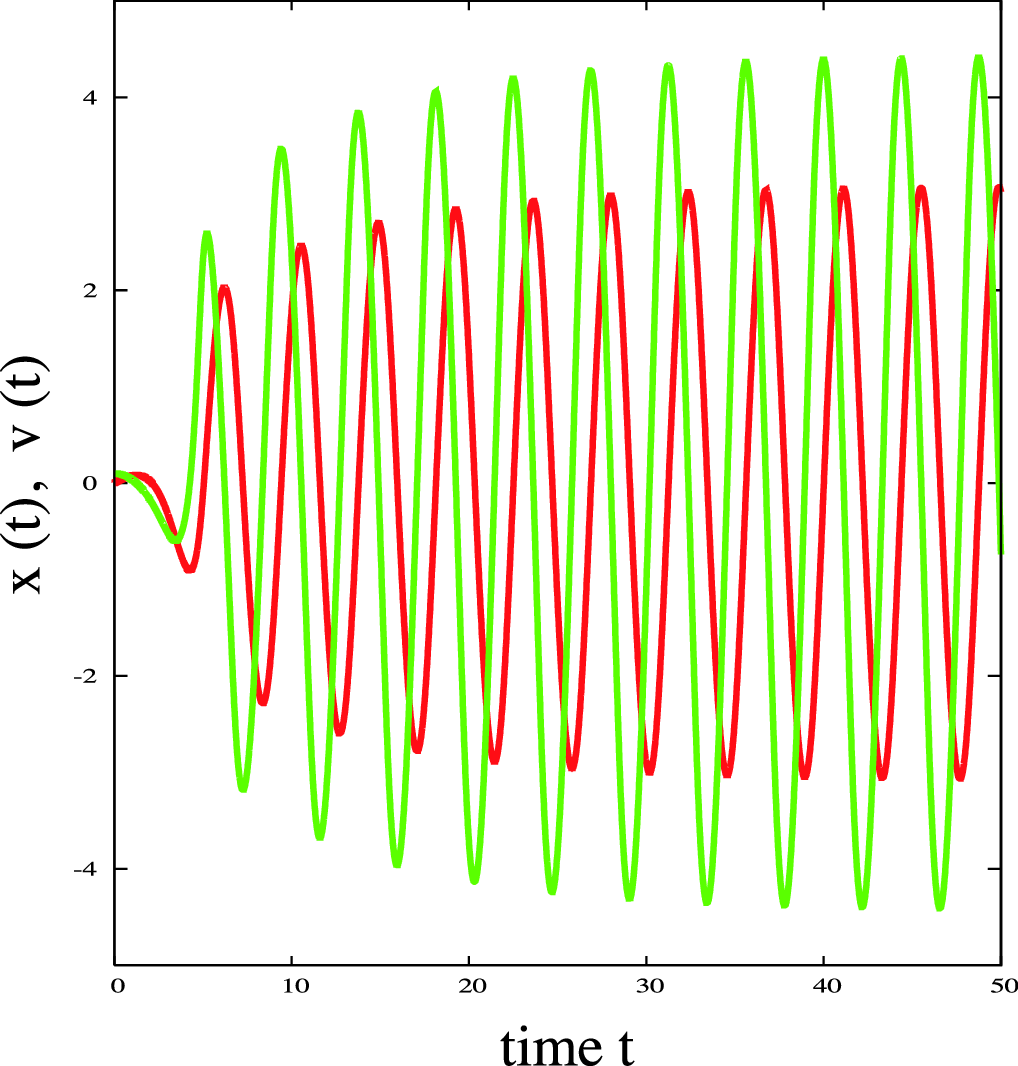}
\caption{\label{lim1Dosc} The coordinate $x(t)$ and the velocity
  $v(t)$ for limit-cycle oscillations of a $1d$ driven oscillatory
  systems. The active particle is driven by a Rayleigh friction force.
  Parameters: $q = 1, d = 0.8, c =0.1, \gamma_0 = 0.1$ .}
\end{center}
\end{figure}

Let us first study the active motion of a 1d driven oscillator without
noise, which for the Rayleigh model is described by \cite{anishchenko_nonlinear_2002}
\begin{equation}
  \label{Raylosc}
  \frac{d x}{dt} = v, \qquad \frac{d v}{d t} + \omega_0^2 x = v (\alpha - \beta v^2),
\end{equation}
where $v_0^2 = \alpha / \beta$ defines the stationary velocity.
Fig \ref{lim1Dosc} shows the results obtained from integration of \eqref{Raylosc}.
The systems exhibits self-excited oscillations, so-called limit cycles - as discussed already by Rayleigh in 1894 \cite{rayleigh_theory_1894}. 
An approximative solution for small driving (small values of $\alpha$) \cite{anishchenko_nonlinear_2002} reads
\begin{equation} 
x(t) = r_0 \sin (\omega_0 t + \Phi), \qquad v(t)=v_0
\cos(\omega_0 t + \Phi) ,
\end{equation} 
where the amplitude of the
oscillations is $r_0 = v_0 / \omega_0$ and the energy is $H_0 = m
v_0^2$. This approximate solution is valid for all active friction
functions which have one zero at $v_0$. Any initial condition if
the $x,v$ space converges to the circle 
\begin{equation} 
  H(x,v) = \frac{1}{2}m
  v^2 +\frac{1}{2}  \omega_0^2 x^2 = 2 v_0^2   
\end{equation} 
This circle is an attracting limit cycle of our system \cite{anishchenko_nonlinear_2002}.
In a next step we will study two-dimensional oscillators. We 
specify the potential $U(\bbox{r})$ as a symmetric
harmonic potential in two dimensions:
\begin{equation}
  U(x_1,x_2) = \frac{m}{2}\omega_0^2 \,(x_1^2 + x_2^2)
  \label{parab}
\end{equation}
First, we restrict the discussion to the deterministic case, which then is
described by four coupled first-order differential equations:
\begin{equation}
  \begin{array}{rcl}
    \dot{x}_1 = v_1, & \qquad &
    \dot{v}_1 = - \gamma\left(v_1,v_2\right) v_{1}-\omega_0^2x_{1}\\ 
    \dot{x}_2 = v_2, & \qquad &
    \dot{v}_2 = - \gamma\left(v_1,v_2\right) v_2 - \omega_0^2 x_{2}
  \end{array}
  \label{2d-det}
\end{equation}
In the 2d case the energy pump of the r.h.s. of the velocity equation \eqref{Raylosc} is generalized to
\begin{equation}
  \label{eq:pum_2d}
  \gamma\left(v_1,v_2\right)\,=\, -\alpha +\beta\,(v_1^2\,+\,v_2^2)\,.
\end{equation}
We can show by simulation and theoretical considerations that two
limit cycles in the four-dimensional phase space are created
\cite{erdmann_brownian_2000}. The projections of both these periodic motions to
the $\{v_1,v_2\}$ plane is the circle
\begin{equation}
  v_1^2 + v_2^2 = v_0^2 = \frac{\alpha}{\beta}\,=\,{\rm const.}
\end{equation}
The projection to the $\{x_1,x_2\}$ plane also corresponds to a circle
\begin{equation}
  x_1^2 + x_2^2 =  r_0^2 = {\rm const.}
\end{equation} 
Due to the condition of equilibrium between centripetal and
centrifugal forces on the limit cycle we have
\begin{equation}
  \label{centri-eq}
  \frac{v_0^2}{r_0} =  r_0 \omega_0^2.
\end{equation}
Therefore the radius of the limit cycle is given by
\begin{equation}
  r_0 = \frac{v_0}{\omega_0}
\end{equation}
From equation~(\ref{centri-eq}) follows
\begin{equation}
  \frac{1}{2} v_0^2 = \frac{\omega_0^2}{2} r_0^2
\end{equation}
This means we have equal distribution of potential and kinetic energy
on the limit cycle \cite{ebeling_active_1999}. As for the harmonic oscillator
in 1-d, both parts of energy contribute the same amount to the total
energy. Therefore the energy of motions on the limit cycle, which is
asymptotically reached, is twice the kinetic energy
\begin{equation}
  \label{Hv}
  H\longrightarrow H_0 =  v_0^2 . 
\end{equation}
The energy is a slow (adiabatic) variable which allows a phase average with
respect to the phases of the rotation \cite{erdmann_brownian_2000}.

Two exact stationary solutions can be easily found. The first cycle in
the four-dimensional phase space reads with arbitrary initial phase
$\Phi$:
\begin{equation}
\label{eq:solution}
  \begin{array}{rcl}
    x_1 = r_0 \cos(\omega_0 t + \Phi) &\quad&
    v_1 = - r_0 \omega \sin(\omega_0 t + \Phi)\nonumber \\\\
    x_2 = r_0 \sin(\omega_0 t + \Phi) &\quad&
    v_2 = r_0 \omega \cos(\omega_0 t + \Phi)\nonumber
  \end{array}
\end{equation}
This means, the particle rotates even for strong pumping with the
frequency given by the linear oscillator frequency $\omega_0$.  One
can check that this is indeed an exact solution of the dynamic
equations.  The trajectory defined by the four equations looks like a
hoop in the four-dimensional phase space. Most projections to the
two-dimensional subspaces are circles or ellipses however there are to
subspaces namely $\{x_1,v_2\}$ and $\{x_2,v_1\}$ where the projection
is like a rod.

\begin{figure}
\psfrag{v_x}{\large $v_x$}
\psfrag{x}[l]{\large $x$}
\psfrag{y}{\large $y$}
  \begin{center}
\includegraphics[width=.5\linewidth,angle=-90]{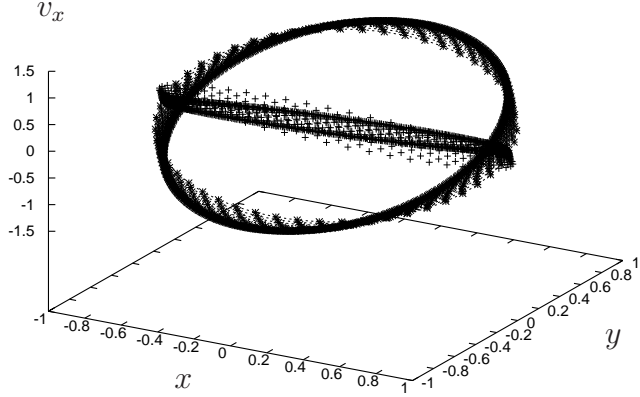}
    \caption{\label{lc234} Stroboscopic plot of the
      2 limit cycles for driven Brownian motion. We show projections of
      solutions for $v_0=1$ to the subspace $\{x_1,x_2,v_1\}$. }
  \end{center}
\end{figure}

The second limit cycle is obtained by inversion of the motion, i.e.
$t \rightarrow -t, v_1(0) \rightarrow - v_1(0), v_2(0) \rightarrow
-v_2(0)$,
$\omega_0 \rightarrow - \omega_0$
which yields
\begin{equation}
\label{eq:solution2}
  \begin{array}{rcl}
    x_1 = r_0 \cos(\omega_0 t - \Phi) &\quad& v_1 = - r_0 \omega \sin(\omega_0 t - \Phi)\nonumber \\\\
    x_2 = - r_0 \sin(\omega_0 t - \Phi) &\quad&
    v_2 = - r_0 \omega \cos(\omega_0 t - \Phi)\,. \nonumber
  \end{array}
\end{equation}
This second cycle forms also a ring in the four-dimensional phase space which is different from the
first one, however both limit cycles have the same projections to the
$\{x_1,x_2\}$ and to the $\{v_1,v_2\}$ plane. The projection to the
$x_1-x_2-$ plane has the opposite direction of rotation in comparison
with the first limit cycle.  The projections of the two rings on
the $\{x_1,x_2\}$ plane or on the $\{v_1,v_2\}$ plane are circles
(Figure~\ref{lc234}). The ring-like distribution intersect
perpendicular the $\{x_1,v_2\}$ plane and the $\{x_2,v_1\}$ plane (see
Figure~\ref{lc234}). The projections to these planes are rod-like and
the intersection manifold with these planes consists of two ellipses
located in the diagonals of the planes (see Figure~\ref{lc234}).

The main effect of noise is the spreading of the
deterministic attractors. Thus, in the presence of noise, the two ring-like embracing limit cycles
are converted into two crossing toroids with cross-sections increasing with increasing noise intensity (see
Fig. \ref{distrib}).

We already examined that in the case of dimension $d=2$ a very useful
representation is obtained in polar coordinates
\cite{schimansky-geier_stationary_2005}.  Following Eq.
(\ref{eq:polar_veloc})  and with 
\begin{equation}
  \label{eq:polar}
  x = r(t) \cos\left(\psi(t)\right),\quad y  = r(t) \sin\left(\psi(t)\right)\,.
\end{equation}
we introduce polar coordinates $r(t), \psi(t),s(t),\varphi(t)$ in the
four-dimensional phase space ($\bbox{r},\bbox{v}$) for the case of
active motion.  One easily obtains the dynamics similar to
Eqs.(\ref{eq:dyn_polar})
    \begin{eqnarray}
  \label{eq:dyn_polar}
  \dot{r} &=& v \cos{\theta}\nonumber, ~~~~~\dot{s} = - \gamma(s) s  -\omega r \cos{\theta}\\\\
  \dot{\theta} &=& \left(\frac{\omega_0^2 r}{s}- \frac sr\right) \sin\theta, ~~\dot{\psi} = \frac sr \;\sin\theta\nonumber
\end{eqnarray}
with $\theta(t)=\varphi-\psi$.

The stationary solutions (Eqs. \ref{eq:solution}, \ref{eq:solution2})
can be easily found. The difference of the two angles $\theta$
approaches two values, $\theta=\pm \pi/2$. These two solution resemble
the two limit cycles with $v_0=\alpha/\beta$, $r_0=v_0 \omega_0$
and two stationary rotations (clockwise and counter clockwise) with
stationary angular velocity $\dot{\psi}=\dot{\varphi}=\pm \omega_0$.

This means, the particle rotates even at strong pumping with the frequency
given by the linear oscillator frequency $\omega_0$.  The trajectory defined
by the above four equations looks like a hoop in the four-dimensional phase
space. Most projections to the two-dimensional subspaces are circles or
ellipses however there are to subspaces namely $\{x_1,v_2\}$ and $\{x_2,v_1\}$
where the projection is like a rod.

In order to construct solutions for stochastic motions we need beside
$H = v_0^2$ other appropriate invariants of motion.  Looking at the
first solution (\ref{eq:solution}) we see, that the following relation
is valid
\begin{equation}
  v_1 + \omega_0 x_2 = 0; \qquad v_2 - \omega_0 x_1 = 0.
\end{equation}
In order to characterize the first limit cycle we introduce the
invariant
\begin{equation}
  J_+ = H - \omega_0 L= \frac{1}{2}(v_1 + \omega_0 x_2)^2 
  +\frac{1}{2}(v_2 - \omega_0 x_1)^2 .
\end{equation}
where we have introduced the angular momentum $L= (x_1 v_2 -x_2 v_1)$.
We see immediately that $J_+ = 0$ holds on the first limit cycle which
corresponds to positive angular momentum. In order to characterize the
second limit cycle from equation~(\ref{eq:solution2}) we use the
invariant
\begin{equation}
  J_- = H + \omega_0 L = \frac{1}{2}(v_1 - \omega_0 x_2)^2 
  +\frac{1}{2}(v_2 + \omega_0 x_1)^2 .
\end{equation}
We see that on the second limit cycle, which corresponds to negative angular
momentum, holds $J_- = 0$.

\subsection{Rotational motors and efficiency}
A two dimensional system in rotational mode may be considered as a
simple model of a rotating motor.  We consider again the 2d
oscillatory system (\ref{2d-det}) this time for the general case of a
friction function having one zero $\gamma(v_0) = 0$. As shown above we
are able to find even an exact solution for the limit cycle
oscillations (\ref{eq:solution}) for the Rayleigh-Helmholtz case. The
solution (\ref{eq:solution}) represents a (mathematically ) positive
rotation.  The motion of the particle on the orbit in distance $r_0$
from the center proceeds with velocity $v_0$ and runs in time $t$ over
a path of length
\begin{equation}
l(t) = v_0 t + l(0)
\end{equation}
on the orbit. On this path the motor is doing work against the
friction $\gamma_0$.  

Generating rotations is connected with angular momentum $\bbox{L} =
{\bbox r}$ x ${\bbox v}$ which satisfies the equation of motion
\begin{eqnarray}
  \label{AngMom}
  \frac{d {\bbox L}}{d t} = -  \gamma({\bbox v}) {\bbox L}
\end{eqnarray}
In the 2d-case the self-generated angular momentum of the 
stationary solution is $L = r_0 v_0$.

Let us assume some torque $M_{ex}$ acting in opposite direction, then
the ``motor'' has to perform some additional work. For simplicity we
assume that there is some opposite force acting along the circumference of
the circle with $M_0 = r_0 F_{ex}$. We assume this force, as some additional friction $F_{ex} = - \Gamma v_{lc}$ acting in the direction opposite to the direction of (rotational) motion.  
We define $v_{lc}\leq v_0$ as the velocity of the motion along the limit cycle, with $v_{lc}=v_0$ for the unperturbed limit cycle.

\begin{figure}
\begin{center}
  \includegraphics[width=.49\linewidth]{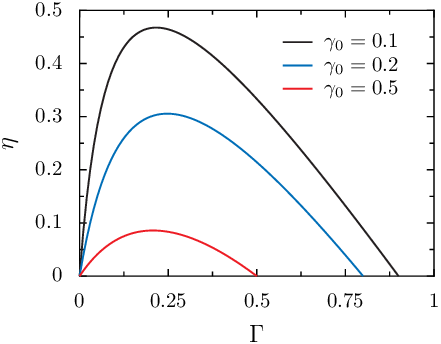}\hfill\includegraphics[width=.49\linewidth]{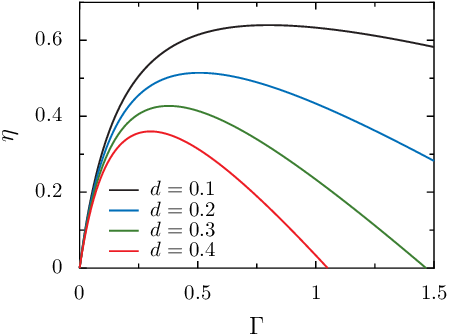}
  \caption{\label{powerSETb} Left: Dependence of the power of the
    "motor" on the external friction $\Gamma$ generated by the load
    force for three different values of the internal motor friction
    $\gamma_0$ and fixed the other parameters $c=0.5, d=0.5, q =
    1$. Right: Dependence of the power of the "motor" on the external
    friction $\Gamma$ generated by the load force for four values of the
    conversion parameter $d$ of the internal motor parameters (other parameters: $\gamma_0=0.2$, $c=0.5$, $q_0=1.0$). 
    We note that the efficiency decreases with increasing parameters $d$. }
\label{motoreffib}
\end{center}
\end{figure}

Let us consider the depot model (see Sec. \ref{sec:depot}).
The velocity along the limit cycle without the additional torque reads
\begin{eqnarray}
  \label{v0}
  v_0^2 = \frac{q_0}{\gamma_0} - \frac{d_0}{c}.
\end{eqnarray}
If we take the additional load into account the velocity along the limit cycle changes to
\begin{eqnarray}
  \label{v1}
  v_{lc}^2 = \frac{q_0}{(\gamma_0 + \Gamma)} - \frac{d_0}{c}.
\end{eqnarray}
The power associated with the additional work is $P = F_{ex} v_{lc}$ and we define
the efficiency as the relation of the generated power $P$ to the flow
of energy into the system $q_0$:
\begin{eqnarray}
  \label{power}
  \eta = \frac{P}{q_0} = \frac{F_{ex} v_{lc}}{q_0}= \frac{\Gamma v_{lc}^2}{q_0}
  = \Gamma\left[ \frac{1}{\gamma_0 + \Gamma} - \frac{d}{c q_0} \right]
\end{eqnarray}
for $F_{ex}v_{lc}\geq0$.
The corresponding dependence on the external additional friction
$\Gamma$ is represented in left Fig. \ref{motoreffib}. For a clear
understanding we should emphasize that $\gamma_0$ is an alway present internal friction, while
$\Gamma$ is connected with the external load and therefore connected
with the work performed against the external force.
The dependence of the power of the motor on the external friction for
different motor parameters is represented in the right panel of
Fig. \ref{motoreffib}.  Depending on the internal motor parameters,
we observe typically a curve with a maximum at a finite load corresponding to maximal
efficiency, which in our examples is around -- or less than -- $50$ percent and is achieved for
\begin{equation} 
  \Gamma_{max} = (c q_0 / d) - \gamma_0 .
\end{equation} 
After the maximum the efficiency decreases with increasing $\Gamma$. Above a maximal value of external friction, corresponding to vanishing velocity along the limit cycle $v_{lc}=0$, the motor stops to work
and the efficiency becomes formally negative.
The curves for the efficiency of rotational motors shown above may
give us some idea about the efficiency of biological motors
responsible for the mobility of animals, whereby the simple model model discussed here reproduces the generic behavior expected to hold for all biological motor systems.
We mention in particular the measurements and discussions of the efficiency of the kinesin motor \cite{harada_phenomenological_2005,zabicki_thermodynamic_2010,zabicki_efficiency_2010}.

\subsection{Stochastic motion in a harmonic external potential}
Since the main effect of noise is the spreading of the deterministic
attractors we may expect that the two hoop-like limit cycles are
converted into a distribution looking like two embracing hoops with
finite size, which for strong noise converts into two embracing tires
in the four-dimensional phase space. In order to get the explicit
form of the distribution we may introduce different variables, like
the amplitude and phase description as used in the previous sections.
Here we introduce the energy and angular momentum as variables and
derive reduced densities.  We remember that throughout the
Stratonovich calculus is used \cite{anishchenko_nonlinear_2002}.

On the basis of the amplitude and phase representation (polar coordinates)
(\ref{eq:polar_veloc}) and (\ref{eq:polar}) we get for the Hamiltonian
\begin{equation}
  \label{Hro}
  H(t) = {1\over 2} s(t)^2 + {1 \over 2} \omega_0^ 2 r(t)^2.
\end{equation}
The angular momentum is given as
\begin{equation}
  \label{eq:angular}
  L(t) =  (x_1 v_2 - x_2 v_1)= s(t) r(t) cos(\theta).  
\end{equation}
Values corresponding to the two limit cycles are
\begin{equation}
  L = + L_0; \qquad L = - L_0; \qquad L_0 = v_0^2 / \omega_0 .
\end{equation}
with $v_0^ 2= \alpha/\beta$.  Both limit cycles are located on
the sphere with $H(t) = v_0^2$.

Considering harmonic oscillators and using equipartition of potential
and kinetic energy (see equation~(\ref{Hv})) we find for motions on
the limit cycle $s^2(t) = H(t) $. Assuming that $s^2 \simeq H$ holds
also near to the limit cycle, the dynamic system with the pump
(\ref{eq:pum_2d}) is converted to a canonical dissipative system with
\begin{equation}
  \gamma(s^2) \simeq \gamma(H) = \gamma_H(H).
\end{equation}
This way we come for the Rayleigh-model to the energy balance
\begin{equation}
  \label{eq:lang_enrgy}
  { {\rm d} \over {\rm d} t} H  = -\gamma_H(H) \,H + \sqrt{2 D_H  H} \xi_H(t)
\end{equation}
where $\xi_H(t)=\xi_1(t)\cos(\phi)+ \xi_2(t) \sin(\phi)$ is again
Gaussian white noise and $D_H=D$. This corresponds to the
Fokker-Planck equation in energy representation
\begin{equation}
  \label{eq:fpe_energy}
    \frac{\partial }{\partial t} P(H,t) = \frac{\partial}{\partial H}
    \left[\left(\gamma_H(H)\,H - D_H \right) P  
    + D_H\, \frac{\partial }{\partial H}\, H\, P\right]
\end{equation}
which stationary solution reads
\begin{equation}
  \label{P0H}
  P_0(H) = {\cal N} \exp\left[- \frac{1}{D_H} 
    \int \gamma_H(H) {\rm d} H \right]. 
\end{equation}
The most probable value of the energy is the energy on the limit
cycle. In case of the Rayleigh model it is
\begin{equation}
  {\tilde H} = H_0 = \frac{\alpha}{\beta} = v_0^2 .
\end{equation}
The stationary distribution can be given in compact form ($H \ge 0$)
\begin{equation}
  \label{P0H_lin}
  P_0(H) = {\cal N}  
  \exp \left[-\frac{\beta}{2 D}\,(H\,-H_0)^2 \right]
\end{equation}
which is a Gaussian at positive energies.

This probability is in fact distributed on the surface of the
four-dimensional sphere. By using equation~(\ref{Hro}) we get for the
Rayleigh- model of pumping in our approximation the following
distribution of the coordinate with $r^2=x_1^2+x_2^2$
\begin{equation}\label{eq:P_rho}
  P_0 (x_1 , x_2) \simeq  \exp{\left[\frac{\alpha \omega_0^2}{D} 
      r^2\left(1-\frac{r^2}{2 r_0^2}\right)\right]}. 
\end{equation}
We see in figure~\ref{eq:P_rho} that the probability crater is located
above the trajectory obtained from simulations of an Active Brownian
particle. This way the maximal probability corresponds indeed to the
deterministic limit cycle.
\begin{figure}[htbp]
  \label{P_rho_pur} 
  \includegraphics[width=.45\linewidth]{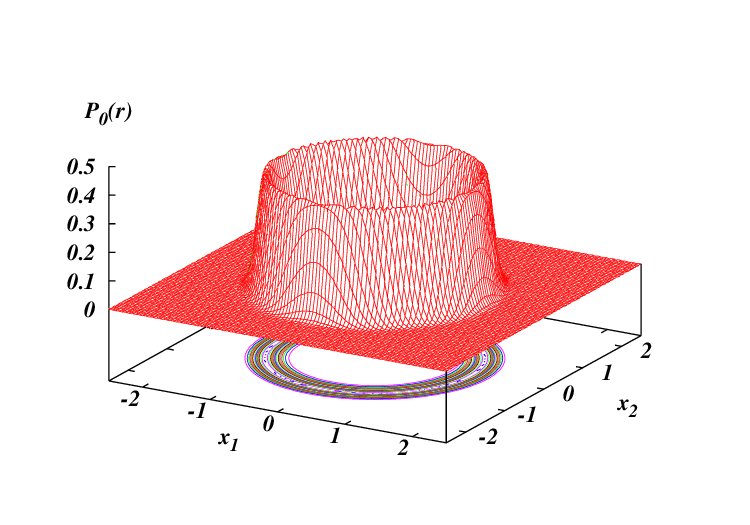}
  \includegraphics[width=.45\linewidth]{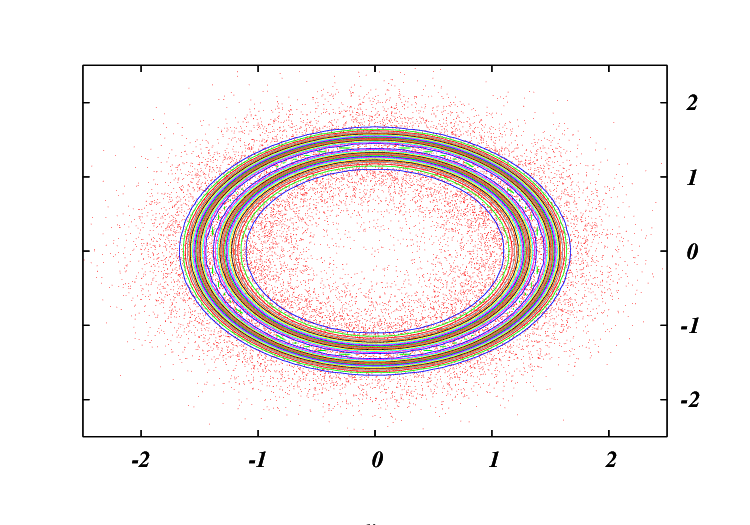}
  \caption{Probability density for the Rayleigh-model represented over
    the $\{x_1,x_2\}$ plane. (a) The probability
    density~(\ref{eq:P_rho}). (b) Contour plot of $P_0(r)$
    superimposed with data points out of simulations of the Active
    Brownian dynamics. Parameters: $\alpha=2$, $D=0.1$ and
    $\omega_0=1$}
\end{figure}

So far we represented only a projection on the $\{x_1,x_2\}$ plane.
The full probability distribution in the four-dimensional phase space
is not constant on the four-dimensional sphere $H = v_0^2$ as
suggested by equation~(\ref{P0H}) but should be concentrated around
the limit cycles which are closed curves on the four-dimensional
sphere $H = v_0^2$.  This means, only a subspace of this sphere is
filled with probability.  The correct stationary probability has the
form of two noisy distributions in the four-dimensional phase space,
which look like hula hoops.  This characteristic form of the
distributions was confirmed also by simulations (see
figure~\ref{P_rho_pur} and \cite{erdmann_brownian_2000}). The projections of the
distribution to the $\{x_1,x_2\}$ plane and to the $\{v_1,v_2\}$ plane
are noisy tori in the four-dimensional phase space. The hula hoop
distribution intersects perpendicular the $\{x_1,v_2\}$ plane and the
$\{x_2,v_1\}$ plane.  The projections to these planes are rod-like and
the intersection manifold with these planes consists of two ellipses
located in the diagonals of the planes.

In order to refine the description we find the distribution of the
angular momenta. We start from the Langevin equation
\begin{equation}
  \frac{{\rm d} }{{\rm d} t} L = - \gamma\left(s^2\right) L 
  + \sqrt{2 D} r\, \xi_L(t)
\end{equation}
\label{ang1}
with $\xi_L(t)=\xi_y(t) \cos(\phi)-\xi_x(t)\sin(\phi)$ being Gaussian
white noise. On the limit cycles it holds
\begin{equation}
  \label{eq:ang_limit}
  L(t) = \pm r(t) s(t), \quad s(t)\,= \omega_0 \,r(t), 
\end{equation}
respectively, the different signs for the different cycles and
$r(t)=r_0$ and $s(t)=v_0$. To find a closed description we assume that
the equations~(\ref{eq:ang_limit}) hold and replace
\begin{equation}
  \label{eq:replace}
  r(t)= \sqrt{{L(t) \over \omega_0}}, \quad  s^2(t)= L(t)\,{\omega_0}, 
  \quad \gamma(s^2)=\gamma\left(L {\omega_0 }\right)= \gamma_L(L) 
\end{equation}
where we have used the positive sign and hence $L > 0$. It follows
\begin{equation}
  \frac{{\rm d} }{{\rm d} t} L = - \gamma_L(L)\, L  + \sqrt{2 D_L L} \, \xi_L(t)
\end{equation}
with
\begin{equation}
  \label{eq:ang_noise}
  D_L=\frac{D}{\omega_0}\,.
\end{equation}
The corresponding Fokker-Planck equation is similar to the energy
representation
\begin{equation}
  \label{eq:fpe_angular}
    \frac{\partial }{\partial t} P(L,t)\, =\,\frac{\partial}{\partial L}
  \left[\left(\gamma_L(L)\,L\, -\,D_L \right) P  
    + D_L\, \frac{\partial }{\partial L}\, L\, P\right].
\end{equation}
Obviously its stationary solution reads
\begin{equation}
  \label{P0L}
  P_0(L) = {\cal N} \exp\left[- \frac{1}{D_L} 
    \int \gamma_L(L) {\rm d} L \right]
\end{equation}
and eventually after introducing the most probable angular momentum
$L_0 = r_0 v_0=H_0/\omega_0$ at the limit cycle the stationary
solution becomes ($L > 0$)
\begin{equation}
  P_0(L) = {\cal N} 
  \exp \left[- \frac{\beta \omega_0^2}{2  D} (L-L_0)^2  \right].
\end{equation}
A corresponding solution can be found for the second cycle by replacing $L_0
\rightarrow -L_0$ for momenta with $L < 0$. Since due to symmetry both values
are provided with same probability one may expect a linear superposition of
the two solutions
\begin{equation}
  P_0(L) = {\cal N}
  \left( 
    \exp \left[
      - \frac{\beta \omega_0^2}{2 D} (L-L_0)^2  \right]
    +\exp \left[
      - \frac{\beta \omega_0^2}{2 D} (L+L_0)^2  \right]
  \right).
\end{equation}

The given method does not provide a complete solution in the
four-dimensional phase space, but gives us a good idea about the
projections on different planes. In order to find a distribution in
the four-dimensional phase space we combine the previously found
distributions and introduce the invariants $J_{+},J_{-}$ which leads
to the following ansatz:
\begin{eqnarray}
 && P_0(x_1,x_2,v_1,v_2) = {\cal N} \exp 
  \left[-\frac{\beta}{2 D} (H - H_0)^2\right]\\
  &&~~~~~~~\times \left(\exp \left[- \frac{\beta}{2 D} 
      J_{+}^2  \right]
    + \exp \left[- \frac{\beta}{2 D}
      J_{-}^2  \right] \right)\nonumber
\end{eqnarray}
We may convince ourselves that this formula agrees with all
projections derived above. Furthermore, it is in agreement with the
general ansatz derived in earlier work from information theory
\cite{ebeling_statistical_2004,ebeling_statistical_2005}. Since our new expression for the stationary
distribution does not contain any parameter characterizing the
concrete potential it may be applied to arbitrary radially symmetric
potentials, in particular we may use it for describing the stationary
distributions for a Coulomb confinement.

\subsection{Dynamics in symmetric anharmonic potentials}
\label{sect:anhrmonic}
Here we will discuss briefly several extensions of the theory
developed in the previous section following \cite{schimansky-geier_stationary_2005}.  At first
we will discuss the case of anharmonic potentials.  For the general
case of radially symmetric but anharmonic potentials $U(r)$ the equal
distribution between potential and kinetic energy $mv_0^2 = a r_0^2$
which leads to $\omega_0 = v_0/r_0 = \omega$ is no more valid. It has
to be replaced by the more general condition that on the limit cycle
the attracting radial forces are in equilibrium with the centrifugal
forces.  This condition leads to $\frac{v_0^2}{r_0} = |U'(r_0)|$.
Then, if $v_0$ is given, the equilibrium radius may be found from the
relation
\begin{equation}
  \label{impl}
  v_0^2 = r_0\, |U'(r_0)|\,.  
\end{equation} 
The frequency of the limit cycle oscillations follows as
\begin{equation}
  \omega_0^2 = \frac{v_0^2}{r_0^2} = \frac{|U'(r_0)|}{r_0} \,.
\end{equation} 
For example, for the case of quartic oscillators with potential $ U(r)
= k r^4/4 $ we get the limit cycle frequency
\begin{equation}
  \omega_0 = \frac{k^{1/4}}{v_0^{1/2}} 
\end{equation} 
Alternatively for attracting Coulomb forces $U(r)\,=\, - \,Z e^2 / r$
the stable radius reads and the limit cycle frequency, respectively,
\begin{equation}
  \label{eq:coulob_radius}
  r_0\,=\,{Z e^2 \over v_0^2 }\,,~~\omega_0=\frac{v_0^3}{Ze^2}
\end{equation}
Integrals of motions follow
\begin{equation}
  \label{eq:H0_L0_Coulomb}
  H_0 = - \frac{1}{2}\, v_0^2\,;\qquad  L_0 = \pm  \frac{Z e^2}{v_0} \,.
\end{equation}
We note that this expression diverges for $v_0 \to 0$ (similarly as in
quantum theory the Bohr radius diverges for $h \to 0$).

If the equation (\ref{impl}) has several solutions, the dynamics might
be much more complicated, e.g. we could find Kepler-like orbits
oscillating between the solutions for $r_0$. In other words we may
find then beside driven rotations also driven oscillations between the
multiple solutions of equation~(\ref{impl}).

An interesting application of the theoretical results given above, is
the following: Let us imagine a system of Brownian particles which are
pairwise bound by a Lennard-Jones-like potential $U(r_1 - r_2)$ to
dumb-bell-like configurations.  Then the motion consists of two
independent parts: The free motion of the center of mass, and the
relative motion under the influence of the potential.  The motion of
the center of mass is described by the equations given in the previous
section and relative motion is described by the equations given in
this section.  As a consequence, the center of mass of the dumb-bell
will make a driven Brownian motion but in addition the dumb-bells are
driven to rotate around there center of mass. What we observe then is
a system of pumped Brownian molecules which show driven translations
with respect to their center of mass.  On the other side the internal
degrees of freedom are also excited and we observe driven rotations
and in general (if equation~(\ref{impl}) has several solutions) also
driven oscillations.  In this way we have shown that the mechanisms
described here may be used also to excite the internal degrees of
freedom of Brownian molecules.

\subsection{Dynamics in asymmetric potentials} 
\begin{figure}[h]
  \begin{center}
    \includegraphics[width=.8\linewidth]{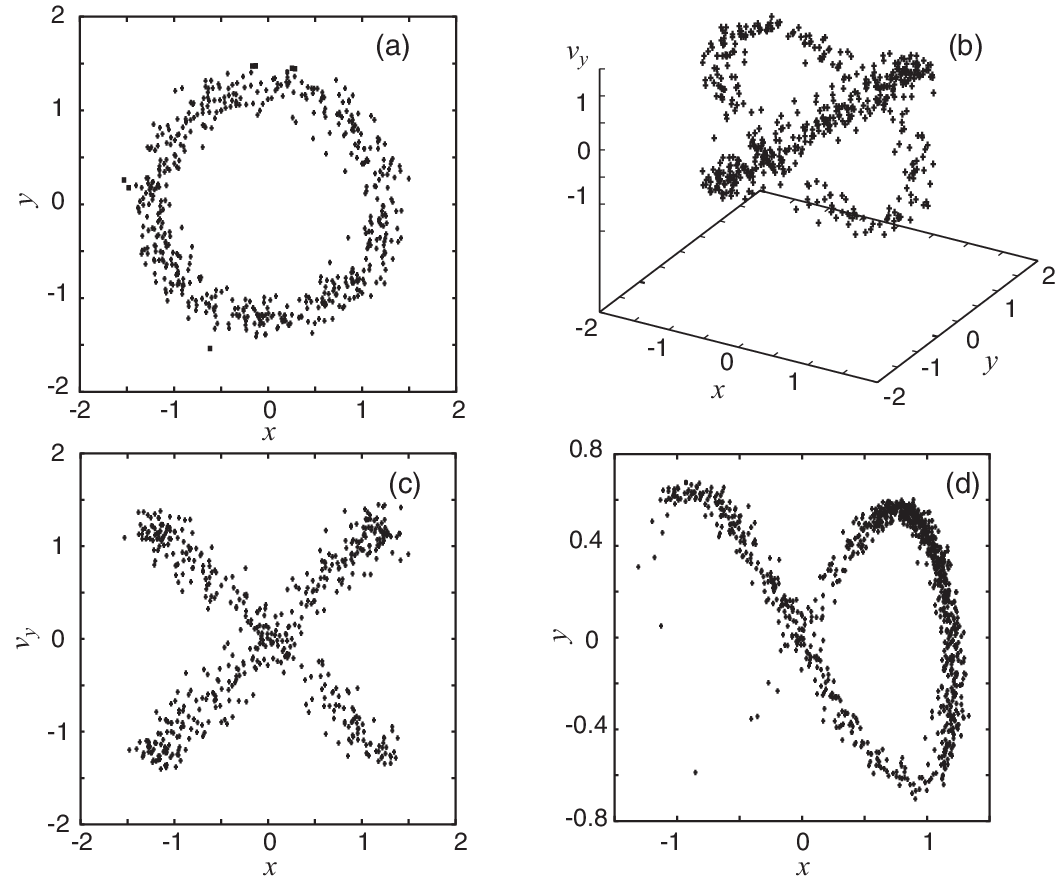}
    \caption{\label{distrib} Distributions according to simulations of
      1000 active Brownian particles with depot pump (Sec. \ref{sec:det_depot}) on limit cycles in symmetric
      (a-c) and asymmetric (d) parabolic potentials.  a) We show
      projections of simulations on the $x-y-$plane. Parameters: $\zeta = 1.5, \gamma_0=1, D=0.01$: (a) 2D
      projections on the subspace $\{x_1, x_2\}$, (b) 3D projections
      on the subspace $\{x_1,x_2,v_2\}$, (c) 2D projections on the
      subspace $\{x_1,v_2\}$, (d) 2D projections on the subspace
      $\{x_1, x_2\}$ for $\omega_2 =2 \omega_1$\label{compsym_unsym} \cite{erdmann_excitation_2002}.}
  \end{center}
\end{figure}

We will study now potential landscapes without radial symmetry and
follow \cite{erdmann_attractors_2005}. Problems which we might be to studied by
help of these models are, for example, the synchronization between the
oscillations along both axis, the existence of Arnold tongues
etc. \cite{erdmann_excitation_2002}.  In the simplest case, we may assume that the
potential is harmonic along the two axes but stretched in an asymmetric way
with different elasticity constants
\begin{equation}
U(x,y) = \frac{1}{2}\omega_1^2 x^2 + \frac{1}{2}\omega_2^2 y^2.
\label{unsympot}
\end{equation}
denoted by $\omega_1 \ne \omega_2$. In order to give some idea about
the influence of asymmetry between the modes led us first present the
result of simulations for 1000 independent active particles with the
depot pump (see Sec. \ref{sec:det_depot}). The distribution of these
particles in the phase space shows us how the probability
distributions are expected to have their maxima (see Fig.
\ref{compsym_unsym})

We also study here a specific case of a simple relation of the
frequencies $\omega_2 = 2 \omega_1$.  The solution of the Hamiltonian
problem (e.g. no dissipation) is given by
\begin{equation}
  \begin{array}{rcl}
    x_1 = r_1 \cos(\omega_1 t - \Phi) &\quad&
    v_1 = - r_1 \omega_1 \sin(\omega_1 t - \Phi)\\\\
    x_2 = r_2 \sin(2 \omega_1 t - \Phi) &\quad&
    v_2 = 2 r_2 \omega_1 \cos(2 \omega_1 t - \Phi)
\end{array}
\end{equation}
where the amplitudes $r_{1,2}$ are determined by the initial
conditions. In the case of active particles, $\delta=\zeta+1>0$ (see Eq. \ref{eq:zeta}), the
dissipative forces drive the trajectories to an attractor which
corresponds to $v_1^2 + v_2^2 \rightarrow v_0^2$ In a good
approximation the attractor is determined by the amplitudes,
\begin{equation}
  r_1 = \frac{v_0}{\omega_1}; \qquad  r_2 = \frac{v_0}{\omega_2}.  
\end{equation}
We see that the smaller amplitude has approximately half the value of
the bigger amplitude. This way we have obtained a limit cycle shaping
like a Lissajous oscillation in the form of an ``8''; a second limit
cycle can be found by inversion of the trajectory. The result of a
stochastic simulation is shown in Fig.  \ref{distrib} (d).  We see
here the projections of two stochastic trajectories (limit cycles);
the projections having the form of an ``8''. In the case of irrational
relations $\omega_2/\omega_1$ as well as for the case of nonlinear
couplings between the modes we find more complicated attractors
\cite{erdmann_excitation_2002}. In order to get analytical expressions for the
distributions we may introduce amplitude--phase representations
\cite{erdmann_brownian_2000}.

\subsection{Transitions bistable potentials}
\begin{figure}[htbp]
  \begin{center}
    \includegraphics[width=.9\linewidth]{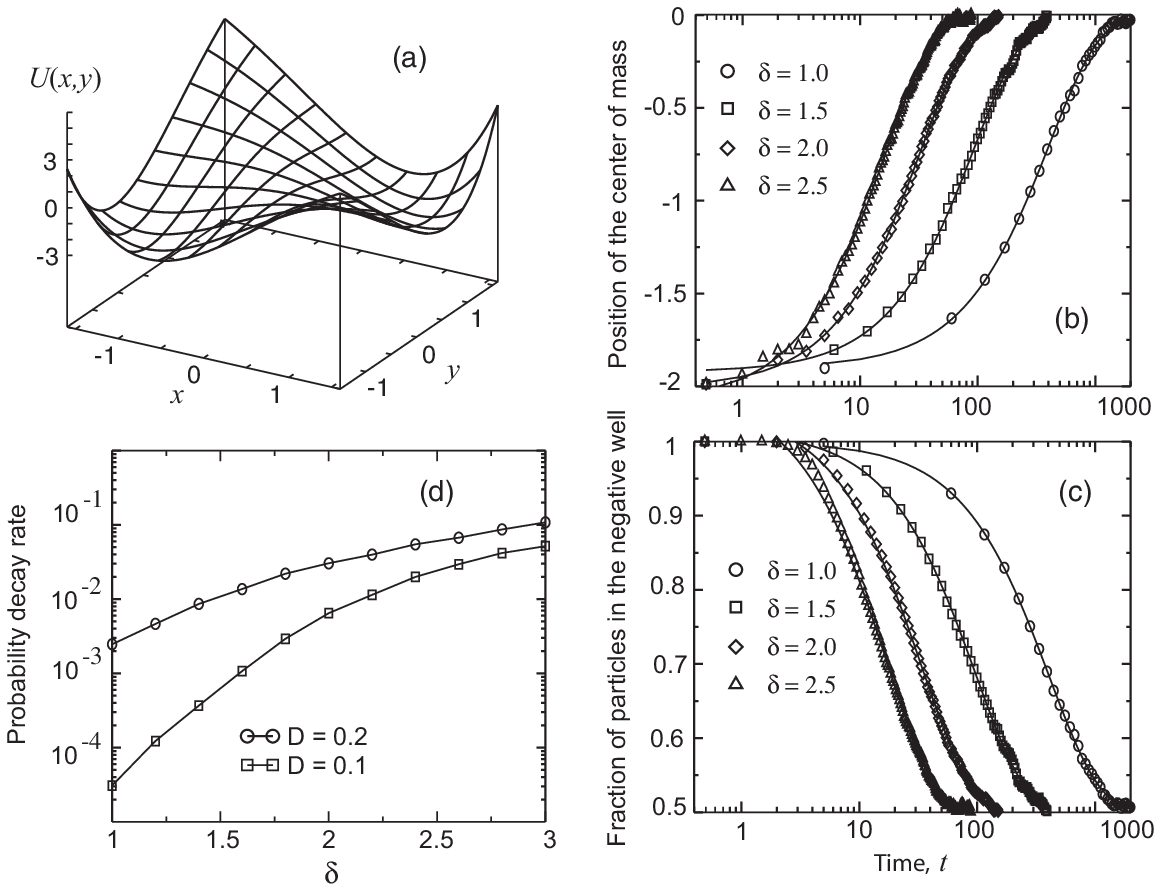}
    \caption{\label{Ubist4} Active Brownian particles in a bistable
      potential in 2 dimensions.  (a) The shape of the potential for
      $\Delta U = 1: a=4, c = 0 $, the minima are located along the
      diagonal.  (b) The position of the center of mass of 10000
      particles initially located in the left well under the influence
      of noise $D=0.1$ as a function of time, for indicated values of
      driving parameter $\delta$. The transitions are enhanced for
      larger driving.  (c) The fraction of particles staying in the
      left well for $D=0.1$ (d) The decay rates of the probability to
      stay in the left well for the noise strength for $D=0.1$ and for
      $D=0.2$. On (b) and (c) symbols shows numerical values, while
      the solid lines are exponential least-square fits due to Eqs.
      (\ref{fit}). Other parameters (see Fig. \ref{gamma54}):
      $\gamma_0=1;v_d=1$ \cite{ebeling_stochastic_2005} }
  \end{center}
\end{figure}

Let us now have a look at bistable situation in a two-dimensional
landscape following \cite{ebeling_stochastic_2005}. In order to investigate the
Kramers problem of transitions between two wells we introduce a simple
model potential with two potential minima placed on the line $x = y$.
The potential is defined by
\begin{equation}
U(x,y) = a (\frac{1}{4}z^4 - \frac{1}{2} z^2 -c z) + \frac{1}{2}\omega^2 (x - y)^2.
\label{bipot}
\end{equation}
Here we introduced $z = (x+y)/2$ being the reaction path, i.e.  the
co-ordinate along the line connecting the minima with the lowest
barrier value between both.  The parameter $c$ determines the
asymmetry of the wells and $a$ determines the height of the well which
is for ($c=0$) given by $\Delta U = a/4$.  The shape of the potential
is presented in Fig.  (\ref{Ubist4}).
In the case of small excitation energy $v_0^2 / 2 \ll \Delta U$, and
weak noise, the dynamics in each well is similar to the parabolic
case.  In other words we may observe two limit cycles in each of the
two wells.  With increasing noise, transitions between the two wells
become possible. For standard Brownian motion this is a well-studied
problem \cite{haenggi_stochastic_1982}.

Here we present numerical simulation of 10000 active particles with
depot pump of energy (see Fig. \ref{gamma54})and provide estimates of
the transition rates (Fig.~\ref{Ubist4}.  Initially particles were put
at $z=Z_0$ in the left potential well.  In Fig. \ref{Ubist4} (b,c) the
temporal evolutions of the position of the center of mass $Z(t)$ and
the fraction of particle in the left well, $p(t)=n_l(t)/N$ are shown,
$n_l$ is the number of particles in the left well and $N=10000$ is the
total number of particles.  As can be seen from the figure, the time
courses can be fitted with exponential laws:
\begin{equation}
\label{fit}
Z(t)=Z_0 \exp\left({-\frac{t}{\tau}}\right), \qquad
p(t)=\frac{1}{2}\left[1+ \exp\left({-\frac{t}{\tau}}\right)\right],
\end{equation}
where $\tau$ is the characteristic transition time. The transition
rate $1/\tau$ is shown in Fig.~\ref{Ubist4}(d) as a function of
$\delta$.  A rough estimate of the transition time is
\begin{equation}
  \tau \propto \exp\left[\frac{\Delta U - (1/2) v_0^2}{D}\right], 
\end{equation} 
where $v_0^2 = (\delta - 1)$ for our model.  Thus, with the increase of
$\delta$ the potential barrier is effectively lowered, providing
higher transition rates which agree with the result for the one-dimensional case \cite{pohlmann_self-organized_1997}.

In other words, the effect of active driving (in comparison to the
passive case) may be estimated by
\begin{equation}
  \frac{\tau_{act}}{\tau_{pas}} \propto \exp\left[- \frac{v_0^2}{ 2 D}\right]
\end{equation}
Due to the nearly exponential dependence on $(\delta - 1)$ the
increase of the rate with the driving strength may be rather large. We
have to note, however, that the exponential ``ansatz'' yield just a
rough estimate for moderate pumping $1 < \delta < 2$. For $\delta > 2$
we observe in Fig.\ref{Ubist4} (d) already a tendency to saturation
(the particles freely penetrate the barrier) and for $\delta < 1$
(transition to the passive case) we observe a more sensitive
dependence on the parameter $\delta$.

%% file: SECTION_swarming.tex
\section{Collective motion and swarming}
\label{sec:swarming}
We apply in this section the model of active Brownian dynamics to
describe the collective motion in active matter systems, such as e.g. swarms of animals.

At first we will characterize the fundamental modes of collective motion of swarms with cohesion
using a quite simple model based on the idea of global coupling of
the individuals in the swarm. 
Global coupling is not based on realistic physical interactions
however it demonstrates already the most important features of
``swarms''. We use here the general notation of ``swarms'' for confined
systems or clusters of particles which can perform collective motion far
from equilibrium.

After the discussion on modes of collective motion induced by global coupling, we will turn to the important class of models of collective motion due to local, short-range velocity alignment between active particles. 
We will start with a description of active Brownian particles with nonlinear friction and velocity alignment. Further on, we proceed with the discussion of a widely used class of self-propelled particles, which correspond to limit of vanishing speed fluctuation. In this context we will focus on different interaction symmetries. 
Starting from the classical Vicsek model of active polar particles with polar interactions, we discuss a 
general classification of models that includes not only so-called active nematics, but also a second important class of models of active polar with apolar interactions.
Throughout this second subsection, we will focus on the onset of collective order for increasing density or decreasing noise in various models. 

Finally, in the end of this section we will introduce alternative mechanisms for swarming and collective motion based on escape and pursuit response or chemotactic behavior of individuals.

\subsection{Dynamics of swarms}
\label{sec:swarms}

\subsubsection{Modes of collective motion with cohesion: translation and rotation}

 The model of global coupling is the
most simple dynamical model of collectively moving swarms of animals.
Global coupling is not based on realistic physical interactions
however it demonstrates already the most important features of
``swarms''. 

Probably one of the first individual based model mathematical model of swarms was introduced by Suzuki and Sakai in 1973 (\cite{suzuki_movement_1973}; see also \cite{okubo_diffusion_2001}). Since then, different models of swarming of animals, many of them motivated by fish schools, have been the subject of biological and ecological investigations (see e.g. \cite{aoki_simulation_1982,huth_simulation_1992,okubo_diffusion_2001,couzin_collective_2002,couzin_effective_2005,bode_making_2010}).  Here we would like to highlight the pioneering paper on swarm dynamics by Hiro-Sato Niwa published in 1994 \cite{niwa_self-organizing_1994}. Niwa, not only introduced a nonlinear friction function in his model, but he is probably the first one who analyzed the stability of the different modes of swarming motion from the point of view of statistical physics. Despite the numerous biologically motivated studies, swarming dynamics is still a rather young field of physical studies. Some recent developments have been covered, for example, by Helbing \cite{helbing_traffic_2001}, who discusses traffic
and related self-driven many-particle systems and the comprehensive books by
Vicsek \cite{vicsek_fluctuations_2001}, Mikhailov and Calenbuhr \cite{mikhailov_cells_2002}, and
Schweitzer \cite{schweitzer_brownian_2003}. 

Based on the simulation studies of the first swarming models \cite{suzuki_movement_1973}, a classification of different modes of swarming motion was performed, which can be also observed in nature (see Okubo and Levin \cite{okubo_diffusion_2001}). According to these authors, three general modes of motion can be identified:
\begin{enumerate}
\item translational motions,
\item rotational excitations and
\item amoeba-like motions.
\end{enumerate}

It is beyond the focus of this review to discuss all the swarming models, 
which were developed in order to account for the rich variety of observations.
Instead, we will study in the following the collective modes and the 
distribution functions of simple models. 
We start here with consideration of finite systems of particles confined
by global coupling which are self-propelled by active
friction and parallelized by small velocity-dependent interactions.
This kind of description may be considered as a simple and to large extent analytically tractable approximation of the
collective dynamics of biological agents \cite{mikhailov_noise-induced_1999,ebeling_swarms_2001,schweitzer_statistical_2001,ebeling_nonequilibrium_2003,erdmann_collective_2003,erdmann_noise_2005,ebeling_statistical_2005}.
More complex models, which incorporate for example local interactions, are discussed in the forthcoming sections as well as in other publications (see e.g. Refs.~ \cite{vicsek_novel_1995,czirok_formation_1996,vicsek_fluctuations_2001,vicsek_collective_2010}).  

We will show here that the collective motion of swarms (large
clusters) of driven Brownian particles resemble very much the typical
modes of motion in swarms of living entities.  The
self-propulsion of the particles is modeled by active friction as
introduced in Sec. \ref{sec:act}. The analysis is
restricted to two dimensional models ($d=2$). The internal driving of 
the system by negative friction yields a dynamical behavior far-from equilibrium. Earlier studies have
shown that such active interacting systems may have many attractors and
that noise may lead to transitions between the deterministic
attractors \cite{mikhailov_noise-induced_1999,schweitzer_statistical_2001}.

\subsubsection{General model of swarms}
\label{sec:gen_mod_swarm}
We proceed with an analysis of the structure of the equations of motion
of a swarm. We will characterize the motion by its attractors in the
phase space and again by the stationary distribution if noise is
included. First we start with the two dimensional problem and harmonic
interactions. In the next section generalizes to three spatial dimensions and
to interaction with Morse potentials.

We consider a system on $N$ active particles at positions $({\bbox
  r}_1(t),\ldots,{\bbox r}_N(t))$ which are moving with velocities ${\bbox
  v}_1(t),\ldots,{\bbox v}_N(t))$. Let the particles be point-like with the mass $m=1$. Since motion is created by forces we postulate the equations of motion for the individuals within the swarm to be
given given by the $N$ (stochastic) Newtonian equations
\begin{eqnarray}
  \label{langev-or2}
  && {{\rm d}\over {\rm d}t} \bbox{r}_i=\bbox{v}_i\,,\\
  && {{\rm d}\over {\rm d}t}\bbox{v}_i=  -\gamma(\bbox{v}_i) \bbox{v}_i  + \bbox{F}_i(\bbox{r}_1,\dots,\bbox{r}_N,\bbox{v}_1,\dots,\bbox{v}_N) + \sqrt{2 D} {\bbox \xi}_i(t), 
\end{eqnarray}
where ${\bbox \xi}_i$ are Gaussian white noise sources acting on every particle, independent from other particles. 

The first term on the right hand side is the dissipative force, which accounts for the individual velocity dynamics (friction/energy pump).
In case of the Rayleigh-Helmholtz model (Eq. \ref{eq:helm_ray}) these forces are read
\begin{equation}
-\gamma(\bbox{v}_i) \bbox{v}_i =  (\alpha-\beta \bbox{v}_i^2)\bbox{v}_i, 
\label{disforce}
\end{equation}
Also the active linear dissipative force with
friction coefficient (\ref{eq:sg_fri}) can be easily generalized to
$N$ particles. Eventually one can define for every particle
$i=1,\dots,N$ an individual energy depot $e_i(t)$ which
will yield a velocity-dependent friction force (\ref{depot_diss_force}).

The second force term $\bbox{F}_i$ accounts for the interactions between the particles, as well as possible external forces acting on individual particles. In general, it can be a function of all positions and velocities of particles forming a swarm.
Here, we assume that  $\bbox{F}_i$ consists of two components: a velocity dependent (dissipative) force and a force due to potential gradients:
\begin{align}
\bbox{F}_i = \bbox{F}_i^\text{va}(\bbox{v}_1,\ldots,\bbox{v}_N) - \nabla U_N(\bbox{r}_1,\ldots,\bbox{r}_N)\,.
\end{align}
The  potential $U_N$ is assumed to consist of an external potential $W(\bbox{r_i})$ and a superposition of pair-wise interaction potential $U(r_{ij})$, with $r_{ij}$ being the distance between particle $i$ and $j$. Hence, the total potential reads:
\begin{equation}
  \label{Upot}
  U_N = \sum_i W(\bbox{r_i}) + \frac{1}{2} \sum_{ij} U (r_{ij}).
\end{equation}

We will start with a parabolic approximations of the interaction
forces generated by the pair-wise potentials \cite{mikhailov_noise-induced_1999,ebeling_swarms_2001,schweitzer_statistical_2001}.
\begin{eqnarray}
  \label{eq:pot_harm}
  U (r_{ij}) \,=\, {\omega^2\over 2} r_{ij}^2, 
\end{eqnarray}
which might hold if the size of the objects is small compared
to the spatial scale of their motion. Harmonic interaction potentials constitute a global coupling between individual particles, which was considered in this context in various publications \cite{niwa_self-organizing_1994,mikhailov_noise-induced_1999,schweitzer_statistical_2001}. It allows to reduce the problem
effectively to the motion in a well formed by harmonic forces which we
considered in Section \ref{sec:harm}. With global coupling the
particles are attracted by an effective spring force to the center of
mass given by
\begin{eqnarray}
  \label{eq:com}
  {\bbox r}_{c.o.m.}(t) = \frac{1}{N} \sum_{j} {\bbox r}_j(t)\,.  
\end{eqnarray}
The effective mean field force acting on the $i$-th particle is defined by
\begin{eqnarray} 
  {\bbox K}_i \,=\, -\nabla_i  U_N(\bbox{r}_1,\ldots,\bbox{r}_N) \,=\, - \omega_0^2 [{\bbox r}_i - {\bbox r}_{c.o.m.}(t)]\,.
  \label{spring}
\end{eqnarray}
Several approaches to include more realistic interactions will be discussed
later on.  Here we mention that the case of constant external forces
was already treated by Schienbein et al.  \cite{schienbein_langevin_1993,schienbein_random_1994}.
Symmetric parabolic external forces were studied in Refs.~
\cite{ebeling_active_1999,erdmann_brownian_2000} and the non-symmetric case is being
investigated in Ref.~\cite{erdmann_excitation_2002}.  More complicated external
fields including short range repulsion were studied in detail in
several publications
\cite{makarov_soliton-like_2000,makarov_dissipative_2001,ebeling_stochastic_2005,schimansky-geier_stationary_2005} and will
also be addressed in Sec. \ref{sec:swarm_3d}.

The dissipative interaction between particles, is assumed to be a velocity-dependent
interactions with the tendency to synchronize the velocities of different particles. 
We assume a simple alignment law \cite{okubo_diffusion_2001}
\begin{equation}\label{eq:va_sigma} 
  \bbox{F}_i^\text{va} (\bbox{v}_1,...,\bbox{v}_N) = -
  \mu \sum_{j} g_N(r_{ij}) (\bbox{v}_i - \bbox{v}_j )\,.
\end{equation}
Here $g(r)$ is a function which accounts for possible distance dependence of the interaction, such as decay of the interaction strength with distance. This models a dissipative force which tends to parallelize the individual
velocities.

In this and the next section, we will consider for simplicity a velocity coupling with an infinite range:  $g_N(r)=1/N$.
Thus, in this model all particles synchronize with the total swarm velocity $\bbox{u}(t)$ which is the
velocity of the center of mass (c.o.m.) of the swarm.
\begin{equation} 
  \label{eq:v_com}
  \bbox{u}(t) = {1 \over N}\sum_{j} \bbox{v_j}.
\end{equation}
In addition we will assume that the alignment strength $\mu$ is vanishingly small. Hence, in the next section our focus lies on the conservative interactions.

Other couplings with finite range $g(r)$ have been studied in numerous previous publications (see e.g. \cite{vicsek_novel_1995,vicsek_fluctuations_2001,chate_collective_2008,peruani_mean-field_2008}). We will discuss the (local) alignment interaction  in detail in Sec. \ref{sec:velal}. There are other types of dissipative interactions, which lead to synchronization of individual velocities as, for example, hydrodynamic interactions \cite{erdmann_collective_2003,baskaran_statistical_2009} or symmetry-breaking contact interactions, such as e.g. inelastic collisions \cite{palsson_model_2000,szabo_phase_2006,peruani_nonequilibrium_2006,grossman_emergence_2008}.  We will discuss local velocity alignment in detail in the forthcoming sections.  Here we restrict our investigation to the simplest model of velocity coupling
- global velocity coupling.

\subsubsection{The model of harmonic swarms with global coupling}
The concept of global coupling has proven to be very useful for the
investigation of stochastic systems \cite{anishchenko_nonlinear_2002}.  For the case of
Rayleigh-driving, the dynamical equations for center of mass and its velocity
have the form \cite{ebeling_swarm_2008}
\begin{eqnarray}
\label{eq:com1}
&&  {{\rm d}\over {\rm d}t} \bbox{r}_{c.o.m.}(t)\,=\,\bbox{u}(t)\,,\\
&&  {{\rm d}\over {\rm d}t}\bbox{u} = \left[\alpha - \beta \bbox{u}^2\right] \bbox{u}
- \frac{\beta}{N-1} \sum_i \left((\delta \bbox{v}_i)^2 \bbox{u}+\left(\bbox{u} \cdot  
    \delta \bbox{v}_i\right) \delta \bbox{v}_i\right) + \sqrt{2 D} \bbox{\xi}_u (t)\,.\nonumber  
\end{eqnarray}
Therein $\delta \bbox{r}_i(t) = \bbox{r}_i(t) - \bbox{r}_{c.o.m.}(t)$ and
$\delta \bbox{v}_i(t) = \bbox{v}_i(t) - \bbox{u}(t)$ are the deviations
of the individual objects from the center of mass, which obey
\begin{eqnarray}
\label{eq:swarm1}
  && {{\rm d}\over {\rm d}t}\delta \bbox{r_i}\,=\, \delta \bbox{v_i}\, \\  
  && {{\rm d}\over {\rm d}t}\delta \bbox{v_i} + \omega_0^2 \delta \bbox{r_i} = \left[\alpha -\mu -\beta \delta \bbox{v}_i^2 -
    \beta \bbox{u}^2 \right] \delta \bbox{v_i}  - 2 \beta ( \bbox{u} \cdot \delta \bbox{v_i}) \bbox{u} + \sqrt{2D} \delta \bbox{\xi}_i (t),  \nonumber
\end{eqnarray}
where the Gaussian white noise acting on the center of mass is defined as
\begin{equation}
  \bbox{\xi}_u(t) = \frac{1}{N} \sum_i \bbox{\xi}_i(t), \quad \mean{\bbox{\xi}_u(t)}=0, \quad \mean{\xi_{uk} \xi_{ul}} = \frac 1N \delta_{kl}\delta(t-t`),~~l,k=x,y\,.  
\end{equation}
The noise agitating the individual objects reads ($i,j=1,\dots,N$)
\begin{align}
  \delta \bbox{\xi}_i = \bbox{\xi}_i -\bbox{\xi}_u, \qquad \mean{\delta \bbox{\xi}_i}=0, \langle \delta \xi_{ik} \delta \xi_{jl} \rangle =\delta_{ij}\delta_{kl} ( 1- \frac{1}{N})\delta(t-t`),
\end{align}
with $k,l=x,y$. We note that these noise terms are again independent of the motion of particles.

Lets first look at dynamics of the center of mass. We start by decoupling is 
from the dynamics of the deviations by averaging with respect to
$\delta \bbox{v}_i$. Approximately we get in this way
\begin{equation}
    {{\rm d}\over {\rm d}t} \bbox{u} = (\alpha_1 - \beta \bbox{u}^2) \bbox{u}  
  + \sqrt{2D} \xi_u(t). 
\end{equation}
Importantly, here $\alpha_1 < \alpha$ is determined by the positive
mean quadratic dispersions of the $\delta\bbox{v}_i$. The
corresponding velocity distribution of the center of mass is
\begin{equation}
  P^{(0)}(\bbox{u}) =
  C \exp\left[ N \,\left(\frac{\alpha_1}{2 D} \bbox{u}^2
      - \frac{\beta}{4 D} \bbox{u}^4 \right)\right]\,.
\label{f(V)}
\end{equation}
This way we find the most probable velocity as $\bbox{u}_1 = u_1
\bbox{e}_v$ with equally distributed and diffusing directions. It owns
values of the speed
\begin{eqnarray}
  \label{eq:swarm_speed}
  u_1=\sqrt{\alpha_1/\beta}
\end{eqnarray}
which are smaller than the speed of an freely moving individual object
which value was $v_0=\sqrt{\alpha/\beta}$.  The shift with respect to
the free mode depends on the noise strength $D$. The solution breaks
down if the dispersion of the relative velocities $\delta
\bbox{v}_i^2$ becomes so large that the $\alpha_1$ becomes negative.
With increasing noise we find a bifurcation to another mode which most
probable value has speed ${u}_2=0$.
\begin{figure}
  \centering \includegraphics[width=6cm, height=6cm]{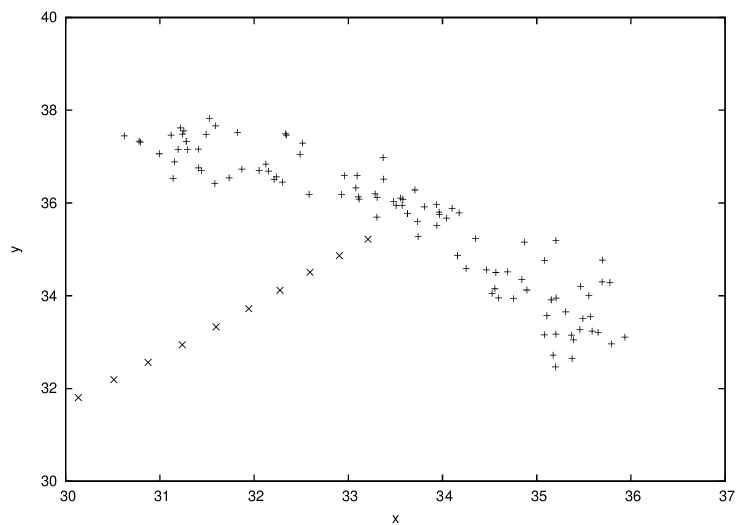}\hfill\includegraphics[width=6cm, height=6cm]{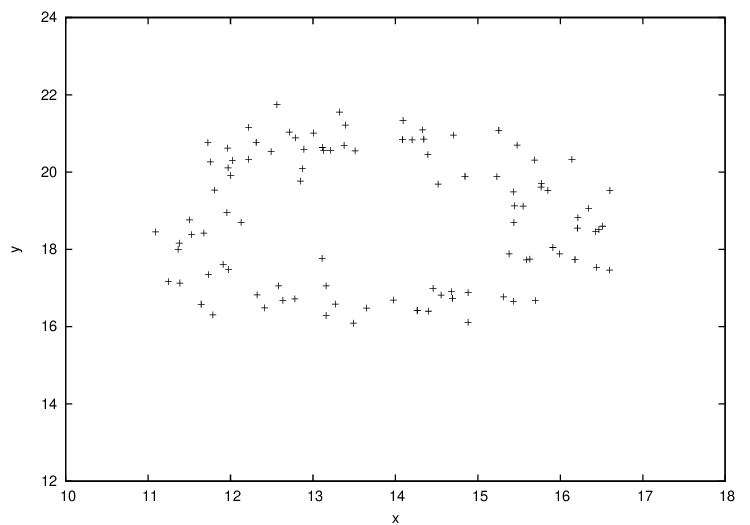}
  \caption{The two basic configurations of a noisy system obtained by
    simulations of of $N=100$ globally coupled active particles. We
    show snapshots in the space of the two coordinates
    ($\omega_0^2=0.2, \mu=0$). Left panel: $D=0.001$,
    translational mode , right panel: $D=0.2$, rotational mode \cite{ebeling_swarm_2008}.}
  \label{trans_rot}
\end{figure}

This corresponds to the findings of Erdmann et al. \cite{erdmann_noise_2005}
which will be discussed more detailed later on.
	Fig. \ref{trans_rot} illustrates the two typical behaviors which can be observed for small noise and for noise intensities above a critical values $D^{crit}$, respectively. Further on we refer to
$u_1\ne 0$ as the translational mode and to the second mode, with vanishing center of mass speed $u_2=0$, as
the rotational mode. We will also see that both
modes can be clearly defined in three spatial dimensions.


\subsubsection{Investigation of the dynamics of the relative motion of active particles}
Here, we discuss the dynamics of the movement of individual particles
relative to the center of mass \cite{ebeling_swarm_2008}. We begin with the analysis of the deterministic case without noise. 
The equations of motion are completely symmetric with respect to the
 particle index $i$. In other words, there are no cross
terms including two different particles as $i,j$.  This is a unique
property of global coupling, the relative dynamics reduces completely
to the binary problem, i.e., to the analysis of the (stochastic) dynamics of pairs of active Brownian particles. Therefore, we
concentrate our investigation on the study of two active particles with
a linear attracting force \cite{erdmann_excitation_2002}.

In this case the center of mass $\bbox{r}_{c.o.m.}$ moves with the
velocity $\bbox{u}$.  The relative motion of an individual particle under the influence of the
interaction force is described by the relative radius vector ${\delta
  \bbox{r}}={\bbox r}_i-{\bbox r}_{c.o.m.}=({\bbox r}_1-{\bbox r}_2)/2$ and the relative velocity ${\delta \bbox{v}}$. 
 The second particle has the same distance and velocity relatively to the center of mass but with opposite sign.  
We may write in Eq. \eqref{eq:swarm1} simply $\delta{\bbox r}, \delta{\bbox v}$ instead
of $\delta{\bbox r}_i, \delta{\bbox v}_i$ and omit the sum and the denominator $1/(N-1)$ in the
equation for the center of mass. 

i) Let us first assume a stable rotational mode. Hence we put
$\bbox{u}=0$ which solves the equation of motion for the center of mass.
Subsequently it follows that $\bbox{r}_{c.o.m.}=\bbox{r}_{c.o.m.}(0)$. in the two
dimensional space the equations of motion for the $\delta \bbox{r}$
and $\delta \bbox{v}$ become symmetric and independent of a direction.
Both assume the form of a van der Pol oscillator for system with a limit cycle
\begin{align}
  \label{eq:vanderpol}
   {{\rm d}\over {\rm d}t} \delta \bbox{r} = \delta {\bbox v}, \quad {{\rm d}\over {\rm d}t} \delta \bbox{v} + \omega_0^2 \delta \bbox{r} =
  \left(\alpha - \mu  -\beta \delta \bbox{v}^2  \right)\delta \bbox{v} \,.
\end{align}
As in the case of an external field (Sec. \ref{sec:rot_ext}) this
system owns two stable limit cycles, see Fig.  \ref{lc234}.  Which of
them will be approached depend on the initial conditions. 
\begin{figure}
  \begin{center}\noindent
    \includegraphics[angle=-90,width=6.5cm]{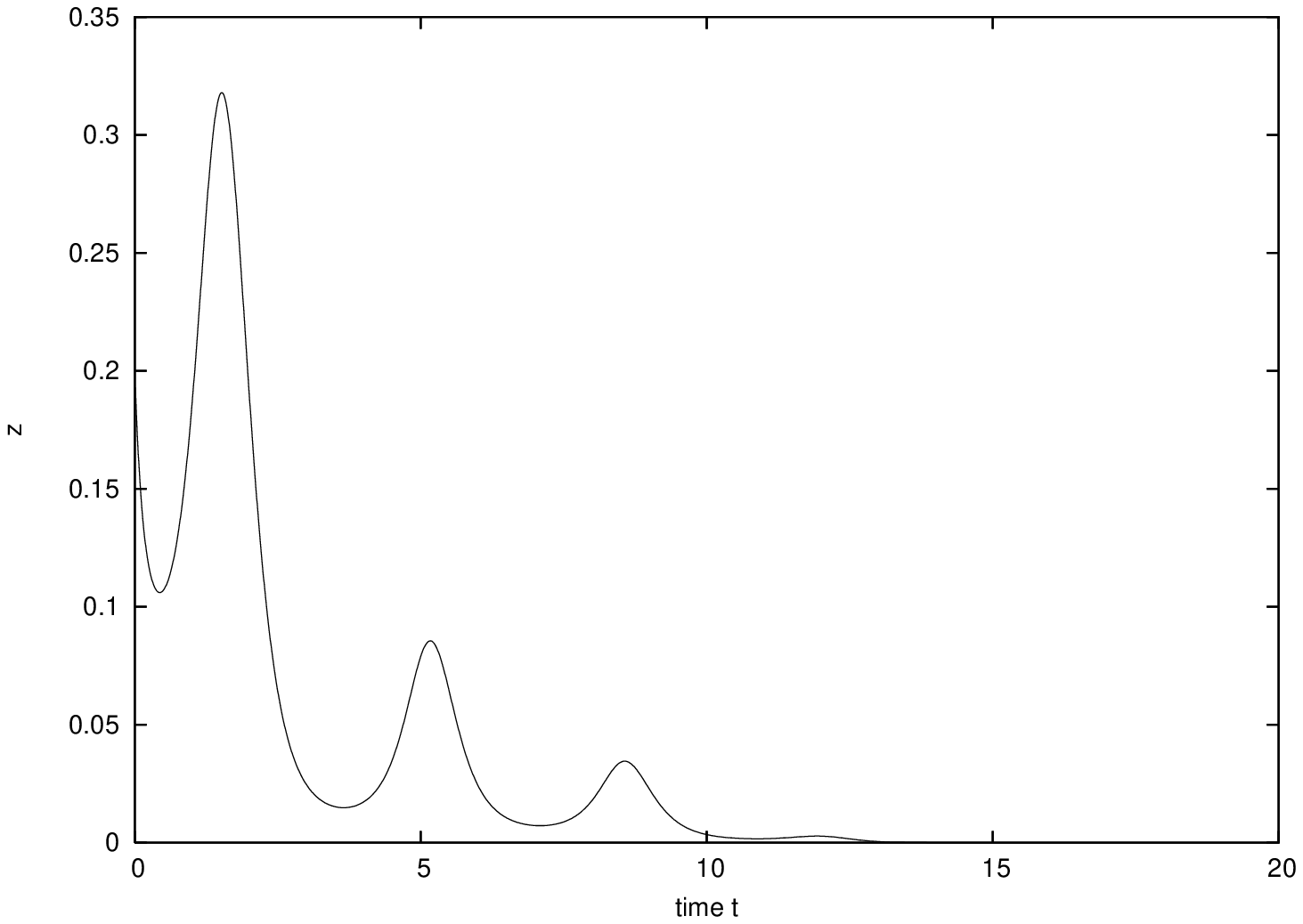}\hfill\includegraphics[angle=-90,width=6.5cm]{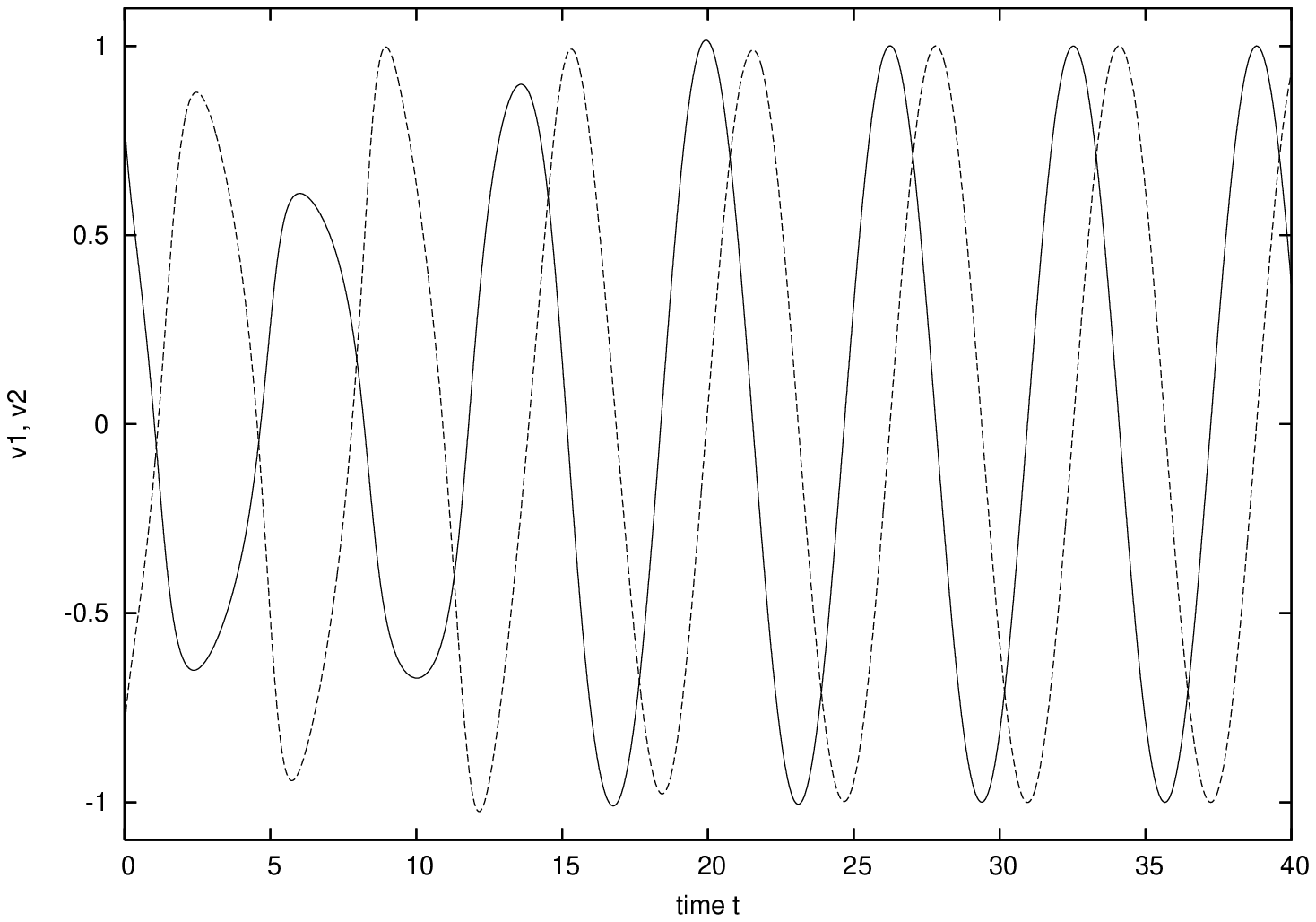}
  \end{center}
  \caption{Dynamics corresponding to the attractor region of the
    rotational motion of the pair: The left panel shows the decay of
    the kinetic energy of the center of mass to zero (rest state), as a function
    of time $z(t) = u^2(t)$ . The right panel shows projections of the
    velocities  $v_1(t), v_2(t)$ as a function of time
    and demonstrates the formation of sustained oscillations \cite{ebeling_swarm_2008}.\label{rotations1}}
 \end{figure}

This attractor is left if we introduce a translational motion with an
overcritical speed value.  In order to understand this we
introduce a small but finite translation $u = v_1 > v_\text{crit}> 0$. If $u$ is sufficiently large, it leads
immediately to a destruction of the rotational symmetry of the limit
cycles, to an elliptic deformation with the longer axis in the
direction perpendicular to the translation. As shown by Erdmann et al.
\cite{erdmann_excitation_2002,erdmann_attractors_2005} the loss of rotational symmetry leads to
leaving an Arnold tongue of stability and consequently to a
destruction of the limit cycles. These authors found that the rotations are indeed
stable only in and near to the plane $u=0$ i.e. for swarms at rest or
near to the resting state. This situation is presented in Fig. \ref{rotations1} for an subcritical initial kinetic energy of the center of mass.

ii) In case of translational mode the vector $\bbox{u} \simeq
\bbox{e}_v$ plays a special role we orientate the coordinate system
relative to it meaning we span ${\delta \bbox{r}} =
(x_{\parallel},x_{\perp})$, ${\delta \bbox v} =
(v_{\parallel},v_{\perp})$ where $x_{\parallel},v_{\parallel}$ are the
components in the direction of $\bbox{u}$ and $x_{\perp},v_{\perp}$ perpendicular to it.
Further we introduce the coordinates $z(t)= u^2(t)$ corresponding to
the twice the kinetic energy of the center of mass. Then we get the
following five differential equations
\begin{eqnarray}
\label{eq_five}
&& {{\rm d}\over {\rm d}t} x_{\parallel} = v_{\parallel},~~{{\rm d}\over {\rm d}t} v_{\parallel}+\omega_0^2 x_{\parallel}  = v_{\parallel} (\alpha - \mu - 3 \beta z - \beta v_{\parallel}^2 - \beta v_{\perp}^2) \,, \nonumber \\
&& {{\rm d}\over {\rm d}t} x_{\perp}= v_{\perp},~~{{\rm d}\over {\rm d}t} v_{\perp}  + \omega_0^2 x_{\perp}= v_{\perp} (\alpha- \mu - \beta z - \beta v_{\parallel}^2 - \beta v_{\perp}^2) \,, \nonumber\\
&& {{\rm d}\over {\rm d}t} z = 2 z (\alpha - \beta z - 3 \beta v_{\parallel}^2 - \beta v_{\perp}^2)\,,.
\end{eqnarray}
supplemented by the definition $\bbox{\dot{r}}_{c.o.m.}=\bbox{u}$. The
qualitative analysis of this system of nonlinear ordinary differential equations shows at
first that the system possesses a stable point attractor at $z=\alpha
/ \beta$, $v_{\parallel}=v_{\perp}=x_{\parallel}=x_{\perp}=0$. With
parabolic attraction without hard core both particle has the same
position and move with the center of mass velocity . Note that this is also
solution for the $N$-particles.

The linear stability analysis provides the eigenvalues $-2 \alpha$ in
$z$-direction, in the parallel direction $\lambda_1 \pm i \sigma_1 $
where $\lambda_1=(2\alpha+\mu)/2$ and $\sigma_1=\sqrt{\lambda_1^2
  -\omega_0^2}$ and in the perpendicular direction $\lambda_1\pm i
\sigma_1 $ with $\lambda_2=\mu/2$ and
$\sigma_2=\sqrt{\lambda_2^2-\omega_0^2}$. The point attractor is
linearly stable provided $\mu > 0$. The linear stability in the
direction $v_{\perp}$ corresponding to the motion perpendicular to the
translation is given only for ${\mu}>0$.  However even at ${\mu}=0$ we
still observe quadratic stability in this particular direction due to
the terms since the $-\beta v_{\perp}^2$ stabilizes the ,motion as can
be seen from the nonlinear equation of motion for $v_{\perp}$ around
the fixed point
\begin{eqnarray}
  {{\rm d}\over {\rm d}t} v_{\parallel} = - \omega_0^2 x_{\perp} - v_{\perp} \left(\mu  +\beta (v_{\parallel}^2 + \beta v_{\perp}^2) \right) \,, \nonumber\\
\end{eqnarray}

Now we suppose that a deviation from the fixed point exist,
which is realized by some small $v_{\perp}^2$ and $v_{\parallel}^2$ we
do not claim that both are stationary but supplement each other to a
stationary values. Afterwards we require that the expression inside th
brackets in the last equation of \eqref{eq_five} vanishes, i.e. it
holds
\begin{eqnarray}
  \label{eq:new}
  \beta z\,=\,\alpha - 3 \beta v_{\parallel}^2 - \beta v_{\perp}^2.
\end{eqnarray}
Since $\dot{z}=0$, the above expression \eqref{eq:new}
represents a translational mode. The stability analysis shows that
perturbations perpendicular to the motion perform harmonic
oscillations around the center of mass if $v_{\parallel}^2$ is vanishing.
Otherwise the start to grow if there is a deviation parallel to the
motion and the translational mode gets oscillating in all variables.

Indeed insertion of \eqref{eq:new} into the dynamics of $v_{\perp}$
yields
\begin{eqnarray}
  \label{eq:newperp}
  {{\rm d}\over {\rm d}t} v_{\perp}  + \omega_0^2 x_{\perp}= v_{\perp} (- \mu + 2\beta v_{\parallel}^2) \,,
\end{eqnarray}
So if $\mu \to 0$ this state is marginally stable if the deviation of
the particle points perpendicular to the center of mass, i.e. $v_\|=0$. Oppositely the deviations
parallel to the direction of the center of mass will be amplified in the
perpendicular direction if
\begin{eqnarray}
  \label{eq:newcond}
  2\beta v_{\parallel}^2 \,>\, \mu \ge 0\,. 
\end{eqnarray}
We see that the alignment increases the stability of parallel motion.
These results illuminate the role of the velocity couplings. Without
the existence of a (positive) velocity coupling, the swarms tends to
show a weak instability in the transversal direction, i.e. it tends to
get broader and broader, remaining concentrated around the center of
mass in the longitudinal direction.

Since this mode is oscillatory with frequency $\omega_0$ we find small
oscillations of the particles perpendicularly to the direction of
motion. It can be also inspected from left Fig. \ref{trans_rot} where
the particle move with a broadened distribution perpendicular to the
center of mass motion.

\begin{figure}
  \begin{center}\noindent
\psfrag{z}{\huge $z$}
\psfrag{time t}{\huge time $t$} 
    \includegraphics[angle=-90,width=6.4cm]{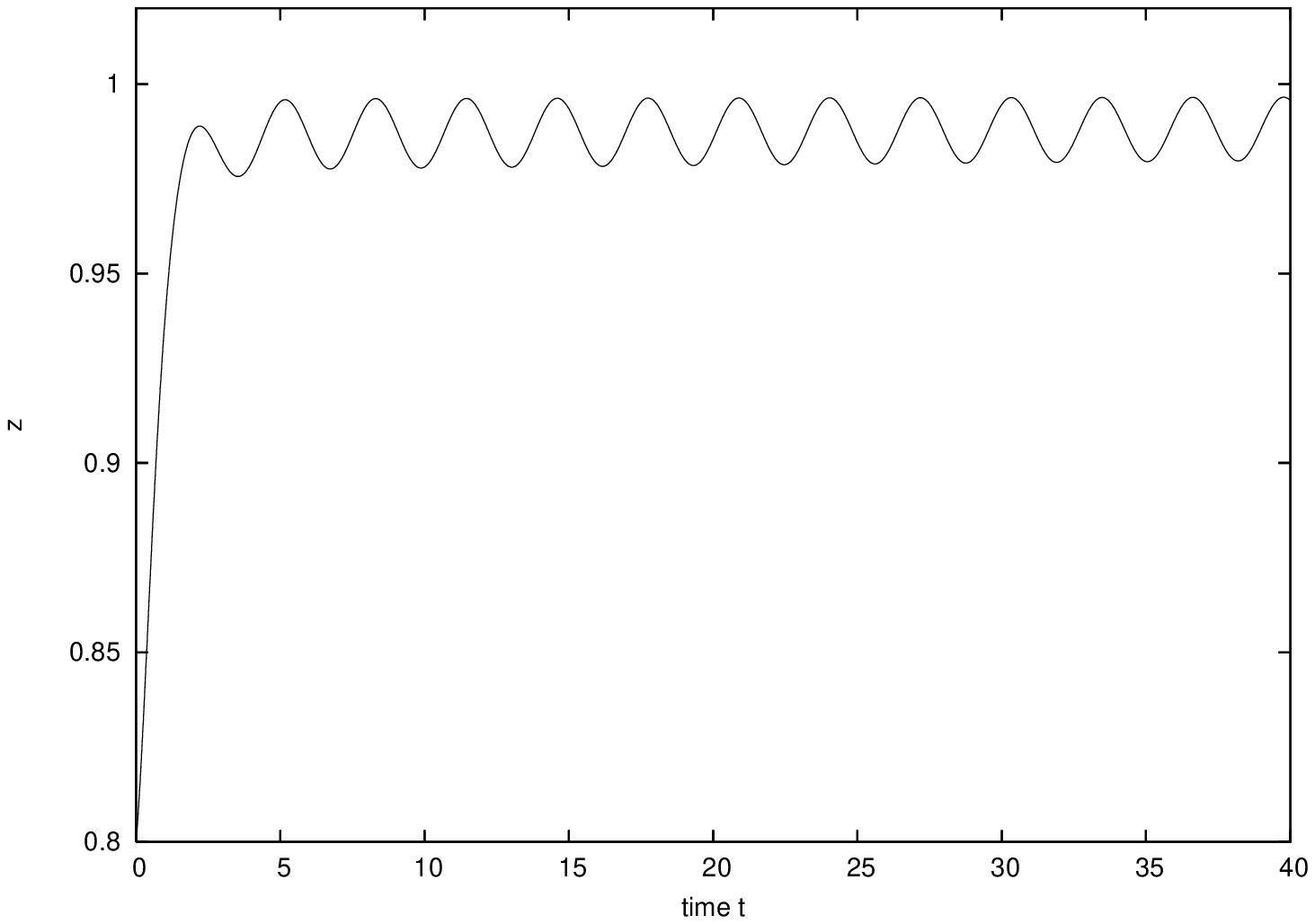}
\psfrag{v1, v2}{\huge $v_\|$, $v_\bot$}
    \includegraphics[angle=-90,width=6.4cm]{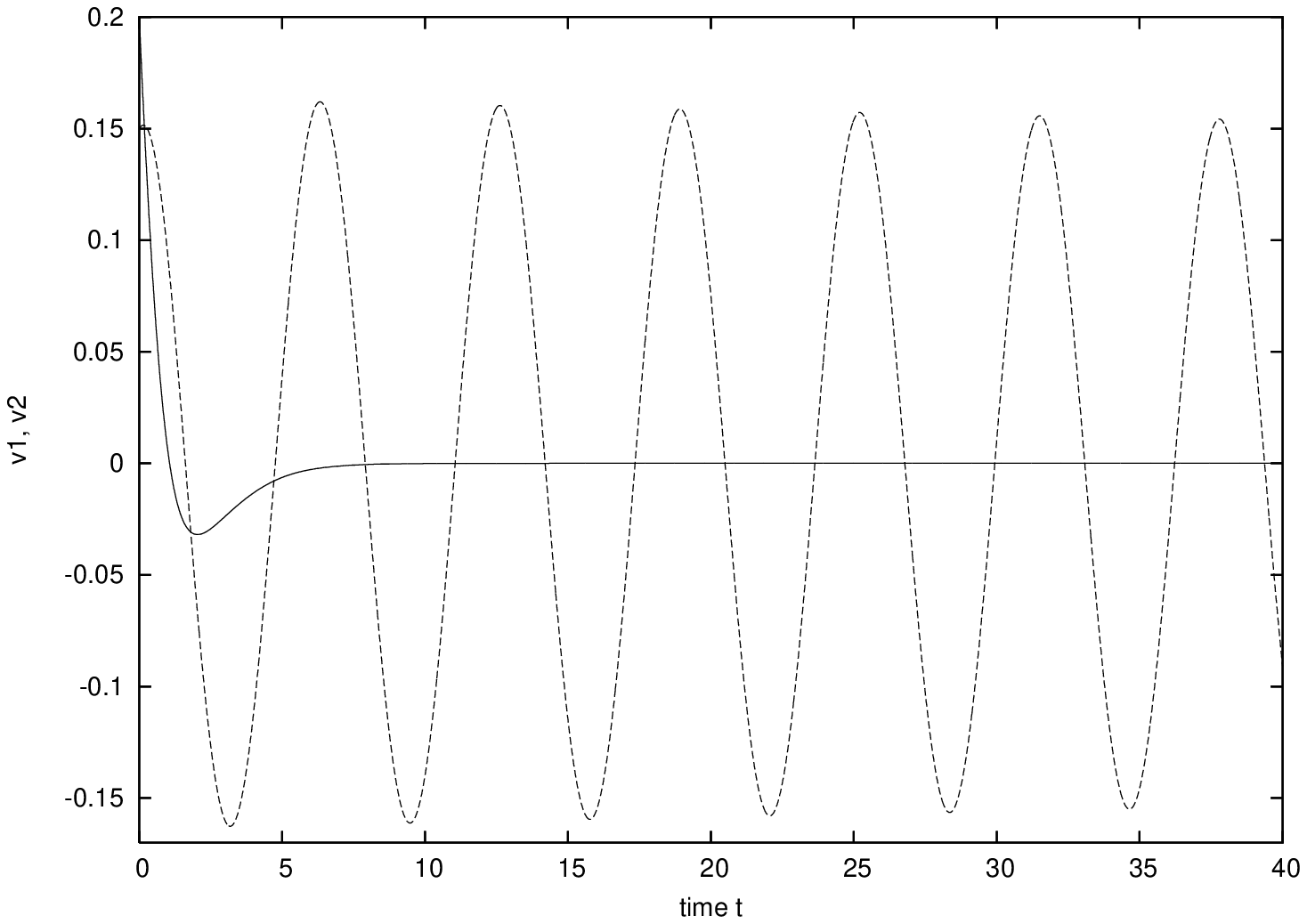}
  \end{center}
  \caption{Solutions of the Eqs. \ref{eq_five}  as a function of time for
    $\mu = 0 $ and initial conditions $z(0)=u^2(0) = 0.8$,
    corresponding to the region of the attractor of translational
    motion. Left panel: The translational (squared) velocity
    $z(t)=u^2(t)$ approaches slowly the maximal value $z=1$. An
    oscillating contribution remains which is due to the neutral
    stability of the $v_{\perp}(t)$ - dynamics.  Right panel: The
    longitudinal component of the velocity $v_{\parallel}(t)$ goes
    quickly to zero and the transversal velocity $v_{\perp}(t)$ decays
    very slowly or remains constant (because of neutral stability) \cite{ebeling_swarm_2008}.
    \label{translations}
    }
  \end{figure}

Details about the dynamics we may obtain from explicit solutions, for
the special set of parameters $\alpha=\beta=1$.  In Fig.
\ref{translations} we show several solutions corresponding to initial
conditions in the attractor region.
We see that typically (for different initial conditions within the
attractor region) the relative velocity perpendicular to the swarm
translation $v_{\perp}(t)$ decays very slowly, and the relative
velocity in the direction of the swam motion $v_{\parallel}(t)$ goes to zero in a very fast way. The
velocity of the center of mass approaches $u_1 < v_0 \alpha/\beta$ in
a rather fast way, however the state becomes unstable and a slow
oscillatory contribution appears.  Including a small amount of
velocity synchronization $\mu > 0$ all oscillatory components in the
translational mode are damped out in the time $1 / {\mu}$.  In the
limit case $\mu = 0$ i.e. no synchronization of the velocities exist.

Note that the relative velocities $v_{\parallel}$ and $v_{\perp}$ are
defined respectively to the velocity of the c.o.m.. Therefore, even if
the parallel velocities vanish the oscillations of particle movement
are also present in the velocities taken respectively to he origin of
coordinate.

\subsubsection{Influence of noise on swarms}
Including noise we expect instead of the point attractor at $u=v_0$
and the two line attractors - the 2 limit cycles in the plane $u=0$ -
that the dynamic systems forms some distributions around the
attractors.  This will obviously lead to some permanent deviations
around the center of mass (c.o.m.) and of its velocity. Here we will find
estimates of the corresponding distributions.  For these we will use
the approximation of independent dynamics of the velocity components
$v_{\parallel}, v_{\perp}$ and of $u$. Above we also found the center of mass velocity
distribution
in Eq. \ref{f(V)} with the most probable velocity given in Eq.
\ref{eq:swarm_speed} with
\begin{equation}\label{alpha1}
  \alpha_1 = \alpha - 3 \beta \langle v_{\parallel}^2\rangle  - \beta \langle v_{\perp}^2\rangle .
\end{equation}
This expression as well as Eq.(\ref{f(V)}) contain still the unknown
constant $\alpha_1$ which is determined by the distributions of the
longitudinal and translational velocities.

For the longitudinal fluctuations around the center of mass of the
swarm near to the stable attractor we get
\begin{equation}
  {{\rm d}\over {\rm d}t} v_{\parallel} + \omega_0^2 x_{\parallel} =
  - 2 \alpha_1 v_{\parallel} + \sqrt{D} \xi_{\parallel} (t),
\end{equation}
where the noise strength of the relative motion is reduced by factor of two.  This follows from the correlators formulated above for $N=2$.
The corresponding stationary Fokker-Planck equation is solved by
\begin{equation}
  P_0(x_{\parallel}, v_{\parallel}) =
  C_0 \exp\left[ -\frac{1} {D} \,
    \left(2 \alpha_1 v_{\parallel}^2 + \omega_0^2 x_{\parallel}^2 \right)\right],
\end{equation}
where $C_0$ is given by the normalization.  The dispersion is given by
\begin{eqnarray}\label{longi}
  \langle v_{\parallel}^2 \rangle   \simeq \frac{D}{4 \alpha_1}.
\end{eqnarray}
The longitudinal dispersion depends on the constant $\alpha_1$ which
is still to be determined.  

For the fluctuations transversal with respect to the c.o.m. of the swarm
the situation is more complicated due to the problems with linear
stability in the $v_{\perp}-$ direction.  Neglecting the correlations with
the longitudinal fluctuations $\langle  v_\parallel v_\perp \rangle$, which are small we find
\begin{equation}
  {{\rm d}\over {\rm d}t} v_{\perp} + \omega^2 x_{\perp} =
  v_{\perp} [(\alpha- \alpha_1 - \mu) - \beta (\delta v_{\perp})^2] + \sqrt{D} \xi_{\perp} (t).
\label{lc}
\end{equation}
We remember that $\alpha_1 < \alpha$ for finite noise.  Therefore the
first term on the right hand side may be positive or negative,
depending on the situation. We have to differ between two cases: 

(i) For the stable solution $\mu > (\alpha- \alpha_1)$, the term with $\beta$ can be neglected for small velocity deviations. We obtain the standard problem of a noisy damped oscillator with the distribution
\begin{equation}
  P_0(x_{\perp}, v_{\perp}) =
  C \exp\left[ -\frac{1} {D} \,
    \left(- (\mu - \alpha - \alpha_1) v_{\perp}^2 + \omega_0^2 x_{\parallel}^2 \right)\right]
\end{equation}
and the dispersion
\begin{eqnarray}
  \langle v_{\perp}^2\rangle   \simeq \frac{D}{2(\mu -\alpha + \alpha_1)} \simeq \frac{D}{2\mu}.
\end{eqnarray}
In the present case the fluctuations of $v_{\parallel}$ and $v_{\perp}$
are rather small and the translational mode
is the most favorable one.

(ii) A different situation is observed in the second case of small (or
zero) contribution from parallelizing interactions.  We consider now
the situation $\mu < (\alpha- \alpha_1)$ in correspondence with the
inequality \eqref{eq:newcond}. 
Then in Eq. \eqref{lc} the term with $\beta$ has to be taken in to account and the dynamics correspond to
a limit cycle, beyond a Hopf-bifurcation.  However this limit cycle is not a standard one, since
$(\alpha- \alpha_1) > \mu $ requires finite noise, i.e. the amplitude
of the limit cycle is noise driven.  

We emphasize that the behavior corresponds to active motion of the
center of mass supplemented by a small active oscillatory motion of individual particles relative to the
center of mass (see Fig. \ref{translations}).  

In order to determine $\alpha_1$, we need to estimate $\langle v_\perp^2 \rangle$. First we remember, that $v_\perp$ is a component of the relative velovity vector $\delta {\bbox v}=({\bbox v}_1 -{\bbox v}_2)/2$. The squared velocity-space diameter of the limit cycle in first approximation is given by $2(\alpha-\alpha_1-\mu)/\beta$. This gives us 
\begin{align}\label{perpdisp}
   \langle v_{\perp}^2 \rangle  \simeq \frac{\alpha - \alpha_1 - \mu }{2 \beta}.  
\end{align}
The longitudinal \eqref{longi} and perpendicular dispersions \eqref{perpdisp} inserted into Eq. \eqref{alpha1} yield: 
and are connected by the relations
\begin{align}
\alpha_1 = \alpha - \frac{3 \beta D}{2 \alpha_1} + \mu. 
\end{align}
It is a quadratic equation in $\alpha_1$ and has the solution
\begin{equation}
  \alpha_1 = \frac{1}{2} (\alpha + \mu) \left[1  -  \sqrt{1 - \frac{6 \beta D}{(\alpha + \mu)^2}} \right].
\end{equation}
We see that the dispersion of $u^2$, corresponding to the small oscillating motion in Fig. \ref{translations},  is maximal for the critical noise
strength
\begin{equation}
  D_{cr} = \frac{(\alpha + \mu)^2}{6 \beta}. 
\end{equation}
This is in good agreement with simulation results in \cite{erdmann_noise_2005}. 
\begin{figure}
  \begin{center}\noindent
    \includegraphics[width=4.5cm,height=4.5cm]{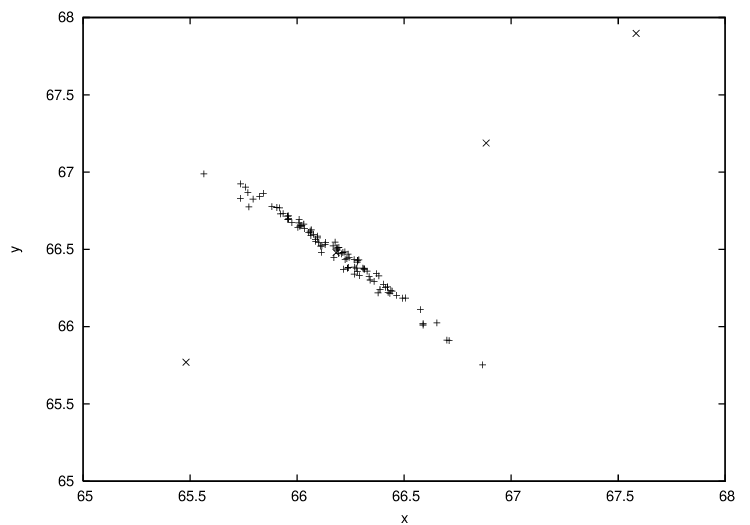}\includegraphics[width=4.5cm,height=4.5cm]{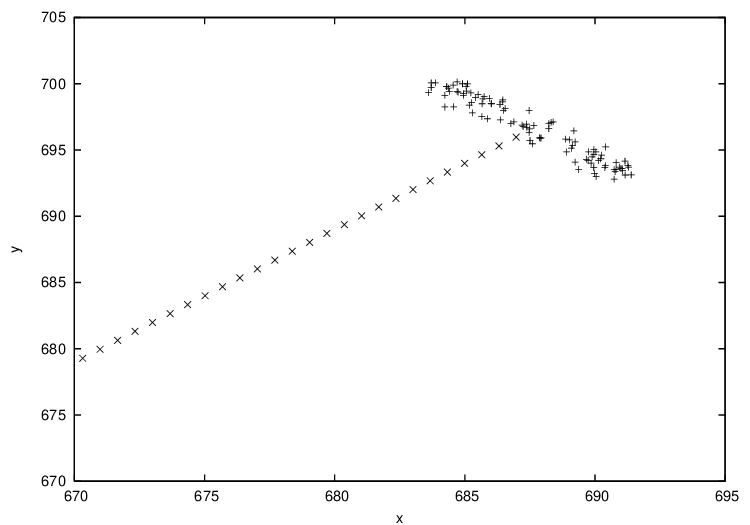}\includegraphics[width=4.5cm,height=4.5cm]{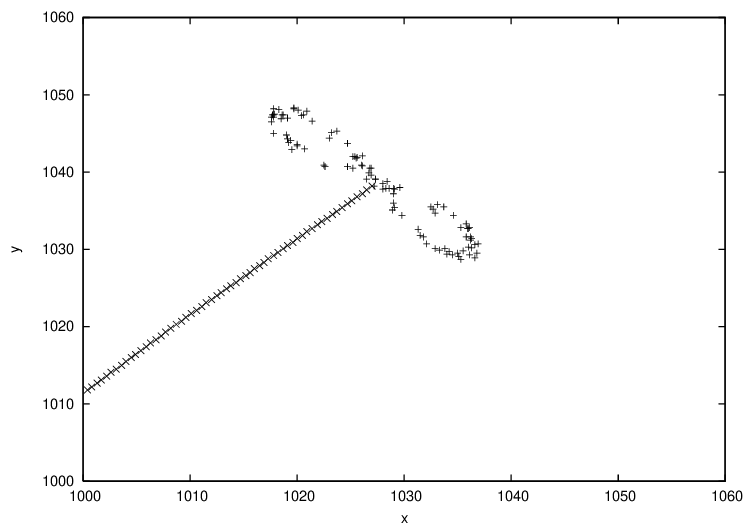}\\
    \includegraphics[width=4.5cm,height=4.5cm]{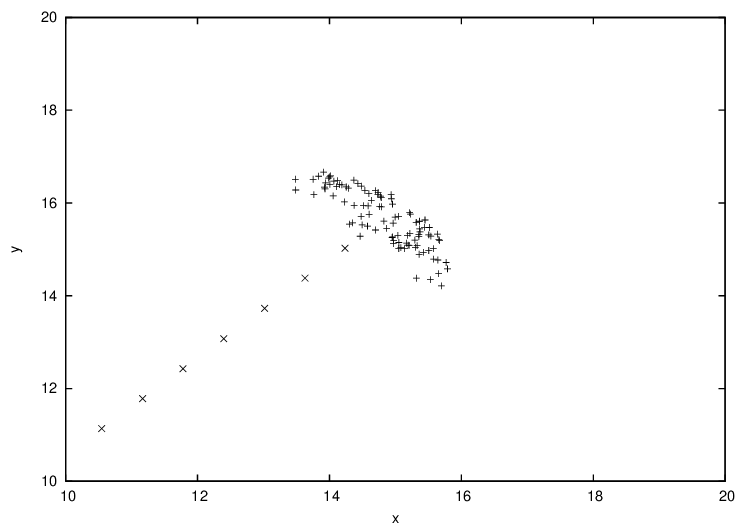}\includegraphics[width=4.5cm,height=4.5cm]{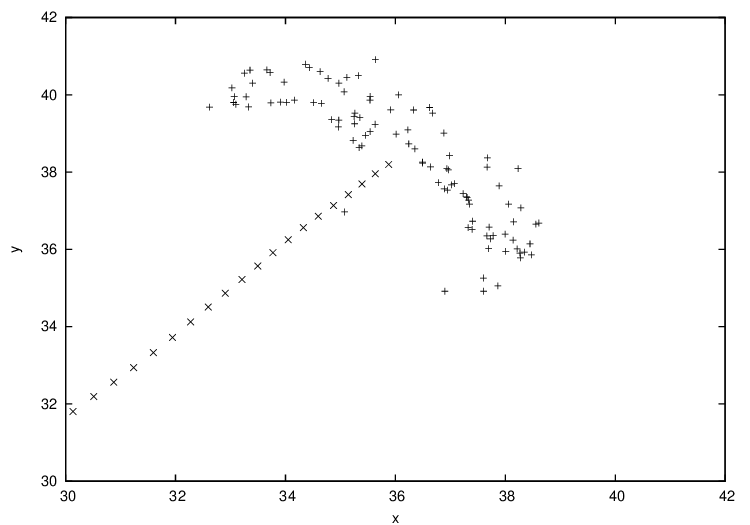}\includegraphics[width=4.5cm,height=4.5cm]{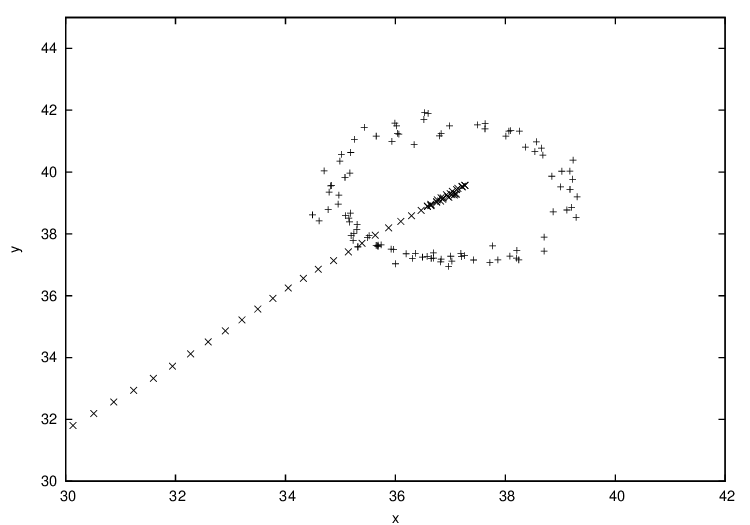}
  \end{center}
  \caption{Upper row: Snapshots of the swarm configurations in space
    of coordinates in the translational mode $u = v_0$ at times
    $t=100,100,1500$ for small noise.  (parameters $\omega_0^2 = 0.01;
    D=0.00001; N=100$). The straight dotted curve represents the
    trajectory of the center of mass.  Lower row: A time sequence
    ($t=20,80,110$) of swarm configurations in space at larger noise
    which end in the rotational mode (parameters $\omega_0^2 = 0.2;
    D=0.001; N=100$). The center of mass becomes eventually resting \cite{ebeling_swarm_2008}.}
  \label{transseq}
\end{figure}

We note however that our simulations show that the translational mode
occurs to be unstable with respect to fluctuations. We observed
that the translational mode for particles with harmonic interactions
in two dimensions is always a transient state. For sufficiently
long simulation times we found for $\mu=0$ that even at such small
noise as $D=0.001$ the system switches to the rotational mode
(see Figs. \ref{transseq}).  We point out that we never observed transitions 
from the rotational mode back to the translational state of a moving
center of mass  even within extremely long simulation times.

\begin{figure}[h]
  \begin{center}\noindent
    \includegraphics[angle=-90,width=6.7cm]{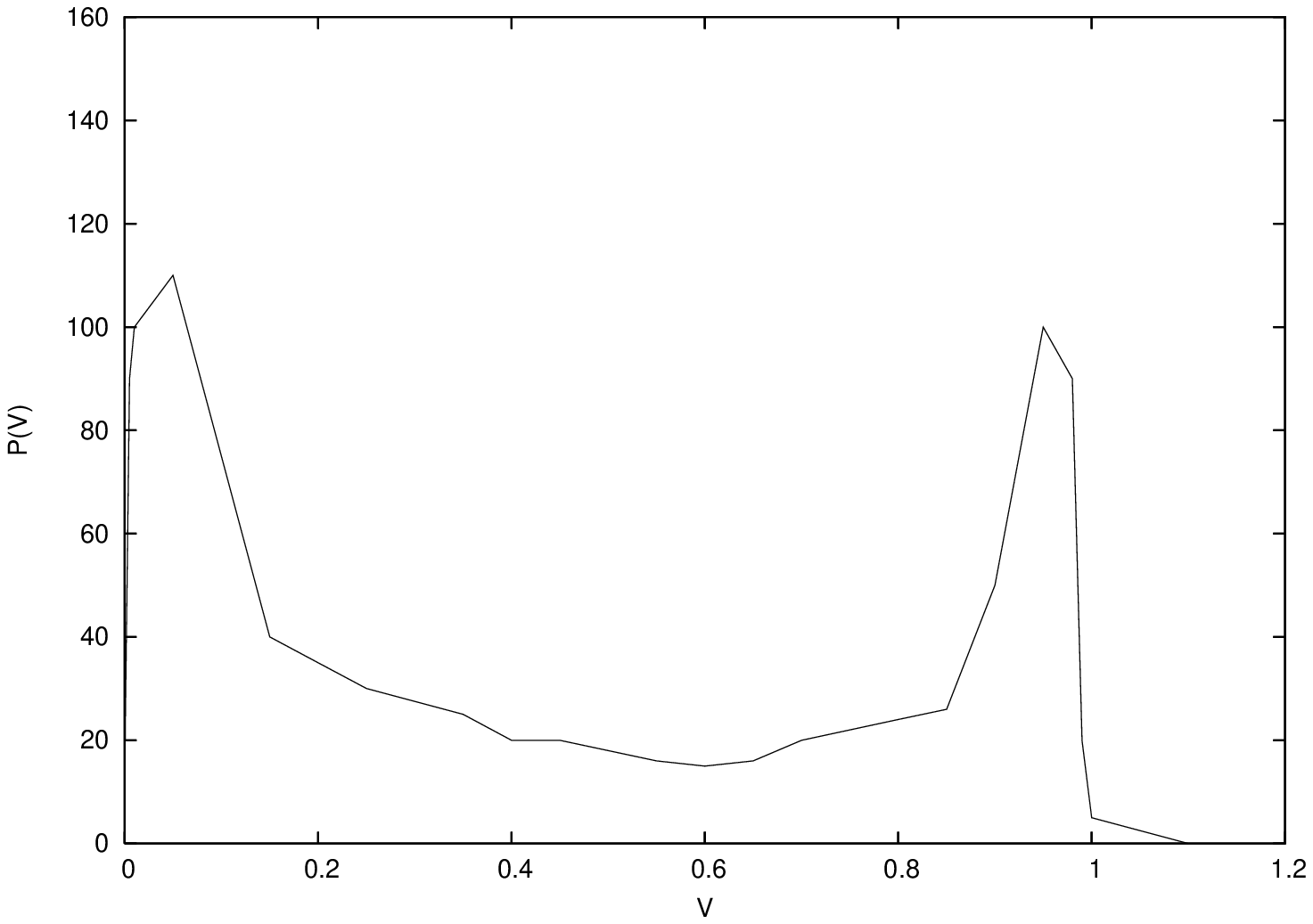}~~~\includegraphics[angle=-90,width=6.7cm]{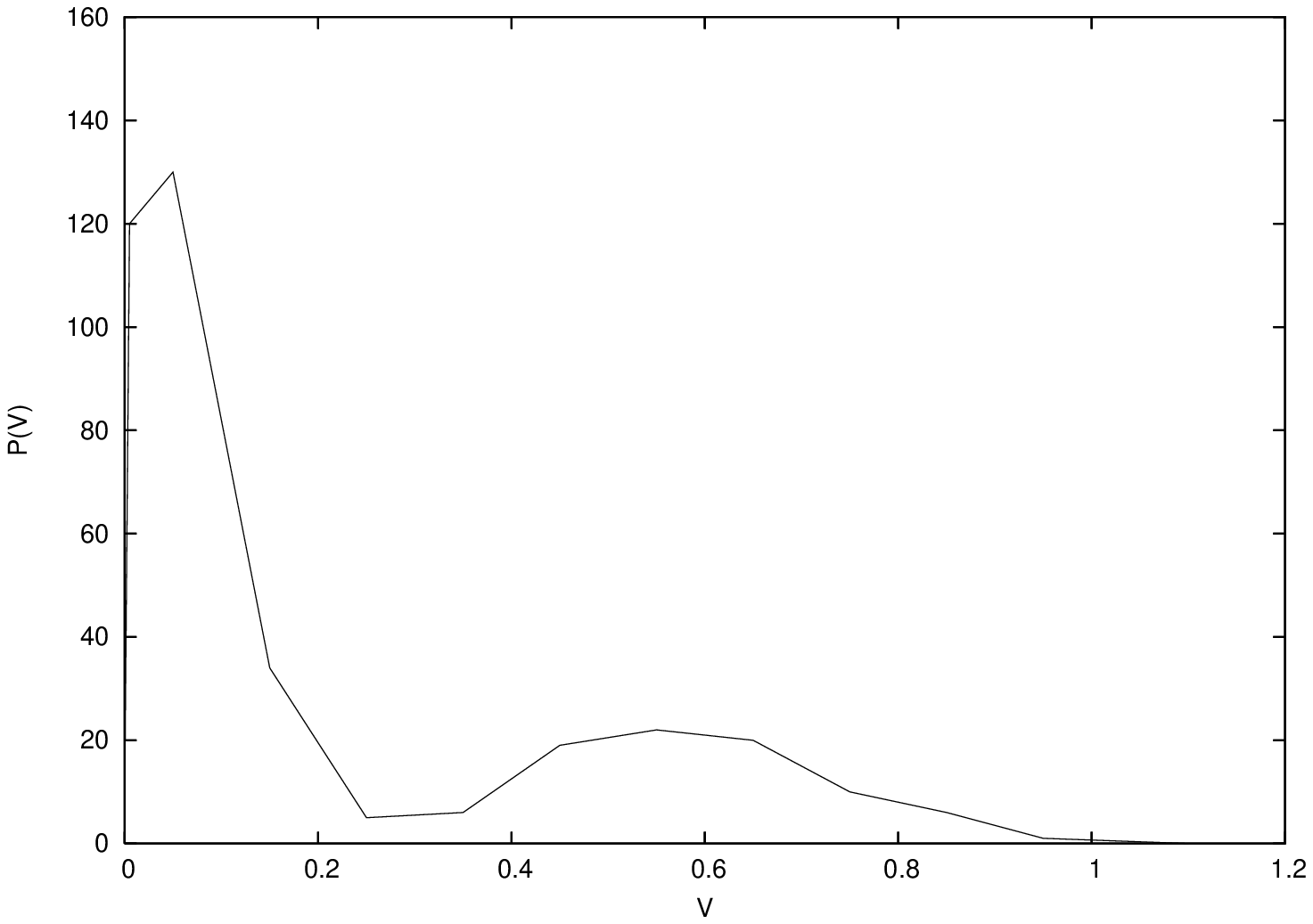}
  \end{center}
  \caption{Relative frequencies $f(|u|)$ of the swarm speed $|u|$
    estimated from simulations along finite times which have started
    in the translational mode.  Left panel: The translational mode
    produces for small noise $D=0.00001$ a narrow maximum near to the
    maximal velocity of the swarm $|u|=v_0$ (here $v_0 = 1$).  Right
    panel: At larger values of noise $D = 0.01$ the translational mode
    broadens and the rotational mode dominates. The latter forms forms
    a big maximum corresponding to more frequent rotational
    configurations \cite{ebeling_swarm_2008}.}
  \label{figf_V}
\end{figure}

This underlines the importance of having at least a small contribution
of velocity synchronization $\mu > 0$, which stabilizes the
translational mode. In the next Section we will discuss bistability between both
modes which is due to a hard core potential in three dimensions. Here
we show frequencies of the center of mass speed, recorded during
finite simulation with translational mode as initial condition in Fig. \ref{figf_V}, which indicate the transient bistable
behavior. The amount of probability around the translation and rotational mode clearly depends on the noise
intensity.

In case of simple harmonic forces the probability is distributed
around two limit cycles corresponding to left or right rotations.
These distributions for the rotational mode are similar to what we
have found for the case of external fields. In the mean field case and
with linear forces the particles do interact effectively with the
resting center of mass, only.  In result there is no tendency to
states where all particles move synchronously along one circle. Such
way the circles does not behave like attractors for the whole swarm
and any decomposition of particles with right and left rotations is
possible.

Summarizing our findings we may state: For two interacting active
particles there exist a translational and a rotational mode. In the
rotational mode the center of the ``dumb-bell'' is at rest and the
system rotates around the center of mass. Only the
internal degrees of freedom are excited and we observe driven
rotations. In the translational mode of the dumb-bell the center of
mass of the dumb-bell performs an active Brownian motion similar to a free
motion of the center of mass.  In this case we may expect a
distribution similar as given \cite{schimansky-geier_stationary_2005}. Larger noise leads to larger deviations from the center of mass, which finally favors transitions to the rotational state.

\subsubsection{Cohesion, repulsion and bistability in 3D swarms} 
\label{sec:swarm_3d}

In this section we present simulations of swarms and their dynamics in
three spatial dimensions. We consider a set of $N$ identical active
particles interacting globally via a pair potential as outlined in the
last section (Sec. \ref{sec:swarms}). The coordinates $\bbox{r}_i$ and velocities
$\bbox{v}_i$ have now three independent components.  

We use a potential which captures fundamental properties of swarming
of animals.  At short range the potential has to be repulsive to avoid
collisions and to prevent the agents from interpenetrating each other.
An attractive force mimics the aim of the individual to stay with the
group. The attraction should be of long range, but eventually approach
zero to account for the limited sensing range of animals. An
exponentially decaying function meets both demands.

Therefore we consider here a generalized Morse potential,
which consists of an attractive and a repulsive part.  Both are
exponential functions with amplitudes $C_a$ and $C_r$ and ranges $l_a$
and $l_r$, respectively.
\begin{equation}
U_{\rm M}^i(\bbox{r}_1, \dots, \bbox{r}_N) = \sum_{k \neq i}^N C_r  \exp\left({\frac{-\left| \bbox{r}_i - \bbox{r}_k \right|}{l_r}}\right) - C_a  \exp\left({\frac{-\left| \bbox{r}_i - \bbox{r}_k \right|}{l_a}}\right).
\label{eq:U}
\end{equation}
The equilibrium distance of two particles is
\begin{equation}
 r_0 = \frac{l_r \ l_a}{l_r-l_a} \ \ln \left(\frac{l_r \ C_a}{l_a \ C_r}\right).
\label{eq:equilibrium}
\end{equation}
For $l_r/l_a < 1$ the potential possesses a minimum
corresponding to short range repulsion and long range attraction. One
can see that for $l_r/l_a < 1$ and $ l_r/l_a > C_r/C_a$ 
the minimum would shift to negative values. Since the
absolute value of the interparticle distance in Eq. \ref{eq:U} is
positive, the potential is attractive everywhere. Therefore we will
concentrate on the parameter space where $l_r/l_a < 1$ and
$l_r/l_a < C_r/C_a$.

The self-organized rotational dynamics of a Hamiltonian system of active particles 
with such a pairwise Morse potential have been studied by Levine {\em et. al} 
\cite{levine_self-organization_2000} and d'Orsogna {\em et. al} 
\cite{dorsogna_self-propelled_2006}. 
Here, we will show that in three dimensions swarms of active Brownian particles interacting via Morse potentials with repulsion and attraction
exhibit noise-induced transitions not only from translation to rotation, as in the purely attractive (harmonic) case, but also the reverse transition. These transitions occur at different noise
intensities, thus leading to a hysteresis curve. This effect was not
observed in two dimensions, here the rotation was stable even
without noise. For comparison we also investigate 
the harmonic forces as introduced above in three dimensions.

%
%
%
%
\begin{figure}
  \begin{center}
    \begin{minipage}[h]{.59\linewidth}
      \vspace*{-1cm}\includegraphics[width=1.\linewidth]{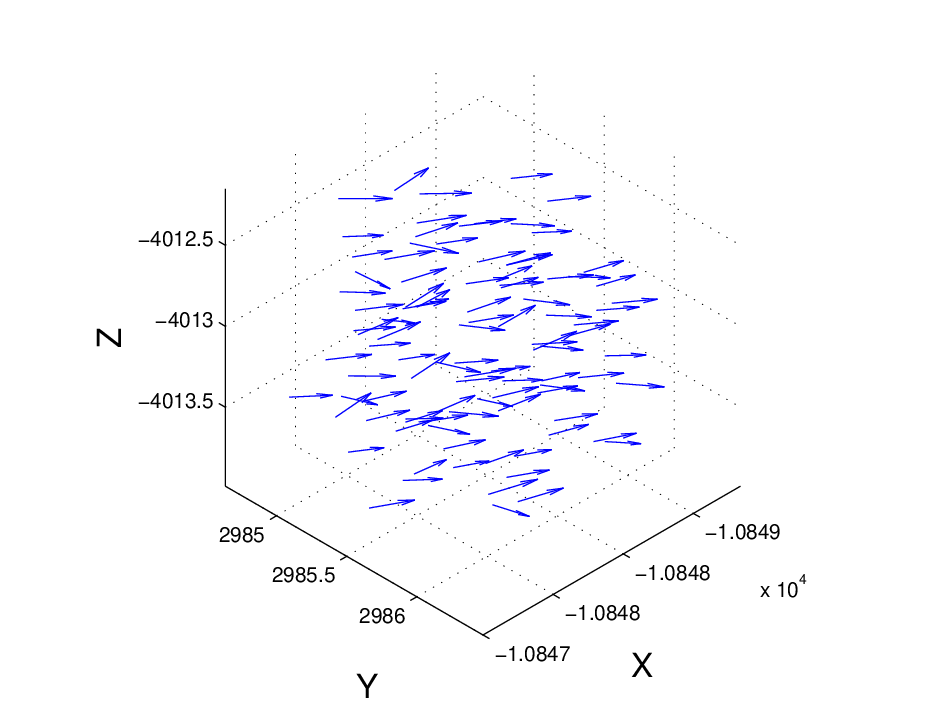}
    \end{minipage}
    \begin{minipage}[h]{0.4\linewidth}
      \includegraphics[width=1.\linewidth]{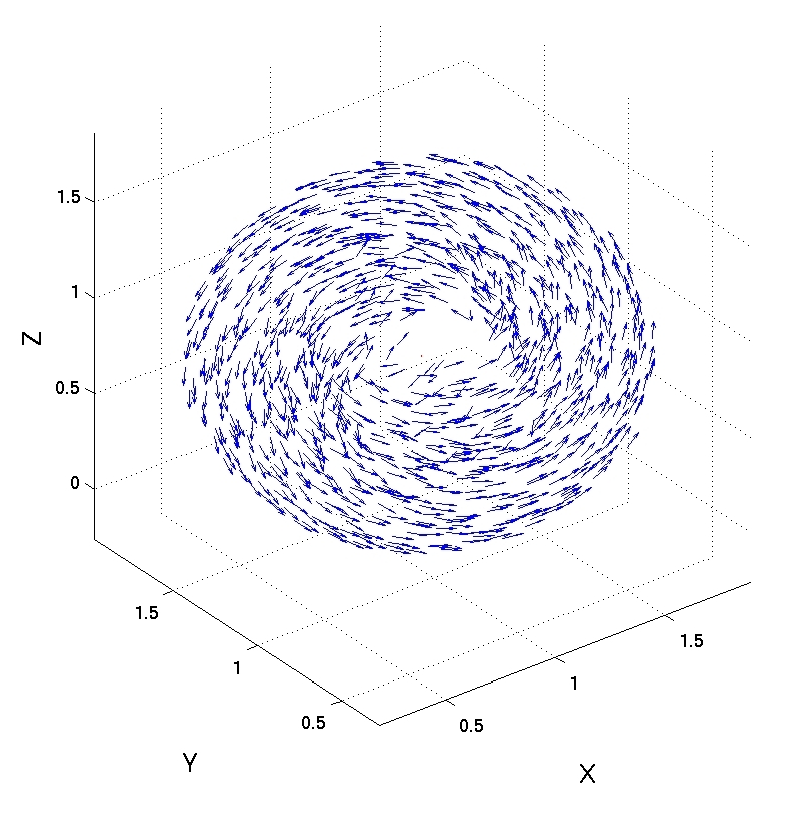}
    \end{minipage}
  \end{center}
  \caption{Rotational (left) and translational (right) mode: Particles
    move either coherently in a torus shape structure or parallel
    with the moving center of mass (c.o.m.) on a straight line. Parameter
    values: $N = 1000, \gamma_1 = 1.6, \gamma_2 = 0.5, C_r = C_sa =
    l_r = 0.5, l_a = 2.0, D = 0$ \cite{strefler_swarming_2008}}
  \label{fig:torus}
\end{figure}

To study the
transition from translation to rotation we prepare the system in the
translational mode (Fig. \ref{fig:torus}, right). In this mode the
particles move parallel with their stationary velocity $v_0 =
\sqrt{\frac{\alpha}{\beta}}$.  The spatial configuration in the
center of mass system corresponds to the equilibrium configuration.
Without noise, there are no fluctuations, the center of mass moves
with $v_0$. Increasing the noise gives rise to fluctuations and leads
to a decreasing velocity of the center of mass \cite{mikhailov_noise-induced_1999,erdmann_noise_2005}. Above a
critical noise value $D_{\rm crit}^{\rm trans}$ the translational
motion breaks down and the particles start to rotate around the center
of mass.

Whereas for a harmonic potential the particles rotate in any direction
on the equipotential sphere, the Morse interaction leads to coherent motion
in a torus shape structure, with the orientation depending on the
initial conditions (Fig. \ref{fig:torus}). The center of mass moves
diffusively, therefore the absolute value of its velocity is not zero
and increases with the noise intensity.

\begin{figure}[h]
  \centerline{
    \includegraphics[width=.49\linewidth]{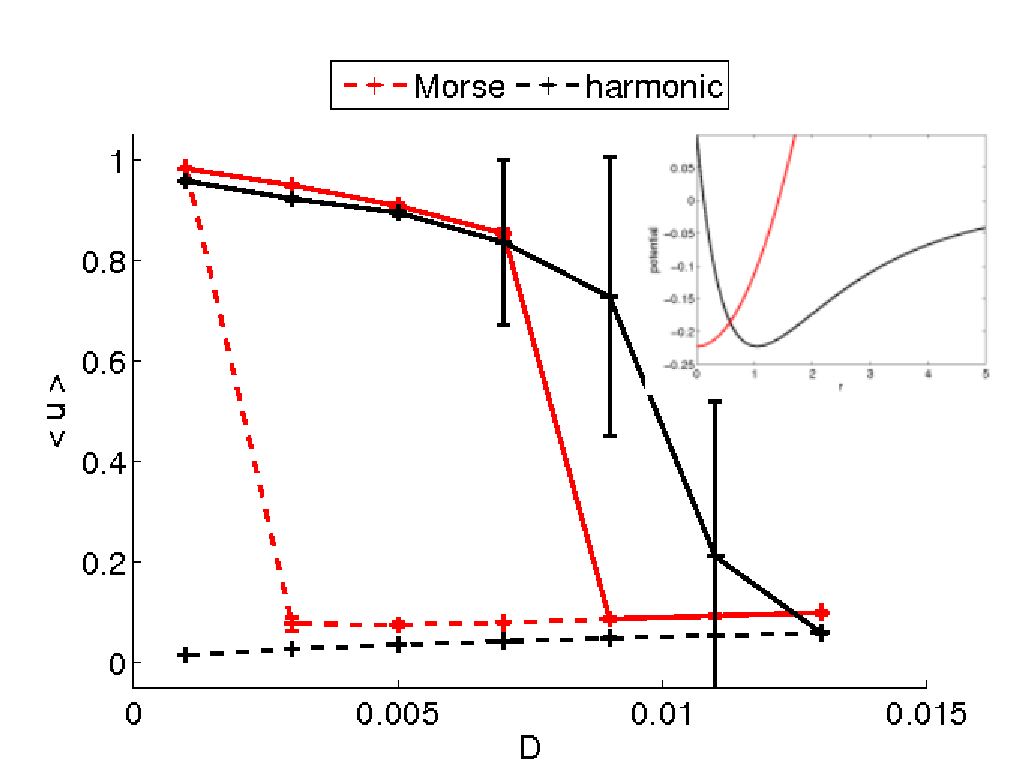}
    \includegraphics[width=.49\linewidth]{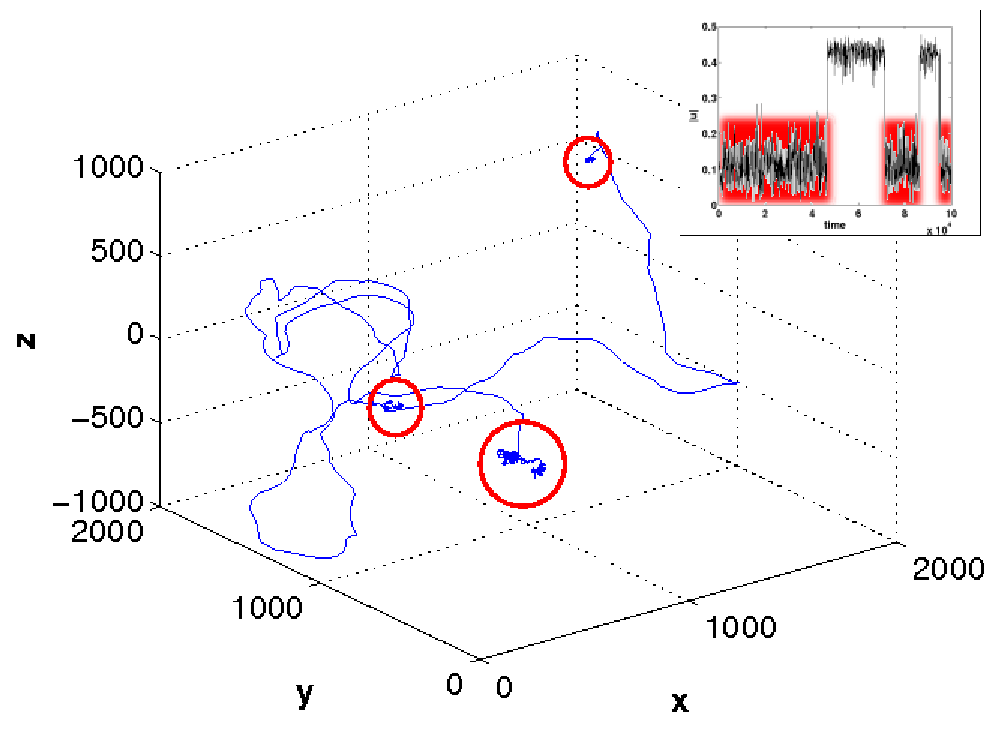}}
  \caption{{\bf Left panel:} Hysteresis of the of the center of
    mass speed $u$ versus noise intensity $D$. Solid lines for
    increasing noise and dashed lines for decreasing $D$. The most
    probable values for Morse potential (red) and its harmonic
    approximation (not shown) are comparable and show instability
    of the rotational state below a critical noise intensity.  
    For harmonic potential, or Morse  potential without a minimum at finite distance, only transitions from    	   translation to rotation can be observed. The latter is stable even without noise.  
    {\bf Right panel:} Example of a trajectory of the center of mass in the bistable
    regime. The inlay shows the corresponding velocity. The rotational
    modes, where the velocity is almost zero, are highlighted red. These
    lead to a diffusive motion of the center of mass within the red
    circles. In between the system displays a stochastic trajectory in
    the translational mode with mean velocity $u \cong v_0 =
    \sqrt{\frac{\alpha}{\beta}}$. 
    \cite{strefler_swarming_2008}}
  \label{fig:hysterese}
\end{figure}

Starting from the rotational state and decreasing the noise intensity,
the system exhibits a transition to the translational mode at a
different critical noise value $D_{\rm crit}^{\rm rot}$. Surprisingly,
this second transition back to translation was not observed in the
two-dimensional case with harmonic attraction. In general, both
transitions occur at different noise values which leads to a
hysteresis curve (Fig.  \ref{fig:hysterese}, left).

The decrease of the center of mass velocity with rising noise
intensity can be shown by considering the equation of motion of the
center of mass velocity. The analysis in three dimensions develops
quite similar to the last Section following Eq.\eqref{eq:com1}
\cite{strefler_swarming_2008}. One solution corresponds to the rotational state,
the second solution to the translational state.  The center of mass
moves (in absence of noise) with the stationary velocity of the
particles $u_0 = \sqrt{\frac{\alpha}{\beta}}$.  Increasing the noise
leads to higher deviations $\delta \bbox{v}$ which decreases the center of
mass velocity.
%
%
%
%
%

For the transition from rotation to translation the critical noise
value decreases with increasing $C_r/C_a$. However, for
$l_r/l_a < {C_r}/{C_a} < 1$ the critical noise value
decreases with decreasing $\frac{C_r}{C_a}$, below ${C_r}/{C_a} =
{l_r}/{l_a}$, no transition to translation takes place.  In this
parameter range the potential does not possess a minimum, i.e. it does
not have a repulsive core and is continuously attractive for all
distances. Even without noise there is no transition from rotation to
translation. Otherwise the transition backwards stays unaffected. From
this fact we conclude that the existence of repulsive forces induces
transitions from rotation to translation.

%

To check this hypothesis, we approximate the Morse potential by a
harmonic potential with equilibrium distance of the shape 
\begin{align}
U_{\rm
  app}(r) = a (r - r_0)^2,
\end{align} 
with 
\begin{align}
a = \left(\frac{l_r C_a}{l_a
    C_r}\right)^{\frac{l_a}{l_a-l_r}} \ \frac{C_r}{2 l_r} \ \left(
  \frac{1}{l_r} - \frac{1}{l_a} \right)
\end{align}    
 and $r_0$ given in Eq. \ref{eq:equilibrium}. We compare it to the
overall attractive harmonic potential $U_{\rm H}(r) = a r^2$ (see Fig.
\ref{fig:hysterese}, left inset). The critical noise value for the
transition from translation to rotation is equal for the Morse
potential and the harmonic approximation (not shown). The
translational mode of the harmonic potential $U_{\rm H}$ is stable for
larger noise values. The main difference occurs in the transition
from rotation to translation. The harmonic approximation $U_{\rm app}$
shows this transition, though at a different noise value than the
Morse potential. In case of the harmonic potential without repulsive
part $U_{\rm H}$ the rotational mode is stable even without noise.
This supports the assumption, that a short-range repulsive part of the
potential is vital for the existence of the transition from rotation
to translation.
%
%
%
%

For both transitions we observe the critical noise value to decrease
with increasing amplitude of the repulsive part of the interaction
potential $C_r$. Yet, the decrease is much faster for the transition
from translation to rotation which leads to a parameter region where
transitions in both directions occur at the same noise value.

In this region, the system alternates between the two states. The
inlay of the right panel in Fig. \ref{fig:hysterese} shows the center of mass velocity $u$
which alternates between translation and rotation (red). The trajectory of the center of mass shows diffusive motion, where the system is rotating, separated by parts with stochastic translational motion (Fig. \ref{fig:hysterese},right panel).

\begin{figure}[tb]
\centerline{
\includegraphics[width=.4\linewidth]{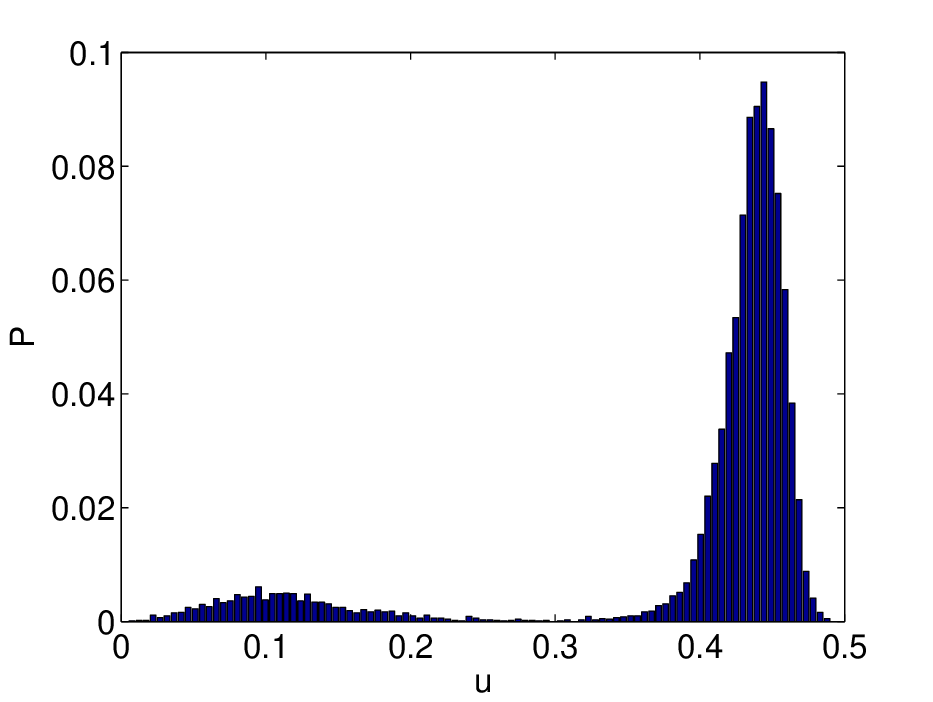}\hfill\includegraphics[width=.4\linewidth]{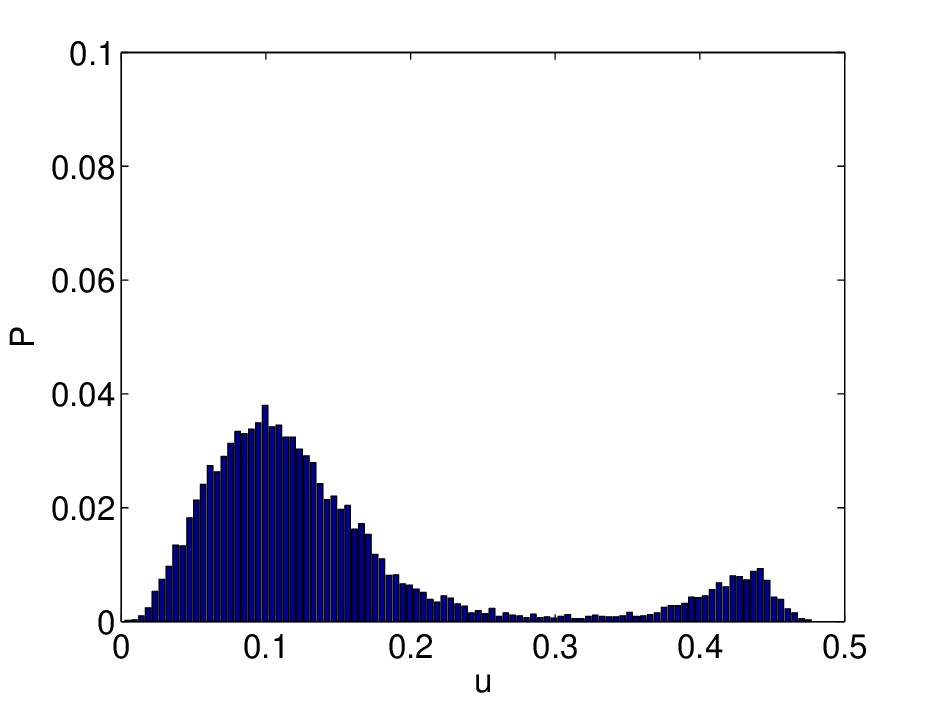}}
\caption{\label{fig:probability} Probability distribution of the speed of center of mass for different noise values of a swarm in 3D with Morse interaction \cite{strefler_swarming_2008}. 
}
\end{figure}

The probability distribution of $u$ depends very sensitively on the
noise intensity (Fig. \ref{fig:probability}). The region where this
oscillatory behavior can be observed is very small. Changing the noise
value by a few percent leads to a shift of the transition
probabilities which is sufficient to destroy the oscillations. It is
sketched in Fig. \ref{fig:probability} where for both graphs the noise
intensity has changed little. Nevertheless, the weights of the two
states in the two distribution changes drastically.

%

\subsection{Collective motion from local velocity alignment without cohesion}
\label{sec:swarming_va}
\subsubsection{Interactions of active particles mechanisms and symmetries}

Active matter may be more generally divided into in externally driven particles, e. g. shaken granular 
matter, and in particles which are intrinsically powered, self-propelled particles, like gliding and 
swarming cells, or chemically active particles, e.g. oscillating yeast cells. 
Self-propelled particles are almost always polar, as their propulsion mechanism determines typically their direction of motion, or, more specifically the leading and trailing edge of the particle are clearly distinguishable (see also Sec. \ref{sec:internalcoord} and \ref{sec:selfdriven_ABP}). 
This requires an initial symmetric breaking in the internal structure of the self-propelled particle, e. g.  
polarization of motile biological cells or  asymmetric design of manufactured granular particles under driving. 
Examples are driven granular particles with non-
symmetric shape or mass distribution \cite{kudrolli_swarming_2008,kudrolli_concentration_2010,deseigne_collective_2010} or different chemical composition at the two ends of a particle, 
like chemically driven running droplets or nano-dimers (Janus particles) \cite{paxton_catalytic_2004,sumino_self-running_2005,ruckner_chemically_2007,tao_design_2008,toyota_self-propelled_2009,valadares_catalytic_2010}. 
To understand the collective dynamics of ensembles of active particles, one needs also to figure out the nature of their  interactions.
Swimming bacteria, flocking birds and schools of fish are often assumed to  have hydrodynamic interactions 
that favor a joint direction of motion. 
Such interactions are analogous to ferromagnetic interactions of spins and try to align the vectors 
describing the motion of the particles; consequently, one speaks of polar interactions. 
In contrast, gliding bacteria or shaken rods tend to align their long axes, but still can either move in a parallel or anti-parallel fashion. Hence, such interactions are classified as apolar. 
Apolar interactions are analogous to nematic interactions that favor alignment, 
e. g. excluded volume interactions of rod-shaped particles.
%

The famous Vicsek model represents an example for  polar particles exhibiting polar interactions \cite{vicsek_novel_1995}. 
Self-propelled rods with volume-exclusion interactions can be identified as polar particles with apolar interactions. 
The so-called {\em active nematics} constitute a third case, namely apolar particles with apolar interactions, that shall not be discussed here.
For recent results on active nematics \cite{toner_hydrodynamics_2005},  such as the analysis of a
 simple Vicsek-type model \cite{chate_simple_2006} and experiments with
 externally driven elongated particles such as rice corns, see instead
e.g. \cite{narayan_nonequilibrium_2006,aranson_swirling_2007,narayan_long-lived_2007} and references therein.


A central question in the study of such active particles with local interactions is if and when long-range order arises. 
The nature of the transition between ordered and disordered types of motion has been investigated thoroughly. 
Below, we will discuss several mean-field theories that neglect spatial variation in the order parameter. 
For the Vicsek model, the mean-field theory predict a continuous second order phase transition to long-range order.
which was reported initially also in numerical simulation \cite{vicsek_novel_1995,czirok_spontaneously_1997}. 
Gr{\'e}goire and Chat{\'e} later found in extensive simulations that the transition to collective motion in the Vicsek model
appeared in fact in a discontinuous way provided the system size in the simulations is large enough 
\cite{gregoire_onset_2004}. 
Their result was contested by later results of Vicsek´s group \cite{nagy_new_2007}, but recent even more extensive studies by Chat{\'e}, Ginelli et al. reaffirm the discontinuous nature of the transition at the onset of collective motion \cite{chate_collective_2008}. This work revealed a crossover from a second to a first order transition at sufficiently large system size, which can depend strongly on the velocity parameter used in the simulations. 
More importantly, their work showed that the appearance of a discontinuous transition is strongly correlated with the 
appearance of segregation bands of high particle density in a low density background that travel in the direction motion
of the self-propelled particles in these bands. 
Such correlations between the transition behavior and spatial patterns of the particle density 
 shall be left aside for the moment and will be treated in Section \ref{sec:pattern} 
that deals with pattern formation aspects of self-propelled particle systems.
An important general lesson is that spatial aspects maybe crucial and need consideration. 
For the moment, we will however exclude them and leave their general discussion to Section \ref{sec:pattern}. 
In the following, we proceed with the discussion of local polar (or ferromagnetic) alignment of active Brownian particles, 
before considering both polar and apolar alignment interactions in the framework of self-propelled particles with Vicsek-type dynamics.

\subsubsection{Active Brownian particles with velocity alignment force}
\label{sec:velal}
We consider now a system of $N$ active Brownian particle interacting only via a local velocity alignment force ${\bbox F}^\text{va}$ (see Eq. \ref{eq:va_sigma} \cite{czirok_formation_1996,okubo_diffusion_2001,romanczuk_collective_2010,romanczuk_mean_2011}), which may be written as
\begin{align}
{\bbox F}^\text{va}_{i} & = \mu \left({\bbox u}_{i,\varepsilon} - {\bbox v}_i \right) 
\end{align} 
with ${\bbox u}_{i,\varepsilon}$ being the mean velocity of the particle gas within within a finite distance $|{\bbox r}_j-{\bbox r}_i|<\varepsilon $ around the focal particle $i$ ($i=1,\dots,N$). The force aligns the velocity of the focal particle to the average local velocity with $\mu$ being the alignment strength, which corresponds in the deterministic case to the inverse relaxation time of ${\bbox v}_i$ towards  ${\bbox u}_{i,\varepsilon}$. For solitary particles with no neighbors or a system in a perfectly ordered state where all particles move with equal velocity the alignment force vanishes as ${\bbox u}_{\varepsilon,i}={\bbox v}_i$. On the other hand for a large number of neighbors moving with random velocities (disordered state), the mean velocity vanishes $|{\bbox u}_{\varepsilon,i}|=u_{\varepsilon,i}\approx 0$ and the velocity alignment force results in additional ``social'' friction $-\mu {\bbox v}_i$.
The force may be considered as a continuous version of the Vicsek model and for self-propelled particles with constant velocity $v_i=const$ it reduces essentially to the polar alignment interaction discussed in the next section \cite{peruani_mean-field_2008}. Another motivation for introduction of such an velocity alignment term comes from hydrodynamic interaction which may result in an alignment term with similar symmetry properties \cite{erdmann_collective_2003,baskaran_statistical_2009}. 

Recently, there has been a number of publication on kinetic description of self-propelled particles with velocity alignment (see e.g. \cite{toner_flocks_1998,toner_long-range_1995,bertin_boltzmann_2006,simha_statistical_2002,simha_hydrodynamic_2002,lee_fluctuation-induced_2010,ihle_kinetic_2011}, see also Sec. \ref{sec:cont_vs_ibm}). In these previous publications,
the authors constructed mesoscopic equations of motion for the density and velocity fields using symmetry and conservation laws. Recently, \cite{bertin_boltzmann_2006} derived corresponding equations with a Boltzmann approach. Further contributions were made by C.-F. Lee \cite{lee_fluctuation-induced_2010}, who studied collective dynamics due to velocity alignment by analyzing an approximated Fokker-Planck equation and by T. Ihle \cite{ihle_kinetic_2011} who derived an Enskog-type kinetic theory for the Vicsek model. 

Here, in contrast to the previous publications, we derive the mean field equations directly from the microscopic Langevin equations. We do not assume a constant speed and analyze the impact of the velocity-dependent friction function on the onset of collective motion. In addition to the density and velocity fields, we consider explicitly the effective temperature field of the active Brownian particle gas.

Our approach is based on the formulation of moment equations of the corresponding probability distribution. 
In general, far from equilibrium the probability distributions are not Gaussian and a correct description requires infinitely many moments (see for example \cite{pawula_approximation_1967,pawula_approximating_1987}). Therefore approximations are necessary in order to obtain a closure of the system of moment equations.

As first example we consider a one-dimensional system with periodic boundary conditions ($d=1$, one-dimensional ring).  
The $n$-th moments of the velocity $ \langle v^n \rangle $ is defined as
\begin{equation}
\langle v^n \rangle = \frac{1}{\rho} \int v^n \psi(x,v,t) dv, \quad n > 0,
\label{eq:moments}
\end{equation}
where $\psi(x,v,t)$ is the one particle distribution describing the probability to find a particle at time $t$, at position $x$ moving with velocity $v$. For $N$ identical particles it is given by simply by the multiple of single particle probability density $\psi(x,v,t)=N P(x,v,t)$ The density of particles is given as 
\begin{equation}
\rho(x,t) = \int d{v}\, \psi(x,v,t) dv = N \int d{v}\, p(x,v,t)\, .
\end{equation}
Multiplying the $n$-th moment with the density and taking the derivative with respect to time, we obtain the dynamics of the moments of velocity
\begin{equation}
\frac{\partial}{\partial t}(\rho \langle v^n \rangle) = \int v^n \frac{\partial \psi}{\partial t} dv.
\label{eq:n-moment}
\end{equation}

The equations of motion for active Brownian particles with Rayleigh-Helmholtz friction and velocity alignment read:
\begin{align}
\dot x_i &= v_{x,i} \\
\dot {v}_{x,i} & = (\alpha-\beta{v}^2_i){v}_{x,i} + \mu (u_{\varepsilon,i} - {v}_{x,i}) +\sqrt{2 D} {\xi}_i \label{eq:RH1d_langevin}.
\end{align}
The Fokker-Planck equation for a single particle in the mean velocity field $u_\varepsilon$ reads 
\begin{align}\label{eq:fpe}
\frac{\partial P}{\partial t} &  = - v_x \frac{\partial}{\partial x} P - \frac{\partial}{\partial v_x} \lbrace (\alpha - \beta v_x^2)v_x + \mu(u_{\varepsilon}-v_x)\rbrace P + D  \frac{\partial^2}{\partial v_x^2} P  . 
\end{align}
Here we omit the index $i$ for simplicity. The velocity ${u}_{\varepsilon}={u}_\varepsilon({x},t)$ is the mean field velocity sensed by the particle. In the continuous description we may express it as an integral over the distribution function:
\begin{align}\label{eq:mf_va_vel}
{u}_{\varepsilon}=\frac{1}{\int_{S_\varepsilon} \rho(x',t) d{x'}}\int_{S_\varepsilon} dx' \int dv' {v'} \psi({x'},{v'},t)  
\end{align}
Here $S_\varepsilon$ represents the spatial neighborhood of the focal particle, which is defined via a metric distance: ${x}\in S_{\varepsilon}$ if $|{x'}-{x}|<\varepsilon$. In the limit $\varepsilon\to0$ together with $N_\varepsilon\to\infty$, $u_\varepsilon$ reduces exactly to the first velocity moment $\langle v \rangle$ \eqref{eq:moments}, whereas for finite $\varepsilon$ it may be seen as an approximation of the first moment under the assumption of a homogeneous density distribution on the corresponding length scale.

Inserting  Eq.~(\ref{eq:fpe}) in to Eq.~(\ref{eq:n-moment}) and using $\lim_{v\to\pm\infty} P(x,v,t)=0$, the terms with partial derivatives with respect to $v$ can be partially integrated, yielding 
\begin{align}\label{eq:dynmoment}
\frac{\partial}{\partial t}(\rho \langle v^n \rangle)   = &  - \frac{\partial}{\partial x} \rho \ \langle v^{n+1} \rangle  
						+ n \ \rho \left[ \alpha \ \langle v^n \rangle - \beta \  \langle v^{n+2} \rangle  
						+ \mu(u_\varepsilon\langle v^{n-1}\rangle - \langle v^{n}\rangle)\right] \nonumber \\
					     & + n  \left(n-1\right) \ D \ \rho \ \langle v^{n-2} \rangle.
\end{align}
We rewrite the velocity of the focal particle as a sum of the local velocity field $u(x,t)$ plus some deviation $\delta{v}$: $v=u+\delta v$. Furthermore, we assume $\langle \delta v^l \rangle=0$ for odd exponents $l$ ($l=1,3,5,\dots$). Thus we obtain for the moments (up to $l=4$):
\begin{subequations}\label{eq:m}
\begin{align}
\langle v \rangle \, &= u \label{eq:m1} \ ,\\
\langle v^2 \rangle &= u^2 + T \label{eq:m2}\ ,\\
\langle v^3 \rangle &= u^3 + 3 \ u \ T  \label{eq:m3}\ ,\\
\langle v^4 \rangle &= u^4 + 6 \ u^2 \ T + T^2 + \theta \label{eq:m4}\ .
\end{align}
\end{subequations}
Here, $T$ is the mean squared velocity deviation $T=\langle \delta v^2 \rangle$, which we will refer to as the ``temperature'' of the active particle gas, whereas $\theta$ is the average of the mean squared temperature fluctuations defined as 
\begin{align}\label{eq:theta}
\theta = \langle \left(\left(v - u\right)^2 - T\right)^2 \rangle= \langle \delta v^4 \rangle - T^2.
\end{align}
Now we can insert Eqs.~(\ref{eq:m}) into Eq.~(\ref{eq:dynmoment}). Considering the dynamics up to $n=2$, after some calculus we arrive at a set of three coupled partial differential equations for the evolution of the density $\rho(x,t)$, the mean velocity field $u(x,t)$, and the temperature field $T(x,t)$:
\begin{subequations}
\begin{align}
 \frac{\partial}{\partial t} \rho  = &  -\frac{\partial}{\partial x} \left(\rho \ u\right) \\
 \frac{\partial u}{\partial t} + u \ \frac{\partial}{\partial x} u = & \ \alpha \ u - \beta \ u \ \left(u^2 + 3 T\right)+\mu(u_\varepsilon-u)
 - \frac{\partial T}{\partial x} - \frac{T}{\rho} \ \frac{\partial \rho}{\partial x} \\
 \frac{1}{2} \left(\frac{\partial T}{\partial t} + u \ \frac{\partial \ T}{\partial x}\right) = & \ (\alpha-\mu) \ T - \beta \ T \left(3 u^2 + T \right)  - \beta \ \theta  + D - T \frac{\partial u}{\partial x}\label{eq:mftemp} 
\end{align}
\label{eq:dgls}
\end{subequations}
Let us consider for simplicity an isotropic system with vanishing gradients in mean velocity $u$ and temperature $T$, which is fully analytically tractable and represents a reasonable approximation of the system dynamics at high particle densities and large $\varepsilon$. In this case, the local velocity in the velocity alignment force equals the constant mean velocity field across the system $u_\varepsilon=u$, and  we end up with the following two ordinary differential equations for the temporal evolution of $u$ and $T$:
\begin{subequations}
\begin{align}
\frac{d u}{d t}  = &\alpha \ u - \beta \ u \ \left(u^2 + 3T\right)\\
\frac{1}{2}\frac{d  T}{d t} =& (\alpha-\mu) \ T - \beta \ T \ \left(3u^2 + T\right) - \beta \theta + D. 
\end{align}
\label{eq:mfT}
\end{subequations}
In order to obtain a closed system of equations, we neglect the temperature fluctuations by setting $\theta=0$ in Eq.~(\ref{eq:theta}), which is a reasonable assumption at small noise intensities. Thus, the above differential equations constitute a two-dimensional dynamical system, with 6 fixed points  (stationary solutions) in the $(u,T)$ phase space, which can be analyzed by means of linear stability analysis.

The stationary solutions ($d u/d t=d T /d t =0$) for $u$ and $T$ read:
\begin{subequations}
\begin{align}
u_{1,2} & = 0,  &  & T_{1,2}  = \frac{\alpha-\mu \pm \sqrt{(\alpha-\mu)^2 +  4 \beta D}}{2 \beta} \, ; \\
u_{3,4} & = \pm \frac{\sqrt{10 \alpha -3 \left(\mu-\Delta\right)}}{4 \sqrt{\beta}}, & &  T_{3,4}  = \frac{2\alpha + \mu - \Delta}{16 \beta} \, ; \\
u_{5,6} & = \pm \frac{\sqrt{10 \alpha -3 \left(\mu+\Delta\right)}}{4 \sqrt{\beta}}, & & T_{5,6}   = \frac{2\alpha + \mu + \Delta}{16 \beta},  
\end{align}
\end{subequations}
with $\Delta = \sqrt{(2\alpha+\mu)^2 - 32 \beta D}$. 

The kinetic temperature $T_j$ has to be positive, therefore,  $T_1$ (positive square root) is the only physically reasonable solution for $u=0$. 

The first solution with vanishing mean velocity $u=0$ describes a disordered phase. For $D=0$, the temperature $T = \frac{\alpha-\mu}{\beta} = v_0^2$ equals the square of the stationary velocity of individual particles. The kinetic energy of all particles consists  only of fluctuations, no systematic translational motion occurs.
\begin{figure}
\begin{center}
  \includegraphics[width=\linewidth]{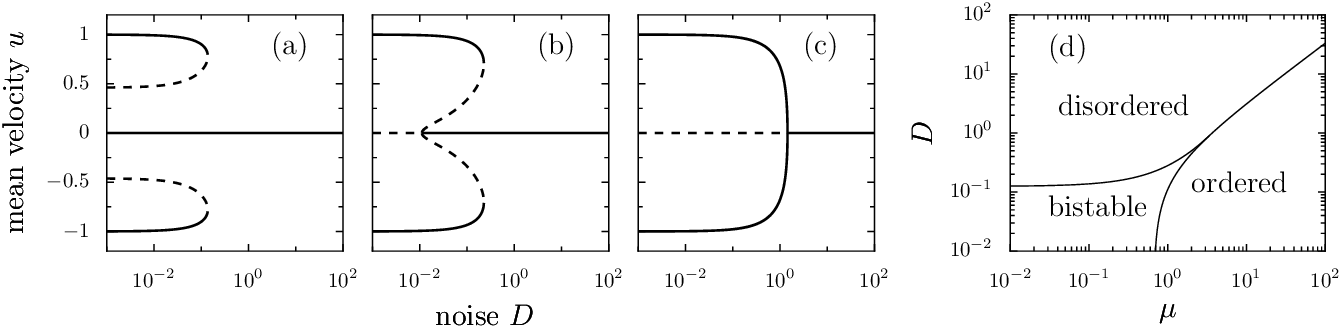} 
\end{center}
\caption{Bifurcation diagram of the mean velocity $u$ vs noise intensity $D$ as predicted from the mean field theory for different velocity alignment strengths ($\alpha=\beta=1$) : {\bf (a)} $\mu<2\alpha/3$ ($\mu=0.1$), {\bf (b)} $2\alpha/3<\mu<10\alpha/3$ ($\mu=0.7$) and {\bf (c)} $\mu>10\alpha/3$ ($\mu=5.0$). {\bf (d)} Phase diagram with respect to velocity alignment $\mu$ and noise intensity $D$ \cite{romanczuk_collective_2010}.} 
\label{fig:ubif}       
\end{figure}
The second pair of solutions corresponds to translational modes which are stable below a critical noise intensity. The two solutions correspond to translational motion with positive or negative velocity $u$, thus, to a collective motion of the particles to the left or right. Without noise, $T_{3,4} = 0$, and the stationary mean velocity reduces to $u_{3,4} = \pm \sqrt{\alpha/\beta}$. Increasing the noise rises the kinetic temperature, and results in a decrease of the mean speed $|u|$.\\
The last solution pair describes unstable modes, for which the temperature decreases and the mean speed increases with increasing noise intensity $D$. 

For low alignment strength $\mu<2\alpha/3$, the disordered phase is always a stable solution; for  $\mu>2\alpha/3$, the linear stability analysis of the mean-field equations predicts the existence of a critical noise intensity 
\begin{equation}
D_{d, \rm crit} = \frac{\alpha(3\mu-2\alpha)}{9 \beta},
\label{eq:D_1}
\end{equation}
which determines the stability boundary of the disordered solution. Starting from large noise intensities where the disordered solution is stable and decreasing the noise below $D_{d, \rm crit}$, we observe a pitchfork-bifurcation, and the disordered phase becomes unstable. Depending on the value of $\mu$, the pitchfork-bifurcation is either sub- or super-critical. For $\mu<10\alpha/3$, the disordered solution becomes unstable through a collision with the two unstable translational solutions, whereas for $\mu>10\alpha/3$ no unstable translational solutions exist and the disordered solution becomes unstable directly through the appearance of the two stable translational solutions (see Fig.~\ref{fig:ubif}). Thus, for $\mu>2\alpha/3$ and $D<D_{d,\rm crit}$, only the translational solutions $u_{3,4}$ are stable. 
\begin{figure}
\begin{center}
  \includegraphics[width=0.75\linewidth]{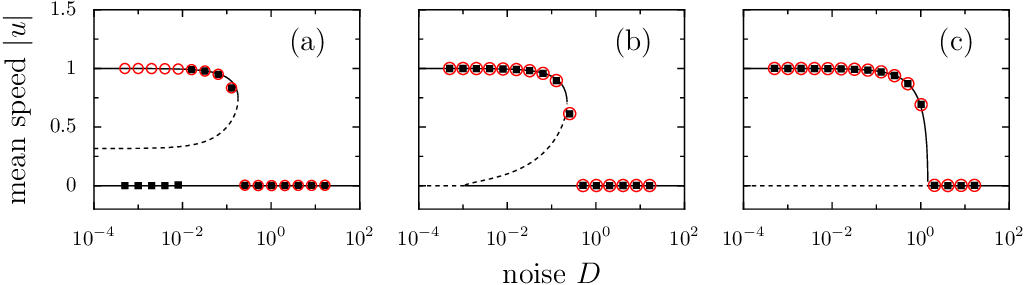}
\end{center}
\caption{(Color online) Comparison of the stationary solution obtained from simulation with theoretical prediction from the mean field theory for different velocity alignment strengths: $\mu=0.4$ {\bf (a)}, $\mu=0.67$ {\bf (b)} and $\mu=5.0$ {\bf (c)}. Other parameters used: particle number $N=8192$, simulation domain $L=500$, velocity alignment range $\epsilon=50$ and $\alpha,\beta=1.0$.  The initial conditions were either the disordered state (black filled squares) or the ordered state (red circles). Solid (dashed) lines show the stable (unstable) stationary solutions of the mean-field equations \cite{romanczuk_collective_2010}.}
\label{fig:usim}       
\end{figure}

For $\mu<10\alpha/3$ there exists a second critical noise intensity which determines the stability of the ordered phase (translational solutions, $u\neq0$). Above the critical noise intensity 
\begin{equation}
D_{o, \rm crit} = \frac{(2\alpha+\mu)^2}{32 \beta}
\label{eq:D_2}
\end{equation}
all translational solutions become unstable through a saddle-node bifurcation (Fig. \ref{fig:ubif} a,b).  

Due to the symmetry of the translational solutions $u_3=-u_4$, we may distinguish the disordered phase and the ordered (translational) phase in simulations by measuring the global mean speed in our simulations:
\begin{align}
\left\langle |u| \right\rangle& =\left\langle \left| \frac{1}{N} \sum_{i=1}^{N} v_i  \right| \right\rangle.
\end{align}
Here, $\langle\cdot\rangle$ denotes temporal average after the system has reached a stationary state.

The stationary speed of the ordered phase versus noise intensity obtained from numerical simulations with ordered state as initial condition are in a good agreement with the theoretical predictions from the mean field theory. But simulations with disordered initial condition reveal an unexpected instability of the disordered solution. At finite $\mu$ the numerical simulations show that at intermediate $D$ the disordered solution $u=0$ becomes unstable via a spontaneous symmetry breaking, which is not predicted by the mean field theory. 

We reproduced the instability of the disordered solution in numerical simulations for intermediate noise strengths for different particle numbers and for global coupling ($\varepsilon=L$). This suggests that it cannot be simply dismissed as a pure finite size effect or an effect of density fluctuations and should be associated with the neglected higher order fluctuations. The latter conclusion is supported by the agreement of the theoretical result and numerics at low $D$, where temperature fluctuations $\theta$ are negligible, as well as with deviations of the mean field temperature from the exact solution for $T$ in the limit $\mu=0$ \cite{erdmann_brownian_2000,romanczuk_collective_2010}.  Due to the nonlinearity of the friction function, the temperature does not increase monotonically with $D$ as predicted by the mean field theory but exhibits a minimum at intermediate noise intensities \cite{romanczuk_collective_2010}. This in turn has a destabilizing effect on the disordered state. Therefore we expect that the extension of the mean field theory to higher orders would account for this effect at the expense of the analytical tractability of the mean field solutions \cite{erdmann_kollektive_2003}.

The same approach can be also extended to two dimensional systems where the mean field velocity and temperature are vectors: ${\bbox u}=(u_x,u_y)=(\langle v_x \rangle, \langle v_y \rangle) $, ${\bbox T}=(T_x,T_y)=(\langle \delta v_x^2 \rangle,\langle \delta v_y^2 \rangle) $. Under the assumption of independent deviations in $x$ and $y$-direction $\langle \delta v_x^n \delta v_y^m \rangle = \langle \delta v_x^n \rangle \langle \delta v_y^m \rangle$ we can derive the following set of mean field equations:
\begin{subequations}
\label{eq:RHva2d_dgls}
\begin{align}\frac{\partial}{\partial t} \rho  = &  -{\nabla} \left(\rho  {\bbox u}\right) \label{eq:continuity}\\
 \frac{\partial u_x}{\partial t} + {\bbox u}  {\nabla}_{\rm r} u_x  =& \  \alpha  u_x - \beta  u_x  \left({\bbox u}^2 + 3 T_x + T_y\right) +\mu(u_{\varepsilon,x}-u_x)\nonumber\\
 &\  - \frac{\partial T_x}{\partial x} - \frac{T_x}{\rho} \ \frac{\partial \rho}{\partial x} \\
 \frac{1}{2} \left(\frac{\partial T_x}{\partial t} + {\bbox u}\nabla_{\rm r} T_x\right) =&\  (\alpha-\mu)  T_x - \beta T_x \left({\bbox u}^2 + 2 u_x^2 + T_x + T_y\right)  - \beta  \theta_x \nonumber \\
 &\  + D - T_x \frac{\partial u_x}{\partial x}. 
\end{align}
\end{subequations}
The corresponding equations for $u_y$ and $T_y$ can be obtained by interchanging the indices. 
We can further simplify the above set of equations by choosing a reference frame, where $u_x=u_{||}=u$ corresponds to the mean field velocity and the orthogonal component vanishes $u_y=u_{\bot}=0$. 

Assuming again the simplest case of a spatially homogeneous system, we obtain the following set of coupled (ordinary) differential equations:
\begin{subequations}\label{eq:MF_RH2dred}
\begin{align}
\frac{d u}{dt} = & \alpha  u - \beta  u  \left(u^2 + 3 T_\| + T_\bot\right)  \\
\frac 12 \frac{d T_\|}{dt} = &  (\alpha-\mu)  T_{\|} - \beta T_\| \left(3 u^2 + T_\| + T_\bot\right)   + D\\
\frac 12 \frac{d T_\bot}{dt} = &  (\alpha-\mu)  T_\bot - \beta T_\bot \left(u^2 + T_\| + T_\bot\right)   + D
\end{align}
\end{subequations}
where $T_\|$ and $T_\bot$ are the temperature components parallel and perpendicular to the mean field direction of motion.
For $u=0$ (disordered state) the components on ${\bbox T}$ can be easily calculated from \eqref{eq:MF_RH2dred}
and the corresponding solution reads:
\begin{subequations}\label{eq:MF_RH2d_dissol}
\begin{align}
u_{1}= & 0 \\
T_{\| ,1}=T_{\bot,1}= 
T_1 = & \frac{\alpha-\mu + \sqrt{(\alpha-\mu)^2 +  8 \beta D}}{4 \beta}  
\end{align}
\end{subequations}
In the case of vanishing noise $D=0$ the ordered solution can be immediately obtained as $u=\sqrt{\alpha/\beta}$ and $T_\|=T_\bot=0$. For $D>0$ the temperature component parallel to the direction of motion is smaller than the perpendicular one: $T_\|<T_\bot$. 

For the general ordered state with $u>0$ and $D>0$ we were not able to obtain explicit stationary solution for $u$, $T_\|$ and $T_\bot$ of the above ODE system \eqref{eq:MF_RH2dred} but the stable and unstable solutions can be determined by a numerical continuation methods as for example provided by the numerical software XPPAUT/AUTO \cite{doedel_auto:_1981,ermentrout_simulating_2002}.  

A possible approach to find an explicit solution is to reduce the dimensionality of the problem. We may use the fact that at a fixed time $t$ we always find a coordinate frame where $u_x=u_y=\tilde u$. In this coordinate frame due to the symmetry of the involved equations we obtain also $T_x=T_y=\tilde T$. Based on this observation we reduce the full problem \eqref{eq:RHva2d_dgls} from a four dimensional system to a two dimensional system in $u$ and $T$, which in the homogeneous case simplifies to
\begin{subequations}\label{eq:MF_RH2dreduced}
\begin{align}
\frac{d \tilde u}{dt} = & \alpha  \tilde u - \beta  \tilde u  \left(2 \tilde u^2 + 4 \tilde T\right) \,, \\
\frac 12 \frac{d \tilde T}{dt} = &  (\alpha-\mu)  \tilde T - \beta \tilde T \left(4 \tilde u^2 + 2 \tilde T\right) + D \,.
\end{align}
\end{subequations}
This gives us a system of equations similar to the problem for $d=1$, with the same structure of stationary solutions but different coefficients.
\begin{figure}
\begin{center}
  \includegraphics[width=0.8\linewidth]{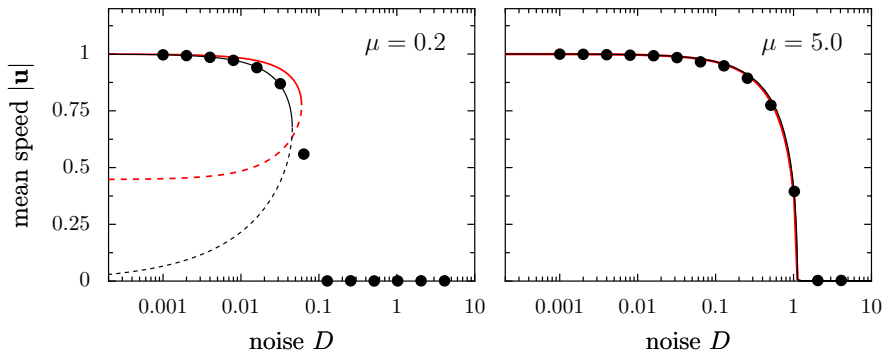} 
\end{center}
\caption[Mean field speed of the RH-model vs noise strength for $d=2$]{Comparison of the mean field speed $|u|$ obtained from Langevin simulations (symbols) of the RH-model in two spatial dimensions at high density with the results of the mean field theory for the homogeneous case. The black lines represent the solutions obtained from the full system of mean field ODE's. The red lines represent the mean field solutions from the reduced system. The stable solutions are shown as solid lines, whereas dashed lines indicate the unstable solution. The simulations were performed with periodic boundary condition and with the disordered state as initial condition. Other parameters: $\alpha=1$, $\beta=1$, $L=200$, $\varepsilon=20$, $N=4096$ \cite{romanczuk_mean_2011}.}
\label{fig:MF_RH2d_U}
\end{figure}

The comparison of the stationary solutions of the reduced systems with the corresponding solutions of the full systems obtained with XPPAUT/AUTO reveals a differences at low velocity alignment strengths $\mu$ (Fig. \ref{fig:MF_RH2d_U}). The velocity of the stable ordered solution of the full system decreases stronger and exhibits an earlier breakdown with increasing $D$. Furthermore from the position of the disordered branch it can be deduced that the basin of attraction of the ordered state at low $D$ for the full system is larger than for the reduced two dimensional system. But at large $\mu$ the differences between the two types of mean field solution vanish and the reduced system \eqref{eq:MF_RH2dreduced} gives a good approximation as shown in Fig. \ref{fig:MF_RH2d_U}. 

The reason for the discrepancy between the two mean field solutions at low $\mu$ is that the performed dimensional reduction throws away all informations about the asymmetry of temperature components parallel and perpendicular to the mean velocity. At large $\mu$ the evolution of the temperature coefficients is dominated by the $-\mu T_k$ term and may in a crude approximation simply be assumed as linear for both components, so that the asymmetry in the temperature components can be neglected.

In general without knowing the temperatures $T_\|$ and $T_\bot$ the mean speed, can be written as   
\begin{align}
|u| = \sqrt{v_0^2 - 3 T_\| - T_\bot}
\end{align} 
where $v_0^2=\alpha/\beta$.
In the limit of large $\mu$ close to the critical noise, where $\alpha,\beta \ll \mu, D$ we may approximate the temperature as  $T_\|=T_\bot=T=D/\mu$ and obtain a simple expression for the ordered state
\begin{align}\label{eq:MF_RH2d_ulimit}
|u| = \sqrt{v_0^2 - \frac{ 4D}{\mu} }.
\end{align}
In this limit the critical noise may be approximated as $D_{d,\rm crit}\approx v_0^2\mu/4=D_\text{crit}$ and the above equation may be rewritten as:
\begin{align}\label{eq:MF_RH2d_ulimit2}
|u| = 2\mu^{-\frac 12 }(D_\text{crit} - D)^{\frac 12},
\end{align}
which is the standard form of the order parameter for a continuous (second order) phase transition. A similar result can be also obtained for self-propelled particles with constant velocity \cite{peruani_mean-field_2008}, corresponding to the limit $\alpha,\beta\to \infty$ with $\alpha/\beta=v_0^2=const.$ with $T_\|=0$ and $T_\bot=D/\mu$. 

The mean-field theory, suggest the possibility of a discontinuous transition from ordered to disordered state (saddle-node bifurcation) as a consequence of a nonlinear friction function at intermediate alignment strenghts. However, for limiting cases, where the nonlinear nature of the friction function is negligible, e.g. for vanishing velocity fluctuations (see next section) or strong velocity alignment, the transition in a homogeneous system is a continuous one.
We should note that for very small alignment strengths $\mu$ the critical noise intensity becomes also very small. In this limit, for arbitrary friction functions, the small passive fluctuation act essentially only on the direction of motion and the speed of individual particles may be assumed as constant. Thus, in this case, the collective mean-field dynamics reduce effectively to a Kuramoto-model of coupled oscillators, which also shows a continuous transition from order (synchronized state) to disorder.  

Recently, it was shown that the spatially homogeneous state is unstable in systems of interacting self-propelled particles \cite{bertin_boltzmann_2006,simha_statistical_2002,simha_hydrodynamic_2002}. Thus, the assumption of a spatially homogeneity should be considered only as an approximation, for sufficiently large $\varepsilon$. For $L\gg\varepsilon$ strong density inhomogeneities appear, such as traveling bands, which affect the global behavior of the system \cite{chate_collective_2008,bertin_microscopic_2009,mishra_fluctuations_2010}. A detailed discussion of the spatial inhomogeneities in self-propelled particle systems will follow in Sec. \ref{sec:pattern}.  

In contrast to $d=1$ no stable disordered solutions at low $\mu$ and low noise intensities $D$ were observed (Fig. \ref{fig:MF_RH2d_U}). A possible explanation can be the reduced basin of attraction of the disordered solution for $d=2$ together with the already discussed instability of the disordered solution due to the neglected temperature fluctuations $\theta_k$ observed already for $d=1$.

A more heuristic explanation for the instability of the disordered solution in two spatial dimensions for low $\mu$ and low $D$, in contrast to the one-dimensional case, is the absence of a velocity potential barrier between different direction of motion: In two dimensions the particles can change their direction of motion by continuous angular drift or diffusion. Thus, any small fluctuation in ${\bbox u}$ in a finite system at vanishing noise ($D\ll1$) will be amplified and eventually will lead to perfect velocity alignment.       

\subsubsection{Mean-field theory for onset of polar and nematic order}
\label{sec:mf_spp}

\begin{figure}[t] 
\centering 
\resizebox{\columnwidth}{!}{\rotatebox{0}{\includegraphics{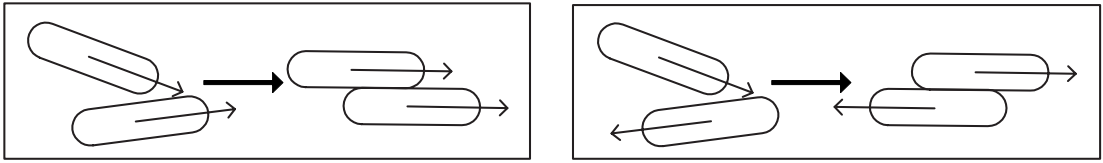}}} 
\caption{Illustration of nematic LC-alignment of polar self-propelled 
 particles illustrated by inelastic collisions of rods. 
 Particles incoming at a small angle (left) align ``polarly'', but those
colliding almost head-on slide past each other, maintaining their nematic alignment
(right).} 
\label{LC-alignment} 
\end{figure} 
The Vicsek-model \cite{vicsek_novel_1995} is considered the simplest model for collective motion of self-propelled particles (SPPs). 
It has become quite popular because it is computationally much more efficient than many other models 
for collective motion of active particles. 
In the Vicsek model, point-like particles moving with a velocity vector of constant magnitude interact 
by aligning their velocity direction to the local average velocity. 
The Vicsek model can be considered as a model of moving spins, in which the velocity of the particles is given by the spin-vector. 
Extending this analogy with spin systems we denote this polar alignment mechanism as ferromagnetic 
(F-alignment). 
Initial simulations of the Vicsek model in 2D  seemed to indicate a second-order
phase transition which leads to long-range orientational order in collectively moving SPPs 
\cite{vicsek_novel_1995,nagy_new_2007}.
Later work pointed towards a first order phase transition to orientational order in sufficiently large 
systems \cite{chate_collective_2008}. 
Theoretical interest was stirred by the that fact that  in analogous equilibrium systems of non-moving 
spins no long-range order is possible \cite{mermin_absence_1966,kosterlitz_ordering_1973}.

F-alignment is one possible alignment mechanism, but clearly not the only one. 
If a system of e. g. self-propelled rods interacts simply by volume exclusion, particles may end up 
moving in the same direction as well as in opposite directions.
Such an apolar interaction mechanism corresponds to interactions in liquid crystals where apolar 
particles get locally aligned  \cite{doi_theory_1988}. 
In analogy to these systems we name this  mechanism hence liquid crystal alignment (LC-alignment). 
An graphical illustration for LC-alignment is given in Fig. \ref{LC-alignment}. 
In a system of SPPs with LC-alignment particles align their velocity to the local average director. 
Orientational order observed in simulations with SPPs with LC-alignment refers to the emergence of a  
global director in the system, while for F-alignment orientational order refers to the appearance of a 
global direction of motion.
%
\begin{figure}

\begin{center}
\includegraphics[width=0.49\textwidth]{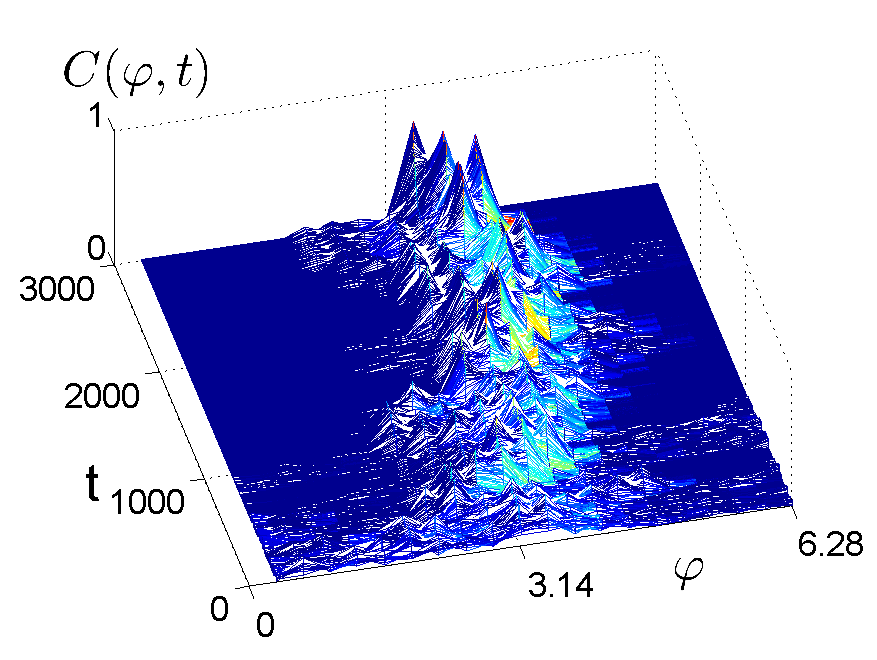}
\includegraphics[width=0.49\textwidth]{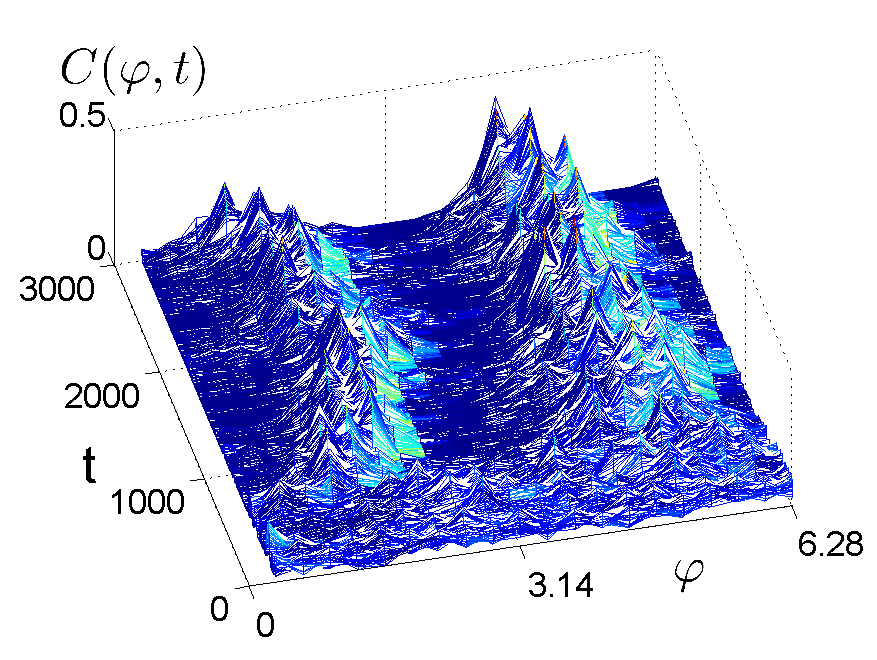}
\end{center}
\caption{Temporal evolution of the velocity direction distribution (angular distribution)
 in simulations of self-propelled particles with very fast angular relaxation, 
where (a) corresponds to the case F-alignment (Vicsek-model), while (b) shows the case of LC-alignment
(Peruani model). 
Number of particles $N=100$, radius of interaction $\varepsilon=2$, 
linear system size $L=42.4$, and noise amplitude $\eta=0.25$. \label{sim_veldistrib}}
\end{figure}
In the following, two alternative continuum models for SPPs with polar, ferromagnetic resp. apolar, 
liquid-crystal like interaction will be introduced and compared to the Vicsek model and a recent 
variation \cite{peruani_mean-field_2008,ginelli_large-scale_2010} with LC-alignment interactions, the Peruani model. 
For the continuum version of the SPP models with local ferromagnetic resp. liquid crystal alignment a 
mean-field theory describing the onset of collective motion is derived and compared to simulation 
results obtained with the Vicsek model. 

We consider point-like particles moving at constant speed in two dimensions  and assume an over-
damped situation such that the state of particle $i$ at time $t$ is given by  its position $\mathbf{r}_i$ 
and its direction of motion $\varphi_i$. The evolution of these quantities follow:
\begin{eqnarray}\label{eq_mot_x}
\dot{\mathbf{r}}_i&=& v_0 \mathbf{e}_{\mathbf{v}}(\varphi_i)  \\
\label{eq_mot_angle} \dot{\varphi_i}&=& - \gamma \frac{\partial
U}{\partial \varphi_i}(\mathbf{r}_i,\varphi_i) +\tilde{\eta}_{i}(t)
\end{eqnarray}
where $\gamma$ is a relaxation constant, and $U$ the interaction potential between particles, and 
hence $\frac{\partial U}{\partial\varphi_i}(\mathbf{r}_i,\varphi_i)$ defines the velocity alignment 
mechanism. 
Moreover, $v_0$ represents the active constant speed of the particles,
the unit vector $\mathbf{e}_{\mathbf{v}}(\varphi_i)$ is again defined
as $\mathbf{e}_{\mathbf{v}}(\varphi_i)=(\cos(\varphi_i),
\sin(\varphi_i))$, and $\tilde{\eta}_{i}(t)$ is an additive white
noise.
In the sense of Sec.  \ref{sec:exter}, it is an active noise acting
perpendicular to the direction of motion.
Eqs. (\ref{eq_mot_x}) and (\ref{eq_mot_angle}) are expressed in terms of first  derivatives. 
In this way, $v_0$ in Eq. (\ref{eq_mot_x}) can be considered as an active force divided by a 
translational friction coefficient.
%
In analogy to spin systems, the ferromagnetic velocity alignment mechanism is given by a potential 
defined as:
\begin{equation} \label{eq:ferromagneticpotential}
U_{F}(\mathbf{r}_i,\varphi_i)=-\sum_{\left|\mathbf{r}_{i}-\mathbf{r}_{j}\right|\leq\varepsilon}
\cos(\varphi_i-\varphi_j)
\end{equation}
where $\varepsilon$ is the radius of interaction of the particles. For the liquid-crystal alignment mechanism, we choose the potential originally introduced by Lebwohl and Lasher on a lattice \cite{lebwohl_nematic-liquid-crystal_1972} which reads:
\begin{equation} \label{eq:liquidcrystalpotential}
U_{LC}(\mathbf{r}_i,\varphi_i)=-\sum_{\left|\mathbf{r}_{i}-\mathbf{r}_{j}\right|\leq\varepsilon}
\cos^2(\varphi_i-\varphi_j)
\end{equation}
One can add a coupling strength coefficient to the expressions (\ref{eq:ferromagneticpotential}) and 
(\ref{eq:liquidcrystalpotential}). 
We assume that the coupling strength is absorbed in $\gamma$ in Eq. (\ref{eq_mot_angle}).
Notice that the potential given by Eq. (\ref{eq:ferromagneticpotential}) exhibits one minimum, while 
Eq. (\ref{eq:liquidcrystalpotential}) has two minima,
which correspond to particles pointing in the same direction and particles pointing in opposite 
directions.

In the limiting case of very fast angular relaxation we obtain 
from Eqs. (\ref{eq_mot_x}) and (\ref{eq_mot_angle}) the updating rules:
\begin{eqnarray}\label{motion_pos}
\mathbf{r}_{i}^{t+\Delta t }&=&\mathbf{r}_{i}^{t} +v_0 \mathbf{e}_{\mathbf{v}}\left(\varphi_i^{t} \right)\Delta t 
\\
\label{motion_vel} \varphi_i^{t+\Delta t }
&=&\arg\left(\sum_{\left|\mathbf{r}_{i}^{t}-
      \mathbf{r}_{j}^{t}\right|\leq\varepsilon}\mathbf{f}(\mathbf{e}_{\mathbf{v}}(\varphi_j^{t}),\mathbf{e}_{\mathbf{v}}(\varphi_i^{t}))\right)+\eta_{i}(t)
\end{eqnarray}
where $\arg\left(\mathbf{b}\right)$ indicates the angle of the vector
$\mathbf{b}$ in polar coordinates, $\eta_{i}^{t}$ is the random
increment of the angle in the interval $\Delta t$ comparable to a
Wiener Process and giving rise to angular diffusion (see Sec. \ref{sec:self_driven_const_s}).
The noise sources are statistically independent for different particles.
In simulations we have used equally distributed random numbers from
$\eta_{i}(t)\varepsilon\left[-\frac{\eta}{2},
  \frac{\eta}{2}\right]$\footnote{Variation of the temporal interval
  $\Delta t$ implies a scaling of $\eta$ with the interval. For the
  case of white noise in Eq. \eqref{eq_mot_x} the variance of
  $\eta_i(t)$ increases linearly $\Delta t$.}.
The term $\mathbf{f}(\mathbf{a},\mathbf{b})$ describes the
interactions and is defined as follows.
For F-alignment, $\mathbf{f}(\mathbf{a},\mathbf{b})=\mathbf{a}$ and Eqs. (\ref{motion_pos}) and 
(\ref{motion_vel}) correspond to the Vicsek model \cite{vicsek_novel_1995}.
For LC-alignment, $\mathbf{f}$ takes the form:
\begin{equation} \label{function_f}
\mathbf{f}\left(\mathbf{a},\mathbf{b}\right) = \left\{
\begin{array}{lcr}
 \mathbf{a} & \mbox{if} & \mathbf{a} \cdot \mathbf{b} \geq 0 \\
-\mathbf{a} & \mbox{if} & \mathbf{a} \cdot \mathbf{b}  <   0
\end{array} \right.
\end{equation}
This interaction was introduced by Peruani {\em et. al.}. Hence we will refer to the corresponding model (Eqs. \ref{motion_pos}, \ref{motion_vel}, \ref{function_f}) as the Peruani model. 
To decide if the described local alignment mechanism give rise to global order, suitable order 
parameters have to be defined.  
One such order parameter that quantifies the direction of alignment and collective motion is the modulus 
of the normalized total momentum (analogous to the magnetization in the XY-model \cite{mermin_absence_1966,kosterlitz_ordering_1973}), 
that we express as:
\begin{eqnarray}\label{eq:orderparam_f}
S^{F} = \left| \frac{1}{N} \sum_{i=0}^{N} \mathbf{e}_{\mathbf{v}} \left(\varphi_i^{t} \right) \right|
\end{eqnarray}
where $N$ stands for the total number of particles in the system. 
The quantity $S^{F}$ has the value $1$ when all particle move in the same direction (perfect  F-
alignment).
In contrast, $S^{F}$ is equal to $0$ in the disordered case in which particles point in any direction 
with equal probability. 
An alternative measure for collective motion is  the velocity direction distribution  in two dimensions 
$C(\varphi)$. 
For high values of the noise, $C(\varphi)$ is flat. 
When the noise is decreased below a critical noise $\eta_c$,  a single peak arises in $C(\varphi)$  
indicating the onset of orientational order, see Fig. \ref{sim_veldistrib}(a). 

On the other hand,  for perfect nematic order or LC-alignment,  {\it i. e.} half of all particles move in 
one direction, and the other half in the opposite direction, $S^{F}$ is identical zero.
Clearly, $S^{F}$  cannot distinguish between a state of LC-alignment and a disordered state. 
To study such LC orientational ordering, one can employ  the order matrix $Q$ of
liquid crystals \cite{doi_theory_1988}. 
For two dimensions one takes the largest eigenvalue $S^{LC}$ of $Q$ as appropriate order parameter:
\begin{equation}\label{eq:orderparam_lc}
S^{LC}=\frac{1}{4}+\frac{3}{2}\sqrt{\frac{1}{4}-\frac{1}{N^{2}}\left\{
\sum_{i,j}^{N}v_{xi}^{2}v_{yj}^{2}-v_{xi}v_{yi}v_{xj}v_{yj}\right\}
}
\end{equation}
where $v_{xi}$ and $v_{yi}$ are defined as $v_{xi}=\cos(\varphi_i)$
and $v_{yi}=\sin(\varphi_i)$. 
The order parameter $S^{LC}$ takes the value $1$ when all particles are aligned along
the same director, and the value $\frac{1}{4}$ in a disordered phase without any preferred orientation.
Alternatively, the velocity direction distribution $C(\varphi)$ can be considered. 
This function displays two peaks separated by $ \pi$ for simulations of the Peruani model low noise 
amplitudes, see Fig. \ref{sim_veldistrib}(b). 

A system of SPPs may be conveniently described through the one
particle density in the phase space $\tilde{\psi}(\mathbf{r},
\mathbf{v}, t)=\psi(\mathbf{r},\varphi,t)$ which equals the one
particle probability density function $P(\mathbf{r}, \mathbf{v}, t)$
multiplied by the particles number $N$.  Then the usual particle
density at a point $\mathbf{r}$ is obtained as
\begin{equation}\label{eq:densitySPP}
\rho \left(\mathbf{r},t\right)=\int_{0}^{2 \pi}\psi\left(\mathbf{r},\varphi,t\right)d\varphi.
\end{equation}
For simplicity, we consider the angular velocity direction distribution: 
\begin{equation}\label{eq:angulardistribution}
C \left(\varphi,t\right)=\int_{\Omega} \psi\left(\mathbf{r},\varphi,t\right){d\mathbf{r}}
\end{equation}
and neglect spatial inhomogeneities in the orientation of the particles. 
We recall that in the individual-based model the kinetic energy is conserved, while 
the momentum is not.
For  F-alignment, the system tends to increase the total momentum, while for LC-alignment the 
tendency is to  decrease it. 
The continuum approach has to reflect that particles always move at constant speed and that number of 
particles is conserved. 


Consequently, the following evolution equation for $\psi(\mathbf{r},\varphi,t)$ is obtained:
\begin{equation}\label{eq_pde_evolution}
\partial_{t} \psi = D_{\varphi}\partial_{\varphi\varphi}\psi-\partial_{\varphi}\left[u_{\varphi}\psi\right]- 
\mathbf{\bigtriangledown}_{\mathbf{r}} \left[\mathbf{u}_{\mathbf{r}}\psi\right],
\end{equation}
where $u_{\varphi}\psi$ and $\mathbf{u}_{\mathbf{r}}\psi$ are
deterministic fluxes, which reflect the local alignment mechanism and
active motion.
The quantity $D_{\varphi}$ refers to the angular diffusion whereby we
have assumed Gaussian noise in Eq. \eqref{eq_mot_angle}. It depends on the square
of the noise amplitude.
The term $u_{\varphi}(\mathbf{r},\varphi)$ describes mean angle
velocity of the alignment interactions of particles with all
neighboring particles which are at a distance less than $\varepsilon$
from their location $\mathbf{r}$. It reads
\begin{equation}\label{eq_f_theta}
  u_{\varphi}\,=\,-\gamma \int_{R(\mathbf{r})}{d\mathbf{r}'}\int^{2 \pi}_{0}{d \varphi'} \frac{\partial 
    U(\mathbf{r},\varphi,\mathbf{r}',\varphi')}{\partial \varphi} \psi(\mathbf{r}',\varphi',t)
\end{equation}
where $U(\mathbf{r},\varphi,\mathbf{r}',\varphi')$ represents the pair
potential between a particle located at $\mathbf{r}$ and pointing in
direction $\varphi$ with another one at $\mathbf{r}'$ and pointing in
direction $\varphi'$. $R(\mathbf{r})$ denotes the interaction
neighborhood.
The above models have the property
$U(\mathbf{r},\varphi,\mathbf{r}',\varphi')=U(\varphi,\varphi')$, i.
e.  within the interaction radius the potential between particles has
equal strength and is homogeneous in space.
Finally, $u_{\varphi}$ represents the "torque" felt by a particle
located at $\mathbf{r}$ and pointing in direction $\varphi$.
The expression for $\mathbf{u}_{r}$ simply reads
\begin{equation}\label{eq_f_x}
  \mathbf{u}_{r}\,= \,v_0 \mathbf{e}_{\mathbf{v}}(\varphi).
\end{equation}
\begin{figure}
\centering
\includegraphics[width=0.49\textwidth]{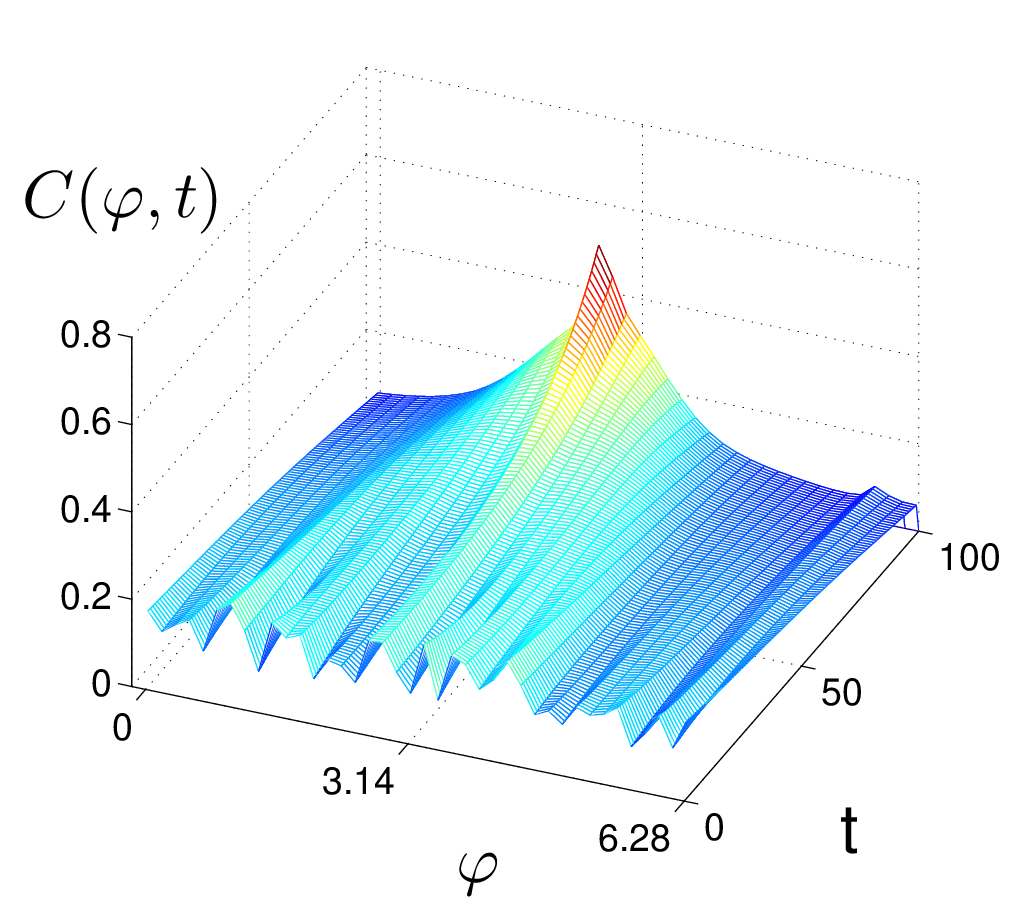}
\includegraphics[width=0.49\textwidth]{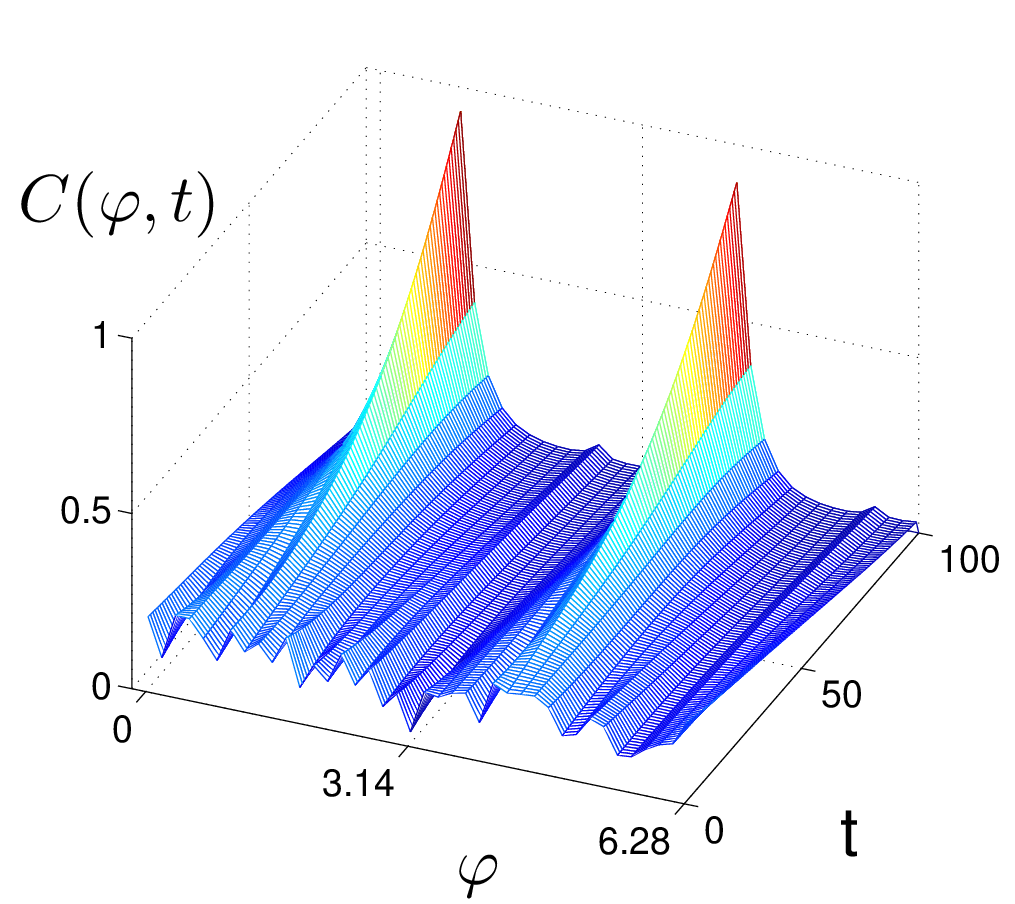}
\caption{Temporal evolution of $C(\varphi,t)$. (top) F-alignment, numerical integration of Eq. 
(\ref{eq:parallel}) with $D_{\varphi}=0.28$. 
(bottom) LC-alignment, numerical integration of Eq. (\ref{eq:parallelantiparallel}) with $D_{\varphi}=0.014$. 
For both, $C^{*}=0.3183$, $\Delta t = 0.001$ and $\Delta
\varphi = 0.16$. The initial condition is a random perturbation around $C^{*}$.
Notice that for F-alignment a single peak emerges, while for LC-alignment the distribution develops 
two peaks. \cite{peruani_mean-field_2008}}
\label{num_example_angle}
\end{figure}
Integrating both sides of Eq. (\ref{eq_pde_evolution}) over the space $\Omega$ and assuming a 
homogeneous spatial distribution of particles $\psi(\mathbf{r},\varphi,t)=C(\varphi,t) \rho_0/N$, where 
$\rho_0$ is defined as $\rho_0=N/L^2$ and  $L$ is the linear size of the system, we obtain an 
evolution equation for  $C(\varphi, t)$:
\begin{align}
\label{eq:angulardistribution2}
\frac{\partial C(\varphi, t)}{\partial
t} = 	& D_{\varphi}\frac{\partial^2 C(\varphi,t)}{\partial \varphi^2} \nonumber \\
	& +\gamma \frac{\pi
\varepsilon^2}{L^2} \partial_{\varphi}\left[ \left\{\int_{0}^{2 \pi} {d\varphi'}
\frac{\partial U(\varphi,\varphi')}{\partial \varphi} C(\varphi',t)  \right\} C(\varphi,t) \right]
\end{align}
The homogeneous angular distribution is a steady state of Eq. (\ref{eq:angulardistribution2}). 
The parameters for the onset of the orientational order and collective motion can be found by 
determining the linear instability of the disordered state. 
We start by analyzing the case of F-alignment.
By dividing both sides of Eq. (\ref{eq:angulardistribution2}) by $\gamma \pi \varepsilon^2 /L^2$, and 
redefining time as $\tau = (\gamma \pi \varepsilon^2 /L^2) t$, and $D_{\varphi}'= D_{\varphi}/ [\gamma \pi 
\varepsilon^2 /L^2]$ one obtains:
\begin{equation}\label{eq:parallel}
\frac{\partial C\left(\varphi,t\right)}{\partial
\tau}=D_{\varphi}'\partial_{\varphi\varphi}C\left(\varphi,t\right)+\partial_{\varphi}\left[\left\{
\int
d\varphi'\sin\left(\varphi-\varphi'\right)C\left(\varphi',t\right)\right\}
C\left(\varphi,t\right)\right]
\end{equation}
Next, a small perturbation of the homogeneous steady state is considered:
\begin{equation}\label{eq:perturb_ang}
C\left(\varphi,t\right)=C^{*}+C_{0}e^{in\varphi}e^{\lambda \tau} \,.
\end{equation}
Note, that ${e^{in\varphi}}$ are eigenfunctions of  the operators emerging from the linearization of Eq. 
(\ref{eq:parallel}) near the homogeneous steady state. 
By substituting  into Eq. (\ref{eq:parallel}) and keeping only terms linear in $C_{0}$, 
we obtain the following  eigenvalues:
\begin{equation}\label{eq:eigenvalue_parallel}
Re(\lambda)=-D_{\varphi}' n^{2}+\pi C^{*}\delta_{n,1}\,.
\end{equation}
The only mode which can become unstable is $n=1$ and the condition for the corresponding instability 
of the homogeneous, disordered state is
\begin{equation}\label{eq:criticaldensity_ferro}
\rho_0>\frac{2 D_{\varphi}}{\gamma \pi \varepsilon^2}\,,
\end{equation}
with $\rho_0 = N/L^2$. 
For a given noise amplitude $D_{\varphi}$, there is a critical particle density above which the 
homogeneous solution is unstable.
Fig. \ref{num_example_angle}(a)  shows that as a result a single peak emerges in $C(\varphi)$
from numerical integration of Eq.(\ref{eq:parallel}) in line with results found in simulations of the 
Vicsek model.

An analogous procedure for LC-alignment yields:
\begin{align}\label{eq:parallelantiparallel}
  \frac{\partial C\left(\varphi,t\right)}{\partial
    \tau}=& D_{\varphi}'\partial_{\varphi\varphi}C\left(\varphi,t\right)\\
  & + \partial_{\varphi}\left[\left\{\int
      d\varphi' 2\cos\left(\varphi-\varphi'\right)\sin\left(\varphi-\varphi'\right) C\left(\varphi',t\right)\right\}
    C\left(\varphi,t\right)\right] \nonumber
\end{align}
As above a small perturbation to the homogeneous distribution $C(\varphi ,t) = C^{*}$ is assumed, see Eq. 
\ref{eq:perturb_ang}.
Again, $e^{in\varphi}$ are eigenfunctions of the linearized operator. 
Substituting Eq. (\ref{eq:perturb_ang}) into Eq. (\ref{eq:parallelantiparallel}) and keeping terms linear 
in $C_{0}$ the following expression for the eigenvalues is obtained:
\begin{equation}\label{eq:eigenvalue_parallel2}
Re(\lambda)=-D_{\varphi}' n^{2}+2 \pi C^{*} \delta_{n,2}
\end{equation}
As for F-alignment, there is only one mode which can become unstable, but this time it is $n=2$. 
The $n=2$-mode exhibits two peaks separated by $\pi$, that correspond to two populations of particles 
moving in exactly the opposite direction.
The homogeneous state is unstable for:
\begin{equation}\label{eq:criticaldensity_LC}
\rho_0>\frac{4 D_{\varphi}}{\gamma \pi \varepsilon^2}
\end{equation}
Again, this inequality defines a critical density  above which the homogeneous solution is no longer 
stable. 
Fig. \ref{sim_veldistrib}(c)-(d) shows the emergence of these two peaks for LC-alignment in 
individual-based  simulations. 
Numerical integration of Eq. \eqref{eq:parallelantiparallel} above this critical density shows again two  
peaks in $C(\varphi,t)$, see Fig. \ref{num_example_angle}(b)
For a given density, there is a critical $D_{\varphi,{c}}$. 
Close to $D_{\varphi,{c}}$ we expect to observe that only one mode dominates $C(\varphi,t)$. As 
already mentioned, $n=1$ is dominant for F-alignment and $n=2$ governs LC-alignment. 
The steady state distribution $C_{st}(\varphi)$ near the instability has the form:
\begin{eqnarray}\label{eq:steady_f}
C_{st}(\varphi) \simeq C^{*} + B_{1} \sqrt{D_{\varphi,{c}} - D_{\varphi}} \cos(\varphi - \varphi_0)
\end{eqnarray}
for F-alignment,  while for LC-alignment it is:
\begin{eqnarray}\label{eq:steady_lc}
C_{st}(\varphi)\simeq C^{*} + B_{2} \sqrt{D_{\varphi,{c}} - D_{\varphi}} \cos(2 (\varphi - \varphi_0))
\end{eqnarray}
where $B_{1}, B_{2}$ are constants and $\varphi_0$ is an arbitrary phase.
In both cases the maximum amplitude of $C_{st}(\varphi)$ close to the $D_{\varphi,{c}}$ grows as 
$\sqrt{D_{\varphi,{c}} - D_{\varphi}}$. Inserting Eq. (\ref{eq:steady_f}) into
Eq. (\ref{eq:orderparam_f}), we obtain the scaling form of the order parameter $S^{F}$:
\begin{eqnarray}\label{eq:scaling_f}
S^{F} \simeq \tilde{B}_{1} \sqrt{\eta_c - \eta}
\end{eqnarray}
where $\tilde{B}_{1} $ is a constant. To obtain the scaling of the order parameter $S^{LC}$, Eq. 
(\ref{eq:steady_lc}) is inserted into  Eq. (\ref{eq:orderparam_lc}):
\begin{eqnarray}\label{eq:scaling_lc}
S^{LC} \simeq \frac{1}{4} + \tilde{B}_{2} \sqrt{\eta_c - \eta}
\end{eqnarray}
where again $\tilde{B}_{2}$ is a constant. The above results confirm our findings for active Brownian particles from Sec. \ref{sec:velal} in the limit of vanishing velocity fluctuations. 
 
It is instructive to compare the results of the mean-field theory qualitatively to the results of 
individual-based simulations with respect to the scaling properties of the order parameters near the 
onset of orientational order. 
In addition, the prediction of the mean-field theory regarding the different critical noise amplitudes 
(resp. critical densities) for F- and LC-alignment can be compared with simulations of the 
Vicsek- and Peruani-model. 

\begin{figure}

\centering
\resizebox{0.7\columnwidth}{!}{
 \includegraphics{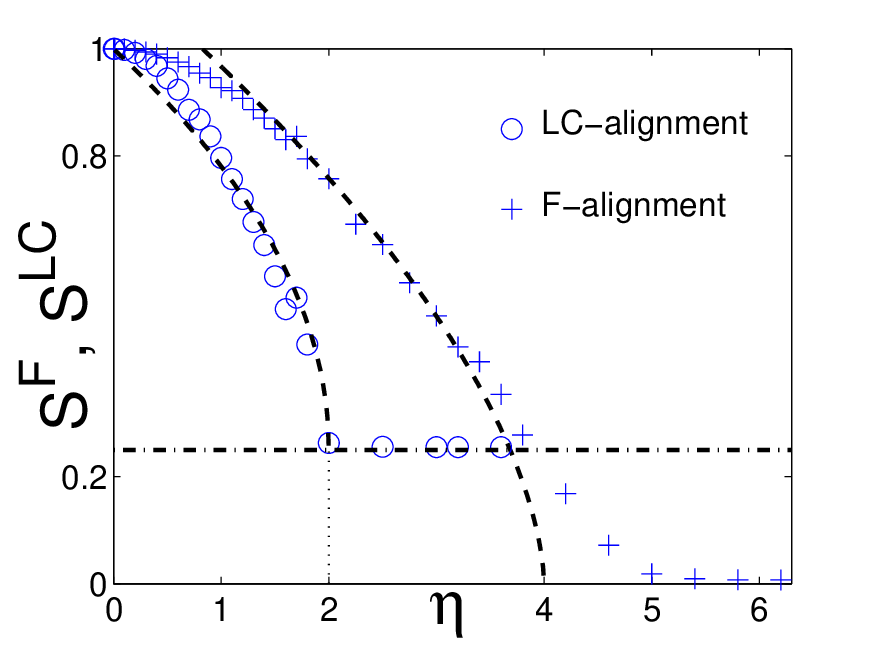} }

\caption{Comparing simulations of particles with F-alignment  
(crosses) and LC-alignment (circles) in the limiting case of very fast angular relaxation. In both cases
$N=2^{14}$ and $\rho=2.0$. Notice that the order parameter for F-alignment is $S^{F}$ while for LC-
alignment is $S^{LC}$ (See text) .
%
The dashed horizontal line indicates the minimum value that $S^{LC}$ could take. 
The dashed curves correspond to the best fit assuming an exponent $0.5$, i.e., $\eta_c$ was the fitting 
parameter \cite{peruani_mean-field_2008}.}
\label{fig_factor_2}
\end{figure}

Fig. \ref{fig_factor_2} shows a comparison between the scaling predicted by the mean-field 
approach for $S^{LC}$ (dashed curve) and the one obtained from individual-based simulations in the limit of very fast angular relaxation (symbols). 
One finds good agreement between the mean-field prediction and the simulations for the scaling of $S$ 
near $\eta_c$ that suggests that individual-based simulations with LC-alignment at high densities 
exhibit a mean-field type transition.
Evidence seems to point towards a mean-field transition if we look at the scaling of the maximum 
amplitude of the angle distribution as function of the angular noise intensity
$\eta$ (see Fig. \ref{fig_factor_2}). 

Finally, Fig. \ref{fig_factor_2} shows that in individual-based simulations with the same parameters 
and different (namely LC- and F-) alignment mechanism, the limit of  fast angular relaxation 
yields $\eta^{LC}_{c}<\eta^{F}_{c}$ as predicted by the mean-field theory. 
Note, however, that the simulations yield  $2 \eta^{LC}_{c} \approx  \eta^{F}_{c}$, while the mean-
field description predicts $\sqrt{2}\eta^{LC}_{c} = \eta^{F}_{c}$. 

In summary, we have derived a mean-field theory for self-propelled particles which accounts for F- and 
LC-alignment. 
This approach predicts a continuous phase transition with the order parameter scaling with an exponent 
one half in both cases.
In addition, the critical noise amplitude below which orientational order emerges is 
smaller for LC-alignment  than for F-alignment, i.e., $\eta^{LC}_{C}<\eta^{F}_{C}$.
These findings were confirmed by individual-based simulations with F- and LC-alignment. 

Furthermore, we have assumed spatial homogeneous density to study the emergence of orientational 
order. 
Thus, the presented approach does not apply to situations where self-propelled particles show 
clustering at the onset of orientational order as will be discussed in the next section (Sec. \ref{sec:pattern}). 
A better understanding of the problem should imply the study of the interplay between 
local orientational order and density fluctuations.

\subsection{Alternative swarming mechanisms}
\subsubsection{Escape and pursuit --- collective motion and group propulsion}
\label{sec:escpur}
Motivated by recent observations on cannibalistic behavior in locusts and crickets \cite{simpson_cannibal_2006,bazazi_collective_2008}, we introduced a model of individuals (active particles) responding to others with escape and pursuit behavior \cite{romanczuk_collective_2009,bazazi_nutritional_2010}. This biologically motivated interactions cannot be expressed simply in terms of an interaction potential as discussed in Section \ref{sec:swarms} and do not include a velocity alignment term as discussed in Sections \ref{sec:swarms} \& \ref{sec:swarming_va}, but constitute selective, velocity dependent attraction and repulsion interactions. Nevertheless they may lead to the onset of large scale collective motion in particular at high densities of individuals.    

The behavioral response of the insects, which are particularly vulnerable to attacks from behind, was assumed in the following way: 
\begin{itemize}
\item If approached from behind by another individual $j$ a focal individual $i$ increases its velocity away from it in order to prevent being attacked from behind. We refer to this behavior as \emph{escape} (e). 
\item If the focal individual "sees" another individual up-front moving away, it increases its velocity in the direction of the escaping individual. We refer to this behavior as \emph{pursuit} (p). 
\item No response in all other cases.
\end{itemize}

The motion of individual particles in two spatial dimension obeys the following Langevin dynamics:
\begin{equation}
\dot{\bbox{r}}_i  =  {\bbox v}_i,  \quad \dot{\bbox{v}}_i  = -\gamma {\text{v}}_i^{a-1}{{\bbox v}}_i+{\bbox F}^s_{i}
	+\sqrt{2 D}\boldsymbol \xi_i \,.\label{eq:dynamicv}
\end{equation}
The first term on the left hand side of the velocity equation (\ref{eq:dynamicv}) is a friction term with coefficient $\gamma$ and an arbitrary power-law dependence on velocity represented by $a=1,2,3,\dots\,$. The social interaction of particles is described by ${\bbox F}^s_i$. The last term is a non-correlated Gaussian random force with intensity $D$. A non-interacting particle (${\bbox F}^s_i=0$) explores its environment by a continuous random walk, where the individual velocity statistics are determined by $\gamma$, $a$ and $D$.     

Furthermore we assume finite-size particles and introduce fully elastic hard-core collisions with a particle radius $R_{hc}=l_r/2$ ($l_r$ - particle diameter) (see e.g. \cite{brilliantov_kinetic_2004}. 

The social force acting on the focal particle $i$ is given as a sum of escape and pursuit force: ${\bbox F}^s_{i}={\bbox f}^e_i+{\bbox f}^p_i$ with
\begin{subequations}\label{eq:ep_forces1}
\begin{align}
{\bbox f}^e_i = \frac{1}{N_e} & \sum_j -\hat {\bbox r}_{ji} K_e(|v_\text{rel}|)\theta(l_s-r_{ji})\theta(r_{ji}-l_{r})\theta(-{\bbox v}_{i}\cdot{\bbox r}_{ji}) \theta(-v_\text{rel}) \\
{\bbox f}^p_i = \frac{1}{N_p} & \sum_j +\hat {\bbox r}_{ji}  K_p(|v_\text{rel}|) \theta(l_s-r_{ji})\theta(r_{ji}-l_{r})\theta(+{\bbox v}_{i}\cdot{\bbox r}_{ji}) \theta(+v_\text{rel})
\end{align}
\end{subequations}
where $\hat{\bbox r}_{ji}=({\bbox r}_{j}-{\bbox r}_{i})/|{\bbox r}_{j}-{\bbox r}_{i}|$ 
{
is the unit vector in the direction of the other individual $j$, and 
$v_\text{rel}={\bbox v}_{ji}\cdot \hat {\bbox r}_{ji} =({\bbox v}_j-{\bbox v}_i)\hat {\bbox r}_{ji}$ 
is the relative velocity of individuals $j$ and $i$.
The functions $K_{e,p}\geq 0$ determine the strength of the interactions. In the following we assume the response functions proportional to the relative speed:
$K_{e,p} = \chi_{e,p}|v_\text{rel}|$,
where $\chi_{p,e}\geq0$ are the corresponding interaction strengths. This choice of the response function leads to stronger response to fast approaching/escaping individuals in comparison to slowly moving ones. 

The product of the $\theta$-functions (step functions), which define the condition for the escape/pursuit interaction to take place can assume only two values: either $1$ or $0$. The product of the first two step functions is identical for both interaction types. It is $1$ only if the individual $j$ is within the social interaction zone defined by the hard-core distance $l_{r}$ and the sensory range $l_s$: $l_{r}< r_{ji} < l_s $. The product of the last two step functions distinguishes the escape and pursuit interaction. The escape interaction takes place only if individual $j$ is behind the focal individual $i$ (${\bbox v}_{i}\cdot{\bbox r}_{ji}<0$) and is coming closer ($v_\text{rel}<0$), whereas the pursuit interaction takes place only if the individual $j$ is in front of the focal individual $i$ (${\bbox v}_{i}\cdot{\bbox r}_{ji}>0$) and is escaping it ($v_\text{rel}>0$).   

The most important property of this asymmetric interactions is their anti-dissipative nature with respect to  kinetic energy. Note that ${\bbox F}^s_i$ leads only to acceleration of individuals and is analogous to the auto-catalytic mechanism proposed in  \cite{bazazi_collective_2008}.

Large scale numerical simulation of a $N$-particle system in a rectangular domain of size $L\times L$ with periodic boundary conditions show that at high density the escape and pursuit interaction leads to collective motion irrespective of the detailed parameter choice as long as the interaction is strong enough to counterbalance the individual fluctuations. Here we use for convenience a dimensionless density $\rho_s=N l_s^2/L^2$ scaled by the interaction range. 
\begin{figure}
\psfrag{r = 2.25}{$\rho_s =2.25$}
\psfrag{r = 1.25}{$\rho_s =1.25$}
\psfrag{r = 0.3}{$\rho_s =0.30$}
\psfrag{p}{p}
\psfrag{e}{e}
\psfrag{p+e}{p + e}
\centering\includegraphics[width=0.6\linewidth]{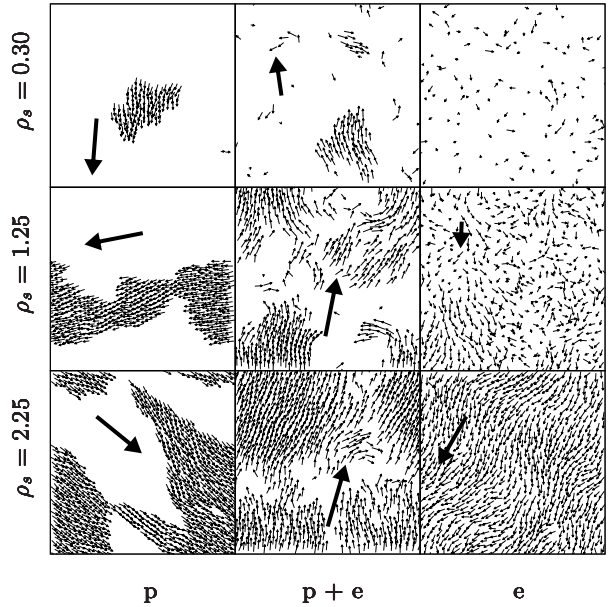}
\caption{Typical spatial configurations and particle velocities (small arrows) for pure pursuit (p), pursuit+escape (p+e) and pure escape (e) cases at different particle densities $\rho_s=0.30$, $1.25$, $2.25$. Mean migration direction and speed $U$ is indicated by large arrows ($U\approx0$ for escape only and  $\rho_s\ll1$) \cite{romanczuk_collective_2009}.\label{fig:ep_snapshot}}
\end{figure}

\begin{figure}
\begin{center}
\centering \includegraphics[width=0.4\linewidth]{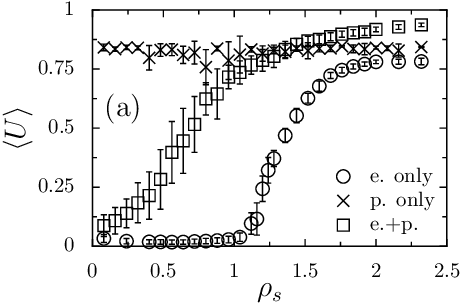}
\includegraphics[width=0.52\linewidth]{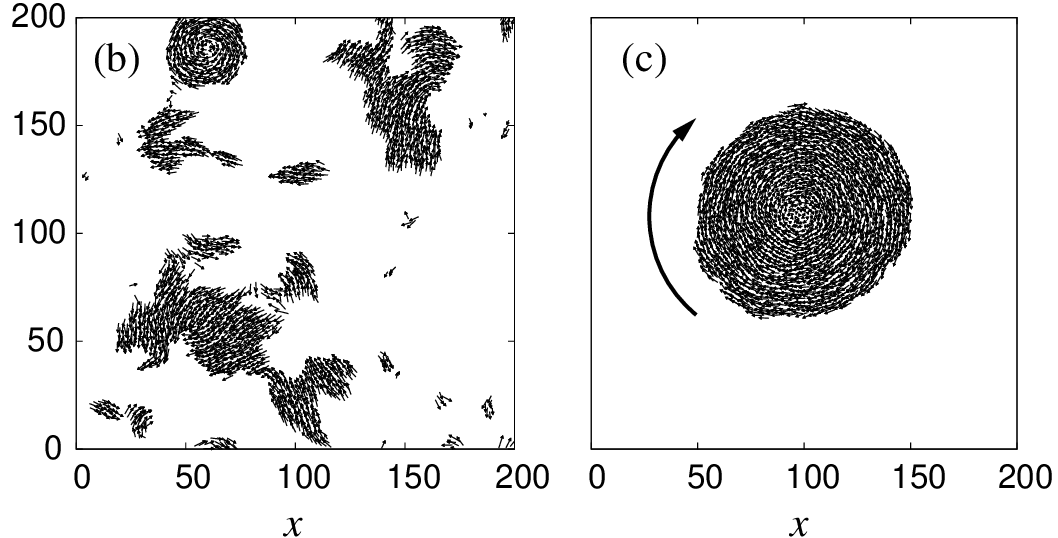}
\end{center}
\caption{(a) Mean velocity $\langle U \rangle$ for escape-only (circles) $\chi_e$=$10$, $\chi_p$=$0$, pursuit-only (crosses) $\chi_e$=$0$, $\chi_p$=$10$ and symmetric escape+pursuit  (squares) $\chi_e$=$\chi_p$=$10$ vs. $\rho_s$ obtained from numerical simulations with periodic boundary conditions ($\gamma$=$1$, $D$=$0.05$, $a$=$3$, $l_{r}$=$2$, $l_{s}$=$4$; only translational solutions were considered; error bars represent one standard. deviation). Vortex formation for pursuit only: Initially a rotating cluster nucleates from collision of two translational clusters at $x\approx50$, $y\approx180$ (b), which then grows through absorption of other translational clusters until a single rotating structure emerges (c). Figure adopted from \cite{romanczuk_collective_2009}.
\label{fig:ep_meanu}}
\end{figure}

At low $\rho_s$ the behavior of the system depends strongly on the relative strength of escape and pursuit. This can be best understood if we consider the extreme cases of pure escape ($\chi_p=0$) and pure pursuit ($\chi_e=0$).

In the escape only case the particles try to keep their distance with respect to individuals approaching from behind. To the front only interactions via the short range repulsion take place. At low $\rho_s$ an escaping particle will eventually move away from the approaching one. The probability of interaction within the characteristic time of velocity relaxation vanishes and the particles perform effectively a disordered Brownian motion. As $\rho_s$ increases the frequency of escape interactions increases and an escaping particle will trigger escape responses from neighboring particles. We observe velocity correlation over several interaction length scales.  Finally at high $\rho_s$ a highly ordered state emerges where all particles are able to correlate their motion. At all $\rho_s$ the density remains almost spatially homogeneous. 

In the pursuit-only case the dynamics change dramatically: At low $\rho_s$ we observe a highly inhomogeneous state initiated by formation of small particle clusters performing coherent collective motion. 
Without escape interaction the density of the clusters is only limited by the hard-core radius. 
At moderate noise intensities the clusters are sufficiently stable and a process of cluster fusion can be observed where larger clusters absorb smaller clusters and solitary particles. 
The typical stationary configuration in a finite system with periodic boundary condition, and moderate noise, is a single cluster performing translational motion (Fig.~\ref{fig:ep_snapshot}). The migration speed $\langle U \rangle$ in Fig.~\ref{fig:ep_meanu}(a) is given by the mean speed of a single cluster $\langle u \rangle=|\sum_{i\in \txr{cluster}} {\bbox v}_i|/N_{\txr{cluster}} $. 
For large clusters $\langle u \rangle$ becomes independent of the cluster size and therefore independent of $\rho_s$. It should be emphasized that this holds only for finite systems with sufficiently strong pursuit interactions. In general there is a finite probability of a large pursuit-only clusters to break up into smaller clusters which increases with cluster size and noise strength. The resulting different clusters will in general move in different direction which results in $\langle U \rangle\to 0$ for $L,N\to \infty$ and constant overall density $\rho_s=const.\ll1$. 

An intriguing feature of the pursuit-only interaction is the possibility of the formation of large scale vortices out of random initial conditions (Fig.~\ref{fig:ep_meanu}b,c), for example via a collision of clusters moving in opposite direction. The vortices may decay either via a collision with a translational cluster or through random fluctuations. For certain parameters the vortices are very stable and may dominate the stationary configurations for the pursuit-only case. 
The emergence of vortex-structures is in particular remarkable because so far they have only been reported for systems of self-propelled particles with confinement, or attracting potential, respectively \cite{czirok_formation_1996,dorsogna_self-propelled_2006,vollmer_vortex_2006}. 
Here, the pursuit behavior has two functions: a propulsion mechanism and an asymmetric attraction.

Both interactions --- escape and pursuit --- lead to collective motion of groups but have an opposite impact on the density distribution. Whereas the escape interaction leads to a homogenization of density within the system, the pursuit interaction facilitates the formation of density inhomogeneities. Thus the combined escape+pursuit case with $\chi_{p},\chi_{e}>0$, is a competition of this two opposite effects. The stability of clusters is determined by the relative ratio of the interaction strengths. In general for the escape+pursuit case at low $\rho_s$ we observe fast formation of actively moving particle clusters with complex behavior: fusion and break up of clusters due to collisions as well as spontaneous break up of clusters due to fluctuations \cite{romanczuk_collective_2009}.

The anti-dissipative nature of the escape and pursuit interactions 
leads to persistent translational motion of interacting Brownian particles. Individual clusters may be considered as self-propelled structures where the propulsion is purely due to social interactions, and constitutes an example of group-propulsion.

The scaling of the speed of individual clusters can be derived by considering the smallest possible cluster: a particle pair. We assume particle $1$ is in front of particle $2$ and  $|{\bbox r}_{12}|<l_s$  at all times.
 The transformation of Eq.~\ref{eq:dynamicv} into polar coordinates with ${\bbox v}_{i}=(v_{i}\cos\delta \varphi_{i},v_i\sin\delta \varphi_i)$, where $\delta \varphi_i$ is defined as the angle between ${\bbox v}_i$ and $\hat {\bbox  r_{12}}$, yields: 
\begin{subequations}\label{eq:pair_polar}
\begin{align} 
\frac{d}{dt} v_1 & = -\gamma v_1^a+\chi_e |v_{12}|\theta(-v_{12}) \cos{\delta \varphi_1}+\sqrt{2 D}\xi_{v,1} \label{eq:v1} \\
\frac{d}{dt} v_2 & = -\gamma v_2^a+\chi_p |v_{12}|\theta(+v_{12}) \cos{\delta \varphi_2}+\sqrt{2 D}\xi_{v,2} \label{eq:v2} \\
\frac{d}{dt} \delta \varphi_1 & = \frac{1}{v_1}\left(-\chi_e |v_{12}|\theta(-v_{12}) \sin{\delta \varphi_1}+\sqrt{2 D}\xi_{\varphi,1}\right) \label{eq:angle1} \\   
\frac{d}{dt} \delta \varphi_2 & = \frac{1}{v_2}\left(-\chi_p |v_{12}|\theta(+v_{12}) \sin{\delta \varphi_2}+\sqrt{2 D}\xi_{\varphi,2}\right). \label{eq:angle2}   
\end{align}
\end{subequations} 
Here $v_{12}=v_1-v_2$ is the relative velocity of the two particles and $\xi_{\varphi,i},\xi_{v,i}$ represents the transformed noise variables. For $-\pi/2<\varphi_1,\varphi_2<\pi/2$ the escape and pursuit interaction leads to an increase of either $v_1$ or $v_2$ in order to harmonize the speed of the slower particle with the faster one. In addition the interaction stabilizes the translational motion along $\hat {\bbox r}_{12}$, i.e. $\langle \delta \varphi_{i}\rangle \to 0$ (Eqs.~\ref{eq:pair_polar}c,d).
After the system relaxes to a stationary state ($\hat {\bbox r}_{12}$ varies slowly in time) we end up with quasi one-dimensional translational motion of the particle pair with slowly diffusing direction of motion defined by $\hat {\bbox r}_{12}$. Please note that for elastic hard-core interaction the total energy and momentum of the particle pair does not change during collisions.

In order to obtain equations of motion for the particle pair in the stationary translational state, we assume that the particle velocities are given by the time averaged mean velocity $u$ of the particle pair plus a small deviation: $v_i=u+\delta v_i$. Furthermore the particles are assumed to have approximately the same heading so that $\delta\varphi_i\ll 1$.
Assuming vanishing mean of speed deviations $\langle \delta v_i \rangle=0$, we obtain for the time evolution of the mean pair velocity:
\begin{align}\label{eq:eppair_avgvel}
\frac{d}{dt} u   = & -\gamma u^a + \frac{\chi_e}{2} \langle |v_{12}| \theta(-v_{12}) \rangle+ \frac{\chi_p}{2}\langle |v_{12}| \theta(+v_{12})  \rangle
\end{align}
In the symmetric case where $\chi_e=\chi_p=\chi$ the social force terms on the right hand side can be summed up and we obtain:
\begin{align}
\frac{d}{dt} u   = & -\gamma u^a + \frac{\chi}{2}\langle |v_{12}| \rangle
\end{align}
There is permanently a social force acting on one of the particles. In order to evaluate the expectation value of the social force we assume that at all times one of the particles moves with the mean velocity $u$ ($\delta v_{1,2} =0$) whereas the velocity of the second particle deviates by $\delta v$ as a result of the stochastic force. We approximate the expectation value $\langle |\delta v| \rangle = \langle | v_{12}| \rangle$, by considering the speed deviations as discrete increments taken from a Gaussian distribution with zero mean and variance $\sigma^2_1=2 D  \tau$ (Wiener process) with $\tau=\chi^{-1}$ being the relaxation time of the escape+pursuit interaction: $\langle |v_{12}| \rangle_{1d} =2\sqrt{\frac{D}{\pi\chi}}$.

In the limit of quasi one-dimensional motion we obtain 
\begin{align}
\frac{d}{dt} u   = & -\gamma u^a + \sqrt{\frac{\chi D}{\pi}},
\end{align}
which results in a stationary pair velocity $u^s_{ep}$
\begin{equation}
 u^{s}_{ep} = \left(\frac{1}{\gamma}\sqrt{\frac{\chi D }{\pi}}\right)^{\frac{1}{a}}. \label{eq:us_pair_ep}
\end{equation}

For symmetric escape and pursuit the particle distance increases slowly due to fluctuations which are not fully compensated by the social force, and the pairs break-up at finite times. Stable pairs, with stable particle distance are only possible for pursuit dominated dynamics and in particular for the pursuit only case $\chi_e=0$. 

In contrast to the escape and pursuit interaction for pursuit only the social force act not all the time but only if the leading (first) particle is faster then the pursuer.
Thus individual velocity deviation from the mean, as well as $\langle |\delta v| \rangle_{1d}$ are larger for pursuit only, in comparison to the escape+pursuit case. We consider now both velocity deviations $\delta v_1$ and $\delta v_2$ as stochastic variables. In analogy to symmetric escape and pursuit we approximate $\langle |v_{12}| \rangle$ by considering discrete Gaussian increments with zero mean and variance $\sigma_2^2=2 \sigma^2_1=4 D  \tau$  (difference of two independent stochastic processes variance $\sigma_1^2$).
The resulting evolution equation for the velocity in the pursuit only case reads:
\begin{align}
\frac{d}{dt} u   = & -\gamma u^a + \sqrt{\frac{\chi_p D}{2\pi}},
\end{align} 
with the stationary speed
\begin{equation}\label{eq:us_pair_op}
 u^{s}_{p} = \left(\frac{1}{\gamma}\sqrt{\frac{\chi_p D }{2\pi}}\right)^{\frac{1}{a}}. 
\end{equation}

The obtained analytical expressions for the averaged pair velocity $\langle u \rangle$ are confirmed by numerical simulations for a wide parameter ranges as shown in Fig. \ref{fig:ep_cluster} (see also \cite{romanczuk_collective_2009}). Deviations become apparent where the pair dynamics deviate strongly from the effective one-dimensional situation as for example for weak coupling strengths $\chi$.

\begin{figure}
\centering \includegraphics[width=1.0\linewidth]{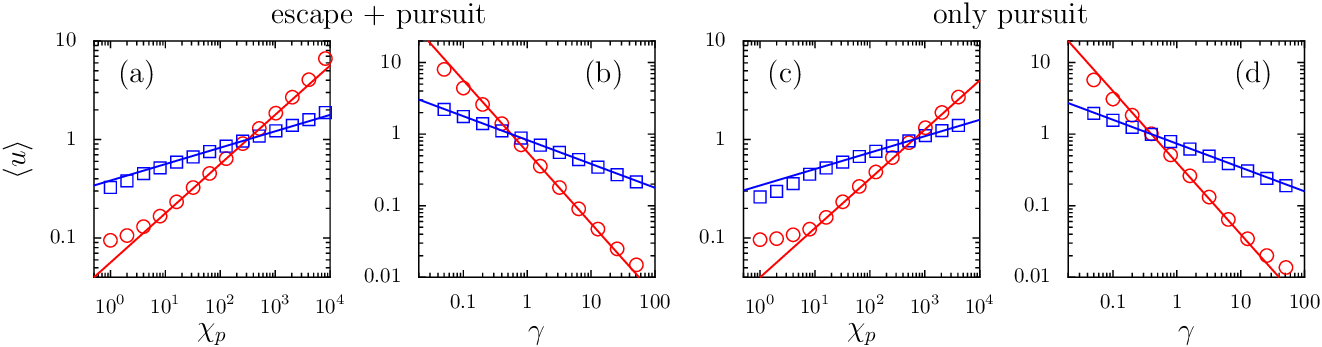}
\caption{(a) Mean pair velocity $\langle u \rangle$ for symmetric escape and pursuit (a,b) and pursuit only (c,d) versus interaction strength (a,c) and friction coefficient (b,d). The symbols represent the results obtained from numerical simulations of interacting particle pairs for linear friction $a=1$ (circles) and nonlinear friction $a=3$ (squares). The solid lines represent the corresponding analytical result Eq. \eqref{eq:us_pair_ep} (a,b) and  Eq. \eqref{eq:us_pair_op} (b,d). 
\label{fig:ep_cluster}}
\end{figure}

Thus we have shown that although isolated particles perform (nonlinear) Brownian motion, the escape and pursuit interactions leads to rectifications of individual fluctuations of interacting particles leading to a collective propulsion with a non-vanishing stationary velocity. Based on Eqs. \eqref{eq:us_pair_ep} and \eqref{eq:us_pair_op} we can consider a pair (or even a cluster) as a single active Brownian particle with a velocity-dependent friction function of the form:
\begin{align}
-\gamma(u)u= \alpha - \gamma u^a 
\end{align}   
with a constant pumping term $\alpha$ accelerating the pair and a friction term $-\gamma u^a$. For $a=1$ this corresponds directly to the Schienbein-Gruler friction introduced in Section \ref{sec:selfdriven_ABP}.

\subsubsection{Chemotactic coupling}
\label{sec:chemotaxis}
Another kind of coupling leading to collective dynamics is known from microbiology. Many microorganisms, as for example different bacteria strains, are able to sense various chemical agents in their environment and bias their motion along their concentration gradients. This ability is referred to as chemotaxis (see \cite{eisenbach_chemotaxis_2004} for a review). Our modern understanding of bacterial chemotaxis was strongly influenced by the systematic investigations  performed by J. Adler, H.C. Berg and their collaborators in the 60's and 70's of the past century \cite{adler_chemotaxis_1965,adler_chemotaxis_1966,adler_chemotaxis_1975,berg_chemotaxis_1972,berg_bacterial_1975,berg_physics_1977,brown_temporal_1974}. Since then a large number of publications has appeared on various experimental and theoretical aspects of chemotaxis (see e.g. \cite{baker_signal_2006,ben-jacob_cooperative_2000,bray_bacterial_2002,de_gennes_chemotaxis:_2004,wadhams_making_2004}). 

Chemotaxis plays an important role for the survival of microorganisms in general, as it enables them to move towards beneficial environments and away from hostile environments. In the case, where individual cells bias their movement towards higher (lower) concentration  of potentially beneficial (dangerous) chemical substances, the corresponding chemical agent is referred to as chemoattractant (chemorepellent).
Effective coupling of individual cells comes from the ability of different microorganisms to produce the respective chemoattractant (chemorepellent) by themselves. By these means bacteria are able to exchange information about favorable or disadvantageous environmental conditions.

The auto-chemotactic interaction was identified as the mechanism responsible for complex spatio-temporal patterns of cell clusters that were observed in colonies of chemotactic bacteria such as Escherichia coli or Salmonella typhimurium \cite{budrene_complex_1991,budrene_dynamics_1995,woodward_spatio-temporal_1995}.

Most models employed for the description of bacterial colonies with chemical cell-to-cell signaling (chemotaxis) are based on the classical Keller-Segel model (KSM) \cite{keller_model_1971,keller_travelling_1971}. It is a continuous model of partial differential equations (PDEs) for the dynamics of the bacterial density $\rho({\bbox r},t)$ and the concentration(s) of the involved chemical agent(s) $c({\bbox r},t)$ (see e.g. \cite{tyson_minimal_1999,matsushita_formations_1999,murray_spatial_2003,hillen_users_2008}). 

Here we discuss briefly models of (active) Brownian particles interacting via a self-generated chemoattractant, as an alternative to a pure PDEs approach \cite{schweitzer_clustering_1994,czirok_formation_1996,romanczuk_beyond_2008}. 

We consider an ensemble of $N$ (active) Brownian particles in two spatial dimensions, each described by individual equations of motion coupled to a self-generated chemoattractant concentration field $c(\bbox x,t)$:
\begin{subequations}\label{langevin}
\begin{align}
\dot {\bbox r}_i & = {\bbox v}_i  \\
\dot {\bbox v}_i & = -\gamma({\bbox v}_i){\bbox v}_i +\kappa(c)\nabla_{{\rm r}} \, c({\bbox r}_i) + \sqrt{2 D} {\boldsymbol \xi}_i \ . \label{v_langevin}
\end{align}
\end{subequations}
The chemotactic force (second term in velocity equation) consists of a chemotactic sensitivity function $\kappa(c)$ which may depend on the concentration $c$ and the gradient $\nabla c$ at the position of the individual particle (cell). Thus depending on the sign of $\kappa(c)$ the force acts either in the direction of the gradient (chemoattractant, $\kappa(c)>0$) or in the opposite direction (chemorepellent, $\kappa(c)<0$). Here we will restrict for simplicity to the discussion of chemotactic coupling via a single chemoattractant. 

The dynamics of the chemoattractant concentration $c$ are assumed to obey a diffusion equation:
\begin{align}
\dot{c}({\bbox r},t) & =  q_0 \sum_{i=1}^{N} \delta({\bbox r}-{\bbox r}_i) 
			- d_c 
			+ D_c\Delta c \label{pde_full}.
\end{align}
The first term describes the production of the chemoattractant by the individual particles with rate $q_0$ at their respective  positions ${\bbox r}_i$. Furthermore the chemical agent $c$ is assumed to decay with the rate $d_c$ and to diffuse with the diffusion coefficient $D_c$.

A simple model of chemotactically interacting Brownian particles with constant friction coefficient $\gamma(v)=\gamma_0=const.$ (Stokes friction) and constant chemotactic response $\kappa(c)=\kappa_0$, was studied by Schweitzer and Schimansky-Geier \cite{schweitzer_clustering_1994} (SSG-Model).  

Depending on the model parameters  they observed formation of spike patterns of the chemoattractant field $c$. The spikes correspond to particle clusters aggregating at high concentration of $c$. The positive feedback between the spike ``height'' and  attraction  on other particles leads to a competition between spikes following an Eigen-Fisher like dynamics. After a certain relaxation time only few spikes ``survive'', but even if the dynamics of the system slows down, in the limiting case of $t\rightarrow \infty$ the only stationary solution  is a single spike of the chemical field or cluster of particles, respectively. This process can be seen as an Ostwald-ripening process known from chemical reactions \cite{schweitzer_stochastic_1988,yao_theory_1993}.

In the case of Stokes friction the particles within a stationary chemotactic cluster perform purely diffusive motion in an effective confining potential caused by the concentration profile of $c$. The introduction of active (self-propelled) motion via a velocity dependent friction function leads in general to complex rotational motion of individual particles within the cluster. In the absence of any additional symmetry breaking interaction no collective rotational modes are possible and the total angular momentum averages to zero.     

Although bacteria are able to sense chemoattractant concentrations over several orders of magnitude via membrane receptors, it is known that at high $c$ they lose the ability to follow the gradient. This can be taken into account in  simple form by the so-called ``receptor law'' \cite{ben-jacob_cooperative_2000,murray_spatial_2003}. The resulting chemotactic force may be written as:
\begin{align}
\kappa(c)=\frac{\kappa_0}{(1+S  c)^2}
\end{align}
with $\kappa_0$ being the sensitivity coefficient and $S$ being the saturation coefficient which takes into account the saturation of chemotactic receptors at high $c$ leading to a decreasing sensitivity with increasing $c$. 

The macroscopic behavior of the model for low $S$ is similar to that observed in the SSG-model but with increasing $S$ (or increasing concentrations $c$) the observed macroscopic patterns and microscopic dynamics within the cluster change.

The decrease of the chemotactic force at high concentrations $c$ for $S>0$, makes particles insensitive towards the gradient of $c$ --- the particles are able to leave the maxima of the field distribution. This behavior has significant impact on the macroscopic pattern formation. At large $S$  any formation of clusters may be inhibited and at moderate values of $S$ a ``smoothing'' of spikes in the concentration profile towards flat spots can be observed \cite{romanczuk_beyond_2008}.  

For the study of pattern formation we neglect the microscopic feature of active motion and consider the particles to behave as normal Brownian particles with a constant friction $\gamma({\bbox v}_i)=\gamma_0= const.$. This approach can be justified by the small stationary velocity of bacteria. With this assumption and in the overdamped limit (Smoluchowski limit) we can derive a simple PDE for the evolution of the particle density $\rho$. The two PDEs for $\rho$ and $c$ give us a reactions-diffusion system with chemotaxis which represent a variant of the Keller-Segel model. A similar system has been studied by Tyson \emph{et al.} \cite{tyson_minimal_1999,tyson_model_1999}: 
\begin{subequations}\label{rdsystem}
\begin{align}
\dot{\rho}({\bbox r},t) & =  \nabla \left( - \frac{\kappa_0 \rho}{\gamma_0(1+S c)^2}\nabla c + D\nabla \rho \right) \label{rd_density}, \\
\dot{c}		& =  q_0 \rho - d_c c + D_c \Delta c.
\end{align}
\end{subequations}
Here, $D=D/\gamma^2_0$ is the spatial diffusion coefficient of the overdamped particles.  

The simplest stationary solution of (\ref{rdsystem}) is the homogeneous solution, given by the averaged particle and chemoattractant concentrations: $\bar{\rho}=N/L^2$ and $\bar c=q_0\bar{\rho}/d_c$.  Analyzing the linear stability of the homogenous solution with respect long wave spatial perturbations we arrive at the condition which has to be fulfilled for stable spatially homogenous solution:
\begin{equation}\label{kappac}
\kappa_0 \leq \frac{D\gamma_0 d_c}{\bar \rho q_0}\left(1+S \frac{q_0}{d_c} \bar \rho \right)^2=\kappa_{c}.
\end{equation}  
For $d_c>0$ and $\kappa_0$  smaller than the critical sensitivity $\kappa_{c}$ all fluctuations around  the homogeneous state decay exponentially. Only if $\kappa_0 \geq \kappa_{c}$ pattern formation on the macroscopic scale can be observed. 

Although this result was derived for overdamped dynamics it can be also applied for actively moving particles with a Rayleigh-Helmholtz friction function, by taking the friction coefficient from the Stokes friction as a single fit parameter \cite{romanczuk_beyond_2008}.  

Not only the macroscopic dynamics but also the microscopic behavior of actively moving particles within a cluster depend strongly on the involved chemoattractant and chemotactic parameters. In the case of strong confinement (e.g. low $S$), the particles perform complex rotational motion, as discussed in Section \ref{sec:rot_ext}. But for large $S$ and low diffusion of the chemoattractant we observe extended clusters where the chemoattractant concentration within a cluster is rather high but approximately constant but drops sharply at the cluster boundary leading to a steep gradient. Inside such clusters the particles perform effectively free motion and are able to sense chemoattractant gradient only at the cluster boundaries which prevents them from leaving the cluster \cite{romanczuk_beyond_2008}.
\begin{figure}
\centering \includegraphics[width=0.4\linewidth]{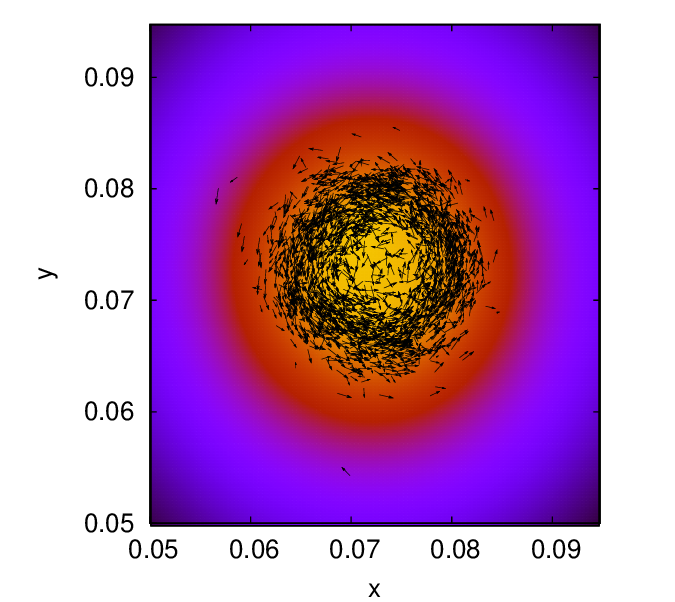}
\centering \includegraphics[width=0.4\linewidth]{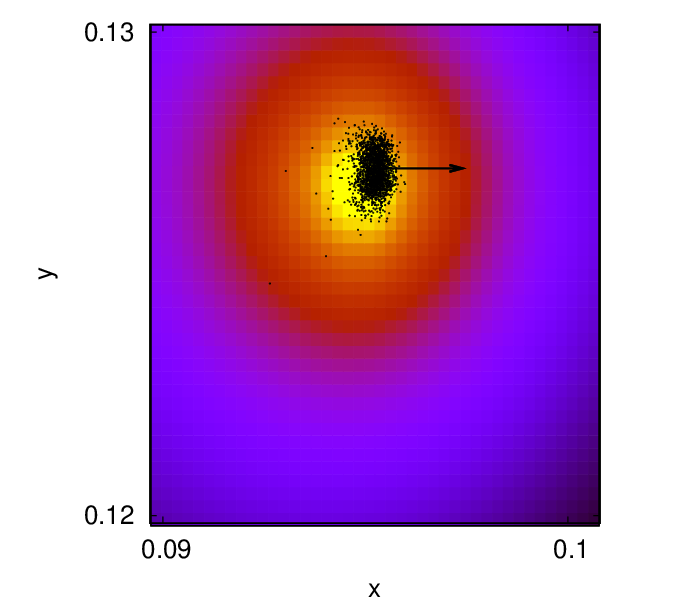}
\caption{Active particles with chemotactic coupling and velocity alignment. The chemoattractant concentration $c$ is indicated by the background color with dark (bright) regions indicating low (high) $c$. Left: Collective rotation at intermediate velocity alignment strengths. Each particle is shown by its velocity vector. Right: A moving cluster for strong velocity alignment. Only the position (symbols) and the mean velocity is shown (vector). 
\label{fig:chemo_collective}}
\end{figure}

%
\begin{figure} 
\begin{center}
	\psfrag{vco}[c]{\; \, $ \langle C_{\txr{vv}}  \rangle$ \normalsize}
        \psfrag{lco}[c]{ $\langle L \rangle$ \normalsize}
	\psfrag{chi}[c]{\large $\chi$ \normalsize}
	\includegraphics[width=0.7\linewidth]{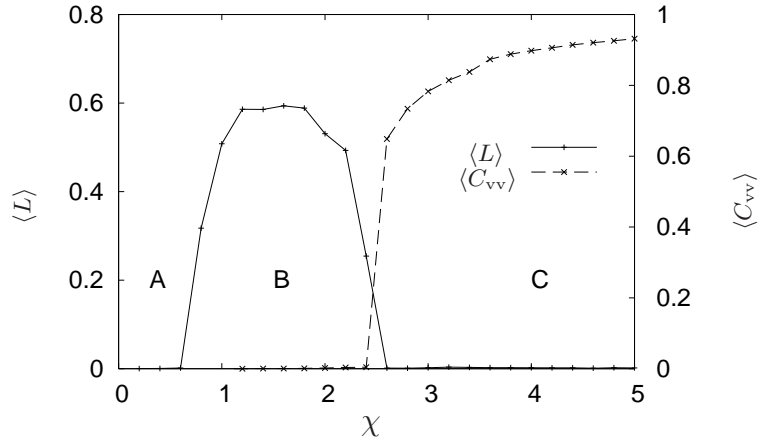}
\caption{The rotational motion $\langle L \rangle$  and parallel motion $\langle C_{\txr{vv}} \rangle$  order parameters for an individual cluster of particles over the velocity-alignment strength $\chi$. Depending on $\chi$ three different states can be identified: ({\bf A}) no collective mode of motion; ({\bf B}) collective rotation; ({\bf C}) collective translation. (simulation parameters: $\kappa_0 = 1\times10^{-5}$~$\txr{mm}^4 \txr{s}^{-2}$  , $S=6\times10^{-5}$~$\txr{mm}^{2}$, $d_c=0.1$~$\txr{s}^{-1}$,$D=1\times10^{-4}\txr{mm}^2\txr{s}$, $N=3000$ and $l_{va}=2\times10^{-3}$~$\txr{mm}$) \label{fig_transition} \cite{romanczuk_beyond_2008}} 
\end{center}
\end{figure}

If the coupling via the self-generated chemoattractant is the only interaction between particles, no collective dynamics within a cluster take place and the center of mass of a stationary cluster performs at most slow diffusive motion.  This changes dramatically if in addition to the chemotactic coupling we include the velocity alignment as introduced in Sec. \ref{sec:velal}. Increasing the alignment strengths leads first to a transition from disordered motion within a cluster to collective rotation within a stationary cluster. This situation corresponds to the dynamics in an external central field \ref{sec:velal}. A further increase in alignment strength enables the particles to escape collectively out of the stationary maximum of $c$. As the particles keep on producing the chemoattractant, they drag a cloud of chemoattractant around them and we observe a compact chemically bounded moving cluster of particles (see Fig. \ref{fig:chemo_collective}).     


%% file: SECTION_pattern.tex
\section{ Pattern formation of active particles}
\label{sec:patt}
\label{sec:pattern}

In this section, various aspects of pattern formation in active matter systems are considered. 
Often pattern formation in active systems like self-propelled or driven particles refers to the formation of substantial spatial inhomogeneities in the density of the particles.  
Phenomena like giant number fluctuations, phase separation and clustering phenomena fall into this class and have all been observed in experiments \cite{narayan_long-lived_2007,peruani_collective_2012,schaller_polar_2010}, predicted or reproduced by theoretical analysis and numerical simulations of appropriate models \cite{mishra_active_2006,peruani_nonequilibrium_2006,ramaswamy_active_2003,peruani_cluster_2011,toner_hydrodynamics_2005}. Another frequent phenomenon in the collective dynamics of active particles is the formation of high-density bands in a low-density environment (segregation bands), that may travel at a constant speed  (as for  the classical Vicsek model) or can be stationary (as for self-propelled rods) \cite{chate_simple_2006,chate_collective_2008,ginelli_large-scale_2010,nagy_new_2007}. 
The high density inside such segregation bands often is associated with polar or nematic alignment  order, while the low density environment typically is disordered. 
The precise nature of these bands (stationary or moving, polar or nematic order) is often determined by symmetries of the system, like the nature of the active particles (polar or apolar) and the specific of their interactions (ferromagnetic or nematic) \cite{chate_collective_2008,ginelli_large-scale_2010}. 
Finally, active particles may also exhibit classical pattern formation phenomena such as  Turing-type patterns or periodic standing and 
 traveling waves of the density, compare e. g.  \cite{baerner_rippling_2002}. 
In comparison to standard pattern formation systems like Benard convection or chemical reaction-diffusion systems where molecular and pattern scale differ by many orders of magnitude, typically the length scales of pattern formation and the particle size in active matter are comparable 
 \cite{baerner_rippling_2002}. 
This allows for agent-based approaches as introduced in Section \ref{sec:swarming}, where the motion and dynamics of each particle is simulated. 
This section is structured as follows: First, we will briefly describe the results of continuum theories and their relation to findings in agent-based models. 
Then, work on clustering of self-propelled particles is reviewed.
As an example, a simple physical model of self-propelled hard rods and simulation results therein are described. 
A theoretical analysis of clustering via a Smoluchowski-type kinetic approach is introduced and applied to simulations with
 hard rods and experiments with myxobacteria. Related phenomena in Vicsek-type models, models of colloids  and granular experiments are also surveyed. 
Second, we describe formation of large-scale coherent structures like traveling and stationary high-density bands in simple models for self-propelled particles, namely the Vicsek model and a variation 
describing polar particles with apolar alignment interactions and their instabilities. 
Due to the rapid development of the studies in collective motion of active (Brownian) particles, 
 we concentrate on detailed descriptions of recent developments. The review in this section  is by no  means exhaustive and 
relevant original and review papers will be cited and may be consulted.  Nevertheless, the examples selected and the discussion will 
address the most important issues. The main message of this section is that different forms of spatial organisation like clusters and 
large-scale patterns, are crucial for a complete understanding of the collective dynamics of active particles. 
Hence, the approaches introduces in this section are necessary to complement the picture sketched in the previous section on swarming.

\subsection{Continuum theories versus agent-based models} 
\label{sec:cont_vs_ibm}
In agent-based models, one encounters often strong finite size effects. The same model may yield clustering and giant fluctuations for small and intermediate particle numbers as well as coherent structures and large-scale patterns for big particle numbers. 
In addition, bands and other large-scale patterns can have instabilities against modes of a definite wavelength, that may be absent in smaller system, but highly relevant in sufficiently large systems. 
Such stability properties of patterns are also quite relevant for the presence of long-range order 
or non-zero global order parameters. 

Alternatively, coarse-grained, hydrodynamic continuum models have been frequently employed to advance the understanding of collective dynamics in active matter. It would be highly desirable to establish links between the two levels of descriptions. 
Such efforts are, however, limited by the fact that continuum models address typically length scales that involve very large number of particles. 
As a result, the study of the corresponding agent-based models is often computationally extremely expensive or even prohibitive.  
Simulations of ``microcospic'' agent-based model are nevertheless very important, since they are necessary to understand the role of the noise and the fluctuations as well as the limitations of the validity of hydrodynamic descriptions. 
Agent-based models also allow for a simultaneous computation of coarse-grained properties like densities or alignment and of characteristic of individual particle motion. 
In other words, they can be used to establish a link between an Eulerian and Lagrange view of active matter systems \cite{parrish_complexity_1999}. 
Often, both Eulerian and Lagrangian properties of active systems are accessible to experiments. 
For example, fluorescence labeling allows for tracking of individual cell motion in bacterial films, while density patterns are conveniently determined by microscopy \cite{welch_cell_2001,igoshin_waves_2004}. 
This enlarges the amount of information necessary for quantitative validation of models substantially. 

In contrast, continuum theories are useful to describe the large-scale behavior of active Brownian particles. 
Shortly after the first publication on the Vicsek model, Toner and Tu carried out pioneering work deriving a continuum theory for active 
systems like the Vicsek model. Their work  analyzes a coarse-grained hydrodynamic description based on the velocity field ${\bf v}$ and 
the particle density $\rho$. These equations are of the general form: 
\begin{eqnarray} \label{toner-tu-1}
\frac{\partial {\bf v}}{\partial t} + \lambda {\bf v} \cdot \nabla {\bf v} + \dots =  ( \alpha - \beta {\bf v} \cdot {\bf v} ) {\bf v} 
+ \Gamma \nabla \nabla {\bf v} - \nabla P(\rho) +  {\bf \xi}   \\
\frac{\partial \rho}{\partial t} + \nabla \cdot \rho {\bf v} = 0.
\label{toner-tu-2}
\end{eqnarray}
Eqs. \ref{toner-tu-1} - \ref{toner-tu-2} capture the phase transition from the disordered state ${\bf v} = 0$ to the ordered state 
$ v = \sqrt{\alpha / \beta}$. 
The analysis of Toner and Tu gave an explanation of the long-range order in two-dimensional systems, that is absent in equilibrium
systems and had been previously observed in the simulations by Vicsek and colleagues. 
Moreover, it allowed the derivation of sound modes and lead to the prediction of giant number fluctuations, that shall be discussed below. 
A detailed discussion of the Toner-Tu field theory can be found in \cite{toner_hydrodynamics_2005,ramaswamy_review_2010}. 
The number fluctuations in the collective dynamics of a many particle system are often expressed as
\begin{equation} 
\Delta N  \propto N^{\phi},
\end{equation}
where $\Delta N$ is the standard deviation of the particle number in a given finite volume.
The quantity  $N$ gives the mean number of particles in this volume. 
Equilibrium systems exhibit normal fluctuations with $\phi = 1/2$, whereas active systems often show $\phi > 1/2$.
Such behavior out of equilibrium is consequently referred to as giant number fluctuations. 
The Toner-Tu field theory predicts $\phi = 7/10 + 1/5d$ for $d < 4$, where $d$ is the spatial dimension of the 
system. The predicted value $\phi = 0.8$ in two dimensions has been recently confirmed in extensive numerical studies of two
agent-based models, namely  the original Vicsek model 
\cite{chate_collective_2008} and the Peruani model \cite{ginelli_large-scale_2010}, in the ordered state without segregation
bands. 
Giant number fluctuations have also been studied for active nematics (=driven apolar particles), 
where an exponent $\phi = 1/2 + 1/d$ was derived from
field theory, see \cite{ramaswamy_active_2003,ramaswamy_review_2010}, where $d$ again refers to the spatial dimension of the
system. In two dimensions, the predicted value of $\phi = 1$ was recovered from extensive numerical simulations of a Vicsek-type
simple agent-based model  \cite{chate_simple_2006}. 
Coincidently, researchers have found in recent experiments with bacteria \cite{zhang_collective_2010,peruani_collective_2012} 
similar exponents ($\phi \approx 0.8$) in two dimensions. 
In the latter case, the interpretation of the giant number fluctuations 
remains controversial because of the simultaneous observation of large coherent clusters of the investigated bacteria. 
Such clustering provides an alternative reason for giant number fluctuations and violates the assumptions of the Toner-Tu field theory. 
A similar controversy appeared in connection with experiments of shaken elongated rice corns that were
devised as a realization of an active nematic and show giant number fluctuations \cite{narayan_long-lived_2007}. 
The initial interpretation of these measurements as confirmation of the predictions of continuum theory were later contested 
\cite{aranson_comment_2008} by pointing out the strong tendency of this system to form clusters.  
In general, the formation of clusters or large-scale patterns like segregation bands is expected to have a strong influence 
on the number fluctuations and goes beyond the range of validity of continuum theories. 

Alternative continuum theories have been recently also been investigated by Marchetti and coworkers based on symmetry consideration 
and expansions \cite{baskaran_statistical_2009,mishra_fluctuations_2010} as well as for a specific approach describing self-propelled
hard rods  \cite{baskaran_hydrodynamics_2008,baskaran_nonequilibrium_2010}. 
Bertin and coworkers have pursued a continuum theory based on a kinetic approach \cite{bertin_boltzmann_2006-1} and have been
able to recover qualitatively  the traveling segregation bands seen in simulations of the Vicsek model 
\cite{bertin_microscopic_2009}.

Altogether, continuum theories have contributed a lot to the understanding of the collective dynamics of active Brownian particles. 
In recent years, large-scale agent based simulations and experiments with, e. g., driven granular particles or moving bacteria, 
have revealed many interesting phenomena that pose new challenges to continuum theorist.
Many open questions revolve around the issue of prediction of spatial inhomogeneities, large-scale patterns and cluster formation which all
have been found to play a dominant role in agent-based simulations and experiments. In the following subsections we will discuss recent 
developments in agent-based models and related experiments.

\subsection{Clustering, segregation and band patterns - phenomena and experiments} 


Examples of large-scale self-organized patterns in systems of 
self-propelled particles  with short-range interactions are found at all scales, from groups of  animals~\cite{parrish_animal_1997,cavagna_scale-free_2010,bhattacharya_collective_2010} and human crowds~\cite{helbing_freezing_2000} down to insects~\cite{buhl_disorder_2006,romanczuk_collective_2009}, bacteria~\cite{zhang_collective_2010}, or actin filaments~\cite{butt_myosin_2010,schaller_polar_2010}. 
Such patterns are also found in non-living system like in driven granular media~\cite{narayan_long-lived_2007,kudrolli_swarming_2008,kudrolli_concentration_2010,deseigne_collective_2010}.
Despite the fact that the interaction mechanisms between
individual elements are of a different nature, it is possible to determine
some common requirement to achieve large-scale (spatial) self-organization. 
Particularly important for the emerging macroscopic patterns are the self-propulsion of the agents, and their velocity alignment mechanism. 
As described above, simple individual-based models like the Vicsek model~\cite{vicsek_novel_1995} have helped to reveal the relevance of these two elements by  reducing the problem to the competition between a local aligning interaction and noise~\cite{chate_modeling_2008}.
Recall, that in two dimensions, self-propelled particles moving at constant speed with a ferromagnetic-like velocity alignment  exhibit at low noise a phase characterized by true long-range polar order which translates into a net flux of particles~\cite{vicsek_novel_1995,chate_collective_2008}. 
In the previous section, we have assumed that the ordered states are spatially homogeneous.
Systematic simulations and theoretical analysis based on coarse-grained continuum models show that these ordered phases often exhibit several remarkable features of spatial organisation.
For ferromagnetic alignment mechanisms,   the  spontaneous formation of elongated high density bands that move at roughly constant speed in the direction perpendicular to the long axis of the band, and anomalous density fluctuations for low noise levels were reported ~\cite{chate_collective_2008,mishra_fluctuations_2010}.
When the alignment is replaced by a nematic velocity alignment, self-propelled particles again display a phase characterized by true long-range nematic order at low noise intensity~\cite{ginelli_large-scale_2010}. 
Interestingly, spontaneous density segregation processes  into bands are observed for these particles as well as anomalous density fluctuations  ~\cite{ginelli_large-scale_2010}. 
It is noteworthy, that self-propelled particles with
display large-scale high-density patterns and segregation in absence of any attracting force~\cite{chate_collective_2008,ginelli_large-scale_2010}. 

At moderate system sizes or particle numbers, self-propelled particles with ferromagnetic or nematic alignment mechanisms often do not show large-scale patterns like segregation bands. 
Nevertheless, collective motion in this regime is still characterized by strong density fluctuations. 
A phenomenon that has attracted a lot of interest recently is the formation of large polar clusters in many experiments and models, that again do not require attractive forces between the particles. 
It seems that the common presence of active motion and effective alignment forces of either ferromagnetic or nematic symmetry are sufficient to facilitate ``condensation'' of self-propelled particles into clusters. 
 This effect has been studied extensively for alignment that results from volume exclusion interactions among self-propelled rods~\cite{peruani_nonequilibrium_2006,peruani_individual_2008,yang_swarm_2010}.
Related findings appeared in models for colloidal rods \cite{wensink_aggregation_2008}, experiments and models of driven granular particles \cite{kudrolli_swarming_2008} and models describing collective motion of sperms \cite{yang_cooperation_2008,yang_swarm_2010}. 
In many instances the appearance of polar clusters was linked to a power-law cluster size distribution \cite{peruani_nonequilibrium_2006,yang_cooperation_2008,yang_swarm_2010}. 
Clustering effects and the emergence of steady state cluster size distributions in self-propelled particle systems were also observed in the Vicsek model~\cite{huepe_intermittency_2004,huepe_new_2008} and the Peruani model \cite{peruani_cluster_2011}. 

In the following subsections, we will first discuss first the formation of polar clusters in a model of self-propelled rods and describe a simple kinetic theory from which the steady state cluster size distribution of self-propelled particles is obtained. 
This approach is a modification of the Smoluchowki kinetic equations that were developed to describe the aggregation of colloids \cite{von_smoluchowski_versuch_1917,chandrasekhar_stochastic_1943}.
Then, we will show that similar phenomena appear and a related kinetic description for the cluster size distribution applies also to models of self-propelled particles with velocity alignment as outlined before in \cite{peruani_individual_2008,peruani_cluster_2011}.  
In the second part of the second of the section, we review results on the formation of large-scale structures in models with velocity alignment. 
Finally, we give an account of results obtained in coarse-grained continuum descriptions of collective motion in active matter specifically designed to account for the phenomena observed in the simulation studies discussed  before. 

\subsection{Clustering, phase separation and giant number fluctuations}
\label{sec:pattern_clustering}

\subsubsection{Clustering of self-propelled hard rods} 

Emergent large-scale patterns of interacting self-driven motile elements are observed in a wide range of biological systems of different complexity: from human crowds,
herds, bird flocks, and fish schools \cite{helbing_traffic_2001,schweitzer_brownian_2003,toner_hydrodynamics_2005} to multicellular aggregates, e.g. of bacteria and amoebae \cite{ben-jacob_cooperative_2000} as well as sperms \cite{riedel_self-organized_2005}.
A recurrent question is how these entities coordinate their behavior to form groups which move collectively.
Specific  models for bacteria like {\it E. coli} as well as for amoebae like {\it D. discoideum} \cite{ben-jacob_cooperative_2000}, have been based on chemotaxis, a long-range cell interaction mechanism according to which individual cells move in response to chemical signals produced by all other cells.
However, in some bacteria there is no evidence for chemotactic cues and cells coordinate their movement by cell-to-cell signalling mechanisms in which physical contact between bacteria is needed  \cite{dworkin_myxobacteria_1993,kaiser_coupling_2003,jelsbak_pattern_2002}.
Consequently, one may ask how such bacteria aggregate in order to communicate.
Another relevant aspect is the influence of the shape of the bacteria.
The shape  has been shown to be essential for individual motion of swimming bacteria \cite{dusenbery_minimum_1997}.
In contrast, the role of the cell shape for collective motion has
remained mostly unexplored.
It has been demonstrated experimentally \cite{kemkemer_nematic_2000}
that migrating elongated amoeboid cells  exhibit alignment effects similar to
those reported in liquid crystals \cite{onsager_effects_1949}. 
A prominent example for collective behavior 
with no apparent long range interactions are the striking patterns
 observed during the life-cycle of gliding myxobacteria, see e.g.
\cite{dworkin_myxobacteria_1993,kaiser_coupling_2003,jelsbak_pattern_2002,myxoexperiments}.
Earlier modeling work has reproduced many of these
patterns in three dimensions assuming either perfect alignment \cite{baerner_rippling_2002}
or a phenomenological alignment force \cite{alber_two-stage_2004,igoshin_waves_2004}.
These models have all considered patterns resulting from exchange of chemical signals,  that are absent in an early stage of the myxobacterial life cycle.
Nevertheless, a trend from initial independent motion
towards formation of larger clusters of aligned bacteria is often observed.

Here, a model of self-propelled rods that have only repulsive excluded volume interactions in two dimensions is considered 
It was found  that the interplay of  rod geometry, self-propulsion
and  repulsive short-range interaction is sufficient to facilitate
aggregation into clusters \cite{peruani_nonequilibrium_2006}. 
%
%
\begin{figure}
%
\begin{center}
\includegraphics[width=0.23\linewidth,height=0.23\linewidth]{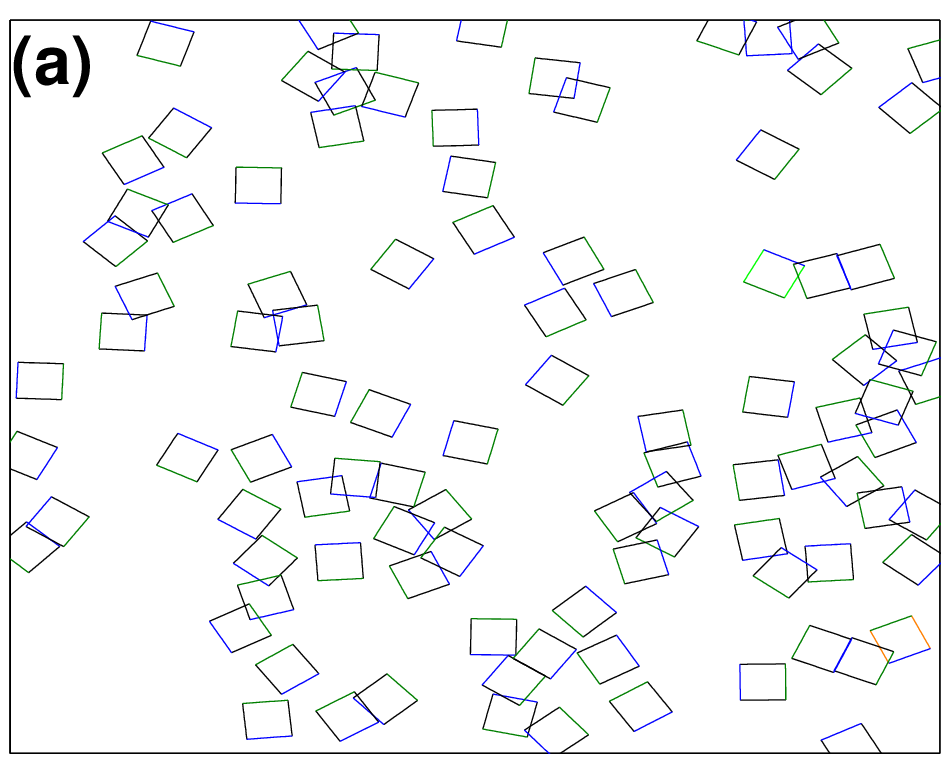}
\includegraphics[width=0.23\linewidth,height=0.23\linewidth]{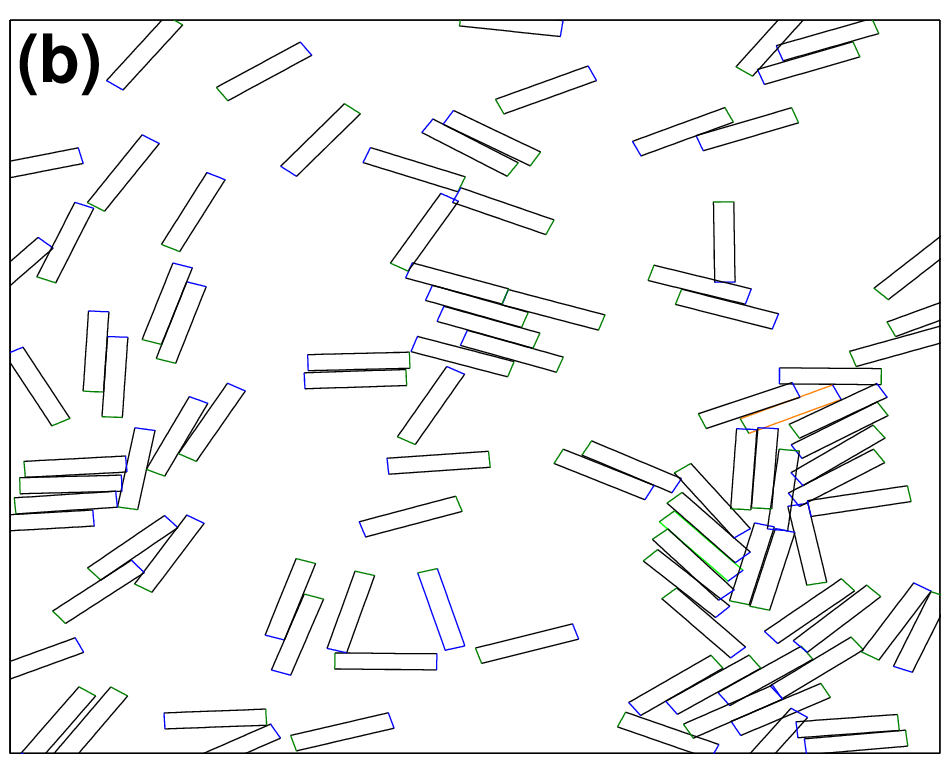}
\includegraphics[width=0.23\linewidth,height=0.23\linewidth]{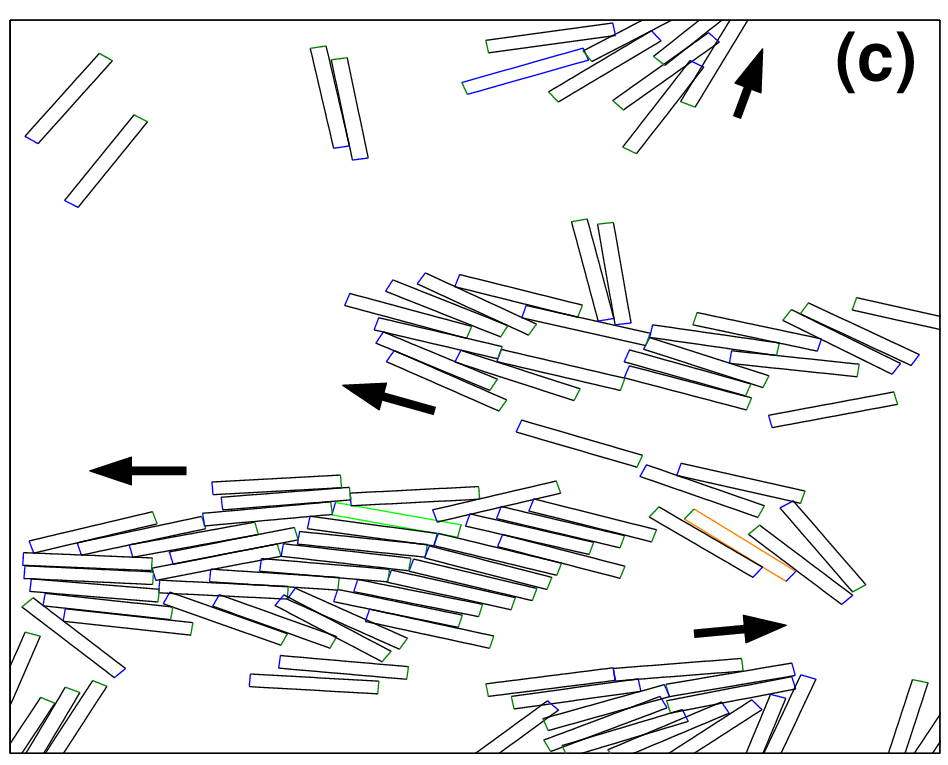} \\
\includegraphics[width=0.23\linewidth,height=0.23\linewidth]{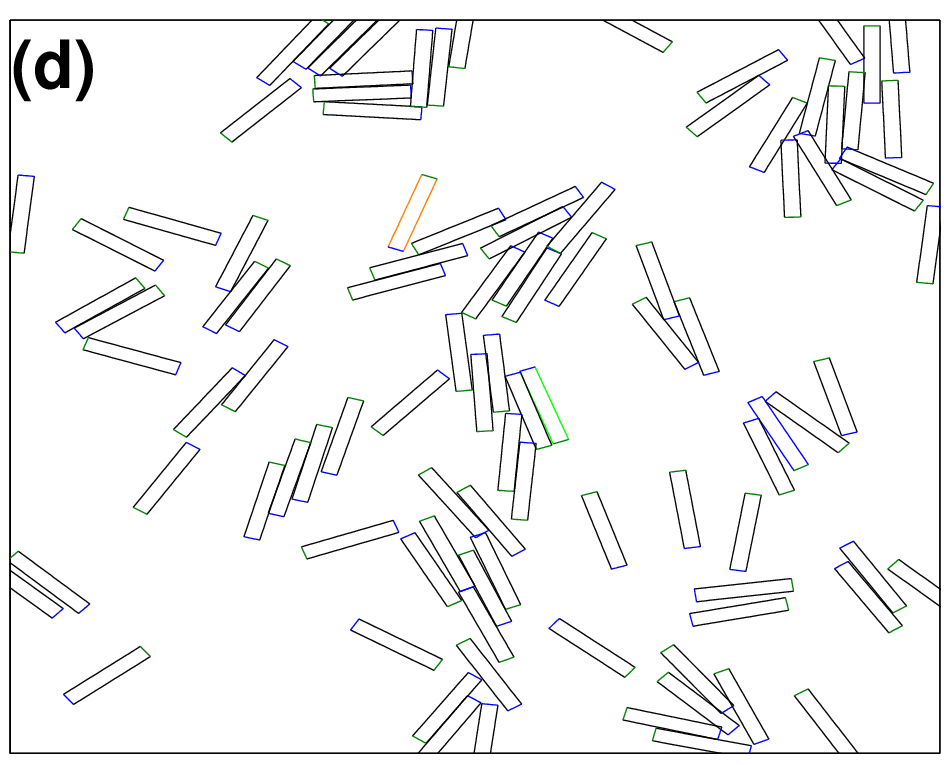}
\includegraphics[width=0.23\linewidth,height=0.23\linewidth]{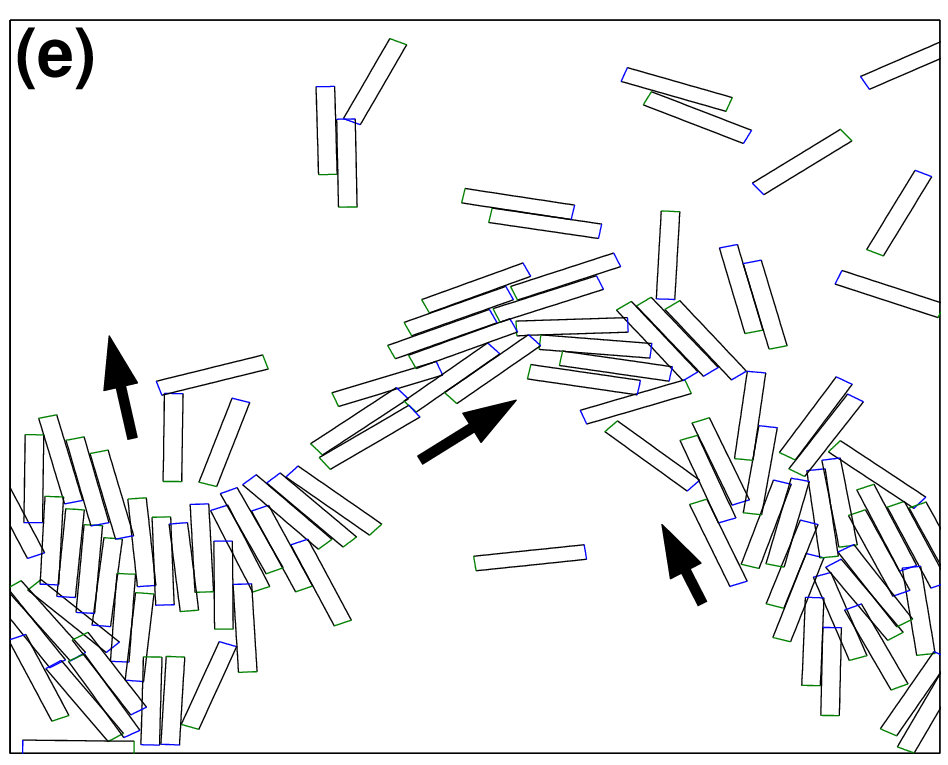}
\includegraphics[width=0.23\linewidth,height=0.23\linewidth]{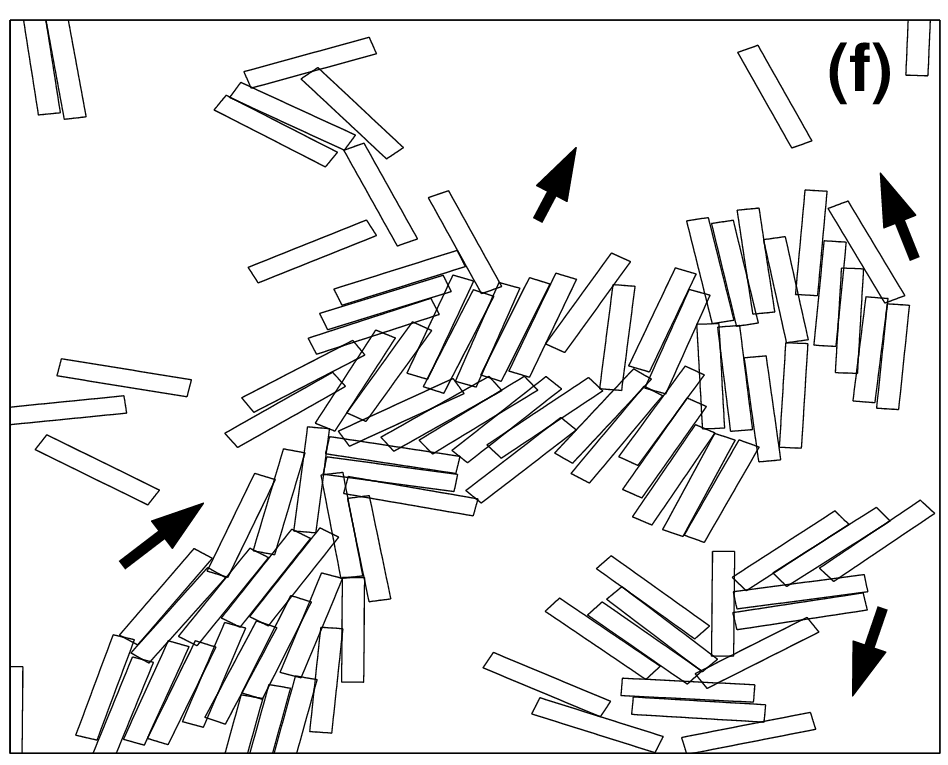}
\end{center}
\caption{Simulation snapshots of the steady states for different
particle anisotropy $\kappa$ and the same packing fraction $\eta$
(a-c), and the same $\kappa$ and different $\eta$ (d-f). Fixing
 $\eta=0.24$: (a) before the transition, $\kappa=1$; (b) almost at
the transition, $\kappa=5$; (c) after the transition, $\kappa=8$.
Fixing $\kappa=6$: (d) before the transition, $\eta=0.18$; (e)
just crossing the transition, $\eta=0.24$; (f) after the
transition, $\eta=0.34$. In all cases, particles $N=100$ and
particle area $a=0.2$. The arrows indicate the direction of motion
of some of the clusters.} \label{fig2}
\end{figure}

Consider $N$ rod-like particles moving on a plane, for a more detailed description see \cite{peruani_nonequilibrium_2006,peruani_individual_2008}.  
Each particle is equipped with a self-propelling force acting
along the long axis of the particle. 
Velocity and angular velocity are proportional to the force and torque,
correspondingly.
The rod-shape of the particles requires  three different friction coefficients which correspond to the resistance exerted by the medium when particles either rotate or move along their long and short axes.
Inertial terms are neglected, i.e. the case of overdamped motion is considered. 
As a result the movement of the {\em i-th} rod is governed
by the following equations for the velocity of its center of
mass and angular velocity:
\begin{eqnarray} \label{update_position}
(v^{(i)}_{\parallel}, v^{(i)}_{\perp}) &=&  \left(
\frac{1}{\zeta_{\parallel}}(F-\frac{\partial U^{(i)}}{\partial
x_{\parallel}}),
 -\frac{1}{\zeta_{\perp}} \frac{\partial U^{(i)}}{\partial x_{\perp}} \right) \nonumber \\
\label{update_orientation} \dot{\varphi}^{(i)} &=&
-\frac{1}{\zeta_{\varphi}}\frac{\partial U^{(i)}}{\partial \varphi}
\end{eqnarray}
where $v^{(i)}_{\parallel}, v^{(i)}_{\perp}$ refer to the
velocities along the long and short axis of the rods,
respectively,
 $\zeta_{i}$ indicates the corresponding friction
coefficients ($\zeta_{\varphi}$ is related to the friction torque),
$U^{(i)}$ refers to the energy of the interaction of the {\em i-th} rod with all other rods, 
and $F$ is the magnitude of the self-propelling force. 
The motion of the center of mass ${{\dot{\bbox r}}^{(i)}= (v_x^{(i)},v_y^{(i)})}$ 
of the {\em i-th} rod is given by 
\begin{eqnarray} 
v_x^{(i)} = v^{(i)}_{\parallel} \cos \varphi^{(i)} +  v^{(i)}_{\perp} \sin \varphi ^{(i)} \nonumber \\
v_y^{(i)} = v^{(i)}_{\parallel} \sin \varphi^{(i)} -  v^{(i)}_{\perp} \cos \varphi ^{(i)} 
\end{eqnarray}
Particles interact by ``soft'' volume exclusion, {\it i. e. } by 
a potential that penalizes particle overlaps in the following way:
\begin{align} \label{potential}
U^{(i)}({\bbox r}^{(i)},\varphi^{(i)},{\bbox r}^{(j)},\varphi^{(j)}) & =  \mu \sum_{j=1,j \neq i}^{N} 
\left( (\gamma - a_{o}^{(ij)})^{- b} - \gamma^{- b} \right)  
\end{align}
where $a_o^{(ij)}=a_{o}({\bbox r}^{(i)},\varphi^{(i)},{\bbox r}^{(j)},\varphi^{(j)})$ 
is the area overlap of the rod $i$ with rod $j$ 
and $\mu$ is the interaction strength.
The simulations were performed placing $N$ identical particles
initially at random inside a box of area $A$ with periodic
boundary conditions. 

\begin{figure}
\centering
\includegraphics[width=0.6\textwidth]{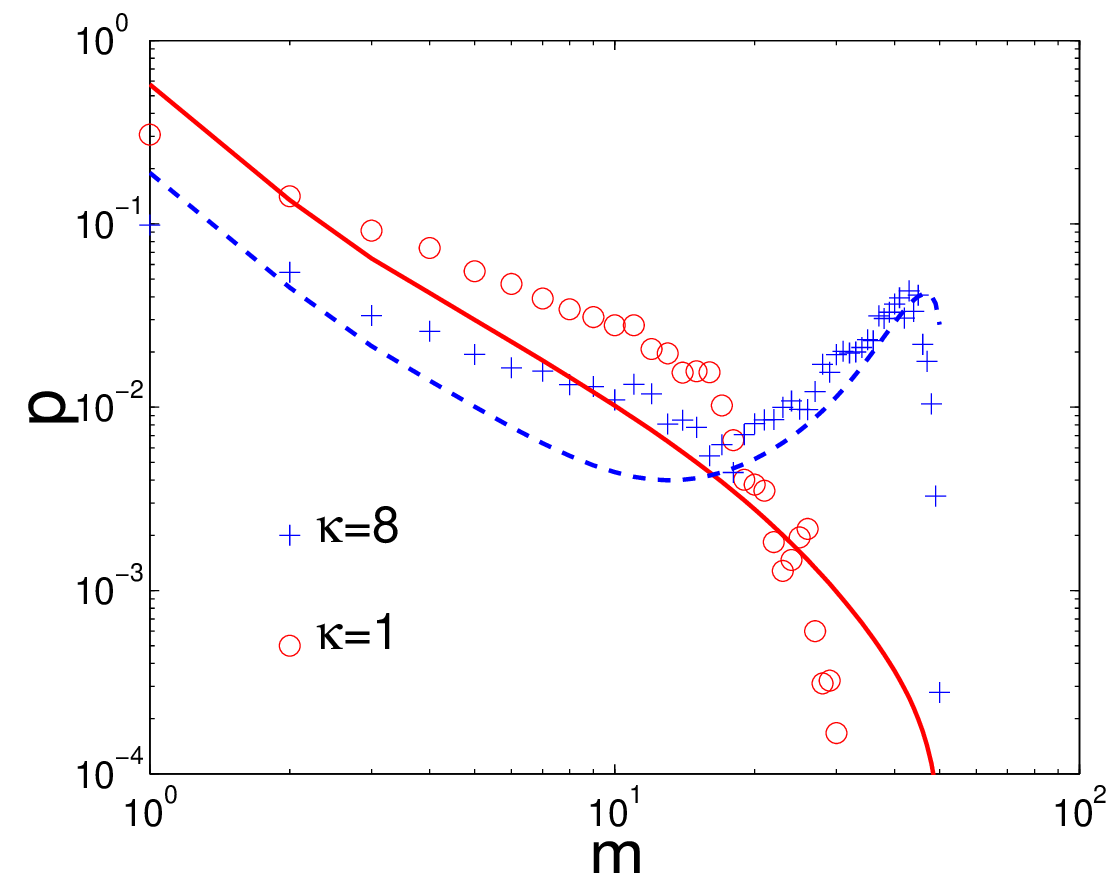}
\caption{
$p(m)$ as function of the cluster size $m$ for $\eta=0.34$. Symbols
show the average over eight IBM simulations for active particles with $N=50$ and
$\kappa=1$ (circles) and $\kappa=8$ (crosses)
. The lines correspond to the mean field theory for $\kappa=1$ (solid) and
$\kappa=8$ (dashed). 
\label{clustersize}} 
\end{figure}

There are two key parameters which control the dynamics of the self-propelled rods: 
i) the packing fraction $\eta$, i.e., the
area occupied by rods divided by the total area ($\eta
= N a/A$, where $N$ is the number of particles in the
system, $a$ is the area of a single particle, and $A$ is the
total area of the box), and ii) the length-to-width aspect ratio
$\kappa$ ($\kappa=L/W$, where $L$ is the length
and $W$ is the width of the rods).
Simulations yield an increase of cluster formation with increasing 
$\kappa$ or $\eta$, see Fig. \ref{fig2}.
Clusters are defined by connected particles that have non-zero overlap area. 
Simulations can be characterized by the weighted cluster size distribution, $p(m)$, which indicates the probability of finding a given particle inside a cluster of mass $m$. 
Fig. \ref{clustersize} shows that for a given $\eta$, a critical $\kappa_{c}$ can be defined as the value of $\kappa$ for which the shape of $p(m)$ changes from unimodal to bimodal. The figure shows also typical shape of $p(m)$ before
clustering, corresponding to low values of $\kappa$, and after clustering, corresponding to large values of $\kappa$. 
The onset of clustering is defined by the
emergence of a second peak in $p(m)$.
The robustness of the model against fluctuations was tested 
by inserting additive noise terms $R_i/\zeta_i$ in Eqs. (\ref{update_orientation}),
which correspond to a switch from purely active to active Brownian particles \cite{erdmann_brownian_2000,erdmann_noise_2005}. 
Clustering is still present in active Brownian rods, albeit the transition
is moved to larger values of $\kappa$ and $\eta$.  
Clustering was absent in all simulations with purely Brownian rods ($F=0$).
%
\subsubsection*{Smoluchowski-type mean field approximation (MFA)} 
The clustering effects in simulations described so far can be  analyzed by deriving kinetic equations for the number $n_j$ of clusters of a given size $j$. 
The equations for $n_j$ contain terms for cluster fusion and fission. 
For the fusion terms  adopted kinetic equations originally derived by Smoluchowski 
for the case of coagulation of colloids \cite{von_smoluchowski_versuch_1917,chandrasekhar_stochastic_1943} were employed, while the fission terms are empirically defined from the typical behavior seen in related simulations \cite{peruani_nonequilibrium_2006}.
The numbers $n_j$ change in time according to 
$\left\{
n_{j}\left(t\right)\right\} _{j=1}^{\infty}$, where
$n_{j}\left(t\right)$ is the number of clusters of mass $j$ at time
$t$.

This description allows 
to consider a single rate constant for all possible collision
processes between clusters of mass $i$ and $j$, as well as a unique
disintegration constant for any cluster of mass $i$.
Four additional crucial assumptions are now made:
i) The total number of particles in the system,
$N=\sum_{j=1}^{N}jn_{j}\left(t\right)$, is conserved.
ii) Only binary cluster collisions are considered. 
Collisions between any two clusters are allowed whenever the sum of the cluster
masses is less or equal to $N$. 
iii) Clusters suffer spontaneous fission only by losing individual 
particles at the boundary one by one, {\it i. e.} a cluster can only decay by a
process by which a $j$-cluster split into a single particle plus a
$\left(j-1\right)$-cluster. 
This is motivated by observations in the above simulations.
iv) All clusters move at constant speed, $\widetilde{v} \approx F/ \zeta_{\parallel}$, 
implying that rods in a cluster have high orientational order and interact only 
very weakly with their neighbors. 
Under all these assumptions the evolution of the $n_j$'s is given
by the following $N$ equations:
\begin{eqnarray}
\dot{n}_{1}&=&2B_{2}n_{2}+\sum_{k=3}^{N}B_{k}n_{k}-\sum_{k=1}^{N-1}A_{k,1}n_{k}n_{1}
\nonumber \\
\dot{n}_{j}&=&B_{j+1}n_{j+1}-B_{j}n_{j}-\sum_{k=1}^{N-j}A_{k,j}n_{k}n_{j}
\nonumber \\
&&+\frac{1}{2}\sum_{k=1}^{j-1}A_{k,j-k}n_{k}n_{j-k} \ \quad \mbox{for} \quad j = 2, .....,N-1
\nonumber \\
\dot{n}_{N}&=&-B_{N}n_{N}+\frac{1}{2}\sum_{k=1}^{N-1}A_{k,N-k}n_{k}n_{N-k} \label{rea}
\end{eqnarray}
where the dot denotes time derivative, $B_{j}$ represents the
fission rate of a cluster of mass $j$, defined by $B_{j}=
(\widetilde{v}/R)\sqrt{j}$, and $A_{j,k}$ is the collision rate
between clusters of mass $j$ and $k$, defined by
$A_{j,k}=(\widetilde{v}\sigma_{0}/A)
\left(\sqrt{j}+\sqrt{k}\right)$. 
$\sigma_{0}$ is the
scattering cross section of a single rod. 
$R$ is the only free parameter and is defined  the 
characteristic length a rod at the boundary moves with the cluster before 
leaving it in a typical fission event.
One sets  $R = \alpha L$ taken into account that longer rods will stay attached to 
cluster for a longer time. 
  
Since $\sigma_{0}$ can be approximated by $\sigma_{0}\approx
L+W=\sqrt{a}\left(\sqrt{\kappa}+\frac{1}{\sqrt{\kappa}}\right)$,
the MFA depends only on the parameters $\kappa$, $a$, $A$, 
$\widetilde{v}$ and $\alpha$. 
If one integrates Eqs. (\ref{rea}) with  parameters used in
IBM simulations and an initial condition $n_{j}\left(t=0\right)=N\delta_{1,j}$, 
their solution yields steady state values $n_{j}^0$  for $t \rightarrow \infty$. 
From these values, we obtain a MFA for the weighted cluster size distribution 
 $p(m)=n_{m}^0 m/N$ for given values of the free parameters $R$ resp. $\alpha$. 
The best agreement between the MFA and the IBM simulations is found for a choice of  $\alpha = 1.0 \pm 0.05$ (see Fig.
\ref{clustersize}).
To understand the relation between the parameters of the model and
clustering effects better, one can rescale Eqs. (\ref{rea}) by introducing a new time
variable: $\tau=t \widetilde{v} / \sqrt{a
\kappa}$. 
The resulting equations depend
only on a dimensionless parameter $P=(\kappa+1)a/A$. 
Note that $\widetilde{v}\neq 0$ is scaled and does not affect the qualitative dynamics of the system. 
In the dimensionless model the parameter $P$ stands for ratio between 
fusion and fission processes and therefore triggers the transition
from a unimodal to a bimodal cluster size distribution.
By numerical solving for the steady state solution of the kinetic equations, one can accurately determine a critical transition parameter $P_{c}$. 
Given the system area $A$, the rod area $a$ and the number of rods $N$, 
this method provides a straightforward way to calculate $\kappa_{c}$:
\begin{equation} \label{kappa_pc}
\kappa_{c}=P_{c}(N)\frac{A}{a}-1
\end{equation}
It was found that in the MFA the critical parameter value $\kappa_c$ for the 
clustering transition does practically not depend on the number of particles as soon as $N > 50$ \cite{peruani_nonequilibrium_2006}. 
One proceeds by assuming that $P_{c}$ is inversely proportional with
$N$ and expresses
 $\kappa_{c}$ as a  function of the packing fraction:
\begin{equation} \label{kappa_eta}
\kappa_{c} = C/\eta-1
\end{equation}
where the constant was  found to be $C \approx 1.46$ by comparison with simulation results \cite{peruani_nonequilibrium_2006}. 
%
%
So, for the range of parameters used in the IBM,  the unimodal shape of the weighted cluster size distribution for  small values of $\kappa$ and $\eta$, and the bimodal shape for large values of the two parameters seen in the IBM  was qualitatively reproduced  in the MFA, see Fig. \ref{clustersize}.
A second interesting feature obtained both in the IBM and MFA  is that the cluster-size distribution changes notably in shape from an function exponentially decaying with size to a function with a power-law behaviour $p(m) \propto m^{-c}$ at small cluster size and  a second peak confined by a cutoff at large cluster sizes. 
The exponent of the power law is, however, substantially larger for the MFA ($c \approx 1.35$) than for the simulation of self-propelled rods ($c \approx 0.95 $). 
The power-law behaviour of the cluster-size distribution characterizing self-propelled rods was also discussed extensively in recent more detailed simulations of self-propelled rods \cite{yang_swarm_2010}, where exponents of the weighted cluster-size distributions are in the range between 0.95 and 1.4. 
Interestingly, also recent experiments with bacteria, namely {\em Bacillus subtilis} \cite{zhang_collective_2010} and a mutant of {\em Myxococcus xanthus} \cite{peruani_collective_2012} report exponents $c$ of the weighted cluster size distribution in the range of 0.85--0.9\, . 
We expect further theoretical and experimental activities, that may finally uncover universal properties in the clustering of self-propelled rods and related systems. 
%


In summary,  one finds non-equilibrium clustering for interacting
self-propelled rod shaped-particles with sufficient packing density
$\eta$ and aspect ratio $\kappa$ in simulations.
The onset of clustering can be defined by a transition from a unimodal to bimodal
cluster size distribution.
This transition is reproduced by a mean-field description of the
cluster size distribution, which yielded a simple criterion, $\kappa
= C/\eta - 1$, for the onset of clustering.
It is instructive to compare these results rewritten in the form
 $\kappa\eta + \eta \approx 1.46$ with the formula for the
 isotropic-nematic transition  $\kappa\eta \approx 4.7 $ found in the two-dimensional version \cite{kayser_bifurcation_1978} of Onsager's mean-field theory for Brownian rods \cite{onsager_effects_1949}.
This shows that actively moving rods can achieve alignment at much lower densities
than Brownian rods resp. particles in equilibrium systems.
The clustering phenomenon is absent in simulations with isotropic self-propelled particles as well as with Brownian rods.
The model of self-propelled rods provides also an alternative explanation for collective behavior of rod-shaped objects - previous swarming models have achieved 
aggregation and clustering  by assuming attractive long-range interactions 
\cite{vicsek_novel_1995,gregoire_onset_2004,erdmann_brownian_2000,erdmann_noise_2005}.  
With respect to biology, the observations made for self-propelled rods offer a physical
explanation for the formation of clusters in many gliding rod-shaped bacteria,
that often precedes the formation of biofilms and the appearance of
more complex patterns. 
%
%


\subsubsection{Clustering of self-propelled particles with velocity alignment} 

In the following, the interplay between orientation ordering by velocity alignment 
and clustering in self-propelled particles with ferromagnetic $F$- or nematic $LC$-alignment is analyzed and briefly discussed. 
Through  simulations evidence was provided  that at high density orientation ordering sets in before clustering \cite{peruani_individual_2008}. 
In contrast,  for low particle densities, the onset of orientation ordering and clustering are closely related and seem to occur at the same value of the noise \cite{peruani_individual_2008,peruani_cluster_2011} for $LC$-alignment.
These findings indicate that the phase transition occurs
rather due to mixing of particles than exclusively to the directed active motion.
%
\begin{figure}
\centering
\includegraphics[width=0.7\textwidth]{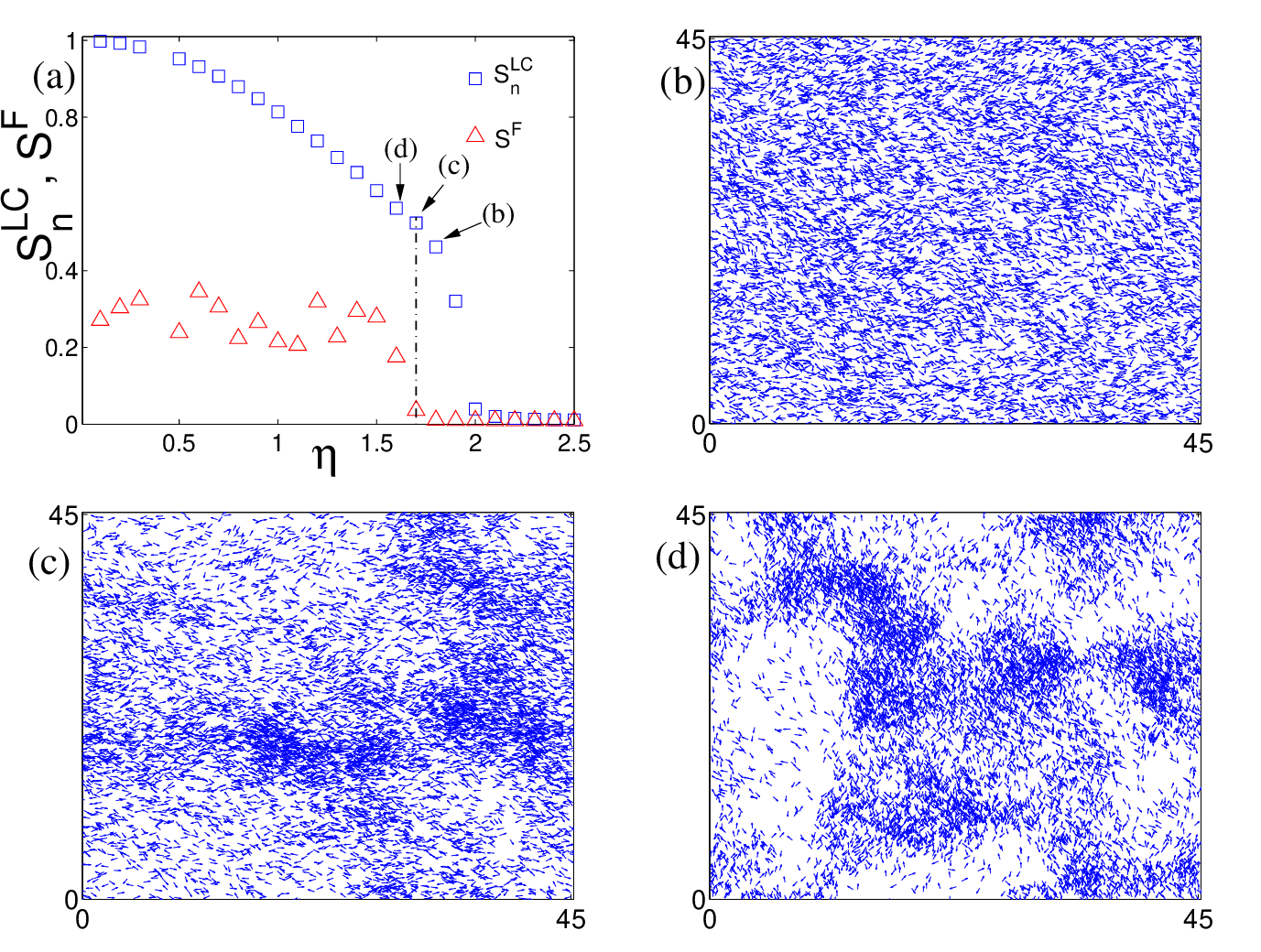}
\caption{Orientational dynamics and clustering at high density. Number of particles $N=2^{14}$ and density $\rho=4$. (a) Orientation order parameter $S$ vs. noise amplitude $\eta$. Symbols are average over $10$ realizations.
The dashed curve corresponds to the scaling predicted by the mean-field while the vertical dot-dashed line indicates the onset of clustering effects. 
(b)-(c) snapshots of the simulations for the values of $\eta$ indicated in (a) at the steady state, time step $2.5$ $10^5$. \cite{peruani_individual_2008} \label{figs_mfa}}
\end{figure}

\begin{figure}
\centering
\includegraphics[width=0.7\textwidth]{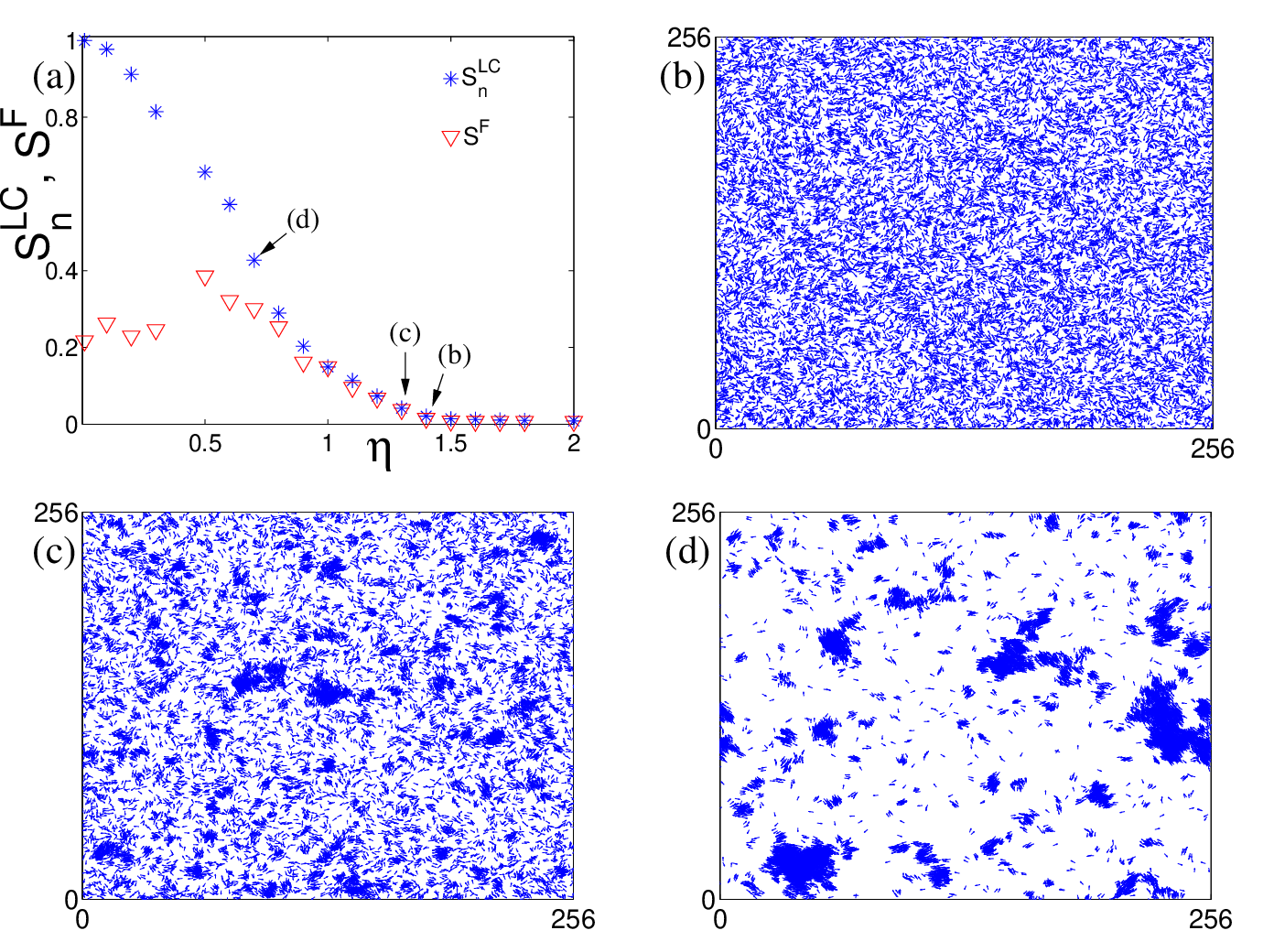}
\caption{Orientational dynamics and clustering at low
density. Number of particles $N=2^{12}$ and density
$\rho=0.25$. (a) Orientational order parameter $S$ vs. directional
noise amplitude $\eta$. Symbols are average over $10$ realizations. 
The dashed curve corresponds to the best fit of the simulation data close to $\eta_c$. 
(b)-(c) snapshots of the simulations for the values of $\eta$ indicated in (a) at the steady state, time step $2.5$ $10^5$ \cite{peruani_individual_2008}.} \label{figs_clustering}
\end{figure}

In Fig. \ref{figs_mfa}(a) the behavior of the orientational order parameters $S^{LC}$ and $S^{LC}$ Â vs. the noise amplitude $\eta$ for high density is shown for a model with LC-alignment. 
A transition from a disordered (isotropic) to an orientational ordered (nematic) state is strongly suggested by the simulation data.
Close to the critical noise amplitude $\eta_c$ the scaling of the curve follows the scaling predicted by the mean field outlined in section 4 \cite{peruani_mean-field_2008}.
The departure from the mean-field prediction occurs exactly when the spatial distribution of particles can not longer be considered homogeneous, compare snapshots Fig. \ref{figs_mfa}(b-d). 
Note, also that the ferromagnetic order parameter is zero as long as the density appears spatially homogeneous. 
At low density the scenario how orientational order emerges is
significantly different. 
Fig. \ref{figs_clustering}(a) shows  that for $\rho=0.25$ the dependency of the orientational order parameter $S$ on the noise amplitude Â $\eta$ is qualitatively different from the one observed at high density. 
The second remarkable difference is that the apparent onset of orientational ordering coincides with the onset of clustering. 
This is also confirmed by  the cluster size distribution which exhibits a power-law distribution near the onset of orientational ordering \cite{peruani_cluster_2011}.
Fig. \ref{figs_clustering}(b) shows that for zero orientational order  the spatial distribution of particle is roughly homogeneous. 
As soon as clustering is observed in the snapshots, see Fig. \ref{figs_clustering}(c), the order parameter starts to deviate from zero. 
As the noise is decreased clustering effects are more pronounced, Fig. \ref{figs_clustering}(d).
The displayed patterns correspond to typical particle configurations in a quasi-steady  steady state. 
In particular, the pattern of cluster does not coarsen as in a phase separation process. 
Here, clusters form and disintegrate in a dynamical way. 
The rate of growth and disintegration of the clusters is highly dependent on the value of $\eta$.
Given a value of $\eta$ the cluster size distribution reaches a steady distribution after an initial transient similar to the behaviour of self-propelled rods described above.
%
%

In the following,  a simple theory to understand the emergence of
steady state cluster size distributions in self-propelled particle systems with velocity alignment. 
The treatment is analogous to the one described above for the cluster-size distribution of self-propelled rods. 
The dynamics of the cluster-size distribution may alternatively be described by deriving
a master equation for the evolution of the probability $p(\mathbf{n}(t))$, where
$\mathbf{n}(t)={n_1(t), n_2(t), ..., n_N(t)}$, with $n_1(t)$ being the number
of isolated particles, $n_2(t)$ the number of two-particle clusters, $n_3(t)$
the number of three-particle clusters, etc. 
This kind of approach has previously been used to understand equilibrium nucleation in gases, where the transition probabilities between states are function of the
associated free energy change ~\cite{schimansky-geier_kinetics_1986,schweitzer_stochastic_1988,schweitzer_stochastics_1988}. 
Here, equations for the time evolution of the values of $n_1(t), n_2(t), ..., n_N(t)$ are derived to describe the cluster dynamics. 
To simplify the notation, one refer to $\langle n_1(t) \rangle$,   $\langle
n_2(t) \rangle$, ...,  $\langle n_N(t) \rangle$ simply as $n_1(t)$, $n_2(t)$,
etc.
The time evolution equations for  the  $n_i(t)$ obey the following form:
\begin{eqnarray}
\dot{n}_{1}&=&2B_{2}n_{2}+\sum_{k=3}^{N}B_{k}n_{k}-\sum_{k=1}^{N-1}A_{k,1}n_{k}n_{1}
\nonumber \\
\dot{n}_{j}&=&B_{j+1}n_{j+1}-B_{j}n_{j}-\sum_{k=1}^{N-j}A_{k,j}n_{k}n_{j}
\nonumber \\
&&+\frac{1}{2}\sum_{k=1}^{j-1}A_{k,j-k}n_{k}n_{j-k} \ \quad \mbox{for} \quad j = 2, .....,N-1
\nonumber \\
\dot{n}_{N}&=&-B_{N}n_{N}+\frac{1}{2}\sum_{k=1}^{N-1}A_{k,N-k}n_{k}n_{N-k} \label{rea_vicsek}
\end{eqnarray}
where the dot denotes the time derivative, $B_{j}$ represents the
 rate for a cluster of mass $j$ to loose  a particle, and is defined as 
\begin{eqnarray}\label{eq:rate_splitting}
B_{j}= \frac{D_\text{eff}(\eta)}{d^2}\sqrt{j} \, ,
\end{eqnarray}
and $A_{j,k}$ is the collision rate
between clusters of mass $j$ and $k$, defined by
\begin{eqnarray}\label{eq:rate_collision}
A_{j,k}= \frac{v_0 2 \epsilon}{a}\left(\sqrt{j}+\sqrt{k}\right) \, ,
\end{eqnarray}
where $a=L^2$ is the area of the two-dimensional space where particles move. 
In Eq. (\ref{eq:rate_splitting}), $d$ denotes,  the maximum distance that two
particles can be separated apart to be considered as {\it connected} and part of the same cluster
The expression  $d^2/D_\text{eff}(\eta)$ is an estimate for the characteristic time a particle spends at the boundary before it moves away from the cluster. 
The splitting rate $B_{j}$ is proportional to the inverse of this characteristic time multiplied by the number of particles on the boundary, which is estimated as $\sqrt{j}$. 
On the other hand, the collision rate $A_{j,k}$ is derived in analogy to the collision rate in kinetic gas theory between two  disk-like particles A and B \cite{peruani_cluster_2011}. 

Numerical integration of  these equations shows that Eq.(\ref{rea_vicsek}) produces 
qualitatively similar distributions as the one observed in individual-based
simulations, see \cite{peruani_cluster_2011}.  
The different curves correspond to various values of the dimensionless
parameter $P$, defined as $P=\frac{2 \epsilon d^2 v_0}{a D_\text{eff}(\eta)}$. 
For small values of $P$, which correspond to large values of $\eta$, the
distribution $p(m)$  monotonically decreases with $m$, while for large
values of $P$, resp. small values of $\eta$, a peak at large cluster sizes
emerges. 
A quantitative comparison between Eq.(\ref{rea_vicsek}) and individual-based simulations is still in a very early stage \cite{peruani_cluster_2011}. 
%
 %
\begin{figure}
\centering
\resizebox{0.55\columnwidth}{!}{\rotatebox{0}{\includegraphics{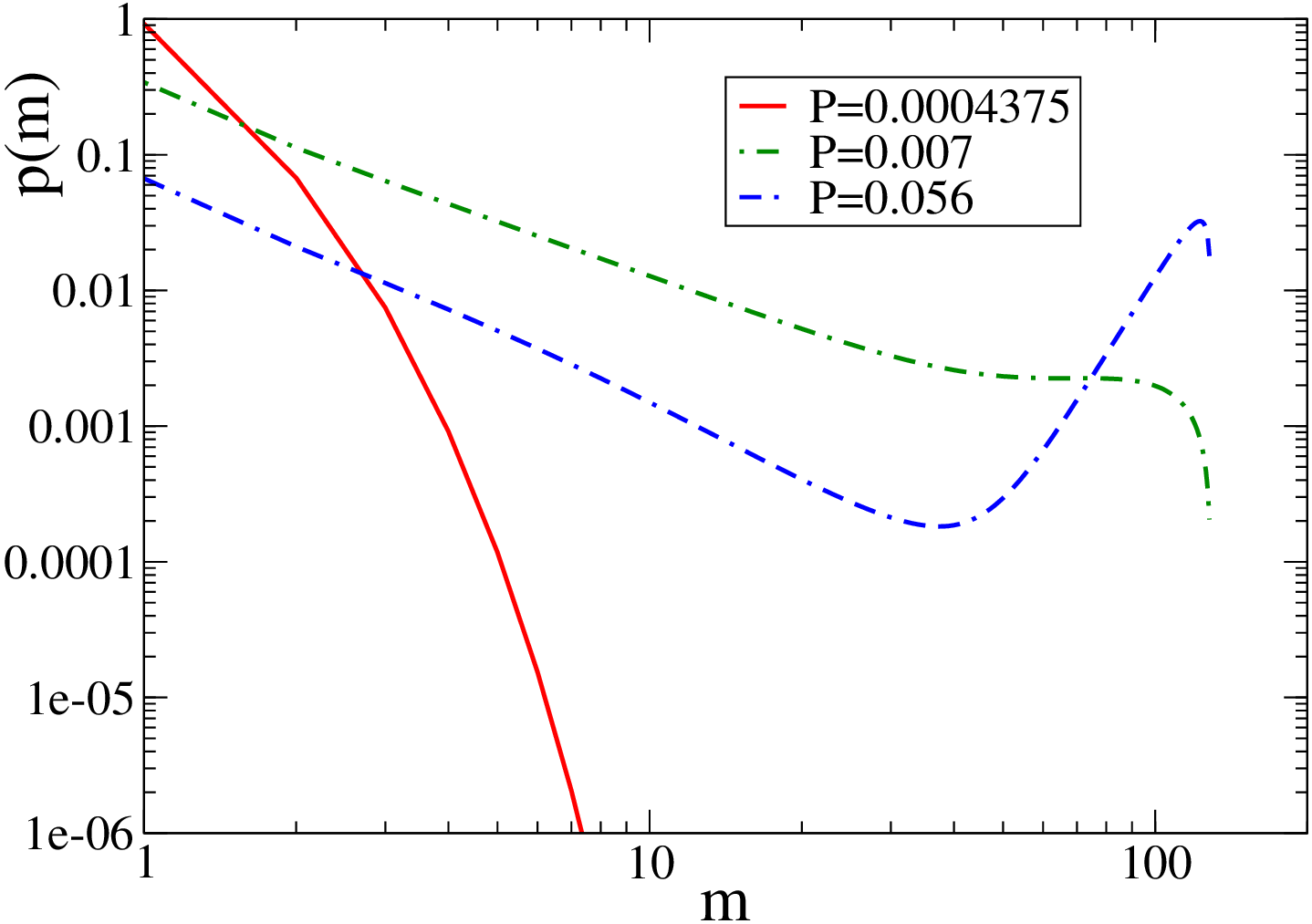}}}
\caption{Steady state cluster size distributions obtained from numerical
  integration of  Eqs.(\ref{rea_vicsek}) with $N=128$ for various values of
  the dimensionless parameter $P$, where $P=\frac{2 \epsilon d^2 v_0}{a D_\text{eff}(\eta)}$. 
Notice the transition from a monotonically decreasing distribution  for small 
values of $P$ to  a non-monotonic distribution with a peak at large cluster
sizes  for large values of $P$.} \label{fig_csd_theo}
\end{figure}
%

Altogether, polars clusters play a fundamental role in the macroscopic dynamics of self-propelled particle systems. 
Orientational order and cluster dynamics are often closely linked. 
The cluster size distribution can be obtained  from  a set of equations of the Smoluchowski type to describe the cluster dynamics in the system. 
Its usefulness awaits further test by comparison to simulations and experiments in systems of self-propelled units.

\subsection{Large-scale segregation bands}
\label{sec:pattern_bands}

Segregation bands are coherent structures typically seen in large-scale simulation of self-propelled particle systems with velocity alignment \cite{vicsek_novel_1995,chate_collective_2008,ginelli_large-scale_2010}. 
These bands often represent high-density zones with a high degree of orientational order embedded in a low-density background with orientational disorder. 
This coupling of orientational order and density goes beyond the mean-field theories describing the onset of orientational order in systems with homogeneous density that were introduced in the previous section. 
Extensive simulation studies have revealed a clear correlation between the appearance of travelling segregation bands and the observation of a first-order transition to orientational order in Vicsek model \cite{chate_collective_2008}, that was first reported in \cite{gregoire_onset_2004}.  
\begin{figure}
\centering 
\includegraphics[width=1.0\textwidth]{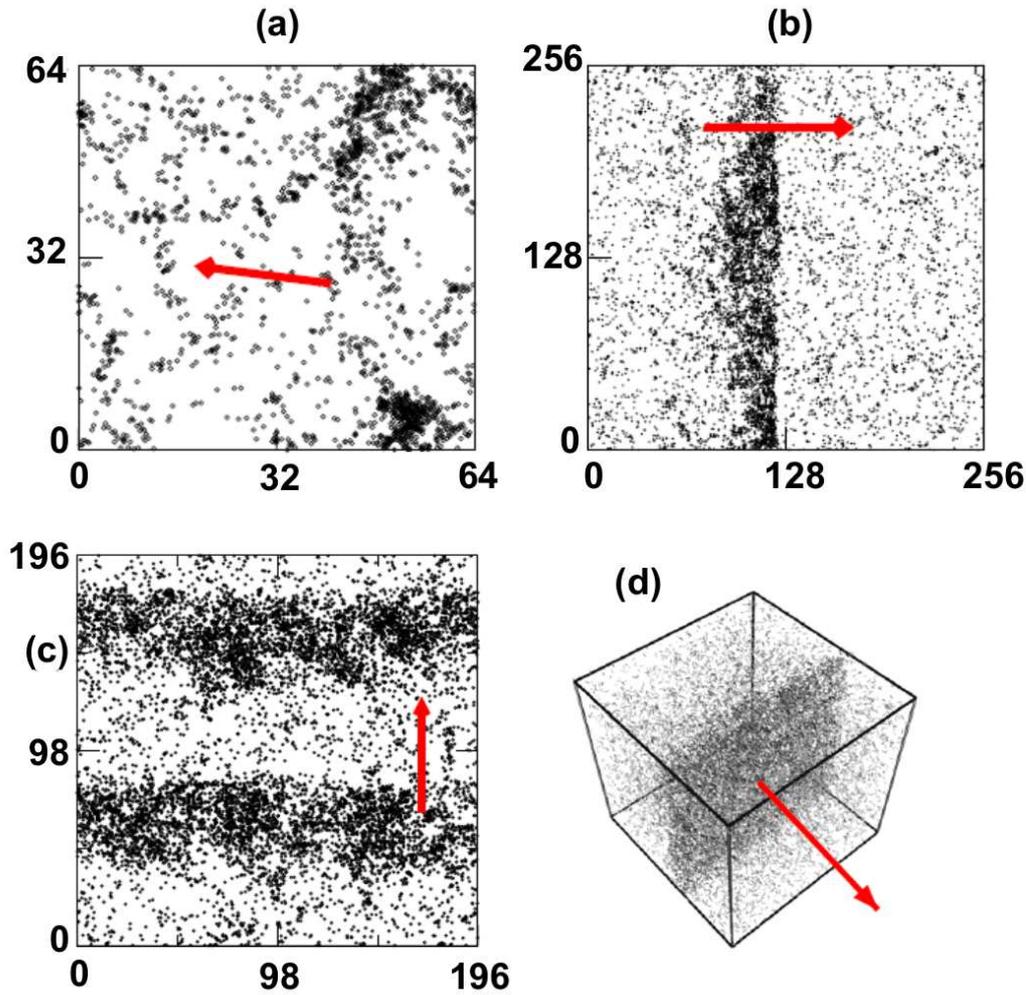}
\caption{ Typical snapshots in the ordered phase obtained from simulations of the Vicsek-model. Points represent the position of individual particles and the red arrow
points along the global direction of motion. (a,b) Vicsek model with angular noise and increasing system size. ( c) bands in simulation of Vicsek-model with angular noise, repulsive force, and periodic boundary conditions. (d) High-density
sheet traveling in a three-dimensional box with periodic boundary
conditions and angular noise. All figures from \cite{chate_collective_2008}. } 
\label{fig:0} 
\end{figure} 

Some examples for the segregation bands in the Vicsek model are displayed in Fig. \ref{fig:0}. 
The segregation bands in the Vicsek-models consist of many particles that are roughly aligned and travel mostly orthogonal to the edge of the band. 
Recently, improved mean-field theories were able to reproduce the simultaneous occurrence of segregation band and orientational order as well as the correct relation between the traveling \cite{bertin_boltzmann_2006,bertin_microscopic_2009,mishra_fluctuations_2010}. 
An exhaustive discussion of these phenomena in the Vicsek model can be found, e. g.  in \cite{chate_collective_2008,vicsek_collective_2010}.

A recent study of the Peruani model addresses the collective properties of self-propelled particles with nematic interaction (LC-alignment). 
Extensive simulations have revealed  long-range nematic order,
phase separation, and space-time chaos mediated by large-scale segregated structures, which we will describe low in greater detail \cite{ginelli_large-scale_2010}.

\begin{figure*}[!t] 
\begin{center} 
\includegraphics[clip,width=\linewidth]{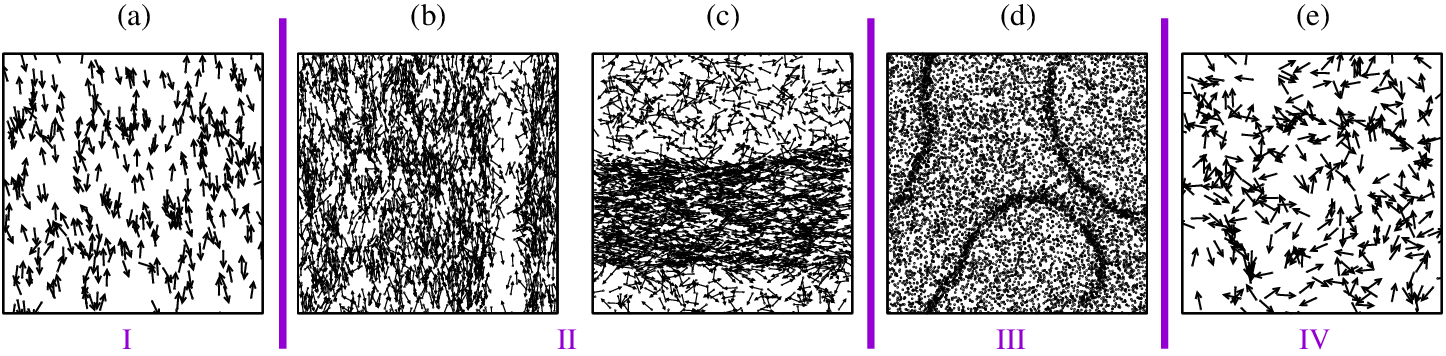} 
\caption{(color online) (a-c)
Typical steady-state snapshots at different noise values 
(linear size $L=2048$). 
(a) $\eta=0.08$, (b) $\eta=0.10$, (c) $\eta=0.13$, (d) $\eta=0.168$, (e) $\eta=0.20$. 
Arrows indicate the polar orientation of particles (except in (d)); only a fraction of 
the particles are shown for clarity reasons. For a movie corresponding to (d) see suppl. material of \cite{ginelli_large-scale_2010}.}
\label{fig:2} 
\end{center} 
\end{figure*} 


The simplicity of this model allowed to deal with large numbers of particles, 
revealing a phenomenology previously unseen in more complicated models sharing 
the same symmetries~\cite{peruani_nonequilibrium_2006,kudrolli_swarming_2008,wensink_aggregation_2008}, where mostly clustering phenomena were reported. 
The two-dimensional simulation study  showed large-scale segregation bands distinctively different from both those of 
the polar-ferromagnetic (F-alignment) case of the Vicsek model and of active nematics. 
Segregation appeared as a phase separation into high and low density areas: in the ordered side, a dense band occupying  a fraction of space along which particles 
move in both directions arises when noise is strong enough. 
Remarkably, the instability marks the order/disorder transition. It vanishes at strong noise,  splitting the disordered phase in two. 
The class of polar particles aligning nematically exhibits thus a total of four phases.
Polar and nematic order in the model with LC-alignment can be characterized by means of the two time-dependent  global scalar order parameters $P(t)=|\langle\exp (i\theta^t_j)\rangle_j|$ (polar) and $S(t)=|\langle\exp (i 2\theta^t_j)\rangle_j|$ (nematic), as well as their asymptotic time averages $P=\langle P(t)\rangle_t$ 
and $S=\langle S(t)\rangle_t$. 
A brief survey of the stationary states observed in a square domain is provided in Figs.~\ref{fig:2}.
Despite the polar nature of the particles, only {\em nematic} orientational 
order arises at low noise, while $P$ always remains near zero.
Both the ordered and the disordered regimes are 
subdivided in two phases, one that is spatially homogeneous (Figs.~\ref{fig:2}(a,e)), 
and one where density segregation occurs,
leading to high-density ordered bands along which the particles move back and forth 
(Figs.~\ref{fig:2}(b-d)).
A total of four phases was observed, labeled I to IV by 
increasing noise strength hereafter. Phases I and II are nematically ordered, 
phases III and IV are disordered.

\begin{figure}[t!] 
\begin{center} 
\includegraphics[clip,width=8.6cm]{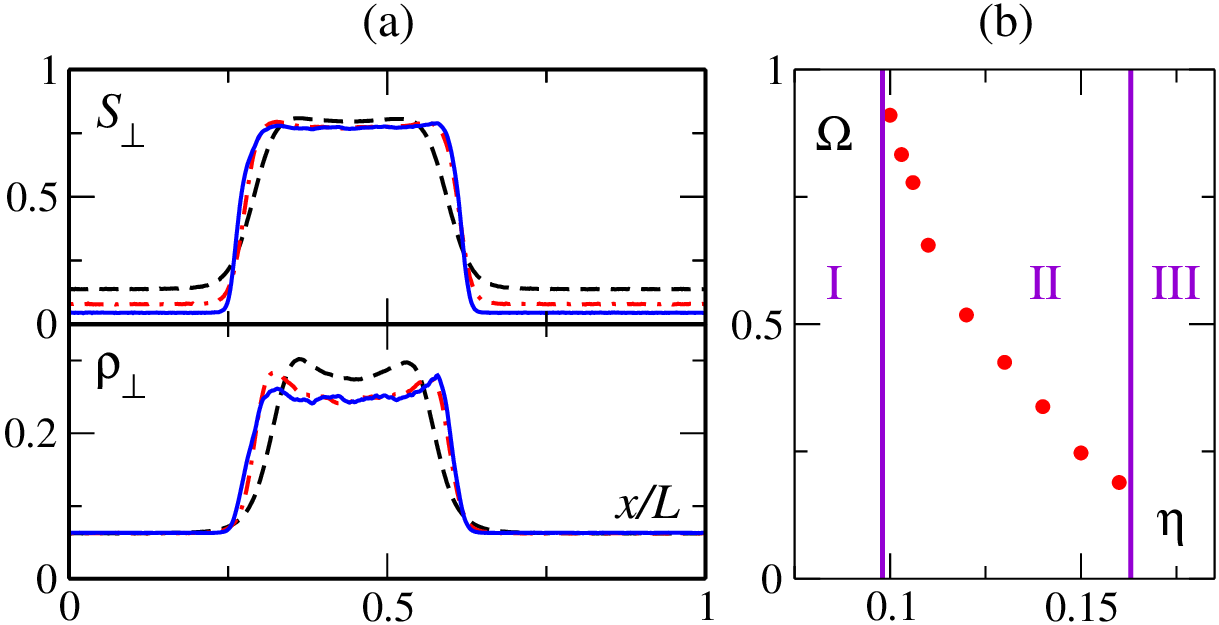} 
\caption{(color online) Phase II (stable bands)
(a) Rescaled transverse profiles in square domains of linear size.
 Data averaged over the longitudinal direction and time, translated to be centered
at the same location. Bottom: density profiles. Top: nematic order parameter profiles. 
(b) Surface fraction $\Omega$ as a function of $\eta$ (defined here as
the mid-height width of the rescaled $S$ profile) \cite{ginelli_large-scale_2010}}
\label{fig:5} 
\end{center} 
\end{figure}

Phase I, present at the lowest $\eta$ values, is ordered and 
spatially homogeneous (Fig.~\ref{fig:2}a). 
Phase II differs from phase I by the presence, in the steady-state, 
of a low-density disordered region.
In large-enough systems,  a narrow, low density channel emerges rather suddenly, like in a nucleation process (Fig.~\ref{fig:2}b). 
It becomes wider at larger $\eta$ values, so that one can then speak of a high-density ordered band, typically oriented along one of the main axes of the box, amidst a low-density disordered background  (Fig.~\ref{fig:2}c). 
Particles travel along the high-density band, turning around or leaving the band
from time to time.
Within the band, nematic order with properties similar to those of phase I 
is found (slow decay of $S$ with system size, giant number fluctuations).
The (rescaled) band possesses a well-defined profile with sharper and sharper edges as $L$ increases  (Fig.~\ref{fig:5}a). 
The fraction area $\Omega$ occupied by the band is asymptotically independent of system size and decreases continuously as the noise strength $\eta$ increases (Fig.~\ref{fig:5}b). 

\begin{figure}[t] 
\begin{center} 
\includegraphics[clip,width=8.6cm]{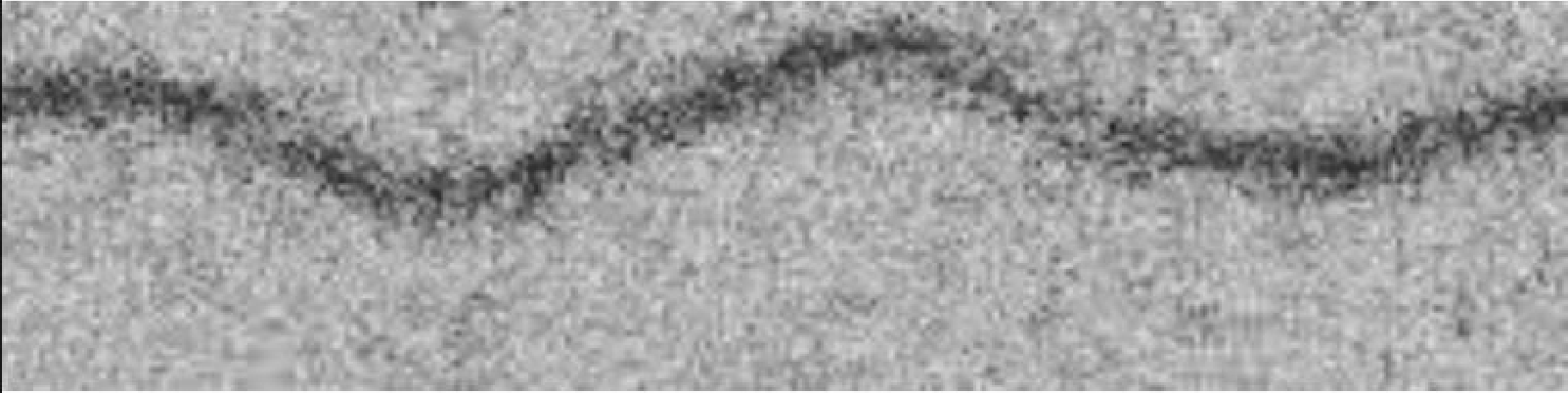} 
\caption{Phase III (unstable bands, $\eta=0168$). Snapshot of coarse-grained density field during the growth of the instability of 
an initially straight band in a 4:1 aspect ratio rectangular domain.} 
\label{fig:6} 
\end{center} 
\end{figure}

In phase III, spontaneous segregation into bands still occurs 
(for large-enough domains), however these thinner bands are unstable 
and constantly bend, break, reform, and merge, in an 
unending spectacular display of space-time chaos (Fig.~\ref{fig:2}d). 
Thus, the transition between phase II and III, located near 
$\eta_{\rm II-III}\simeq 0.163(1)$, is the order-disorder transition of the model. 
It resembles a long wavelength transversal instability of the band
(see for instance Fig.~\ref{fig:6}).

Increasing further the noise strength, the segregated bands vanish,
leaving phase IV, an ordinary disordered phase, spatially-homogeneous,
and with very short correlations in space and time (Fig.~\ref{fig:2}e).
Near the transition point, at $\eta_{\rm III-IV}\simeq 0.169(1)$, 
the nematic order parameter $S(t)$ 
exhibits bistability between a low amplitude, fast fluctuating state 
(typical of phase IV) and a larger amplitude, slowly fluctuating one 
typical of phase III \cite{ginelli_large-scale_2010}. 
This suggests a discontinuous disorder-disorder transition  between phase III and IV for the Peruani model similar to the above cited results for the Vicsek-model. 

Note also that the above results, and in particular the space-time chaotic motion of the  spontaneously segregated bands (phase III) (see supp. material in \cite{ginelli_large-scale_2010}), are reminiscent of the streaming and swirling regime which characterizes the aggregation of myxobacteria \cite{jelsbak_pattern_2002,kaiser_coupling_2003,zusman_chemosensory_2007}. 
The model results suggests that no adhesion or chemical signaling is needed for such behavior to emerge, LC alignment mediated by volume-exclusion interaction is sufficient to obtain complex patterns. 
These results may therefore be relevant for the 
collective dynamics of gliding bacteria, biofilms and other cells with 
friction and moderate adhesion. 

At a more general level, the findings reveal unexpected emergent behavior among   the simplest situations giving rise to collective motion. 
The described model of self-propelled polar objects
aligning nematically stands out as a member of a universality class
distinct from both that of the Vicsek model \cite{vicsek_novel_1995,toner_flocks_1998,chate_collective_2008} 
and that of active nematics \cite{chate_simple_2006}.
Thus, in the out-of-equilibrium context of self-propelled particles, the symmetries of
the moving particles and of their alignment interactions must be considered separately and are both relevant ingredients.
A major lesson of the discovery of large-scale separation bands of different symmetries for the models with F- and LC-alignment is that mean-field and other continuum theories have to be extended to allow for density instabilities.
Improved continuum models should eventually reproduce the behaviour found in large-scale simulation, first promising steps in this direction has been provided by Bertin et al. \cite{bertin_microscopic_2009} and Mishra et al. \cite{mishra_fluctuations_2010}.

%% file: SECTION_concl.tex
\section{Discussion, Outlook and Conclusions}

The first systematic studies of systems far from equilibrium and their surprising self-organizing capabilities date back more than half a century back and are connected to such distinguished scientists as {\em Alan Turing}, {\em Ilya Prigogine} or {\em Hermann Haken}. The ideas and concepts formulated and introduced by this scientific pioneers inspired generations of statistical physicists and applied mathematicians to pursue their research on corresponding problems with applications to physics, chemistry and biology.     

Recently, a subclass of such far from equilibrium systems, characterized by active motion of individuals units, received growing attention. The continuously increasing number of publications on such systems, are probably the best proof for an intense and ongoing research activity.
The individual active units may be of very different type and the relevant spatial length scales may span over many orders of magnitude starting from the nanometer scale governing the motion of individual molecular motors to kilometers in the case of large collectively moving swarms of insects, such as desert locusts. 
Despite this apparent heterogeneity of active systems, the common fundamental properties and universal dynamical features, suggest the formulation of generic models of active motion within the framework of dynamical and stochastic systems. The mathematical description based on the concept of individual active particles, allows on the one hand the detailed understanding of the dynamics of individual units constituting an active system and on the other hand enables us in a simplified setting to derive coarse-grained equations and to study the large-scale behavior of ``active matter'' systems. Here, the focus lies on the rather simple models, which in the ideal case provide qualitative insights to the universal dynamics and allow often for analytical treatment at the costs of quantitative predicitive power of specific active matter realizations.

In Sections \ref{sec:brown} and \ref{sec:selfdriven} we have introduced the mathematical framework for the description of self-propelled motion of individual Brownian particles and analyzed the behavior of a number of different models. 
We discussed the concept of active Brownian motion and its description via velocity-dependent friction functions based on the assumption of an internal degree of freedom of individual particles (energy depot) or as an effective description of ensembles of coupled active particles (molecular motors). 
Here, we did not restrict ourselves to the case of Gaussian fluctuations, but discussed different types of stochastic forces, such as dichotomous Markov noise or shot-noise, and their impact on the system dynamics. 
Based on the non-equilibrium nature of the studied dynamics, we addressed the question of the different impact of active and passive fluctuations and have shown that active fluctuations, which are correlated with the direction of motion of individual particles lead to characteristic deviations of the corresponding speed and velocity distributions, independent on model details, such as a particular choice of the friction function.

Furthermore we analyzed the diffusive motion of free active particles (Sec.~\ref{sec:sing}) and their dynamics in external confinements (Sec.~\ref{sec:rot_ext}). The detailed analysis of the individual dynamics reveals surprising features of active Brownian motion such as a giant diffusion regimes or optimal noise values, which maximize or minimize the spatial diffusion.  

In the last two sections we have extended our analysis to ``swarms'' and ``gases'' of interacting active particles. 
In Sec. \ref{sec:swarming}, we identified for example the fundamental stationary modes of collective motion of active Brownian particles with attracting interactions (swarms) and discussed novel results on the complex behavior of swarms with attraction and repulsion in three spatial dimensions. 

An important class of interactions studied in the literature is the so-called velocity-alignment, which we put a particular emphasis on in Sec. \ref{sec:swarming}. 
We have shown how, starting from microscopic Langevin equations of active Brownian particles with polar velocity alignment, we can derive systematically the corresponding mean field equations. We then focus on the onset of order in the special class of minimal models of self-propelled particles, with polar and apolar alignment interaction, motivated by the well known Vicsek-model. 

Finally, we have discussed pattern formation in such minimal models, which show features such as clustering and formations of large scale density inhomogeneities in Sec. \ref{sec:pattern}. This patterns are intrinsically connected to the active motion of the interacting units and the onset of large-scale collective motion, and are up to date under intense investigation.

Despite our focus on qualitative understanding, the comparison with experimental results of active matter systems and corresponding  modelling approaches, which were motivated by those, must not be neglected.

\subsubsection*{Individual Dynamics - Experiments and Models}
The most obvious example of autonomous self-propelled motion, which does not require external driving, is the motile behavior of biological agents. Here, the single-celled motile organism, such as certain bacteria or eukaryotic cells (e.g. Dictyostelium discoideum) are probably what comes closest to the concept of an ``active Brownian particle'' as discussed in this review. However, being ``simple'' in comparison to higher organisms, does not mean that the corresponding mechanisms of active motion is not complex. 
In fact, in recent decades we witnessed a burst in scientific advances which pushed forward our understanding of cell motility. The active translocation of cells is driven by complex dynamics of the intracellular actin cytoskeleton (see e.g. \cite{mitchison_actin-based_1996}). It is a fascinating field of interdisciplinary research already for decades, which continues to thrive and continuously motivates theoretical investigations  (see e.g \cite{kruse_self-propagating_2001,gholami_velocity_2008,enculescu_modeling_2010} and recent reviews \cite{flaherty_mathematical_2007,juelicher_active_2007}). 

Bacteria, for example, can exhibit different motility types to propel themselves under different environmental conditions \cite{eisenbach_chemotaxis_2004,kearns_field_2010}. Probably the best understood bacterial motility type is swimming in a liquid medium due to the action of rotating flagella \cite{berg_symmetries_1996,sowa_bacterial_2008}. The typical bacterial swimming motion consists of straight runs interrupted by short reorientation events. This so-called ``run \& tumble'' motion has been successfully modelled by a random walk approach \cite{berg_random_1993,codling_random_2008}, where usually constant speed and Poissonian distribution of reorientation events is assumed. However, a closer look at some real trajectories reveals that in between the tumbles a bacterium does not move in a perfect straight line with constant speed \cite{mittal_motility_2003}. Brownian fluctuations as well as possible fluctuation in the driving force lead to more complicated dynamics. Thus, a more realistic model can be obtained by a combination of a Langevin description, as discussed in this review, together with tumbling dynamics, as suggested by Condat et al. \cite{condat_randomly_2005}. Yet the authors considered in their model only passive fluctuations and a consideration of active fluctuations (Sec. \ref{sec:SG}) will yield different results. 

Swimming of bacteria, as well as other microorganisms (e.g. microalgae \cite{drescher_direct_2010}), takes place at extremely low Reynolds numbers ($\sim 10^{-6}$) and a large number of theoretical studies investigated the active motility of swimmers using an overdamped hydrodynamical description (Stokes equation) which takes into account interactions between the active swimmer and the fluid. This approach yields some interesting preditions on long-ranged hydrodynamical coupling between different swimmers as well as between cells and surfaces (see e.g. \cite{hernandez-ortiz_transport_2005,lauga_swimming_2006,alexander_dumb-bell_2008,pooley_hydrodynamic_2007,hernandez-ortiz_dynamics_2009,putz_cuda_2010}, or a recent review in \cite{lauga_hydrodynamics_2009}). However, recent a recent experiments with {\emph E. Coli} conducted by Drescher et al. \cite{drescher_fluid_2011}, show that in most cases the effects of long-ranged hydrodynamic interactions are negligible in comparison to the intrinsic stochasticity in the motion of bacteria (i.e. rotational diffusion). Thus, the authors conclude that the collective dynamics of bacteria might be quite similar to the dynamics exhibited for example by granular systems. Based on this surprising results, Langevin equations may be an interesting modeling alternative, as they allow for simple implementation of various interactions (see Section and \cite{czirok_formation_1996,ben-jacob_cooperative_2000,romanczuk_beyond_2008}). 

A common objection against bacterial equations of motion of the Langevin type are negligible inertial effects due to the extremely low Reynolds numbers at which the dynamics takes place. However, second-order equations of motion may offer a suitable effective description in the presence of additional time scale(s) in the propulsion mechanism and the resulting bacterial response to external signals, as for example to gradients of chemical agents in their environment (chemotaxis). Furthermore, Langevin equations can provide a reasonable framework for modelling other motility types observed in bacteria, such as twitching or gliding. In particular in cases where the shortest time-scales -- where the details of the specific propulsion mechanism play an important role -- are not of interest.

We have only briefly discussed chemotactic behavior in the context of collective dynamics in Section \ref{sec:chemotaxis}. For a more detailed discussion on chemotaxis, we refer the interested reader to the rich literature on this fascinating topic \cite{berg_physics_1977,manson_bacterial_1992,eisenbach_chemotaxis_2004,wadhams_making_2004,iglesias_navigating_2008}.

In recent years, there has been a number of empirical studies analyzing the motion of various eukaryotic cells crawling on substrates \cite{selmeczi_cell_2005,dieterich_anomalous_2008,li_persistent_2008,selmeczi_cell_2008,boedeker_quantitative_2010,li_dicty_2011}. All these studies have shown that the Ornstein-Uhlenbeck model of persistent Brownian motion is not sufficient to explain the empirical observation, but differ in their conclusions. Dieterich et al. \cite{dieterich_anomalous_2008} suggest the description of the cell migration by a fractional Klein-Kramers equation, in order to explain the observed anomalous dynamics. However, this conclusion have been questioned as the apparent super-diffusive behavior may be a consequence of too short observation times with respect to possible long relaxation times of the direction of motion of individual cells \cite{peruani_self-propelled_2007,campos_persistent_2010,li_dicty_2011}.

The analysis of the deterministic and stochastic accelerations derived from the tracking data in \cite{selmeczi_cell_2005,selmeczi_cell_2008,boedeker_quantitative_2010,li_dicty_2011}, show similar behavior  for different cells types: 1) a negative linear drift which corresponds to constant Stokes friction, and 2 ) fluctuations strengths increasing with the speed of the cell, which indicates multiplicative noise in the corresponding Langevin equation.

Researchers around Henrik Flyvbjerg and Edward C. Cox, have carefully analyzed the experimental trajectories of different cell types and proposed different models of persistent motion using Langevin equations based on their empirical findings. For human epidermal cells, they formulated stochastic integro-differential equation of motion with a kernel representing a short-ranged memory of the cell \cite{selmeczi_cell_2005,selmeczi_cell_2008}. Hereby, they motivated their choice with the double-exponential decay of the cell velocity autocorrelation. This however can be also explained by a simper model with independent fluctuations in the velocity and direction of motion as proposed by Peruani and Morelli \cite{peruani_self-propelled_2007} and discussed in Sec. \ref{sec:diff_independent}. The same group also analyzed the motion of Dictyostelium discoideum amoeba, and have shown that its motion consists of two components: a persistent stochastic motion and a fast oscillatory contribution. This oscillatory dynamics at short time-scales can be linked to the specific propulsion mechanism. The cells move forward by growing protrusions at their leading edge, so-called pseudopods, attaching them to the substrate and ``pulling'' themselves forward by contracting their trailing edge. A repeated right-left-right-left formation of such pseudopods leads to a zig-zag motion responsible for the oscillatory contribution \cite{selmeczi_cell_2008,li_persistent_2008}. Similar observations have been reported previously by Shenderov and Sheetz \cite{shenderov_inversely_1997}, and just recently by Yang et al. \cite{yang_zigzag_2011}. In fact, oscillatory dynamics are not restricted to Dictyostelium amoeba,but have been reported by Barnhart et al. \cite{barnhart_bipedal_2010} in motile fish keratocytes. The authors use a modelling approach to show, that such oscillation may be explained by the  effective elastic coupling of the different parts of the cells, in particular the leading and trailing edges.

Only recently, Zaburdaev et al. \cite{zaburdaev_langevin_2011} investigated the swimming motion of the parasite African tryponosome. Based on similar analysis of empirical data as discussed above for eukaryotes, the authors propose a Langevin model of its motion. In their model, the authors distinguish a slow velocity component, characterized by a constant speed and subject only to rotational diffusion and an additive fluctuating component with linear relaxation. Essentially, this corresponds to an active Brownian particle, with a Schienbein-Gruler friction function as discussed in Sec. \ref{sec:SG} with active and passive fluctuation terms, corresponding to the angular diffusion of the slow component and the vectorial noise in the fast component, respectively. 

The Langevin models based on velocity-dependent friction function as discussed in this review, can not account for the specific details of active motion of crawling cells such as the observed oscillating dynamics on short time-scales. However, they yield a simple description at longer temporal and spatial scales, where such fast and typically small oscillatory motion can be neglected or considered as an additional quasi-stochastic contribution to the dynamical behavior. 

We have introduced and discussed in Sec. \ref{sec:internalcoord} and \ref{sec:motorcontrol} ``internal coordinates'', a co-moving coordinate system determined by the propulsion and symmetries of the active particle. Based on empirical investigation of cell trajectories, Li et al. emphasize in \cite{li_dicty_2011} the importance of using such a coordinate system attached to the cell instead of the laboratory reference frame. 


Zooplankton species, such as Daphnia, represent another example of actively moving biological agents, which are currently investigated due their ecological importance (see e.g. \cite{strickler_observing_1998,c.okeefe_swimming_1998,ordemann_motions_2003,erdmann_active_2004,garcia_optimal_2007,dees_stochastic_2008,menden-deuer_inherent_2010}). Daphnia motion consists of straight hops interrupted by turning events with exponentially distributed turning angles \cite{erdmann_active_2004,garcia_optimal_2007} and can be modelled as a persistent random walk (see Sec. \ref{sec:sing} and \cite{komin_random_2004,garcia_optimal_2007}). Frank Moss and coworkers advocated the theory that the specific motion pattern of Daphnia is the results of evolutionary adaptation to foraging in finite food patches. Although this hypothesis is difficult to prove, it is at least consistent with various empirical observations \cite{garcia_optimal_2007,dees_physical_2010}.  

An example of non-living active Brownian particles are the autonomously moving micro- and nanoscale particles which convert chemical energy into kinetic energy of motion (see e.g.  
\cite{ismagilov_autonomous_2002,paxton_catalytic_2004,howse_self-motile_2007,ke_motion_2010,mirkovic_nanolocomotion_2010,kolmakov_designing_2011} 
or a recent reviews in \cite{ebbens_pursuit_2010,gibbs_catalytic_2011}).
One general mechanism of self-propelled motion of these objects is so-called self-phoresis. In general, phoresis refers to the effective transport of colloids due to boundary layer forces induced by external fields \cite{anderson_colloid_1989}. For example, diffusiophoresis refers to a drift experienced by a colloidal particle subject to concentration gradient across its interface. Self-phoresis is thus the phenomenon, where the change in the environment which leads to phoretic drift (e. g. formation of a concentration gradient) is induced by the particle itself.  Self-propulsion can be achieved by an breaking the symmetry in the ability to catalyze some ``fuel'' substance on its surface. For example platinum is a catalyst for the decomposition of hydrogen peroxide (H$_2$O$_2$) into oxygen and water. A particle half-coated with platinum in a fluid containing H$_2$O$_2$ will perform self-propelled motion with a speed depending on the concentration of H$_2$O$_2$ \cite{howse_self-motile_2007,valadares_catalytic_2010,ke_motion_2010}. This was theoretically investigated using simplified models by Golestanian et al  \cite{golestanian_propulsion_2005,golestanian_designing_2007}, as well as detailed molecular dynamics simulations by Kapral and co-workers \cite{ruckner_chemically_2007,tao_design_2008}. 

The first experimental realization of such systems, triggered a surge of research on catalytic self-propelled particles. For example, it was shown that these objects can transport cargo \cite{sundararajan_catalytic_2008}, or can be effectively controlled by external (e.g. magnetic) fields \cite{tierno_autonomously_2010}.

The propulsion direction set via the self-phoresis asymmetry, defines the preferred direction of motion (heading)  \cite{ke_motion_2010,ebbens_direct_2011} as introduced in Sec. \ref{sec:internalcoord} \& \ref{sec:SG}. Due to their small size, the particles are subject to Brownian fluctuations, but in addition we expect that non-thermal fluctuations associated with the non-equilibrium propulsion will contribute to the stochasticity in their motion. Thus, these object, as well as related autonomous swimmers  appear as very good candidates for polar active particles with passive and active fluctuations as introduced in Sec. \ref{sec:SG}. In addition, it is possible that for non-sperical but axis-symmetric particles, a mis-alignment of the self-phoresis symetry axis with the ``body axis'' will introduce systematic torques into the dynamics of active particles as discussed in Sec. \ref{sec:stoch_dich_angles} and in \cite{weber_active_2011}.  

Another interesting application of such chemically driven particles is the fabrication of nano-rotors by attaching the particles to a substrate \cite{fournier-bidoz_synthetic_2005,mirkovic_nanolocomotion_2010}, which can be described by a simple models of confined active Brownian particles discussed in Sec. \ref{sec:rot_ext}. 

Interestingly, the Cartesian velocity histograms of both, freely moving cells and artificial self-propelled particles are non-Gaussian and show increasing counts at low velocities (low speeds) \cite{li_persistent_2008,boedeker_quantitative_2010,ke_motion_2010}. In Sec. \ref{sec:active_stokes} and \ref{sec:SG}, we have shown that this behavior is generic in the presence of active fluctuations. Hereby, the deviation from a Gaussian probability distribution function at low velocities increases with increasing strength of active fluctuations. In fact, similar effect can be observed for oscillatory self-propulsion as discussed in Sec. \ref{sec:depot_intosc}  \cite{romanczuk_active_2011}. 

It appears that the concept of Active Brownian particles, is even better suited as a general model of such autonomously moving artificial objects than for the motion of biological agents due to the simpler propulsion mechanism. We believe it may contribute significantly to the understanding of the active diffusion of these self-propelled particles, as well as their behavior in external fields.

\subsubsection*{Collective Dynamics - Experiments and Models}
Experimental examples for collective dynamics and pattern formation of active particles are often found during the life cylces of bacteria or other microorganism like Dictyostelium discoideum (Dd) \cite{ben-jacob_cooperative_2000}. 
While aggregation and patterns in Dd cells are mostly driven by chemotaxis, i. e. motility responses to a chemoattractant generated by the cells itself, other organisms like myxobacteria \cite{kaiser_coupling_2003} provide intriguing examples of self-organization of active, self-propelled objects. 
In contrast to processes controlled and influenced strongly by biochemical communication between the active agents (= cells), more recently experiments with swimming and gliding bacteria have been performed that focus on the interaction between active motion and simple physical interactions. 

Swarming behavior in colonies of up to a thousand swimming bacteria {\em Bacillus subtilis} have been investigated by Zhang and coworkers \cite{zhang_collective_2010}. 
Their setup allowed to locate the individual cells and determine their directions of motion. 
As a result, Zhang et al. could determine the number fluctuations as well as the cluster size statistics of the bacterial swarms.  
They found cluster size distributions similar to the ones reported for the self-propelled rod models  for densities below the nonequilibrium clustering transition in Section 7 above, i. e. the identified distributions of a form $P(n) \propto n^{-b} e^{-n/n_C}$ where $n$ denotes the cluster size and $n_C$ a density-dependent fit parameter. 
The exponent $b$ was density independent and took on a value of $b = 1.85$. 
In parallel, Zhang {\em et al.} report giant number fluctuations for the standard deviation of the particle numbers $\Delta_N \propto N^\alpha$  with an exponent $\alpha$  near 0.75 for small mean cell numbers $N$. These giant number fluctuations are presumably linked to the formation of larger clusters of cells moving in the same direction. 
In more recent work, Peruani {\em et al.} analyzed the collective motion of up to two thousand gliding bacteria in a two-dimensional monolayer. The organism under study was a mutant species of {\em Myxococcus Xanthus} that does no possess flagella and does not exchange biochemical signals relevant for the control of their motility  \cite{peruani_collective_2012}. Experiments were conducted over a large range of densities. As a result a transition to nonequilibrium clustering was found at a critical coverages of around 16 percent. The cluster size distribution at the transition became a power law with exponent $b = 1.88$ and a pronounced second maximum was found at large cluster sizes for densities above the critical value. In addition giant number fluctuations with exponent $\alpha \approx 0.8$ were found. 
 It is striking that quite different bacterial systems (gliding myxobacteria and swimming Bacillus subtilis) exhibit similar clustering dynamics and related giant number fluctuations.  If we recall that the cluster size distribution $P(n)$ discussed here is related to the weighted cluster size distribution $p(n) = n P(n)$, we immediately get a relation $b = 1 + c$, where $c$ is the exponent of the cluster size distribution discussed in Subsection 7.3 above.  The exponents $c \approx 0.9$ \cite{peruani_nonequilibrium_2006} and a $c$ in the range of $0.9 -- 1.3$ \cite{yang_swarm_2010}. This suggests that collective dynamics of some bacterial species may be indeed well described by self-propelled rod-shaped particles with volume exclusion interactions.

Another experimental system, wherein spectacular experimental phenomena were observed, is a motility assay with actin filament, ATP and immobilized molecular motors \cite{schaller_polar_2010}.  Below a critical density of ca. 5 filaments per square micrometre, a disordered phase with no preferred orientation is observed in these experiments. Above, the critical density different ordered patterns are observed. First moving clusters of filaments are observed similar to the observation presented for self-propelled rods. Above a second threshold of 20 filaments per square micrometre traveling density bands are observed similar to the ones found in the Vicsek model, i. e. high density traveling waves (or bands) composed of aligned filaments that travel individually in the some direction of motion as the whole wave. These bands are reminiscent of the observation in simulation presented in Fig. 71.  Schaller et al. were able to reproduce their experimental findings in a cellular model that was based on similar assumptions as the simple models for collective motion discussed in Section 6 and 7 \cite{schaller_polar_2010}.  In parallel, swirling vortex-like motion states were also observed in the whole above the first critical density of ca. 5 filaments per square micrometre. 
In a related study with a motility assay that contained filaments, ATP, motors and crosslinking proteins, K\"ohler {\em et al.} report pattern formation characterized by a broad distribution of cluster sizes \cite{koehler_structure_2011} . With time a coarsening of these clusters and a related higher percentage of large clusters was observed. A detailed quantitative analysis of the cluster sizes was not carried out. Hence, only future experiments will show if the cluster size distributions will follow similar laws as the ones reported above for similar of collective motion of active particles. 
A class of physical systems that realize self-propelled particles are ensembles of driven granular particles with asymmetric shape and weight distribution. Kudrolli {\em et al.} studied cylindrical rod-shaped particles with strongly asymmetric weight distribution and reported local ordering as well as a strong tendency to aggregation and persistent swirling motion in which velocities are strongly correlated with particle orientation \cite{kudrolli_swarming_2008}. The findings agree quite well with simulations of self-propelled system, where Wensink et al. have also reported aggregation of rods near the boundary \cite{wensink_aggregation_2008}.

Another experimental realization of collective motion in granular particles was designed by Deseigne {\em at al.} \cite{deseigne_collective_2010}, who studied vibrated polar discs (= discs with asymmetric material density) in a system with a petal-shaped boundary. This particular shape avoids the aggregation of particles at the boundary and leads to more homogeneous distributions of the moving particles. The experiments then yields orientational order and giant number fluctuations with exponent $\alpha\approx 0.725$. This is again in the ballpark of the observation for the simple models discussed in Section 7 as well as similar to the values measured for bacterial swarms above.

Altogether, simulations of models describing the motion of active Brownian particles and their interactions are in good agreement with many recent findings for living systems such as bacteria, in-vitro biological systems like motility assays and driven granular matter. One is tempted to look for universal properties of such systems. There is however a large plethora of patterns that cannot be reproduced from the simple assumption discussed in extenso in this review. For such systems, the models analyzed and described may nevertheless provide good starting points in model developments. 

Bacterial colonies and social amoeba as discussed above, are by far not the only biological systems exhibiting collective motion on large scales. In fact, colloquially a ``swarm'' is typically associated with collective motion of higher organisms such as, flocks of birds, schools of fish, or the devastating mass migration of insects (e.g. desert locusts). Up to recently, experimental data on the individual behavior and collective motion patterns was rather limited \cite{okubo_diffusion_2001,krause_living_2002}. As a consequence most models of collective motion motivated by the different swarming phenomena relied on more or less empirically based assumptions and qualitative matching of the model behavior with empirical observations (see e.g. \cite{aoki_simulation_1982,reynolds_flocks_1987,couzin_collective_2002,parrish_self-organized_2002,bode_making_2010} and further references in Sec. \ref{sec:swarming}). Most of these models share the following three interactions mechanisms: 1) short-ranged repulsion responsible for collision avoidance, 2) long-ranged attraction ensuring group cohesion and 3) directional alignment (typically at intermediate distances) facilitating collective motion of the group. These type of models is usually referred to as zone-models. Many of these models assume for simplicity constant speed of individuals.

However, new experimental methods, such as automated digital video recordings, allow the collection of large data sets. This enables not only the statistical analysis of collective motion in nature \cite{buhl_disorder_2006} but allows also to infer the interaction rules between individuals \cite{ballerini_interaction_2008,cavagna_starflag_2008,lukeman_inferring_2010,katz_inferring_2011,herbert-read_inferring_2011}. 

Based on careful analysis of experimental recordings of starling flocks and the resulting correlation functions, Ballerini {\em et al.} \cite{ballerini_interaction_2008} have shown that each individual responds on average  to the behavior with a fixed number of its nearest neighbors. Thus, they conclude that the interaction between individual birds is governed primarily by the topological distance and not the metric distance as assumed in most theoretical models also those discussed in Sec. \ref{sec:swarming}. This feature of the interactions offers a simple explanation for large density differences observed between different flocks and may also contribute to the robustness of cohesive bird flocks with respect to external pertubations (e.g. predators).  In a follow-up work, Cavagna et al. \cite{cavagna_scale-free_2010} report scale-free fluctuations in starling flocks, which may indicate the operation of the flock close to criticality. Being close to a critical point may be advantageous for a swarm or flock acting as a collective information processing system \cite{sumpter_information_2008,mora_are_2011,vanni_criticality_2011}. 

Lukeman et al. \cite{lukeman_inferring_2010} have collected and analyzed data on collective behavior of surf scoters swimming on the water surface. The authors fit their data to a zone model, identify the best parameter values and argue that the standard zone model has to be complemented by an additional interaction to the front.  

Very recently two studies have appeared, which infer the social interaction rules between fish \cite{katz_inferring_2011,herbert-read_inferring_2011}. Both studies report that speed modulation is the primary response to close by individuals in front or in the back. This speeding up/slowing down as well as turning behavior of the fish are consistent with an attraction-repulsion behavior but show no clear evidence for directional alignment. In addition Katz et al. \cite{katz_inferring_2011} report that the three fish interaction is neither given by a superposition nor an averaging of pairwise-interactions, whereas Herbert-Read et al. \cite{herbert-read_inferring_2011} stress the importance of the interactions with the nearest neighbor.  

In the light of these recent results, the active Brownian particle concept appears as a promising starting point for the development of more realistic models of collective motion as it 1) naturally accounts for a variable velocity of individuals and acceleration/deceleration due to effective forces, and 2) exhibits collective motion for simple attraction/repulsion interactions without need of alignment terms as discussed in Sec \ref{sec:swarms} and \ref{sec:swarm_3d} . In this context, we should note that the escape \& pursuit interaction introduced in Sec. \ref{sec:escpur}, represents a special case of general position and velocity-dependent attraction/repulsion interaction \cite{romanczuk_thesis_2011}.

As stated in Sec. \ref{sec:escpur}, the escape \& pursuit interactions were directly motivated by the empirical evidence for cannibalism being the driving mechanism of collective motion in certain insect species (Mormon crickets, desert locusts) \cite{simpson_cannibal_2006,bazazi_collective_2008,bazazi_social_2010}. The simple Brownian agent model \cite{romanczuk_collective_2009} can be easily modified to account in more detail for movement patterns of individual insects. A parametrization of such an extended escape \& pursuit model with experimental results results not only in good agreement with the observations but enables us also to make specific predictions on the impact of the nutritional state of individuals on the onset of collective motion \cite{bazazi_nutritional_2010}. A generalized version of the model \cite{romanczuk_thesis_2011} can account for different types of individual interactions ranging from pure avoidance behavior, via escape \& pursuit to pure attraction and allows the evolution of escape and pursuit behavior and collective motion \cite{guttal_cannibalism_2012}.

Finally, there are important biological question in the context of collective motion of higher organisms, which go far beyond the scope of this review (see e.g. \cite{okubo_diffusion_2001,krause_living_2002,sumpter_collective_2010}). For example, some recent publications investigate the impact of different information available to individuals \cite{couzin_effective_2005,couzin_uninformed_2011}, or the evolution of heterogeneous behavioral strategies within a group \cite{guttal_social_2010}.     

\subsubsection*{Outlook \& Final Remarks}

A major challenge for the future is the comparison of novel experimental results to predictions of the different theoretical models discussed here as well as their underlying assumptions. For example, the statistical properties of fluctuations in the dynamics of active particles can be measured and analyzed in order to refine our description of active noise terms. Furthermore, the issue of variable speed and corresponding fluctuations needs certainly to be addressed in the future.
Even if the general framework of Active Brownian particles accounts for variable speed, there remain many open questions on important details. The corresponding theoretical results rely often on approximations and rather simple  assumptions, which may not be justified. For example,  possible non-trivial correlations between velocity fluctuations and changes in the direction may strongly influence the theoretical results. 

In collective dynamics it is import to distinguish between universal and system specific properties. One would expect that minimal model systems should provide the answer. However, the unresolved connection between the giant number fluctuations and clustering in those systems reveals the conceptual difficulties which have to be addressed in the future. 
From the more biological perspective, in the light of the new measurement discussed above, the question arises about new models which are able to account for the observed interactions between individuals. Here, it appears that simple physically motivated pairwise ``forces'' might be not sufficient, and that a more biological centered ansatz, based on the sensory and cognitive capabilities of individuals, can be very promising.

In general, statistical physicists possess a large inventory of methods for describing and analysing complex systems, which is being continuously developed, on the background of a long experience in applying these methods to natural phenomena. 
In this spirit, the development of the mathematical description of natural active matter systems and their analysis, is essential for our understanding of the dynamical behavior of these systems, and gives us important insights into their role and function in the biological and ecological context. 

In conclusion, this review gives an overview over the theoretical foundations and concepts of active (Brownian) particles systems and discusses recent developments in the field of statistical physics applied to active particle systems far from equilibrium.  We are aware that an all-encompassing review of the field is not feasible here, given the large number of publications from different disciplines and the intense ongoing research activity. Nevertheless, the focus on theoretical concepts and recent developments can be seen as complementary to other reviews on the topic \cite{helbing_traffic_2001,schweitzer_brownian_2003,vicsek_collective_2010} and we hope it will be of interest to researchers in statistical physics who would like to broaden their knowledge in this rapidly developing research field. 

\section{Acknowledgments}

We acknowledge the fruitful collaboration on various problems discussed in this review with our former co-authors: G. Cebiroglu, I.D. Couzin, U. Erdmann, E. Gudowska-Nowak, L. Haeggqwist, N. Komin, F. Peruani, T. P{\"o}schel, T. Riethm{\"u}ller, F. Schweitzer, I.M. Sokolov, J. Strefler, C. Weber. All of them have actively contributed to a better understanding of active particles. In particular, we  would also like to thank F. Peruani and C. Weber for providing valuable assistance in the preparations of the manuscript. Further, we have strongly benefited from many scientific discussions with our friends and colleagues V.S. Anishchenko, P. H{\"a}nggi, H. Malchow, A.S. Mikhailov, A.B. Neiman and Yu.M. Romanovsky.
 
The authors will always remember the stimulating discussions with the late Frank E. Moss, to whom this work is dedicated.
